\documentclass[a4paper,12pt,titlepage,english]{book}

\usepackage{amsmath,amssymb,url,cite,ifpdf,slashed,multirow,tabularx}
\ifpdf                                     % For pdfLaTeX
  \usepackage{graphicx}                    %   usepackage without driver option
\else                                      % For (p)LaTeX + dvipdfmx
  \usepackage[dvipdfmx]{graphicx}          %   usepackage with driver optiont
\fi

\usepackage[dvipdfmx,colorlinks=true,linkcolor=blue,citecolor=red,urlcolor=blue]{hyperref}

\usepackage{feynmp}
\usepackage{color}

\usepackage[charter]{mathdesign}
\def\thefootnote{\ifnum\c@footnote>\z@\textasteriskcentered\@arabic\c@footnote\fi}
\makeatletter
\renewcommand{\footnoterule}{%
\kern-3\p@
\hrule width 0.4\columnwidth
\kern 2.6\p@}
\def\thefootnote{\ifnum\c@footnote>\z@\@arabic\c@footnote\fi}
\makeatother
\allowdisplaybreaks

\newcommand{\TeV}{\,{\rm TeV}}
\newcommand{\GeV}{\,{\rm GeV}}
\newcommand{\MeV}{\,{\rm MeV}}

\newcommand{\invfb}{\,{\rm fb^{-1}}}

\def\be{\begin{equation}}
\def\ee{\end{equation}}
\def\beq{\begin{eqnarray}}
\def\eeq{\end{eqnarray}}

\def\({\left(}
\def\){\right)}
\def\<{\langle}
\def\>{\rangle}

\newcommand{\Order}{\mathop{\mathcal{O}}}

\newcommand{\Slash}[1]{{\ooalign{\hfil/\hfil\crcr$#1$}}}

\makeatletter
\newcommand{\@authornote}[2]{{\def\thefootnote{\fnsymbol{footnote}}\setcounter{footnote}{#1}#2\setcounter{footnote}{0}}}
\newcommand{\authornotemark}[1]{\@authornote#1{\addtocounter{footnote}{-1}\footnotemark}}
\newcommand{\authornotetext}[2]{\@authornote#1{\footnotetext{#2}}}
\makeatother
\begin{document}

%%% TITLE
\begin{titlepage}

%\begin{flushright}
%UT--**--**\\
%Month, Year
%\end{flushright}

\begin{center}

\Large Ph.D Dissertation \\
\vspace{4cm}
{\Huge \bf
Prospects for Slepton Searches in \\
\vspace{0.5cm}
Future Experiments
} \\

\vspace{3cm}
{\large
Takahiro Yoshinaga
} \\

\vspace{1cm}
{\large
{\it 
Department of Physics,  University of Tokyo, Tokyo 113-0033, Japan
}}
\end{center}

\end{titlepage}
%%%%%%%%%%%%%%%%%%%%%%%%%%%%%%%%%%%%%%%%%%%%%

%%%%%%%%%%%%%%%%%%%%%%%%%%%%%%%%%%%%%%%%%%%%%
\begin{titlepage}

\begin{center}

\mbox{} \\
\vspace{5cm} 
{\large To Suzuka \\}
\end{center}

\end{titlepage}
%%%%%%%%%%%%%%%%%%%%%%%%%%%%%%%%%%%%%%%%%%%%%

%%% Contents
\tableofcontents

%%% List of Figures
\listoffigures 

%%% List of Tables
\listoftables

%%% Praface
\chapter*{Preface}
%!TEX root = ../Dthesis.tex

\textbf{This dissertation is based on the following works by author accomplished during the Ph.D course at the Graduate School of Science, the University of Tokyo.}

\begin{enumerate}
\item[\bf{[A]}] 
T.~Kitahara, and T.~Yoshinaga, "Stau with Large Mass Difference and Enhancement of the
Higgs to Diphoton Decay Rate in the MSSM," 
\href{http://dx.doi.org/10.1007/JHEP05(2013)035}{{\em JHEP} {\bf1305} (2013) 035}, 
\href{http://arxiv.org/abs/1303.0461}{{\tt arXiv:1303.0461 [hep-ph]}}, 
{\color[named]{Salmon}{\small Reference}}\cite{Kitahara:2013lfa}.
\item[\bf{[B]}] 
M.~Endo, K.~Hamaguchi, T.~Kitahara, and T.~Yoshinaga, “Probing Bino contribution to muon $g-2$,"
\href{http://dx.doi.org/10.1007/JHEP11(2013)013}{{\em JHEP} {\bf1311} (2013) 013}, 
\href{http://arxiv.org/abs/1309.3065}{{\tt arXiv:1309.3065 [hep-ph]}},
{\color[named]{Salmon}{\small Reference}}\cite{Endo:2013lva}.
\item[\bf{[C]}] 
M.~Endo,  T.~Kitahara, and T.~Yoshinaga, "Future Prospects for Stau in Higgs Coupling to Di-photon,"
\href{http://dx.doi.org/10.1007/JHEP04(2014)139}{{\em JHEP} {\bf1404} (2014) 139}, 
\href{http://arxiv.org/abs/1401.3748}{{\tt arXiv:1401.3748 [hep-ph]}}, 
{\color[named]{Salmon}{\small Reference}}\cite{Endo:2014pja}.
\end{enumerate}

\noindent
\textbf{Among the other works by Author, the following work is referred in this dissertation.}

\begin{enumerate}
\item[\bf{[D]}] 
M.~Endo, K.~Hamaguchi, S.~Iwamoto, and T.~Yoshinaga,~“Muon $g-2$ vs LHC in Supersymmetric Models,"
\href{http://dx.doi.org/10.1007/JHEP01(2014)123}{{\em JHEP} {\bf1401} (2014) 123}, 
\href{http://arxiv.org/abs/1303.4256}{{\tt arXiv:1303.4256 [hep-ph]}}, 
{\color[named]{Salmon}{\small Reference}}\cite{Endo:2013bba}.
\end{enumerate}

%%% Abstract
\chapter*{Abstract}
%!TEX root = ../Dthesis.tex

Muon $g-2$ anomaly, which is mismatch between the theoretical and the experimental values of the anomalous magnetic moment of muons, 
provides a sensitive probe of new physics.
The supersymmetry (SUSY) is one of candidates for new physics models to solve the muon $g-2$ anomaly.
The minimal supersymmetric extension of the standard model is called the MSSM.
The MSSM contains superpartner of the SM particles, called superparticles.
If masses of the superpartners relevant to the muon $g-2$ are $\mathcal{O}(100\GeV)$, 
the SUSY contributions to the muon $g-2$ become sizable and the anomaly can be solved.

There are two representative SUSY contributions to the muon $g-2$, the chargino and the neutralino contributions, respectively.
Particularly, when the discrepancy of the muon $g-2$ is dominated by the neutralino contributions,  
it is found that the discrepancy can be explained if only the Bino and sleptons are light. 
This setup is "minimal" to solve the muon $g-2$ anomaly.

In this dissertation, we discuss the "minimal" SUSY models where only the Bino and sleptons are light, while the other superparticles are decoupled.
When masses of relevant superparticles are $\mathcal{O}(100)\GeV$ and the left-right mixing of the sleptons are large, 
the SUSY contributions to the muon $g-2$ are enhanced.
In this model, size of left-right mixing is important. 
If it was allowed to be arbitrarily large, 
the sleptons could be extremely heavy while keeping the SUSY contribution,
thereby escaping any collider searches.
However, it is constrained by the vacuum stability condition of the slepton--Higgs potential. 
Therefore, there are upper bounds on masses of sleptons and Bino.

The searches for the sleptons rely on the slepton mass spectrum.
When the slepton soft masses are universal, the upper bound on the smuon mass becomes 330 (460) $\GeV$ 
in order to solve the muon $g-2$ anomaly at the $1\sigma $ ($2\sigma $) level. 
It is within the reach of LHC and ILC. 
Further, light staus might affect the Higgs coupling to di-photon, which is sensitive to new physics. 
We also study future prospects of the stau which contributes to the Higgs coupling.
If the stau is heavier than the smuon, 
the bound can be as large as 1.4 (1.9) $\TeV$. 
Such non- universal slepton mass spectrum generically predicts too large LFV/CPV. 
We show that the models are expected to be probed by LHC/ILC and LFV/CPV complementarily in future.

%%%%%%%%%%%%%%%%%%%%%%%%%%%%% Contents

\chapter{Introduction}
\label{chp:introduction}
%!TEX root = ../Dthesis.tex

\section{Overview}
\label{sec:overview}

The Standard Model (SM) of the elementary particle physics is a successful model
which describes almost all phenomena of the elementary particles.
Only the Higgs boson was the last piece of the particle contents of SM for a long time. 
On 4th July 2012, the ATLAS and the CMS collaborations claimed that the Higgs like new boson was discovered at last.  
If this boson is the SM Higgs boson, this discovery means that the matter content of SM is completed.

However, we physicists do not consider the SM as a fundamental theory of elementary particles.
In fact, new physics models are suggested from some viewpoints, 
for example, by the Hierarchy problem, 
absent of the dark matter, the dark energy, and the mechanism which generated current baryon asymmetry of the Universe, 
the mismatch between the theoretical and the experimental values of the anomalous magnetic moment of muons 
(muon $g-2$ anomaly\cite{Bennett:2006fi, Roberts:2010cj, Hagiwara:2011af}), 
and so on.
We are in the middle of a trip toward the ultimate theory, 
above the suggestions and/or problems are useful guidelines to clarify new physics.

Particularly, the muon $g-2$ anomaly is promising because it provides a sensitive probe for new physics contributions.
New particles which have masses of $\mathcal{O}(100)\GeV$ are suggested to explain the discrepancy, 
and they are within kinematical reach of future collider experiments, such as the Large Hadron Collider (LHC) 
and the International electron-positron Linear Collider (ILC).
Furthermore, new muon $g-2$ experiments are planed at Fermilab\cite{Venanzoni:2014ixa} and J-PARC\cite{Mibe:2010zz}, 
and the statistical significance is expected to become 7--8 $\sigma $ in next 3--5 years if the central value of experimental would remain the same.
In other words, if the muon $g-2$ anomaly is true, it can be confirmed at $\gtrsim 5\sigma $ level.
Hence, it is important to investigate new physics models which are motivated by the muon $g-2$ anomaly, just now. 
In this dissertation, we regard the anomaly as basic guideline to search for new physics.

The supersymmetry (SUSY), which is the symmetry between fermions and bosons,   
is a good candidate for new physics models to solve the muon $g-2$ anomally.\footnote{
In fact, as we will see in Sec.~\ref{sec:motivation}, SUSY is often motivated by the hierarchy problem.
We consider the muon $g-2$ anomaly as particularly important problem in this dissertation. 
Since SUSY can naturally provide a solution to the anomaly, 
we take SUSY models as good candidates for new physics models.
For a detail, see Sec.~\ref{sec:gm2mssm}}
The SM can be extended to supersymmetric models, 
and this model includes new particles, which is associated with SUSY, called SUSY particles (superparticles). 
The minimal supersymmetric extension of SM, called the minimal supersymmetric standard model (MSSM), 
has scalar quarks (squarks) and scalar leptons (sleptons) as  bosonic partners of the SM fermions.
Also the fermionic partners of the Higgs and gauge boson are called the Higgsinos and the gauginos.
As we will see in Sec.~\ref{sec:gm2mssm}, if the superparicles couple with the muon, such as smuons, electroweak gauiginos, and Higgsinos 
are around $\mathcal{O}(100)\GeV$, 
the SUSY contributions to the muon $g-2$ become sizable and the muon $g-2$ anomaly is solved.

Superparticles have been searched for at the LHC.
Since none of them has been discovered, 
colored superparticles, such as gluinos and squarks, are considered to be heavier than about $1\TeV$.
Further, the Higgs boson mass of $126\GeV$ suggests that the stops are as heavy as 
$\mathcal{O}(1-10)\TeV$\cite{Okada:1990vk, Ellis:1990nz, Haber:1990aw, Ellis:1991zd} 
and/or have a large trilinear scalar coupling\cite{Okada:1990gg}. 
These two facts contradict with the suggestion from the muon $g-2$ anomaly, 
which relevant superparticles should have masses of $\mathcal{O}(100)\GeV$.
In fact, some of the representative SUSY-breaking mechanism, such as the minimal supergravity (mSUGRA) 
and the gauge mediation (GMSB) cannot simultaneously satisfy (i) explaining the $126\GeV$ Higgs, 
(ii) avoiding the constraints from LHC experiment, and (iii) solving the muon $g-2$ anomaly\cite{Endo:2011gy}.  
From the viewpoint of model building, Some of extended MSSM have been proposed by some theoretical groups.

In this dissertation, we take model-independent approach. 
The above conflict implies that the colored superparticles are heavy enough to explaine the Higgs mass and avoid current LHC bound, 
while the superparticles relevant to the muon $g-2$ are light ($\mathcal{O}(100)\GeV$).
The candidates of light superparticles depend on dominant SUSY contributions to the muon $g-2$.
There are two representative SUSY contributions as follows, 
\begin{enumerate}
\item[(1)] Chargino--muon sneutrino contribution (see the left panel of Fig.~\ref{fig:gm2mass}),
\item[(2)] Neutralino--smuon contribution (see the right panel of Fig.~\ref{fig:gm2mass}).
\end{enumerate}
When the discrepancy of the muon $g-2$ is explained the chargino--muon sneutrino contribution mainly, 
the Bino, Wino, Higgsino, and sleptons should be light. 
On the contrary, if the neutralino--smuon contribution are dominant, the Bino and sleptons must be relatively light.
In order to test whether the SUSY contributions is the origin of the muon $g-2$ anomaly, 
it is important to investigate the phenomenogy of these two cases as a first step.

The case (1) has been studied in Ref.~\cite{Endo:2013bba}.
Authors considered the muon $g-2$ motivated SUSY models, 
where only the Bino, Winos, Higgsinos, and sleptons are light, 
while the other superparticles are decoupled.
In this case, the searches for (Wino-like) electroweak gauginos are promising. 
It has been shown that they can be discovered in near future in most of the parameter regions, 
where the SUSY contributions to the muon $g-2$ are dominated by chargino--muon sneutrino diagrams.

The case (2) is "minimal" setup  to explain the muon $g-2$.
In the case (1), the Wino is also required to be light, 
because the chargino contribution to the muon $g-2$ is suppressed in large Wino mass regions.
On the other hand, the neurtalino contribution does not depend on the Wino mass, 
%It is enhanced when masses of the Bino and sleptons are $\mathcal{O}(100)\GeV$, 
%and the left-right mixing of the slepton (see \eqref{eq:gminus2bino} of Sec.~\ref{subsec:setup}) are large.
and its phenomenology has not studied sufficiently yet.

%Note that the searches in Ref.~\cite{Endo:2013bba} rely on the assumption that Wino is light.
%This assumption is not always necessary to explain the muon $g-2$ discrepancy, 
%when the SUSY contributions is mainly from neutralino-smuon diagrams
%(see also Fig.~\ref{fig:gm2mass} of Sec.~\ref{sec:gm2mssm}).
%Among the neutralino contributions, Bino-smuon diagram as Fig.~\ref{fig:gm2gauge} (b) of Sec.~\ref{sec:gm2mssm} is particularly important in the next step.
%Eventually, we notice that the models in which only the Bino and the sleptons are light, are the "minimal"  SUSY models 
%to solve the muon $g-2$ anomaly.
  
%%%%%%%%%%%%%%%%%%%%%%%%%%%%%%%%%% Figure
\begin{figure}[t]
 \begin{center}
 \includegraphics[width=15cm]{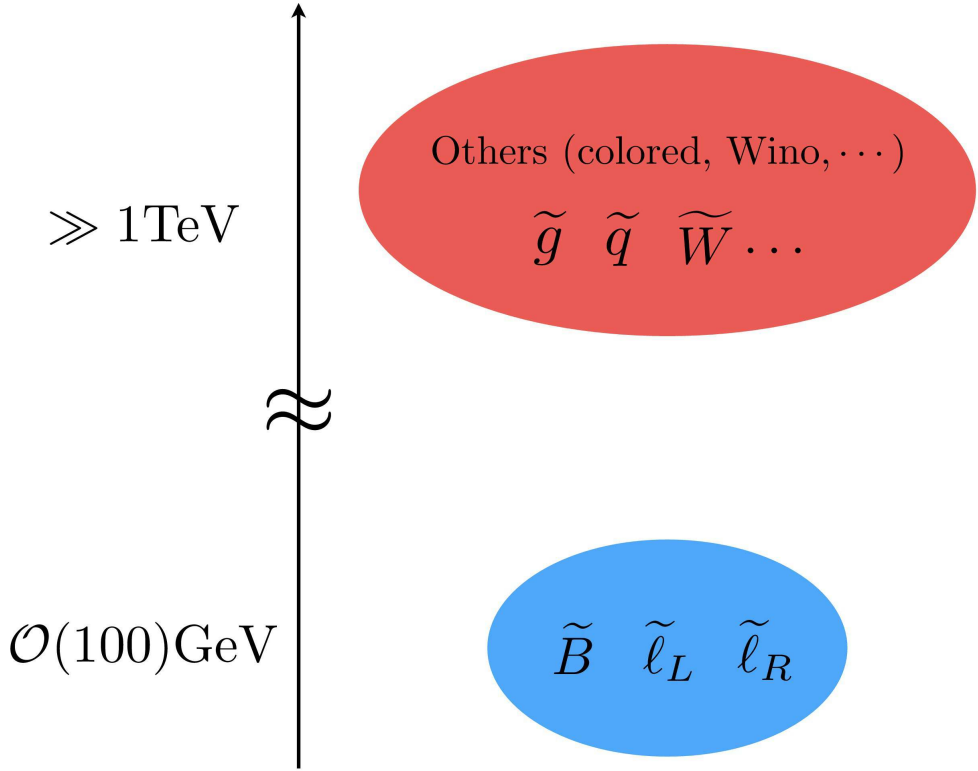} \hspace*{2mm}
 \end{center}
 \caption[Mass spectrum of the "minimal" SUSY model]
 {Mass spectrum of the "minimal" SUSY model, where 
 only the Bino and the sleptons are light, while the other superparticles are decoupled.}
 \label{fig:massspec}
\end{figure}
%%%%%%%%%%%%%%%%%%%%%%%%%%%%%%%%%%  
In this dissertation, we discuss the phenomenology of the "minimal" SUSY model, in which the following superparticles are light, 
\beq
 \widetilde{B},~\widetilde{\ell }_L ,~\widetilde{\ell }_R,  \label{eq:intromodelparam1}
\eeq
while the other superparticles are decoupled,\footnote{
 In this dissertation, masses of the heavy superparticles set to 30\TeV~for simplicity. 
 The results of this dissertation are almost independent on their mass spectrum except for the Wino.
 When the Wino mass are $\lesssim 1\TeV$, the chargino--muon sneutrino contribution causes a large theoretical uncertainty.  
 } as seen in Fig.~\ref{fig:massspec}.
Here, $\widetilde{B}$ denotes the Bino and $\widetilde{\ell }$ is the sleptons.
This study is based on our study\cite{Kitahara:2013lfa, Endo:2013lva, Endo:2014pja}.
The model parameters are 
\beq
 M_1,~m^2_{\widetilde{\ell }_L},~m^2_{\widetilde{\ell }_R},~m^2_{\widetilde{\ell }_{LR}}, \label{eq:intromodelparam2}
\eeq
where $M_1$ is the Bino mass,
and $m^2_{\widetilde{\ell }_L}$ and $m^2_{\widetilde{\ell }_R}$ are soft SUSY-breaking masses of the left- and right-handed sleptons, respectively. 
$m^2_{\widetilde{\ell }_{LR}}$ is off-diagonal components of the slepton mass matrices (left-right mixing of the sleptons).
Naively if masses of  Bino and sleptons are order of 100\GeV~and  the left-right mixing is large, 
the muon $g-2$ anomaly can be solved.

\begin{figure}[t]
 \begin{center}
 \includegraphics[width=17cm]{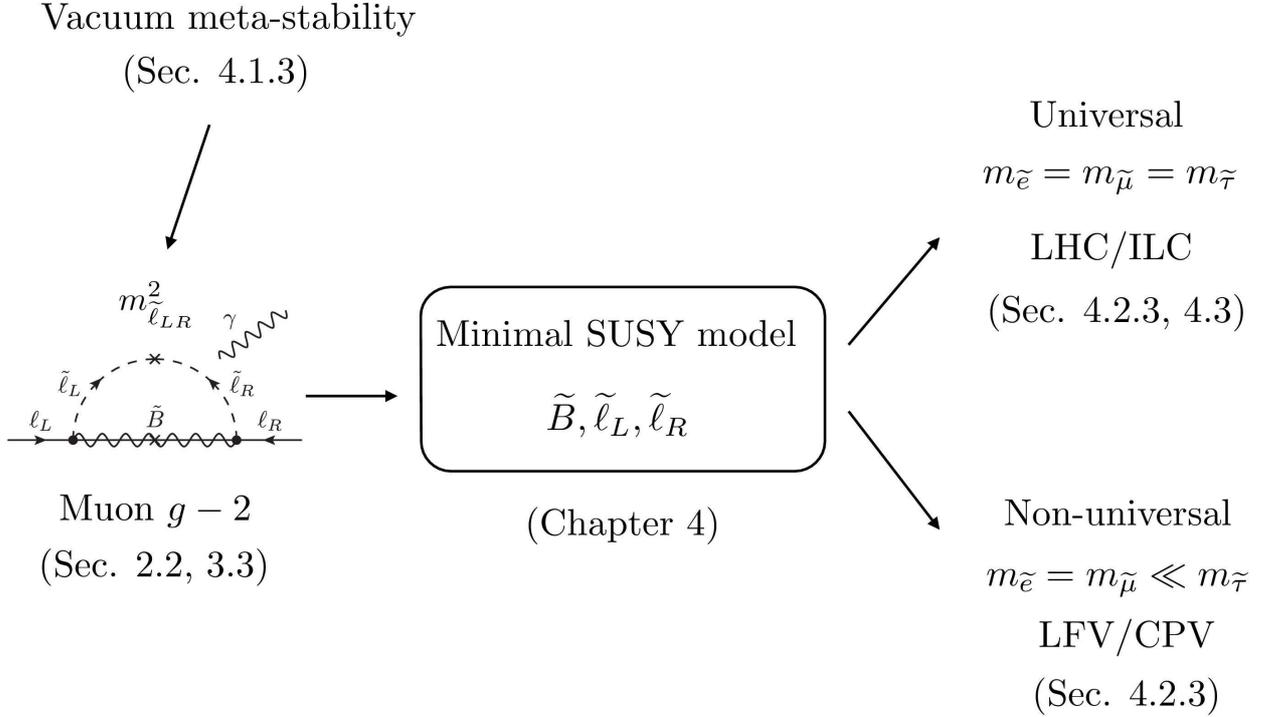}
 \end{center}
 \vspace{-30pt}
 \caption[Overview]{Overview of this dissertation.}
 \label{fig:overview}
\end{figure}

In this model, size of the left-right mixing is important.
If it was allowed to be arbitrarily large, the sleptons could be extremely heavy while keeping the contribution to the muon $g-2$, 
thereby escaping any collider searches.
However, too large left-right mixing spoils stability of the electroweak vacuum.
Therefore, the slepton masses are bounded from above by the vacuum meta-stability condition.
We will discuss future prospects of slepton searches based on the mass bounds.
The overview is summarized in Fig.~\ref{fig:overview}.

\section{Outline of the dissertation}

%%%%%%%%%%%%%%%%%%%%%%%%%%%%%%%%%% Figure
%%%%%%%%%%%%%%%%%%%%%%%%%%%%%%%%%% 
The outline of this dissertation is as follows.
In Chapter \ref{chp:foundation},  we review the electromagnetic dipole moments and the Higgs coupling.
First, we summarize the electromagnetic form factors in Sec.~\ref{sec:EMDM}.
They induce the magnetic and the electric magnetic moments, respectively.
The lepton flavor-changing version corresponds to the lepton flavor violating processes.
In Sec.~\ref{sec:gm2exp}, we review the anomalous magnetic moment of the muon, 
which has discrepancy between the theoretical and experimental values.
Then, the lepton flavor and CP violations are highly suppressed in SM. 
In SUSY models, they are sensitive to off-diagonal components of the slepton mass matrices, 
and are enhanced in case of non-universal slepton masses.
Sec.~\ref{sec:LFVexp} and \ref{sec:EDMexp} give a review of the lepton flavor violation and the electric dipole moment, respectively.
The Higgs couplings are also promising.
Particularly, since the Higgs coupling to di-photon is from loop diagrams, 
it is sensitive to new physics contributions, such as staus.   
In Sec.~\ref{sec:kappaexp}, we review the Higgs coupling to di-photon.

In Chapter \ref{chp:susy}, we review the supersymmetry, which is symmetry between fermions and boson.
In Sec.~\ref{sec:motivation}, we introduce the supersymmetry.
Then, we briefly review the minimal supersymmetric standard model (MSSM) in Sec.~\ref{sec:mssm}.
New particles included in MSSM are called the SUSY particles (superparticles).
If they have the masses of TeV scale,
it is expected that they affect some observables indirectly.
Particularly, observables corresponding to the electromagnetic and the weak interactions are sensitive to new physics contributions.
In this dissertation, we focus on the observables in which the non-colored superparticles are expected to have large contributions.
The SUSY contributions to the muon $g-2$, the LFV,  the electron EDM, and Higgs coupling to di-photon are reviewed in Sec.~\ref{sec:gm2mssm}, \ref{sec:LFVmssm}, and  \ref{sec:HiggsCmssm}, respectively.
Finally, we discuss the current status for superparticle searches in Sec.~\ref{sec:cstatus}.

Chapter \ref{chp:main} is the main part and based on Refs.~\cite{Kitahara:2013lfa, Endo:2013lva, Endo:2014pja}. 
Future prospects for the slepton searches are discussed.
First, we introduce the "minimal" SUSY model in Sec.~\ref{sec:foundation}.
Then, we show that the slepton masses are bounded from above by vacuum mata-stability condition in slepton-Higgs potential. 
The searches for sleptons depend on slepton mass spectrum as follows, 
(1) universal mass spectrum, $m_{\widetilde{e}} = m_{\widetilde{\mu }} = m_{\widetilde{\tau }}$, 
and (2) non-universal mass spectrum,  $m_{\widetilde{e}} = m_{\widetilde{\mu }} \ll  m_{\widetilde{\tau }}$, respectively.
The former case is discussed in Sec.~\ref{subsec:universal} and \ref{sec:stau}. 
The latter case is studied in Sec.~\ref{subsec:non-univ}.

Chapter \ref{chp:summary} is devoted to conclusion.
The vacuum decay rate is summarized in Appendix \ref{app:vacuum}. 
The corrections to the Bino-muon-smuon coupling are discussed in Appendix \ref{app:bino}.
The cross section of slepton productions is summarized in Appendix \ref{app:slepprod}.

\chapter{Foundation}
\label{chp:foundation}
%!TEX root = ../Dthesis.tex

%%%%%%%%%%%%%%%%%%%%%%%%%%%%%%%%%%%%%%%%%%%%%%%%%%%%%%%%%%%%% Intro

This chapter is a review of measurements which are sensitive to new physics contributions.
In this dissertation, we pick up the anomalous magnetic moment of the muon (muon $g-2$), 
the charged lepton flavor violation (cLFV), the electric dipole moment (EDM), and the Higgs coupling to di-photon.
First, we start from the dipole contributions of leptons, generated by the electromagnetic form factor of leptons in Sec.~\ref{sec:EMDM}.
They contribute to the muon $g-2$, cLFV, and EDM, respectively.
In Sec.~\ref{sec:gm2exp}, we review the muon $g-2$, 
which has a discrepancy between the experimental and the theoretical values.
The discrepancy is expected to be confirmed by future muon $g-2$ experiments.
%, such as Fermilab and J-PARC.
Although processes of LFV and CP violation (CPV) are extremely suppressed in standard model, 
some of new physics models might induce large LFV and CPV.
They are also expected to be probed in future experiments.
In Sec.~\ref{sec:LFVexp} and \ref{sec:EDMexp}, we review LFV and EDM.
For a detailed review of the muon $g-2$, LFV, and EDM, see textbook\cite{Roberts:2010zz}. 

The Higgs couplings are also promising to probe new physics contributions.
Particularly, since the Higgs coupling to di-photon is loop-induced, 
it is sensitive to new physics contributions. 
Sec.~\ref{sec:kappaexp} is a review of the Higgs coupling to di-photon.

%%%%%%%%%%%%%%%%%%%%%%%%%%%%%%%%%%%%%%%%%%%%%%%%%%%%%%%%%%%%%

%%%%%%%%%%%%%%%%%%%%%%%%%%%%%%%%%%%%%%%%%%%%%%%%%%%%%%%%%%%%% Sections

\section{Electromagnetic Dipole Moments}
\label{sec:EMDM}
%!TEX root = ../Dthesis.tex
%This section is based on Sec.~2.2 of \cite{Roberts:2010zz}.

The dipole contributions of leptons, which induce the electric and magnetic dipole moment and flavor-changing processes, 
not only have been precisely measured, but also are sensitive to new physics contributions.
In this section, we summarize the contributions of electromagnetic dipole moments in field theory approach.

\subsection{Electromagnetic Form Factors}
\label{subsec:EMFF}

First, we consider the matrix element of the electromagnetic current $J_{\mu }^{\text{em}} = e Q_{\ell } \overline{\ell } \gamma _{\mu } \ell $,
between initial and final states of a spin $1/2$ fermion $\ell $,
with momenta $p$ and $p^{\prime }$, respectively.
Here, we define a transfer momentum as $q \equiv p^{\prime } - p$.
The matrix element is written as 
\beq
 \langle \ell (p^{\prime })| J_{\mu } ^{\text{em}} | \ell (p) \rangle = e Q_{\ell }\overline{u}_{\ell } (p^{\prime }) \Gamma _{\mu } u_{\ell }(p), \label{eq:form}
\eeq
where $Q_{\ell }$ is the electric charge of the lepton,\footnote{
We define $e > 0$ and $Q_{\ell } \mp 1$ for leptons (anti-leptons).} and $\overline{u}_{\ell }$ and $u_{\ell }$ are Dirac spinor fields of the lepton. 
The matrix $\Gamma _{\mu }$ has the general Lorentz structure as \cite{Roberts:2010zz}\footnote{
This definition is slightly different from Ref.~\cite{Roberts:2010zz}.
Correspondences of \cite{Roberts:2010zz} to our definitions are 
\beq
 \Gamma _{\mu }   &\leftrightarrow& eQ_{\ell } \cdot \Gamma _{\mu } , \notag \\
 F_2(q^2)  &\leftrightarrow&  F_2(q^2) / 2m_{\ell }, \notag
\eeq
where the left hand sides are from \cite{Roberts:2010zz}, our definitions are the left hand sides.}
\beq
 \Gamma _{\mu } = F_1(q^2) \gamma _{\mu } + i \frac{F_2(q^2)}{2m_{\ell }} \sigma _{\mu \nu } q^{\nu } 
 - F_3(q^2) \sigma _{\mu \nu } q^{\nu} \gamma _5 + F_A(q^2) (\gamma _{\mu }  - 2m_{\ell } q_{\mu } ) \gamma _5, \label{eq:form_general}
\eeq
where $\gamma _{\mu }$ is the gamma matrices, $\sigma _{\mu \nu } = (i/2) [\gamma _{\mu }, \gamma _{\nu }]$, 
and $F_i(q^2)~(i = 1,2,3, A), $ are called the charge, anomalous magnetic dipole, electric dipole and anapole form factors,\footnote{
The anapole form factor $F_A(q^2)$ corresponds to parity violations and is generated by electroweak loop physics.
In this dissertation, we will not discuss anapole induced interactions.} 
respectively.
The form factors become real since the current  $J_{\mu }^{\text{em}}$ is required to be hermitian.

The static charge and dipole moments are defined at $q^2 = 0$ as follows,
\beq
 F_1 (0) &=& 1  : \text{electric charge}, \label{eq:form_charge} \\
 F_2 (0) &=&  a_{\ell } :  \text{anomalous magnetic moment}, \label{eq:form_mdipole} \\
 F_3 (0) &=& \frac{d_{\ell }}{e} : \text{electric dipole moment}. \label{eq:form_edipole}
\eeq
To lowest order, $F_1(0) = 1$ and $F_{2,3}(0) = 0$.
Non-vanishing contributions to $F_{2, 3}$ are generated by radiative levels.

The dipole contributions are  also evaluated in effective Lagrangian approach.
The effective Lagrangian, which provide to $F_{2,3}(0)$ is given by 
\beq
 \mathcal{L}_{\text{eff}} &=& \left( \frac{e}{4 m_{\ell }}  F_2 \cdot \overline {\ell } \sigma _{\mu \nu } \ell  
 + \frac{i}{2} F_3 \cdot \overline{\ell } \sigma _{\mu \nu } \gamma _5 \ell \right) F^{\mu \nu} , \label{eq:effH} \\
 F_{\mu \nu } &=& \partial _{\mu } A_{\nu} - \partial _{\mu } A_{\mu },  \label{eq:fieldstrength}
\eeq
where $F_{\mu \nu }$ is the field strength and $A_{\mu }$ is the electromagnetic field.
The interactions in Eq.~\eqref{eq:effH} are called dimension 5 operators.
They are generally not allowed in classical Lagrangian since they spoil renormalizability at quantum field theory.
Nevertheless, they arise at the radiative levels, if no symmetry forbid them.
Both the anomalous magnetic and electric dipole moments must be finite and are calculable in principle in term of other parameters of the theory. 
In Sec.~\ref{sec:gm2exp} and \ref{sec:EDMexp}, 
we will summarize the current status and the future prospects for them.

\subsection{Transition dipole moments}
\label{subsec:TDM}

Flavor-changing transition amplitudes between distinct fermions are composed of 
flavor off-diagonal part of electromagnetic current $J_{\mu }^{\text{em}}$ 
and induce $\ell _{i} \to \ell _{j} + \gamma $ processes.
They can be parameterized in analogy with Eqs. \eqref{eq:form} and \eqref{eq:form_general} as follows\cite{Roberts:2010zz}
\beq
  \langle \ell _j (p^{\prime })| J_{\mu } ^{\text{em}} | \ell _i(p) \rangle 
  &=& e Q_{\ell } \overline{u}_{\ell _j} (p^{\prime }) \Gamma ^{ij}_{\mu } u_{\ell _i}(p), \label{eq:form_FV} \\
  \Gamma ^{ij}_{\mu } &=& \left( q^2 g_{\mu \nu } - q_{\mu } q_{\nu} \right) \gamma ^{\nu } 
  \left[ F^{ij}_{E0} (q^2)  + \gamma _5 F^{ij}_{M0} (q^2) \right] \notag \\
  &&+ i \sigma _{\mu \nu } q^{\nu } \left[ F^{ij}_{M1} (q^2)  + \gamma _5 F^{ij}_{E1} (q^2)   \right] . \label{eq:form_general_FV}
\eeq
The matrix $\Gamma ^{ij}_{\mu }$ is composed of two parts.
The first two form factors $F^{ij}_{E0}$ and $F^{ij}_{M0}$ correspond to chirality-conserving processes at $q^2 = 0$.
They are part of dimension 6 operators in effective Lagrangian.
For example, they mainly contribute to $\mu \to e\bar{e}e$ process.
In this dissertation, we are interested in scenarios in which the contributions to the dipole moments are dominant, 
thus we will not discuss the dimension 6 operators.\footnote{
Note that the dimension 5 operators also contribute to such as $\mu \to e\bar{e}e$ process even if we neglect the dimension 6 operators.}

The two form factors in second line of Eq.~\eqref{eq:form_general_FV}, $F^{ij}_{M1}$ and $F^{ij}_{E1}$, change the chirality. 
They  contribute to flavor-changing processes analog of magnetic and electric dipole moments whose contributions are included in dimension 5 operators.
They correspond to $\ell _i \to \ell _j + \gamma $ decays, e.g., $\mu \to e \gamma $.
In this dissertation, we assume that dipole contributions are dominant.  

Flavor-changing processes can be also calculated by effective Lagrangian approach.
The effective Lagrangian which give rise to $F^{ij}_{E1}$ and $F^{ij}_{M1}$, is represented as \cite{Cho:2001hx}
\beq
 \mathcal{L}_{\text{eff}} &=& \frac{e}{2} \overline{\ell }_i \sigma _{\mu \nu }
  \left( [F^{ij}_{E1} - F^{ij}_{M1}] P_L + [F^{ij}_{E1} + F^{ij}_{M1}] P_R \right) \ell _j F^{\mu \nu } + \text{h.c.}, \notag \\
  &\equiv& e \frac{m_{\ell _i}}{2} \overline{\ell }_i \sigma _{\mu \nu }
  \left( A^L_{ij} P_L + A^R_{ij} P_R \right) \ell _j F^{\mu \nu } + \text{h.c.}, \label{eq:eff_LFV}
\eeq
where we define the coefficients $A^L_{ij} $ and  $A^R_{ij} $ as 
$A^L_{ij} = (F^{ij}_{E1} - F^{ij}_{M1}) / m_{\ell _i}$ and $A^R_{ij} = (F^{ij}_{E1} + F^{ij}_{M1}) / m_{\ell _i}$, respectively. 
Further, $m_{\ell _i} > m_{\ell _j}$ is defined and $m_{\ell_{j}} \simeq 0$ is assumed for simplicity.

\section{Muon $g-2$}
\label{sec:gm2exp}
%!TEX root = ../Dthesis.tex

%This section is based on Sec.~2.3 of \cite{Roberts:2010zz}.

The form factor $F_2(0)$ in Eq.~\eqref{eq:form_mdipole} equals to the anomalous magnetic moment of the lepton $a_{\ell }$, 
as mentioned in Sec.~\ref{subsec:EMFF}. 
Particularly, the muon anomalous magnetic moment,  $a_{\mu }$ is called the muon $g-2$.
It gives a sensitive probe for new physics contributions as follows\cite{Roberts:2010zz}, 
\begin{description}
\item[Experiment:]
Muons can be copiously produced in a fully polarized state and 
the lifetimes are long enough to precisely measure their precession frequencies in a magnetic field \cite{Bennett:2006fi}.
\item[Theory:]
The SM prediction has been precisely evaluated by several group \cite{Melnikov:2006sr}, and 
the muon is heavier than the electron enough to be sensitive to new physics contributions.   
\end{description}
In this section, we briefly review the theoretical prediction and the experimental value of the muon $g-2$.

\subsection{Definition}

The $g$-value, $g_{f}$ is a dimensionless quantity which characterizes the magnetic moment.
It is proportional to the spin magnetic moment $\overrightarrow{\mu }_{{\rm spin}}$, 
and is defined as 
\beq 
  \overrightarrow{\mu }_{{\rm spin}} = g_{f} \times  \frac{e Q_{f}}{2m _{f}} \overrightarrow{S}, \label{eq:magneticm}
\eeq 
where $m_{f}$, $e Q_{f}$, and $\overrightarrow{S}$ are the mass, electric charge, and spin of the particle $f$, respectively.
The quantity $\mu _{B} \equiv e / 2m_{f} $ is called the Bohr magneton. 
The interaction Hamiltonian $H_{\text{int}}$ between the particle and the magnetic field $\overrightarrow{B}$ is expressed as
\beq
 H_{{\rm int} } \supset - \overrightarrow{\mu }_{{\rm spin}} \cdot \overrightarrow{B}.  \label{eq:hamiltonian}
\eeq

The $g$-value for leptons $g_{\ell }$ is predicted as 2 in Dirac theory.
However, radiative corrections shift it slightly. 
The shift from 2 is known as the anomalous magnetic moment
\beq
 g_{\ell } = 2 (1 + a_{\ell }). \label{eq:a_magnetic_moment}
\eeq
In particular, $a_{\mu }$ is called the muon $g-2$.

Eq.~\eqref{eq:a_magnetic_moment} can be checked by field theory based approach in Sec.~\ref{sec:EMDM}.
First, we consider a lepton scattering from a static vector potential $A^{\text{cl}}_{\mu } = (0, \overrightarrow{A_{\text{cl}}})$.
The scattering matrix element $\mathcal{M}$ is evaluated as\cite{Peskin:1995ev}
\beq
 \mathcal{M} =  e Q_{\ell} \left[ \overline{u}_{\ell } 
 \left(  F_1(q^2) \gamma _{i} + i \frac{F_2(q^2)}{2m_{\ell }} \sigma _{i \nu } q^{\nu }  \right)  u_{\ell } \right] 
 \widetilde{A}_{\text{cl}}^{i}, \label{eq:gm2static1}
\eeq
where $\widetilde{A}_{\text{cl}}$ is the Fourier transform of $\overrightarrow{A_{\text{cl}}}$ and 
the electric dipole form factor $F_3$ is neglected. 
The spinor $u_{\ell }(p)$ is expressed as\footnote{
The representation obeys in Ref.~\cite{Peskin:1995ev}.
}
\beq
 u_{\ell } (p) = \begin{pmatrix}
                        \sqrt{p \cdot \sigma} \xi \\
                        \sqrt{p \cdot \overline{\sigma }}\xi   \\
                     \end{pmatrix} 
                     \simeq \sqrt{m_{\ell}}
                     \begin{pmatrix}
                      \left( 1- \frac{ \overrightarrow{p}\cdot \overrightarrow{\sigma }}{2m_{\ell }}\right) \xi  \\
                      \left( 1+ \frac{\overrightarrow{p}\cdot \overrightarrow{\sigma }}{2m_{\ell }}\right) \xi  \\
                     \end{pmatrix} ~~(\text{non-relativistic limit}). 
\eeq
Then, we extract linear contributions in $q^i$ from Eq.~\eqref{eq:gm2static1} as
\beq
 \mathcal{M} \simeq  - (2m_{\ell }) e Q_{\ell } \xi ^{\prime \dagger} 
 \left( \frac{-1}{2m_{\ell }} \sigma ^k \left[ F_1(0) + F_2(0) \right]\right) \xi \widetilde{B}^{k}~~~(q \to 0),  \label{eq:gm2static2}
\eeq
where $\widetilde{B}^k = i \epsilon ^{ijk} q^i \widetilde{A}^j_{\text{cl}}$ is the Fourier transform of the magnetic field.

The amplitude $\mathcal{M}$ can be interpreted as the Born approximation to the scattering of the electron from a potential (For detail, see textbook \cite{Peskin:1995ev}).
The interaction Hamiltonian which corresponds to such potential is given by 
\beq
 H_{{\rm int} }  = - \langle \overrightarrow{\mu } \rangle \cdot \overrightarrow{B}, \label{eq:bornint}
\eeq
where 
\beq
 \langle \overrightarrow{\mu } \rangle = 2\left[F_1(0) + F_2(0)  \right] \times \frac{e Q_{\ell }}{2m_{\ell }} 
 \xi ^{\prime \dagger} \frac{\overrightarrow{\sigma }}{2} \xi .
 \label{eq:magneticborn}
\eeq
The factor $\xi ^{\prime \dagger} (\overrightarrow{\sigma }/  2 ) \xi $ can be interpreted as the spin of the leptons, $\overrightarrow{S}$.
Comparing Eq.~\eqref{eq:magneticborn} with Eqs.~\eqref{eq:form_charge} and \eqref{eq:form_mdipole}, the coefficient $2\left[F_1(0) + F_2(0)  \right]$ becomes 
\beq
  2\left[F_1(0) + F_2(0)  \right] = 2 (1 + a_{\ell }) = g_{\ell }.
\eeq
It is just the $g$-value.

\subsection{The Standard Model prediction of the muon $g-2$}
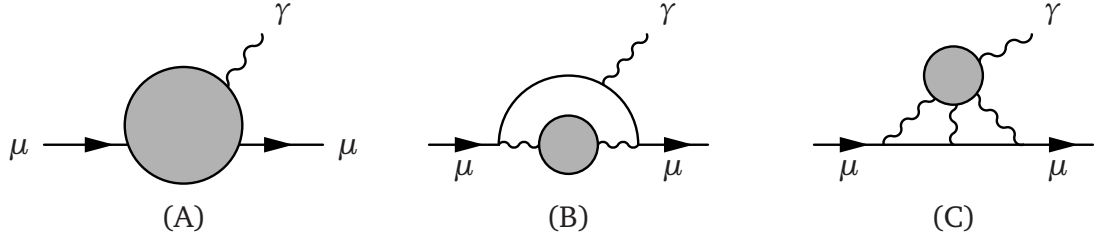
\begin{figure}[t]\begin{center}
 \begin{fmffile}{feyns/g_2-SM}
  \begin{minipage}[t]{0.3\textwidth}
   \begin{center}
    \begin{fmfgraph*}(110,70)
\fmfleft{d1,p1,d2,d3,d4}\fmfright{d5,p2,d6,d7,gc}
\fmf{fermion,label=,l.s=right}{p1,x1}
\fmf{vanilla,tension=1}{x2,x1}
\fmf{phantom,tension=1}{x2,x4,xx,x3,x1}
\fmf{fermion,label=,l.s=right}{x2,p2}
\fmf{phantom,tension=5}{gc,ge}
\fmf{phantom,tension=1}{ge,d1}
\fmfposition
\fmf{phantom}{gc,yy,d1}
\fmf{phantom}{yy,gb,xx}
\fmf{phantom,left,tag=1,l.s=right}{x3,x4}
\fmfipath{p[]}
\fmfiset{p1}{vpath1(__x3,__x4)}
\fmfi{photon}{point length(p1)/3*2 of p1 -- vloc(__ge)}
\fmflabel{$\gamma$}{ge}
\fmflabel{$\mu$}{p1}
\fmflabel{$\mu$}{p2}
%\fmfblob{0.3w,d.shape=circle,d.filled=30}{gb}
\fmfv{d.size=0.4w,d.shape=circle,d.filled=30}{gb}
    \end{fmfgraph*}
    \\(A)
   \end{center} 
  \end{minipage}
  \begin{minipage}[t]{0.3\textwidth}
   \begin{center}
    \begin{fmfgraph*}(110,70)
\fmfleft{d1,p1,d2,d3,d4}\fmfright{d5,p2,d6,d7,gc}
\fmf{fermion,label=$\mu$,l.s=right}{p1,x1}
\fmf{photon}{x2,xx,x1}
\fmf{fermion,label=$\mu$,l.s=right}{x2,p2}
\fmf{phantom,tension=5}{gc,ge}
\fmf{phantom,tension=1}{ge,d1}
\fmfposition
\fmf{vanilla,left,tag=1,l.s=right}{x1,x2}
\fmfipath{p[]}
\fmfiset{p1}{vpath1(__x1,__x2)}
\fmfi{photon}{point length(p1)/3*2 of p1 -- vloc(__ge)}
\fmflabel{$\gamma$}{ge}
\fmfv{d.size=0.2w,d.shape=circle,d.filled=30}{xx}
    \end{fmfgraph*}
    \\(B)
   \end{center} 
  \end{minipage}
  \begin{minipage}[t]{0.3\textwidth}
   \begin{center}
    \begin{fmfgraph*}(110,70)
\fmfleft{d1,p1,d2,d3,d4}\fmfright{d5,p2,d6,d7,gc}
\fmf{fermion,label=$\mu$,l.s=right}{p1,x1}
\fmf{vanilla}{x2,xx,x1}
\fmf{fermion,label=$\mu$,l.s=right}{x2,p2}
\fmf{phantom,tension=5}{gc,ge}
\fmf{phantom,tension=1}{ge,d1}
\fmfposition
\fmf{phantom,left,tag=1,l.s=right}{x1,x2}
\fmfipath{p[]}
\fmfiset{p1}{vpath1(__x1,__x2)}
\fmfipair{mid}
\fmfiset{mid}{point length(p1)/2 of p1}
\fmfi{photon}{mid -- vloc(__ge)}
\fmfi{photon}{mid -- vloc(__xx)}
\fmflabel{$\gamma$}{ge}
\fmfi{photon,left}{mid -- vloc(__x1)}
\fmfi{photon,left}{mid -- vloc(__x2)}
\fmfiv{d.shape=circle,d.filled=30,d.size=0.2w}{mid}
    \end{fmfgraph*}
    \\(C)
   \end{center} 
  \end{minipage}
  \caption[The Feynman diagram which contributes to Muon $g-2$.]
{(A) The Feynman diagram which contributes to the anomalous magnetic moment of the muon.
(B) The hadronic vacuum-polarization contributions to the muon $g-2$. 
(C) The hadronic light-by-light contributions to the muon $g-2$.}
  \label{fig:gm2SM}
 \end{fmffile}
\end{center}
\end{figure}

The SM prediction of the muon $g-2$ has been precisely evaluated through many efforts of several groups of theorists.
Fig.~\ref{fig:gm2SM} (A) shows the Feynman diagram which contribute to the muon $g-2$.
The theoretical uncertainty reaches to about 0.4 ppm.
In this section, we briefly review the SM prediction of the muon $g-2$.

\subsubsection{QED Contribution}
\label{subsec:gm2qed}

The quantum electromagnetic dynamics (QED) contributions are the dominant contributions to the muon $g-2$ (99.993\%), 
and come from the diagrams with leptons and photons.
They have been calculated analytically up to $\mathcal{O}(\alpha ^3)$,
and recently the full five-loop ($\mathcal{O}(\alpha ^5)$) contribution has been calculated \cite{Aoyama:2012wj, Aoyama:2012wk}.
The present QED value is reported as
\beq
  a_{\mu }(\text{QED}) &=& (11658471.8951 \pm 0.0080) \times 10^{-10}, \label{eq:gm2qed}
\eeq
where the uncertainties are from the lepton mass ratios, the $\mathcal{O}(\alpha ^4, \alpha ^5)$ terms. 
The value of $\alpha $ is taken from $^{87}$Rb atom, $\alpha ^{-1}$(Rb) $= 137.035999049 (90)$.
The current uncertainty from QED contributions is far below the experimental error
and is less important for the discrepancy between the theoretical and the experimental value of the muon $g-2$.
 
\subsubsection{Electroweak Contribution}
\label{subsec:gm2ew}

The electroweak (EW) contributions to the muon $g-2$ are from diagrams with the $W$, $Z$, and Higgs bosons. 
The one-loop contribution is evaluated as
\beq
 a^{1\text{-loop}}_{\mu }(\text{EW}) &=& \frac{G_F}{\sqrt{2}} \frac{m_{\mu }^2}{8\pi ^2} 
 \left[ \frac{10}{3} + \frac{1}{3} \left( 1 - 4 \sin ^2 \theta _W \right) ^2 - \frac{5}{3} 
 + \mathcal{O}\left( \frac{m^2_{\mu }}{m^2_Z} \log \frac{m^2_Z}{m^2_{\mu }} \right)  \right] \notag \\
 && +  \frac{G_F}{\sqrt{2}} \frac{m_{\mu }^2}{8\pi ^2} \frac{m^2_{\mu }}{m^2_h} 
 \int ^1 _0 dx \frac{2x^2 (2-x)}{1-x+\frac{m^2_{\mu }}{m^2_h} x^2} \notag \\
 && = 19.48 \times 10^{-10}, \label{eq:gm2ew1} 
\eeq
where $G_F = 1.16637(1) \times 10^{-5} \GeV$ is the Fermi constant, 
and $\sin ^2 \theta _W \equiv 1- m^2_W / m^2_Z \sim 0.223$ is the Weinberg angle.
It was calculated by several groups\cite{Bardeen:1972vi,Jackiw:1972jz,Bars:1972pe,Fujikawa:1972fe}, 
after the Glashow-Weinberg-Salam model was shown to be renormalizable.
In one-loop level, only $W$ and $Z$ boson contributions are relevant at a measurable level.
On the other hand, the Higgs boson contribution is $\mathcal{O}(10^{-11})$, i.e., negligible, due to the small Yukawa coupling of the Higgs boson to muons.

The EW contribution was calculated at the two-loop level, 
which is negative\cite{Czarnecki:1995sz,Peris:1995bb, Czarnecki:1995wq, Czarnecki:2002nt}, 
with including leading log effects at three-loop level.
Recently, it has been re-evaluated using recent LHC result of the Higgs mass\cite{Gnendiger:2013pva}.
The total EW contributions to the muon $g-2$ are 
\beq
 a_{\mu }(\text{EW}) &=& (15.36 \pm 0.1) \times 10^{-10}, \label{eq:gm2ew} 
\eeq
where the uncertainty comes from hadronic effects in the second-order EW diagrams with quark triangle loops,
along with unknown three-loop contributions\cite{Czarnecki:2002nt, Knecht:2002hr,Vainshtein:2002nv ,Knecht:2003xy }.
The leading logs for the next-order term are shown to be small\cite{Czarnecki:2002nt,Gnendiger:2013pva}.

\subsubsection{Hadronic Contribution}
\label{subsec:gm2had}

The hadronic contributions are from QCD interaction.
They are composed of the two types as follows:
(1) Hadronic Vacuum Polarization contribution (HVP), 
and (2) Hadronic Light-by-Light contribution (HLbL), respectively.
HVP originates from hadronic quantum corrections in the photon propagator.
It is shown in Fig.~\ref{fig:gm2SM} (B).
The lowest order of HVP is included at $\mathcal{O}(\alpha ^2)$, 
the higher order correction of HVP and the HLbL are of $\mathcal{O}(\alpha ^3)$.

Unlike the QED contribution, HVP cannot be calculated from original QCD,
because hadronic physics in low energy, i.e., non-perturbative QCD regime, contribute to HVP. 
Nevertheless, thanks to analyticity of the vacuum polarization, 
the leading effect of HVP can be evaluated via the dispersion relation as\cite{Roberts:2010zz} 
\beq
 a_{\mu } (\text{HVP-LO}) = \frac{1}{4 \pi ^3} \int ^{\infty} _{4 m^2_{\pi }} ds 
 ~K(s) ~ \sigma ^0 (s)_{e^+ e^- \to \text{hadrons}}, \label{eq:gm2hvp0}
\eeq
where $\sigma ^0 (s)_{e^+ e^- \to \text{hadrons}}$ is cross sections for $e^+e^- \to \text{hadrons}$.
They are determined by measured values as input.
The kinematical function $K(s)$ is given by\cite{Roberts:2010zz}
\beq
 K(s) &=& x^2 \left( 1 - \frac{x^2}{2} \right) + (1 + x )^2 \left( 1 + \frac{1}{x^2} \right) 
 \left[ \log (1 + x) - x + \frac{x^2}{2} \right] \notag \\
  &&+ \frac{1+x}{1-x} x^2 \log x, \label{eq:Ks} \\
  x &=& \frac{1 - \sqrt{1-\frac{4m^2 _{\mu }}{s} }}{1 + \sqrt{1-\frac{4m^2 _{\mu }}{s} }}. \label{eq:xsub}
\eeq

In the lowest order, 
two groups have analyzed by using $e^+ e^- \to \text{hadrons}$ data\cite{Davier:2010nc, Hagiwara:2011af} as
\beq
 a_{\mu }(\text{HVP-LO}) &=& (692.3 \pm 4.2) \times 10^{-10} \text{\cite{Davier:2010nc}}, \label{eq:gm2hvp1} \\
a_{\mu }(\text{HVP-LO}) &=& (694.91 \pm 4.27) \times 10^{-10} \text{\cite{Hagiwara:2011af}}. \label{eq:gm2hvp2} 
\eeq

The most recent result is evaluation of the next-to-leading order\cite{Hagiwara:2011af},
which can be also calculated from dispersion relation as
\beq
  a_{\mu }(\text{HVP-HO}) &=& -(9.84 \pm 0.07) \times 10^{-10}. \label{eq:gm2hvp3} 
\eeq

Another hadronic contribution originates though a light-by-light scattering process.
It is shown in Fig.~\ref{fig:gm2SM} (C).
HLbL cannot be determined from data, unlike HVP.
It must be calculated using low energy hadronic models that correctly reproduce property of QCD.
In this dissertation, we quote two results as follows:
\beq
 a_{\mu }(\text{HLbL}) &=& (10.5 \pm 2.6) \times 10^{-10}, \label{eq:gm2hlbl1} \\
 a_{\mu }(\text{HLbL}) &=& (11.6 \pm 4.0) \times 10^{-10}. \label{eq:gm2hlbl2}
\eeq
The former is obtained in Ref.~\cite{Prades:2009tw}, the latter is from Ref.~\cite{Jegerlehner:2009ry,Nyffeler:2009tw}.

\subsection{Current Status of the muon $g-2$}
\label{subsec:gm2status}

The complete Standard Model prediction of the muon $g-2$ is 
\beq
 a_{\mu }(\text{SM}) = a_{\mu }(\text{QED}) + a_{\mu }(\text{EW}) + a_{\mu }(\text{HVP}+\text{HLbL}). 
\eeq
As for the hadronic contribution, we take the result in Ref.\cite{Hagiwara:2011af}.\footnote{
As for HLbL, Ref.~\cite{Prades:2009tw} is quoted in Ref.~\cite{Hagiwara:2011af}.} 
Combining Eqs.\eqref{eq:gm2qed}, \eqref{eq:gm2ew}, \eqref{eq:gm2hvp2}, \eqref{eq:gm2hvp3}, and \eqref{eq:gm2hlbl1}, 
the SM value is 
\beq
 a_{\mu }(\text{SM}) &=& (11659182.8 \pm 4.9) \times 10^{-10} \label{eq:gm2sm}.
\eeq

On the other hand, the muon $g-2$ has now been measured to a precision of $0.54$ ppm
by the experiment E821 at the Brookhaven AGS \cite{Bennett:2006fi, Roberts:2010cj}.
The values is summarized as
\beq
 a_{\mu } (\text{exp}) &=& (11659208.9 \pm 6.3) \times 10^{-10},  \label{eq:gm2exp}
\eeq
and the discrepancy between the experimental and the theoretical value becomes 
\beq
 \Delta a_{\mu } \equiv a_{\mu } (\text{exp}) - a_{\mu } (\text{SM}) = (26.1 \pm 8.0) \times 10^{-10}. \label{eq:g-2_deviation}
\eeq
It is found that the discrepancy is as large as $3\sigma $ level. 
In this dissertation, we call this problem the muon $g-2$ anomaly, 
and assume that the discrepancy suggests new contributions from new physics beyond the SM.

%%%%%%%%%%%%%%%%%%%%%%%%%%%%%%%%%% Table
\begin{table}[tb]
\begin{center}
  \begin{tabular}{|c|c|} \hline
    Contributions & Value ($\times 10^{-10}$) \\ \hline \hline
    QED  &  $11658471.8951 \pm 0.0080$ \\  
    EW &  $15.36 \pm 0.1$ \\ 
    HVP (LO) \cite{Davier:2010nc} &  $692.3 \pm 4.2$   \\ 
    HVP (LO) \cite{Hagiwara:2011af} & $694.91 \pm 4.27$    \\ 
    HVP (HO) \cite{Hagiwara:2011af} & $-9.84 \pm 0.07$  \\ 
    HLbL \cite{Prades:2009tw} & $10.5 \pm 2.6$    \\ 
    HLbL \cite{Jegerlehner:2009ry,Nyffeler:2009tw} & $11.6 \pm 4.0$   \\ \hline
    Total SM \cite{Davier:2010nc} & $11659180.2 \pm 4.2_{\text{H-LO}} \pm 2.6_{\text{H-HO}} \pm 0.2_{\text{other}} (\pm 4.9_{\text{tot}})$ \\
    Total SM \cite{Hagiwara:2011af} & $11659182.8 \pm 4.3_{\text{H-LO}} \pm 2.6_{\text{H-HO}} \pm 0.2_{\text{other}} (\pm 5.0_{\text{tot}})$ \\ \hline 
    Discrepancy \cite{Davier:2010nc} & $28.7 \pm 8.0$ \\
     Discrepancy \cite{Hagiwara:2011af} & $26.1 \pm 8.0$ \\ \hline \hline
  \end{tabular}
   \caption[Summary of the Standard Model prediction.]
   {Summary of the Standard Model contributions to the muon $g-2$.
   The lowest-order hadronic contribution and the hadronic light-by-light contribution is quoted by using two recent evaluations.
   We take the result in Ref.~\cite{Hagiwara:2011af}.}
   \label{tab:sum_gm2}
\end{center}
\end{table}

%%%%%%%%%%%%%%%%%%%%%%%%%%%%%%%%%%

\subsection{Future Prospects of the muon $g-2$}

%%%%%%%%%%%%%%%%%%%%%%%%%%%%%%%%%% Figure
\begin{figure}[t!]
 \begin{center}
 \includegraphics[width=10cm]{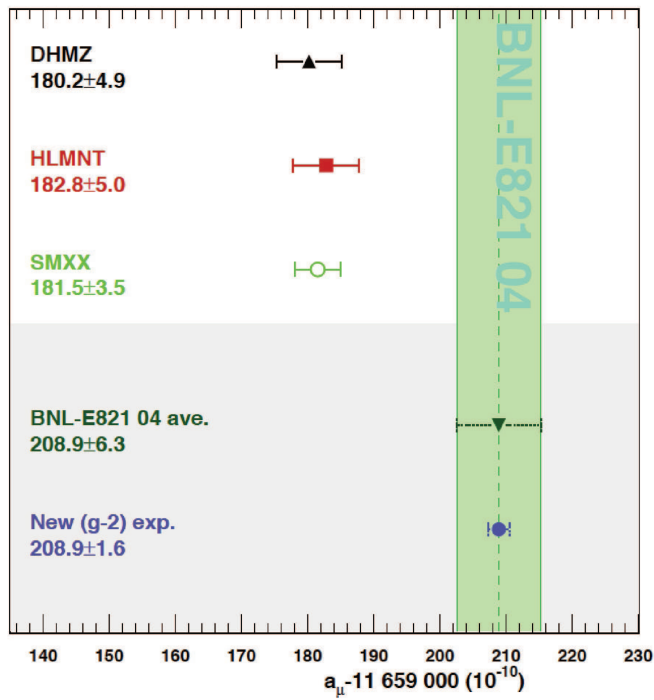} \hspace*{2mm}
 \end{center}
 \caption[The current status and the future prospect for the experimental and the theoretical values of the muon $g-2$.]
 {The current status and the future prospect for the experimental and the theoretical values of the muon $g-2$.
 The figure provide comparisons between $a_{\mu }(\text{exp})$ and $a_{\mu }(\text{SM})$.
 "DHMZ" is from Ref.~\cite{Davier:2010nc}, "HLMNT" is from Ref.~\cite{Hagiwara:2011af}.
 "SMXX" means the same central value with a reduced error as expected by the improvement on the hadronic cross section measurement.
 "BNL-E821 04 ave." is the current experimental value, and "New ($g-2$) exp." is the same central value with a reduced error as
 planned by the future $(g-2)$ experiments at Fermilab\cite{Venanzoni:2014ixa} and J-PARC\cite{Mibe:2010zz}.
 This figure is from Ref.~\cite{Blum:2013xva}.}
 \label{fig:gm2_future}
\end{figure}
%%%%%%%%%%%%%%%%%%%%%%%%%%%%%%%%%% 

The theoretical uncertainty on the standard model value is about 0.4 ppm, as seen in  Eq.~\eqref{eq:gm2sm}.
We summarize the standard model contributions in Tab.~\ref{tab:sum_gm2}.
From this,  it is found that the theoretical uncertainty by the hadronic contribution is dominant.
The lowest-order hadronic contribution is determined from the $e^+ e^- \to \text{hadrons}$ cross section data.
It will be improved by new data from Novosibirsk and BESIII \cite{Blum:2013xva}.
The theoretical uncertainty is expected to be reduced by a factor 2, i.e., $\sim$0.2 ppm.

The hadronic light-by-light contribution might be also improved.
Two approaches are planed to reduce the uncertainty.
One is the measurements of $\gamma ^*$ physics at 
KLOE\cite{AmelinoCamelia:2010me, Babusci:2011bg} and BESIII.
For example, the reaction $e^+e^- \to e^+e^- \gamma ^* \gamma ^* \to e^+e^- \pi ^0$ 
gives a useful information for the form factor $F_{\pi ^0 \gamma ^* \gamma ^*}$,
which is the model parameter of HLbL.
The other is the calculation on the lattice, 
which could produce a meaningful result by 2018\cite{Blum:2013xva}.
The goal is to compute HLbL to 10 \% accuracy, or better during the next 3-5 years.\footnote{
Improvement for HVP is also planned by the lattice QCD.
It is expected to be reduced to 1 or 2 \% within the next few years.}

The new $g-2$ experiments are planed at Fermilab\cite{Venanzoni:2014ixa} and J-PARC\cite{Mibe:2010zz}.
The experimental uncertainty is expected to be reduced about 0.1 ppm.
Fig.~\ref{fig:gm2_future}~shows the current status and the future prospects for the theoretical and the experimental values.
It is found that the statistical significance would become 7-8 $\sigma $ in near future if the central value would remain the same.
Thus, if the muon $g-2$ anomaly is true, it can be confirmed at more than 5 $\sigma $ levels.

The sensitivity to the muon $g-2$ anomaly will be improved during the next 3-5 years.
Therefore, detailed investigation new physics models which are motivated by the muon $g-2$ anomaly are important, just now.
Several models have been invented to solve this anomaly.
The supersymmetry can be a solution, 
if the masses of new particles couple to the muon, are order of 100\GeV.
In this dissertation, we concentrate on the supersymmetric standard model.
The new contributions from supersymmetric models are given in Sec.~\ref{sec:gm2mssm}.
Searches for supersymmetric models which solve the muon $g-2$ anomaly are discussed in Chapter.~\ref{chp:main}.

\section{Lepton Flavor Violation}
\label{sec:LFVexp}
%!TEX root = ../Dthesis.tex

New physics contributions to flavor changing neutral current  (FCNC) processes tend to be much larger than the SM predictions. 
Particularly, muon FCNCs are generically enhanced when the muon $g-2$ discrepancy is explained by new physics contributions. 
Therefore, the size of FCNC have been strongly constrained from lepton flavor experiments.
In this section, we summarize theoretical foundations and status for the experimental searches.

\subsection{Motivation}

Particle flavors have often provided hints of new physics that was not accessible at the stage.
For example, the Glashow-Iliopoulos-Maiani mechanism (GIM mechanism)\cite{Glashow:1970gm} was suggested to explain 
the absence of quark FCNCs, and predicted the existence of charm quark, when only three quarks were discovered yet.
Kobayashi and Maskawa hypothesized three generations of quarks\cite{Kobayashi:1973fv} 
to explain the CP violation observed in neutral Kaon mixing and decays,
and after predicted large CP violations in B meson (or extremely heavy quark, i.e., the top quark)\cite{Bigi:1981qs}. 
This hypothesis was confirmed by discovery of top quark and the B factory experiments.
Further, the neutrino oscillations, which are evidence tiny neutrino mass, 
could imply the existence of extremely heavy Majorana neutrinos 
(Seesaw mechanism\cite{Sawada:1979gf,GellMann:1980vs, Minkowski:1977sc,Mohapatra:1979ia})
and Leptogenesis\cite{Fukugita:1986hr}, 
a scenario which the CP violation in the neutrino sector might be origin of asymmetry between the matter and the anti-matter in the Universe.
 
Neutrinos oscillate almost freely among three lepton generations.
On the other hand, no FCNC processes have ever been observed yet in charged leptons.
In the standard model,\footnote{
Here, the "standard model" includes neutrino masses.} 
they are extremely suppressed because neutrino masses are much smaller than the EW scale.
For example, a process in which a muon converts into an electron and a photon, $\mu \to e \gamma $,
can proceed through a neutrino oscillation $\nu _{\mu } \to \nu _e $ at the one-loop level.
The branching ratio of $\mu \to e \gamma $ process is evaluated as follows, 
\beq
 B (\mu \to e \gamma ) = \frac{3 \alpha }{32 \pi } \left| \sum _{i} (V_{MNS})^*_{\mu i} (V_{MNS})_{e i}  \frac{m^2_{\nu _{i}}}{m^2_W} \right| ^2,
 \label{eq:LFVSM}
\eeq 
where $V_{MNS}$ is the Maki-Nakagawa-Sakata (MNS) matrix\cite{Maki:1962mu} and $m_{\nu _i}$ are neutrino masses. 
It is found that the branching ratio is negligibly small ($\leq \mathcal{O}(10^{-54})$) because of tiny neutrino masses. 
In this way, the lepton FCNCs are highly suppressed in SM.

On the other hand, some of New Physics scenario tends to 
introduce relatively large charged lepton flavor violation (cLFV) 
since no fundamental symmetry forbid it.
Particularly, muon FCNCs are more sensitive to new physics contributions.
Typical new physics contribution to the branching ratio of $\mu \to e \gamma $ is estimated as
\beq
 B (\mu \to e \gamma )_{\text{NP}} \sim \frac{48 \pi ^3\alpha }{G^2_F} \frac{\left| \epsilon _{\mu e} \right| ^2}{\Lambda ^4_{\text{NP}}}, \label{eq:NPLFV}
\eeq
where $\Lambda _{\text{NP}}$ is a typical mass scale of new physics, and $\epsilon _{\mu e}$ characterizes a size of $\mu $--$e$ flavor violation.
Then, using $B (\mu \to e \gamma ) < 5.7 \times 10 ^{-13}$\cite{Adam:2013mnn}, 
we can estimate the constraint as
\beq
 \Lambda _{\text{NP}} \gtrsim 600 \TeV \times \sqrt{\epsilon _{\mu e}}. \label{eq:LFVconst}
\eeq
If the muon $g-2$ anomaly is solved by new physics contributions, masses of new particles should be order of 100\GeV.
In this case, the size of $\epsilon _{\mu e}$ is constrained as
\beq
 \epsilon _{\mu e} \lesssim 2 \times 10^{-8}~\text{for}~ \Lambda _{\text{NP}}\sim 100\GeV.
\eeq
In this way, muon FCNCs provide a sensitive probe of new physics.
Therefore, we concentrate cLFV processes of muon in this dissertation.

\subsection{Flavor violating muon decays}
\label{subsec:FVMD}

In this section, we provide formulae for muon FCNC processes in model independently.
We assume that dipole moment contributions are dominant.  
The starting point is the effective Lagrangian \eqref{eq:eff_LFV} in Sec.~\ref{subsec:TDM}.
Here, we discuss representative three processes, $\mu \to e \gamma $, $\mu \to e\bar{e}e$, and $\mu - e$ Conversion in a Muonic Atom, respectively.

\subsubsection{$\mu \to e\gamma $ Decay}

One of the most popular LFV processes is the $\mu \to e \gamma $ decay.
The leading contributions proceed through the dipole-type operators.
Using the effective Lagrangian \eqref{eq:eff_LFV}, 
the branching ratio of $\mu \to e \gamma $ is evaluated as\cite{Roberts:2010zz}
\beq
 B (\mu \to e \gamma ) \equiv \frac{\Gamma (\mu \to e \gamma )}{\Gamma (\mu \to \text{all})} 
 \simeq \frac{48\pi ^3 \alpha }{G^2_F} \left(  |A^L_{12} |^2 + |A^R_{12} |^2 \right) \label{eq:mueg0},  
\eeq
where $A^L_{12}$ and $A^R_{12}$ are Wilson coefficients of  the dipole operators. 
The total decay rate of the muon $\Gamma (\mu \to \text{all})$ is well approximated as
\beq
 \Gamma (\mu \to \text{all}) \simeq \Gamma(\mu \to e \nu \bar{\nu }) \simeq \frac{G^2_F m^5_{\mu }}{192 \pi ^3}. \label{eq:totalmu}
\eeq

\subsubsection{$\mu \to e\bar{e}e$ Decay}

The dipole operators also contribute to $\mu \to e\bar{e}e$,  $\mu -e$ conversion processes
via virtual photon effects at $q^2 \neq 0$.
In this case, the branching ratio of $\mu \to e\bar{e}e$ can be estimated in term of that of $\mu \to e \gamma$.
The ratio between $\mu \to e\bar{e}e$ and $\mu \to e \gamma$ processes is evaluated as \cite{Roberts:2010zz}
\beq
  \frac{B(\mu \to e \bar{e} e  )}{B(\mu \to e \gamma  ) } \simeq  \frac{\alpha }{3\pi } \left[ \log \left( \frac{m^2_e}{m^2_{\mu }} \right) - \frac{11}{4} \right]  
  \simeq 0.006. \label{eq:muto3e}
\eeq

\subsubsection{$\mu -e$ Conversion in a Muonic Atom}

The conversion of a muon captured by a nucleus into an electron has been one of the most powerful methods to search for cLFV. 
The ratio of $\mu -e$ conversion is defined as
\beq
R(\mu N \to e  N) \equiv \frac{\Gamma (\mu N \to e N)}{\Gamma (\mu N \to \text{all})}, \label{eq:mueconv0}
\eeq
where $\Gamma (\mu N \to \text{all})$ is the total decay width.
The dipole contributions in this process correspond to that of $\mu \to e \gamma $ as a function of the mass number (A) and the atomic number (Z), 
as with $\mu \to e\bar{e}e$ decay.
The relation is estimated as\cite{Roberts:2010zz}
\beq
  \frac{B(\mu \to e \gamma  )}{R(\mu N \to e  N)} \simeq  \frac{96\pi ^3 \alpha }{G_F^2 m^4_{\mu }}\frac{1}{3 \times 10^{12}~B(A,Z)}
  \simeq \frac{428}{B(A,Z)}, \label{eq:mueconv}
\eeq
where $B(A,Z)$ is a nucleus-dependent factor that includes atomic and nuclear effects, 
and is evaluated as 1.1, 1.8, and 1.25 for Al, Ti, and Pb, respectively\cite{Czarnecki:1998iz,Kitano:2002mt}.

\subsection{Current Status and Future Sensitivity}
\label{subsec:LFVexp}

%%%%%%%%%%%%%%%%%%%%%%%%%%%%%%%%%% Table
\begin{table}[tb]
\begin{center}
  \begin{tabular}{|c|c|c|} \hline
    Process & Current bound & Future sensitivity \\ \hline \hline
    $\mu \to e \gamma $  & $5.7 \times 10^{-13}$ (MEG)\cite{Adam:2013mnn} & $6 \times 10^{-14}$ (MEG)\cite{Baldini:2013ke}  \\ 
    $\mu \to e\bar{e}e$  &$10^{-12}$ (SINDRUM I)\cite{Bellgardt:1987du}  & $10^{-16}$ (Mu3e)\cite{Blondel:2013ia}\\
    $\mu - e$ conversion & $7\times 10^{-13}$ (SINDRUM II)\cite{Bertl:2006up}  & $3 \times 10^{-17}$ (COMET/Mu2e)\cite{Kuno:2013mha,Abrams:2012er} \\
     & & 2 $\times 10^{-19}$ (PRISM/PRIME)\cite{Kuno:2012pt} \\ \hline \hline
    \end{tabular}
   \caption[Summary of the flavor violating muon decays.]
   {Summary of the flavor violating muon decays in Sec. \ref{subsec:LFVexp}. 
   The left column shows muon FCNC processes.
   The middle column is current experimental limits for each of process.
   The right column represents sensitivities of future experiments.}
   \label{tab:sum_LFV}
\end{center}
\end{table}
%%%%%%%%%%%%%%%%%%%%%%%%%%%%%%%%%%

Signals of cLFV have been searched by many experimental groups.
The current sensitivity to flavor violating muon decays has reached to about $10^{-13}$.
The best upper bound on $B(\mu \to e \gamma  )$ is obtained by the MEG experiment at PSI, 
$B(\mu \to e \gamma  ) < 5.7 \times 10^{-13}$ at 90\% CL\cite{Adam:2013mnn} using $\sim 1/2$ of data taken in summer 2013.
The final result of MEG is expected during 2014.
Further, preparations for an upgrade of MEG experiment (MEG Upgrade) are underway.
The sensitivity is expected to be $B(\mu \to e \gamma  ) = 6 \times 10^{-14}$ at 90\% CL\cite{Baldini:2013ke} in 3 years of running. 

Searches for $\mu \to e\bar{e}e$ decay are also promising.
The current limit is obtained in SINDRUM, $B(\mu \to e\bar{e}e) < 1.0\times 10^{-12}$ at 90\% level\cite{Bellgardt:1987du}.
Using Eq.~\eqref{eq:muto3e}, we find that it is transformed to $1.7 \times 10^{-10}$ in term of $\mu \to e \gamma $ decay, 
and is much weaker than the limit at MEG. 
New measurements is required that an experimental sensitivity is $\mathcal{O}(10^{-16})$ at least 
to be competitive with existing limits and other planned measurements.
Recently, the Mu3e experiment proposed at PSI.
In the first (second) phase of Mu3e, the sensitivity is expected to reach to $10^{-15}$ ($10^{-16}$)\cite{Blondel:2013ia}.  
This sensitivity is comparable to that of MEG Upgrade.
 
The $\mu - e$ conversion is expected to provide the ultimate sensetivity to cLFV. 
The present experimental bound is $R(\mu N \to e N) < 7 \times 10^{-13}$ at 90\% CL\cite{Bertl:2006up} set by the SINDRUM II experiment, 
in which the muonic gold, i.e., $N = \text{Au}$ is used. 
As with $\mu \to e\bar{e}e$ decay, the sensitivity to $\mu -e$ conversion is expected to be upgraded by two projects.
The one is the Mu2e experiment at Fermilab, the other is the COMET experiment at J-PARC.
COMET will follow a staged approach, COMET phase-I and phase-II, respectively.
The expected sensitivities of COMET phase-I (II) reach to $3 \times 10^{-15}$ ($3 \times 10^{-17}$) at 90\% CL with Al\cite{Kuno:2013mha}.
Mu2e experiment provides almost same sensitivity, $5.4 \times 10^{-17}$ at 90\% CL with Al\cite{Abrams:2012er}.
Further, the PRISM project is being developed in Japan.
The sensitivity is expected to reach to $2 \times 10^{-19}$ at 90\% CL with Ti\cite{Kuno:2012pt}.
These results are summarized in Tab.~\ref{tab:sum_LFV}.

Finally, let us estimate the constraint for mass scale of new physics again, using the expected sensitivity of PRISM project.
Using Eq.~\eqref{eq:mueconv} and $B(A.Z) = 1.8$ for Al, the sensitivity is transformed to $4.8 \times 10^{-17}$  in term of $\mu \to e \gamma $ decay.
In the same way in \eqref{eq:LFVconst}, the constraint for new physics mass is estimated as 
\beq
 \Lambda _{\text{NP}} \gtrsim 6000 \TeV \times \sqrt{\epsilon _{\mu e}}. \label{eq:LFVconstpros}
\eeq
It is found that sensitivity to new physics scale is improved by order of magnitude in the long-term future.
Similarly, the sensitivity to size of $\mu $--$e$ flavor violation $\epsilon _{\mu e}$ is also estimated as $\epsilon _{\mu e} \lesssim 3 \times 10^{-10}$ in case that 
$ \Lambda _{\text{NP}} = 100\GeV$.

In Sec.~\ref{subsec:non-univ}, we will discuss the supersymmetric models with non-universal slepton mass spectrum, 
which generically induces large lepton FCNCs.
Then, it is found that such models are tightly constrained by lepton flavor experiments.

\section{Electric Dipole Moment}
\label{sec:EDMexp}
%!TEX root = ../Dthesis.tex

There is no experimental evidence for the electric dipole moment (EDM),  
despite nearly a half-century of search.
However, laboratories have attempted to find the EDM by new approaches, 
the searches become more important than ever before. 
In this section, we summarize current status of the EDM experiment.
Particularly, we focus on the electron EDM, 
because it provides a sensitive probe for new physics models where we will consider in this dissertation. 

\subsection{Motivation}
\label{subsec:motivationEDM}

Unless both parity (P) and time reversal (T) invariance are violated, no EDM can exist. 
The interaction Hamiltonian between the lepton $\ell $ and the electric field $ \overrightarrow{E}$ is defined as\cite{Roberts:2010zz}
\beq
 H_{{\rm int} } \supset - d_{\ell } \overrightarrow{E} \cdot \frac{\overrightarrow{S}}{S},  \label{eq:hamiltonianEDM}
\eeq
where $\overrightarrow{S}$ is spin of the lepton.
Under a parity transformation, the axial vector $\overrightarrow{S}$ remains invariant, 
but the polar vector $\overrightarrow{E}$ changes sign.
On the other hand, under a time reversal transformation, $\overrightarrow{E}$ is invariant but $\overrightarrow{S}$ changes sign.
In this way, $H_{{\rm int} }$ is not invariant under P and T transformations. 
Further, assuming CPT invariance, T violation is equivalent to CP violation.
Since we know that CP invariance is violated in the decay of neutral Kaon and B mesons, 
the existence of EDM appears quite possible.

However, EDM is highly suppressed in the standard model.
Since the SM has only one CP phase in CKM matrix, 
EDM of the lepton is induced by the CKM phase in the quark sector via a diagram with a closed quark loop.
Then, a non-vanishing effect appears first at the four-loop level\cite{Pospelov:1991zt}.
Therefore, the lepton EDM in the SM is extremely suppressed.
In electron case, it is estimated as
\beq
 d^{\text{CKM}}_e \lesssim 10^{-38} e ~\text{cm}. \label{eq:EDMSM}
\eeq
This is far from the current experimental sensitivity, $\sim 10^{-27\sim 29} e ~\text{cm}$.

New physics models generically contain complex parameters, which induce CP violating phase.
The new particles which have complex parameters can contribute to EDM at two-loop or one-loop levels.
It is difficult to justify a suppression mechanism of these CP phases.
Hence, EDM tends to be enhanced by new physics effects, 
the observation of a non-vanishing EDM would be heralded as a discovery of new physics.

Let us estimate the new physics contribution.
We assume that a typical mass scale of new physic is $\Lambda _{\text{NP}}$ 
and $\phi _{\text{NP}}$ parameterizes the size of CP violating phase.
Just like the muon FCNCs, the typical new physics contribution to the lepton EDM can be naively given by\cite{Roberts:2010zz}
\beq
 d_{\ell } \simeq \frac{m^2_{\ell }}{\Lambda ^2_{\text{NP}}} \frac{e}{2m_{\ell }} \tan \phi _{\text{NP}}, \label{eq:estimateEDM}
\eeq
where the Wilson coefficient of effective Lagrangian is assumed to be $\mathcal{O}(1)$.
Then, using current bound of the electron EDM $d_e < 8.7 \times 10^{-29} e \text{ cm}$ at 90\% CL\cite{Baron:2013eja}, 
we can estimated as 
\beq
 \Lambda _{\text{NP}} \gtrsim 240\TeV \times \sqrt{\tan \phi _{\text{NP}}} .\label{eq:estimateEDM2}
\eeq
If new particles with $\mathcal{O}(100)\GeV$ masses exist, and they could contribute to both the muon $g-2$ and the electron EDM, 
the size of CP phase is constrained as $\phi _{\text{NP}} \lesssim 1.7 \times 10^{-7}$ for $\Lambda _{\text{NP}} = 100\GeV$.
In this way, the electron EDM also provides a sensitive probe for new physics.

\subsection{Current Status and Future Sensitivity }
\label{subsec:futureEDM}

%%%%%%%%%%%%%%%%%%%%%%%%%%%%%%%%%% Table
\begin{table}[t]
\begin{center}
  \begin{tabular}{|c|c|c|} \hline
    Material & Current bound (90\% CL, unit of $e$ cm) & Future sensitivity \\ \hline \hline
    YbF  & $1.05 \times 10^{-27}$ \cite{Hudson:2011zz} & $10^{-30}$\cite{Kara:2012ay}  \\ 
    Fr &  & $10^{-29}$ \cite{Sakemi:2011zz}\\
    ThO & $8.7 \times 10^{-29}$\cite{Baron:2013eja} & 
    $1\times 10^{-28}/ \sqrt{(\text{day})}$ \cite{Vutha:2009ux, Campbell:2013ota} \\
    WN  & & $10^{-30}$ / day \cite{Kawall:2011zz} \\ \hline \hline
    \end{tabular}
   \caption[Summary of the electron EDM.]
   {Summary of the electron EDM in Sec. \ref{subsec:futureEDM}.
   The left column shows atoms or molecules which are used to measure the electron EDM.
   The most severe bound is obtained by ThO molecule\cite{Baron:2013eja}. 
   The right column is future sensitivities, where "day" represents the running time in days.
   If "day" is $\mathcal{O}(10)$, the future sensitivity will become about $10^{-31}e$ cm.}
   \label{tab:sum_EDM}
\end{center}
\end{table}
%%%%%%%%%%%%%%%%%%%%%%%%%%%%%%%%%%

We cannot directly measure only the contribution of the electron EDM, $d_e$,  
but search for a paramagnetic atom or molecule EDM and then interpret the results in term of the unpaired electron EDM.
Naively, this approaches seem to be impossible due to Schiff"s theorem\cite{Schiff:1963zz}.\footnote
{
This is the theorem that the atom or molecule cannot exhibit a linear Stark effect to first order in non-relativistic limit.
For details, see Ref.~\cite{Schiff:1963zz} or textbook\cite{Roberts:2010zz}.} 
However, this theorem fails when relativistic effects are taken in account as shown by Sanders\cite{Sandars1965194, Sandars1966290}.
Further, for paramagnetic atoms or molecules, since the effective electric field is proportional to $Z^{2-3}$, 
where $Z$ is atomic number, the EDMs are enhanced.
Hence, a paramagnetic atom or molecule EDMs with large $Z$ are expected to provide a more sensitive probe for new physics effects.

The bound for electron EDM was obtained by the ytterbium fluoride (YbF) molecules, $d_e < 1.05 \times 10^{-27} e \text{ cm}$ at 90\% CL\cite{Hudson:2011zz} until quite recently.
The new bound is reported by ACME collaboration, and 
is obtained by the polar molecule thorium monoxide (ThO), $d_e < 8.7 \times 10^{-29} e \text{ cm}$ at 90\% CL\cite{Baron:2013eja}.

The sensitivity is expected to be improved by future experiments.
For atoms, the future sensitivity with the Fr atom will reach to $d_e = 10^{-29} e \text{ cm}$\cite{Sakemi:2011zz}.
For molecules, the YbF molecule and the WN ion are promising.
The sensitivity with the YbF is $d_e = 10^{-30} e \text{ cm}$\cite{Kara:2012ay}.
The experiment with the WN ion can probe down to $d_e = 10^{-30} e \text{ cm}$,
where the systematic limit is at the level of $10^{-31} e \text{ cm}$\cite{Kawall:2011zz}.
These results are summarized in Tab.~\ref{tab:sum_EDM}.
 
Finally, let us estimate the constraint for mass scale of new physics again, using the expected sensitivity by WN ion.
In the same way in \eqref{eq:estimateEDM2}, the constraint for new physics mass is estimated as 
\beq
  \Lambda _{\text{NP}} \gtrsim 7000\TeV \times \sqrt{\tan \phi _{\text{NP}}}. \label{eq:estimateEDM3}
\eeq
Therefore, it is found that sensitivity to new physics scale is improved by order of magnitude in the future, 
and is comparable to that of muon FCNCs.
Similarly, the sensitivity to CP phase $\phi _{\text{NP}}$ is also estimated as 
$\phi _{\text{NP}} \lesssim 2 \times 10^{-10}$ for $\Lambda _{\text{NP}} = 100\GeV$.

\section{Higgs Coupling}
\label{sec:kappaexp}
%!TEX root = ../Dthesis.tex

In 2012, a new scalar boson was discovered at the LHC experiment.
Its observed properties are consistent with the SM Higgs boson within current experimental uncertainty.
In addition, so far, no extra particles has ever been discovered yet.
Therefore, it has been found that the SM provides a good description for the elementary particle physics at electroweak scale,
in both the gauge interaction and the electroweak symmetry breaking sector.
 
It is important to investigate Higgs couplings to SM particles.
If new physics exist near the electroweak scale, they affect the couplings indirectly.  
Particularly, loop-induced Higgs couplings, i.e., the Higgs boson coupling to di-photon, di-gluon or $Z\gamma $ strongly constrain the new physics,
and they are called the Higgs oblique corrections\cite{Gori:2013mia}, named after the electroweak oblique corrections.
In the SM, these couplings are prevented by the gauge symmetry at the tree level and are induced by quantum corrections.
Therefore, new physics contributions might be probed indirectly by measuring the loop-induced Higgs coupling in future.

In particular, the Higgs coupling to di-photon, which corresponds to the electromagnetic interactions, is important.
In the SM, it is dominated by one-loop contributions of the electroweak gauge bosons and the top quark.
If new physics contain charged particles which couple to the Higgs boson, 
they contribute to  the Higgs coupling to di-photon at radiative levels.
Hence, it is sensitive to the new physics contributions with charged particles.
If such new particles exist, it is expected that deviations from the SM prediction are observed.

In this dissertation, we parameterize the Higgs coupling to $\gamma \gamma $ particles and its deviation from the SM prediction as
\beq
 \kappa _{\gamma } = \frac{g_{h\gamma \gamma }}{g_{h\gamma \gamma}(\text{SM})} = 1 + \delta \kappa _{\gamma }, \label{eq:defkappag}
\eeq
where $g_{h \gamma \gamma }$ is the Higgs coupling to $\gamma \gamma $ particles and new physics contributions are represented by $\delta \kappa _{\gamma }$.
The explicit formula for  $\kappa _{\gamma }$ will be given in Sec.~\ref{sec:HiggsCmssm}.

The Higgs coupling to di-photon $\kappa _{\gamma }$ has been measured at LHC experiment.
The measured uncertainty is 15\% ($1\sigma$) at ATLAS\cite{Aad:2013wqa} and 25\% at CMS\cite{CMS:yva}, respectively.
The results are consistent with the SM prediction, though they are not yet precise enough to probe new physics contributions.
In future, the LHC accumulate the luminosity $\mathcal{L} \sim 300\invfb$ at $\sqrt{s} = 14\TeV$, 
and further upgrade is proposed for $\mathcal{L} \sim 3000\invfb$ at High-Luminosity LHC (HL-LHC).
The accuracies are expected to be about 7\% (5\%) at  300$\invfb$ (3000$\invfb$).\footnote{
They are dominated by systematic uncertainties.The accuracies could be improved by reducing them.}

%%%%%%%%%%%%%%%%%%%%%%%%%%%%%%%%%% Figure
\begin{figure}[t]
 \begin{center}
 \includegraphics[width = 10cm]{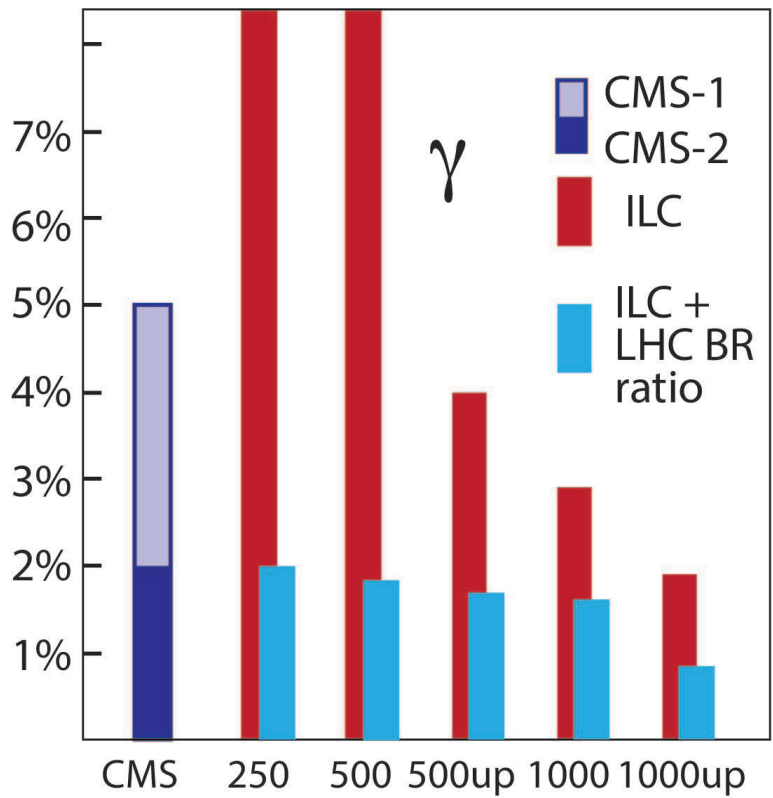} 
  \end{center}
 \caption[Joint analysis of $\kappa_{\gamma }$ at LHC and ILC.]
 {The accuracy of the Higgs coupling to di-photon $\kappa _{\gamma }$.
 The horizontal axis is luminosity of ILC and the vertical axis is the accuracy of $\kappa _{\gamma }$. 
 CMS-1 (2) are the future prospects of $\kappa _{\gamma }$ at CMS for $\mathcal{L} = 3000\invfb$. 
 CMS has presented a less conservative set of estimates, using two scenarios. 
 ILC (red) is the accuracy of $\kappa _{\gamma }$ at ILC 
 and ILC + LHC BR ratio (aqua blue) is  $\kappa _{\gamma }$ 
 combined with the measurement of $B(h \to \gamma \gamma )/ B(h \to ZZ^{*} )$ by ATLAS.
 Using joint analysis of HL-LHC and ILC, the accuracy of $\kappa _{\gamma }$ will be improved well.
 This figure is from\cite{Peskin:2013xra}.}
 \label{fig:jointana}
\end{figure}
%%%%%%%%%%%%%%%%%%%%%%%%%%%%%%%%%% 

Recently, it is argued that the sensitivity to the Higgs coupling to di-photon can be improved well, 
once the international electron-positron linear collider (ILC) will be constructed\cite{Peskin:2013xra}.
At LHC, the ratio $B(h\to \gamma \gamma ) / B(h\to ZZ^*)$ will be precisely measured, 
where the $Z^*$ means $Z$ boson at off-shell.  
On the other hand, at ILC, the Higgs couplings to $Z$ boson and di-photon can be measured at (sub) percent levels\cite{Asner:2013psa}. 
The joint analysis of HL-LHC and ILC enable us to realize the accuracy of $\kappa _{\gamma }$ of about 2\%\cite{Peskin:2013xra}.
Here, it is assumed that the uncertainty of $B(h\to \gamma \gamma ) / B(h\to ZZ^*)$ is 3.6\% from HL-LHC and ILC runs 
at $\sqrt{s} = 250\GeV$ and $\mathcal{L} = 250\invfb$.
The direct measurement of $\kappa _{\gamma }$ at ILC is not so precise that of LHC, due to lack of  luminosity.
If ILC accumulate more luminosity, e.g., $\mathcal{L} = 2500\invfb$ at $\sqrt{s} = 1\TeV$, 
the accuracy of $\kappa _{\gamma }$ can become 1.9\% \cite{Peskin:2013xra, Asner:2013psa} at ILC.
the joint analysis can be better than 1\%\cite{Peskin:2013xra}.
They are so precise that the new charged particles could be sufficiently probed by measuring $\kappa _{\gamma }$.
These discussions are summarized in Table.~\ref{fig:jointana}.

In Sec.~\ref{sec:stau}, we consider a situation that an excess of $\kappa _{\gamma }$ is measured in HL-LHC and ILC,
and consider new physics models that explain the anomalous excess.
Although many models have been proposed, we focus on the supersymmetry (SUSY), which includes a scalar partner of the tau lepton (stau).
Then we will study the properties of staus that are responsible for the $\kappa _{\gamma }$ excess,
and discuss whether the stau contribution to $\kappa _{\gamma }$ can be probed at ILC.

%%%%%%%%%%%%%%%%%%%%%%%%%%%%%%%%%%%%%%%%%%%%%%%%%%%%%%%%%%%%%

\chapter{Supersymmetric Standard Model}
\label{chp:susy}
%!TEX root = ../Dthesis.tex

The chapter is a review of the supersymmetry.
The supersymmetric extension of the standard model provides a beautiful solution to the hierarchy problem of the standard model. 
In Sec.~\ref{sec:motivation}, we introduce the supersymmetry motivated by the hierarchy problem.
The minimal extension model is called the minimal supersymmetric standard model (MSSM).
In Sec.~\ref{sec:mssm}, we provide MSSM Lagrangian.
The supersymmetric model includes superpartners of standard model particles associated with supersymmetry, 
which is called superparticles.
If superparticles have masses of $\mathcal{O}(100)\GeV$, 
they are not only discovered by collider experiments, but also affect low energy measurements through radiative corrections. 
Particularly, searches for superparticles without color charge are promising in near future experiments.
In Sec.~\ref{sec:gm2mssm}--\ref{sec:HiggsCmssm}, 
we summarize measurements which non-colored superparticles could provide large contributions. 
Sec.~\ref{sec:cstatus} is a summary on current status of supersymmetry.
The prospects for searches of sleptons will be discuss in Chapter.~\ref{chp:main}.

\section{Motivation}
\label{sec:motivation}
%!TEX root = ../Dthesis.tex

The Standard Model (SM) provides a very successful description of phenomena of the elementary particles at the electroweak scale $m_W$.
Although no additional structures have ever discovered yet, 
physicists do not consider the SM as the ultimate particle theory, but a low energy effective theory of a fundamental theory, 
whose typical energy scale is characterized by $M_{\text{UV}}$.
For example, the reduced Planck scale $M_P = (8 \pi G _{\text{Newton}}) = 2.4 \times 10^{18}\GeV$, 
which characterizes a quantum gravity, is one of important fundamental scales.
Once we take above standpoint, the Higgs sector includes an "unnaturalness" in the SM.

This unnaturalness comes from the squared Higgs mass parameter $m^2_H$.
First, we introduce the SM Higgs field $H$ with a potential 
\beq
 V = m^2_H \left| H \right| ^2 + \lambda  \left| H \right| ^4, \label{eq:potentialSM}
\eeq
where $\lambda $ is the quartic Higgs self-coupling.  
The Higgs field has a vacuum expectation value (VEV), $\langle H \rangle = \sqrt{-m^2_H / 2 \lambda}$ at the minimum of the potential,
if $\lambda > 0$ and $m^2_H < 0$.
Since the VEV is known as about 174\GeV,  the mass parameter $m^2_H$ must be order of $-(100\GeV)^2$.

%%%%%%%%%%%%%%%%%%%%%%%%%%%%%%%%%%%%%%%%%%% Diagram
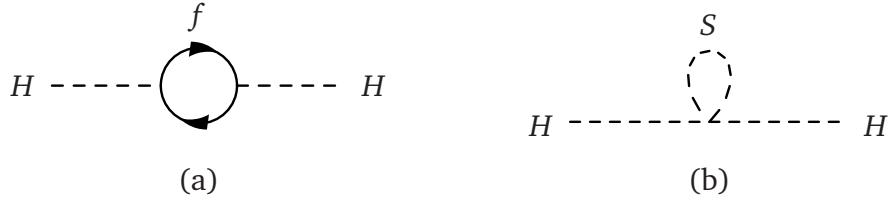
\begin{figure}[t]
\begin{center}
\begin{fmffile}{feyns/Higgscorr}
 \begin{minipage}[t]{0.4\textwidth}
 \begin{center}
 \begin{fmfgraph*}(110,50)
 \fmfleft{d1,h1,d2}\fmfright{d3,h2,d4}
 \fmf{phantom}{d1,d3}\fmf{phantom}{d2,d4}
 \fmf{dashes,tension=0.7}{h1,v1}
 \fmf{fermion,left,tension=0.5,label=$f$}{v1,v2}
 \fmf{fermion,left,tension=0.5}{v2,v1}
 \fmf{dashes,tension=0.7}{v2,h2}
 \fmflabel{$H$}{h1}
 \fmflabel{$H$}{h2}
 \fmfv{label= ,l.a=-120}{v1}
 \fmfv{label= ,l.a=-60}{v2}
% \fmfdot{v1,v2}
 \end{fmfgraph*}
 \\(a)
 \end{center}
 \end{minipage}
 \begin{minipage}[t]{0.4\textwidth}
 \begin{center}
 \begin{fmfgraph*}(110,50)
 \fmfleft{d1,h1,d2,d3,d4}
 \fmfright{d6,h2,d7,d8,d9}
 \fmf{phantom}{d1,d6}\fmf{phantom}{d4,d9}
 \fmf{dashes,tension=0.7}{h1,v}
 \fmf{dashes,right,tension=1.3,label=$S$}{v,v}
 \fmf{dashes,tension=0.7}{v,h2}
 \fmflabel{$H$}{h1}
 \fmflabel{$H$}{h2}
 \fmfv{label= }{v}
% \fmfdot{v}
 \end{fmfgraph*}
 \\(b)
 \end{center}
 \end{minipage}
\end{fmffile}
\caption[The Feynman diagrams corresponding to the quantum correction of Higgs mass parameter.]
{
The radiative correction to the Higgs squared mass parameter $m^2_H$, 
where (a) is from a Dirac fermion $f$, and (b) is from a scalar $S$.
}
\label{fig:higgs1loopcorrection}
\end{center}
\end{figure}

%%%%%%%%%%%%%%%%%%%%%%%%%%%%%%%%%%%%%%%%%%%%%%%%%%%%%

The squared Higgs mass parameter receives quantum corrections from diagrams in Fig.~\ref{fig:higgs1loopcorrection}.
When the Higgs field couples to a Dirac fermion $f$, the correction as Fig.~\ref{fig:higgs1loopcorrection} (a) become 
\beq
 \Delta m^2_H = -\frac{\left| \lambda _f \right| ^2  }{8\pi ^2} M^2_{\text{UV}} + \mathcal{O}(\log( M^2_{\text{UV}} ) ).
 \label{eq:hierarf}
\eeq
On the other hand, 
\beq
 \Delta m^2_H = \frac{\lambda _s }{16\pi ^2} M^2_{\text{UV}} + \mathcal{O}(\log( M^2_{\text{UV}} ) ), 
  \label{eq:hierars}
\eeq
is from a scalar $S$ in Fig.~\ref{fig:higgs1loopcorrection} (b).

In the SM, top quark has strong coupling with Higgs field, $\lambda _f \sim 1$.
However, scalar fields other than Higgs does not exist.
Hense, we need a fine tuned cancellation between the bare mass and the corrections which are proportional to $M^2_{\text{UV}}$ as
\beq
 m^2_H(\text{physical}) = m^2_H(\text{bare}) + \Delta m^2_H, \label{eq:physhiggs}
\eeq
to realize $m^2_H = -(\mathcal{O}(100)\GeV )^2$.
If $M_{\text{UV}} = M_P$, size of the quantum correction is some 30 orders of magnitude and is extremely larger than $m_W$.
This unnaturalness is known as the hierarchy problem\cite{Weinberg:1975gm, Weinberg:1979bn, Gildener:1976ai, Susskind:1978ms,tHooft:1980ps}, 
and is one of motivations introducing models beyond the SM.

Supersymmetry (SUSY) \cite{Haag:1974qh}, a symmetry between fermions and bosons, provides a beautiful solution for the problem, 
and introduces two scalar boson for one Dirac fermion, and vice versa.
Further, SUSY guarantees that couplings of scalar partners with the Higgs field are same as those of Dirac fermions, 
$\lambda _s = \left| \lambda _f\right| ^2$.
Thanks to new contributions of superparticles, the quadratic divergence of the Higgs mass can cancel as 
\beq
 \Delta m^2_H = -\frac{\left| \lambda _f \right| ^2  }{8\pi ^2} M^2_{\text{UV}} + 2 \times \frac{\lambda _s }{16\pi ^2} M^2_{\text{UV}} 
 +  \mathcal{O}(\log( M^2_{\text{UV}} ) ) \to \mathcal{O}(\log( M^2_{\text{UV}} ) ). \label{eq:hierarsusy}
\eeq
In this case, the tuning is not only so fine, but also we need not care quantum corrections at any energy scales.   
In this way, the supersymmetric extended models of the SM can solve the problem of the SM. 
Hereafter, we adopt SUSY models as candidates of new physics.

\section{The MSSM}
\label{sec:mssm}
%!TEX root = ../Dthesis.tex

%%%%%%%%%%%%%%%%%%%%%%%%%%%%%%%%%%%%%%%%%%%%%%%% Table
\begin{table}[htb]
\begin{center}
{\renewcommand\arraystretch{1.5}
  \begin{tabular}{|c|ccc|cc|cc|} \hline
  \multicolumn{8}{|c|}{Matter and Higgs fields} \\ \hline
    & SU(3)$_c$ & SU(2)$_L$ & U(1)$_Y$ & $B$ & $L$ & spin 0 &  spin 1/2 \\ \hline
    $Q_i$ & \bf{3} & \bf{2} & 1/6 & 1/3 & 0 & $\left( \widetilde{u}_L, \widetilde{d}_L \right)$ &  $\left( u_L, d_L \right)$ \\ \hline
    $\overline{U}_i$ & $\overline{ \bf{3}}$ & \bf 1 & -2/3 & -1/3 & 0 & $\widetilde{u}^*_R$ &  $u^{\dagger }_R$ \\ \hline
    $\overline{D}_i$ & $\overline{ \bf{3}}$ & \bf 1 & 1/3 & -1/3 & 0 & $\widetilde{d}^*_R$ &  $d^{\dagger }_R$ \\ \hline
    $L_i$ & \bf 1 & \bf 2 & 1/2 & 0 & 1 & $\left( \widetilde{\nu }_{\ell }, \widetilde{\ell }_L \right)$ &  $\left( \nu _L, {\ell }_L \right)$ \\ \hline
    $\overline{E}_i$ & \bf 1 & \bf 1 & 1 & 0 & -1 & $\widetilde{\ell }^*_R $ &  $\ell ^{\dagger }_R$ \\ \hline 
    $ H_u$ & \bf 1  & \bf 2 & 1/2 & 0 & 0 & $\left( H^+_u, H^0_u \right)$ & $\left( \widetilde{H}^+_u, \widetilde{H}^0_u \right)$ \\ \hline
    $ H_d$ & \bf 1  & \bf 2 & -1/2 & 0 & 0 & $\left( H^0_d, H^-_d \right)$ & $\left( \widetilde{H}^0_d, \widetilde{H}^-_d \right)$ \\ \hline
    \multicolumn{8}{|c|}{Gauge fields} \\ \hline
        & SU(3)$_c$ & SU(2)$_L$ & U(1)$_Y$  & $B$ & $L$ & spin 1/2 &  spin 1 \\ \hline
         $G$ & \bf 8 & \bf 1 & 0 & 0 & 0 & $\widetilde{g}$ &  $g$ \\ \hline
          $W$ & \bf 1 & \bf 3 & 0 & 0 & 0 & $\widetilde{W}$ &  $W$ \\ \hline
           $B$ & \bf 1 & \bf 1 & 0 & 0 & 0 & $\widetilde{B}$ &  $B$ \\ \hline
  \end{tabular}}
   \caption[Particle contents of the MSSM.]{Particle contents of the MSSM.}
   \label{tab:mssm}
   \end{center}
\end{table}
%%%%%%%%%%%%%%%%%%%%%%%%%%%%%%%%%%%%%%%%%%%%%%%%

The minimal supersymmetric standard model (MSSM) is the minimal supersymmetric extension of the standard model.
The gauge symmetries of the MSSM are SU(3)$_C \times $SU(2)$_L \times$U(1)$_Y$, just like the SM. 
The MSSM provides rich phenomenology since it includes many new particles coupled with SM particles and new parameters.
In this section, we introduce the MSSM superpotential and the soft supersymmetry-breaking terms.

The superpotential for the MSSM is given by\cite{Martin:1997ns}
\beq
W= \mu H_u H_d
      - \left(Y_u\right)_{ij}  H_u Q_i \overline{U}_j
      + \left(Y_d\right)_{ij} H_d Q_i \overline{D}_j
      + \left(Y_e\right)_{ij} H_d L_i \overline{E}_j,
\label{eq:superpotential}
\eeq
where $H_u$, $H_d$, $Q$, $L$, $U$, $D$ are chiral superfields in Tab.~\ref{tab:mssm}.
Unlike the SM, we need two Higgs fields $H_u$ and $H_d$ 
because of holomorphy of superpotential and anomaly cancellation condition.
Parameters $Y_{u, d, e}$ are the Yukawa coupling matrices whose indices correspond to the generation of the fermions.
All of the gauge (color and weak isospin) indices in Eq. \eqref{eq:superpotential} are suppressed.
The first term of right hand side, $\mu (H_u) _{\alpha } (H_d) _{\beta} \epsilon ^{\alpha \beta }$ is called $\mu $ term, 
where $\epsilon ^{\alpha \beta }$ is the invariant tensor for SU(2)$_L$ and indices $\alpha , \beta =1,2$.
Similarly, the term $\left(Y_u\right)_{ij}  H_u Q_i \overline{U}_j$ is written out as 
$\left(Y_u\right)_{ij} H_{u\alpha } Q_{i\beta a } \overline{U}^{a}_j \epsilon ^{\alpha \beta }$, 
where $a = 1,2,3$ is a color index which is lowered (raised) in the $\bf 3$ $( \bf \bar{3})$ representation of SU(3)$_C$.

The superpotentilal (\ref{eq:superpotential}) is minimal in the sense that it is sufficient to produce a phenomenologically viable model.
The most general gauge-invariant and renormalizable superpotential includes not only  Eq. (\ref{eq:superpotential}) 
but also the terms\cite{Martin:1997ns}
\beq
 W_{\Delta L = 1}  &=&  \frac{1}{2} \lambda ^{ijk} L_i L_j \overline{E}_k + \lambda ^{\prime ijk} L_iQ_j \overline{D}_k + \mu ^{\prime } L_i  H_u, 
 \label{eq:deltaL1} \\
 W_{\Delta B = 1}  &=&  \frac{1}{2} \lambda ^{\prime \prime ijk} U_i \overline{D}_j \overline{D}_k,
 \label{eq:deltaB1}
\eeq
where $i = 1,2,3$ correspond to generation of the fermion, and $B, L$ are baryon and total lepton numbers, respectively.
The chiral superfield $Q_i$ has $B = + 1/3$, $\overline{U}_i, \overline{D}_i$ have $B = -1/3$ and $B = 0$ for all others.
The lepton number assignments are $L = +1$ for $L_i$, $L = -1$ for $\overline{E}_i$ and $L=0$ for all others.
Hense, the terms in Eq. \eqref{eq:deltaL1} violate lepton number by 1 unit 
and those in Eq. \eqref{eq:deltaB1} violate baryon number by 1 unit.

%%%%%%%%%%%%%%%%%%%%%%%%%%%%%%%%%% Figure
\begin{figure}[t]
 \begin{center}
 \includegraphics[width = 12cm]{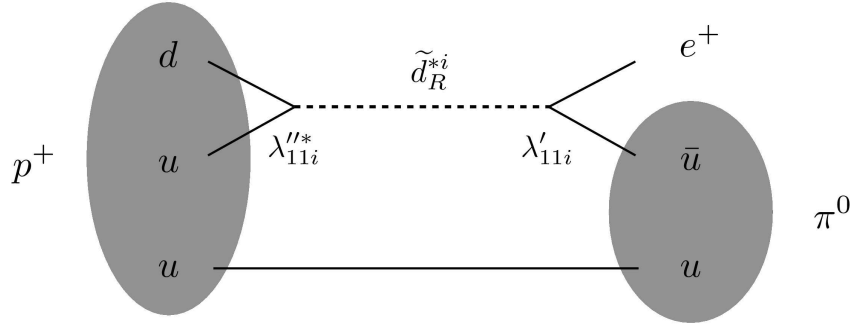} 
  \end{center}
 \caption[Proton Decay]{Feynman diagram for proton decay occurred by the interactions $U\overline{D}\overline{D}$ and $LQ\overline{D}$.
 These interactions cause dangerous contributions to proton decay. To forbid such a interaction, we have to impose the R parity.}
 \label{fig:protond}
\end{figure}
%%%%%%%%%%%%%%%%%%%%%%%%%%%%%%%%%% 

The $B$- and $L$- violating processes have not been discovered yet.
The most severe constraint comes from no observation of proton decay.
If the couplings $\lambda ^{\prime }, \lambda ^{\prime \prime }$ were not suppressed, 
the lifetime of proton would becomes extremely short.
For example, Fig.~\ref{fig:protond} is the Feynman diagram with squark exchange for $p^{+} \to e ^{+} \pi ^{0}$ decay.
The decay rate is roughly estimated as
\beq
 \Gamma _{p \to e^+ \pi ^0} &\sim& m^5_{\rm{proton}} \sum _{i = 2, 3}
  \left| \lambda ^{\prime 11i } \lambda ^{\prime \prime 11i }  \right| ^2  / m^4_{\widetilde{d}_i} \notag \\
   &=& \frac{\sum _{i = 2, 3} \left| \lambda ^{\prime 11i } \lambda ^{\prime \prime 11i }  \right| ^2}{2.9 \times 10 ^{-20}~\text{yr}} 
   \left( \frac{1 \TeV}{m_{\widetilde{d}_i}} \right) ^4 \notag \\
   &<& \left( 8.2 \times 10^{33}~\text{yr}\right) ^{-1}.
   \label{eq:protondecay}
\eeq
The proton decay is strongly constrained for the Super-Kamiokande experiment \cite{Nishino:2009aa}.
When $\Gamma _{p^+ \to e^+ \pi ^0}$ is compared with the experimental constraint, 
one finds $\left| \lambda ^{\prime 11i } \lambda ^{\prime \prime 11i }  \right| \lesssim 10^{-27}$.

To solve above unnaturalness, we usually impose conservation of the $R$-parity\cite{Farrar:1978xj}, 
which is a discrete $Z_2$ symmetry defined as 
\beq
  P_R = (-1)^{3(B-L) + 2s}, 
  \label{eq:rparity}
\eeq
where $s$ is the spin of the particle.
If we impose the $R$-parity conservation in the MSSM, 
the terms in Eqs. (\ref{eq:deltaL1}), (\ref{eq:deltaB1}) are forbidden.

The $R$-parity takes values of $P_R = \pm 1$.
The $R$-parity odd particles are called "superparticles".
If $R$-parity is exactly conserved, there are no mixing between superparticles ($P_R = -1$) and SM particles ($P_R = +1$).
Further, every interaction in the MSSM contains the even number of superparticles.
These have three important phenomenological consequences\cite{Martin:1997ns}:
\begin{itemize}
\item The lightest particles with $P_R = -1$ is called "lightest supersymmetric particle (LSP)", and must be stable.
If LSP is electrically neutral and interacts weakly with ordinary particles, it can be candidate for non-baryonic dark matter
that seems to be required by cosmological consequences.
\item Each superparticle other than LSP decays into states with the odd number of LSPs.
\item Suparparticles can be produced in even numbers at collider experiments.
\end{itemize}
In this dissertation, we concentrate on the MSSM with $R$-parity.

We can construct the full supersymmetric Lagrangian using kinetic terms and the superpotential given in \eqref{eq:superpotential}.
However, it is not sufficient to explain phenomena of the Nature.
The most important constraint is that none of superpartners of the SM particles has been discovered yet.
For example, if SUSY is unbroken, the selections $\widetilde{e}_{L,R}$ must have same mass of the electron, $m_e \simeq 0.511\MeV$.
A similar discussions can apply to each of the other sleptons, squarks, and gauginos, respectively.
Such superparticles should be already detected if they exist.
Hense, SUSY must be a broken symmetry in our vacuum.

A hint to describe the nature of SUSY breaking, is the hierarchy problem as introduced in Sec.~\ref{sec:motivation}.
The hierarchy problem is solved in supersymmetric models 
because SUSY guarantees cancellation of the quadratic divergence to the Higgs mass.
If broken SUSY theories still provide a solution to the hierarchy problem, 
the relationship between dimensionless couplings that hold in an unbroken SUSY theory must be maintained.
Therefore, we are led to consider "soft" supersymmetry breaking.
This means the effective Lagrangian of the MSSM is written in the form\cite{Martin:1997ns}
\beq
 \mathcal{L} _{\text{MSSM}} = \mathcal{L} _{\text{SUSY}} + \mathcal{L} _{\text{soft}} , \label{eq:MSSMlag}
\eeq 
where $\mathcal{L} _{\text{SUSY}}$ is the SUSY invariant Lagrangian and 
$\mathcal{L} _{\text{soft}}$ contains the SUSY breaking interactions with coupling parameter whose mass dimension is "positive".
In this case, the cancellation of the quadratic divergence is maintained.

Then, we provide the soft SUSY breaking terms.
They are composed of the mass term of the gauginos, the sfermions and the Higgs bosons, and the trilinear term of the scalar bosons as follows\cite{Martin:1997ns}\footnote{
In general SUSY models, $\mathcal{L}_{{\rm soft}}$ includes the tadpole terms, such as $t^i \phi _i$.
They do not occur because there is no singlet field in the MSSM.}
\beq
 \mathcal{L}_{{\rm soft}} &=&  
 - \frac{1}{2} \left( M_1 \widetilde{B}\widetilde{B} + M_2 \widetilde{W}\widetilde{W} + M_3 \widetilde{g}\widetilde{g} + \rm{h.c.} \right) \notag \\
  &&+ \left[ (a_u)_{ij} H_u \widetilde{Q}_i \widetilde{\overline{u}}_j -  (a_d)_{ij} H_d\widetilde{Q}_i \widetilde{\overline{d}}_j  
   - (a_e)_{ij} H_d\widetilde{L}_i \widetilde{\overline{e}}_j + \rm{h.c.} \right] \notag \\
  && - \left[
      \left(m^2_{\widetilde{Q}}\right)_{ij}  \widetilde {Q}^{\dagger}_i \widetilde Q_j 
      + \left(m^2_{\widetilde{L}} \right)_{ij}  \widetilde {L}^{\dagger}_i \widetilde L_j 
      + \left(m^2_{\widetilde{U}} \right)_{ij}  \widetilde {\overline{u}}^{\dagger}_i \widetilde{\overline{u}}_j 
      + \left(m^2_{\widetilde{D}} \right)_{ij}  \widetilde {\overline{d}}^{\dagger}_i \widetilde{\overline{d}}_j
      + \left(m^2_{\widetilde{E}} \right)_{ij}  \widetilde {\overline{e}}^{\dagger}_i \widetilde{\overline{e}}_j  \right] \notag \\
   && - \left[ m^2_{H_u} H_u^* H_u +  m^2_{H_d} H_d^* H_d + (b H_u H_d + \rm{h.c.}) \right],
\label{eq:softbreaking}
\eeq
where $M_a$, $m^2_{\widetilde{f}}~(f = Q, L, U, D, E)$, and $m^2_{H_{u,d}}$ are the masses of the gauginos and the scalar bosons (Higgs bosons and sfermions), 
$a_{u,d,e}$ are trilinear scalar couplings called A-term, and $b$ is quadratic coupling of the Higgs boson called-B term.
Note that $H_{u,d}$ is not the chiral superfields but the scalar fields. 
Each of $a_{u, d, e}$ is a complex $3 \times 3$ matrix. 
They are in one-to-one correspondence with the Yukawa couplings of the superpotential.
Each of $m^2_{\widetilde{f}}$ is a Hermitian $3 \times 3$ matrix.
In this dissertation, we parametrize the A term and B term as
\beq
  \left( A_{u,d,e} \right) _{ij} \equiv \frac{\left( a_{u,d,e} \right) _{ij}}{\left(Y_{u,d,e}\right)_{ij}}. ~~~~~~ B \equiv \frac{b}{\mu }.  \label{eq:abterm}
\eeq

The soft SUSY breaking terms $\mathcal{L}_{\text{soft}}$ introduce many new parameters, which do not exist in the SM.
There are 105 parameters of masses, phases, and mixings in the MSSM \cite{Dimopoulos:1995ju} 
that cannot rotate away by redefining the phases and the bases for the fermion supermultiplets.
These parameters introduce new flavor/CP violations, which might easily contradict with current experiments.
These problems are called the SUSY flavor/CP problems. 
Thus, in principle, SUSY breaking induces a tremendous arbitrariness in the Lagrangian.

The size of the flavor/CP violations depends on the origin of the SUSY breaking.
In this dissertation, we do not enter the detailed models of SUSY breaking, 
but consider the MSSM with the soft SUSY breaking terms as  a effective theory at TeV scale.
Then, we investigate the constraints of low energy measurements to probe the possible type of models.
In subsequent chapter, we discuss the SUSY contributions to several measurements 
and introduce the current status of SUSY models.

\section{Muon $g-2$ in the MSSM}
\label{sec:gm2mssm}
%!TEX root = ../Dthesis.tex

As mentioned in Sec.~\ref{sec:gm2exp}, 
the muon $g-2$ has more than 3$\sigma $ discrepancy between the experimental and the theoretical values.
The size of the discrepancy is order of $10^{-9}$.
It is difficult for typical new physics models to explain the discrepancy.
First, let us introduce this difficulty.
Naive new physics contributions to the muon $g-2$ are expressed as
\beq
 \Delta a_{\mu } (\text{NP}) \sim \frac{\alpha _{\text{NP}}}{4 \pi } \frac{m^2_{\mu }}{m^2_{\text{NP}}}, 
 \label{eq:npcontrib}
\eeq
where $\alpha _{\text{NP}}$ and $m_{\text{NP}}$ are typical coupling and mass of new particle.

Since the size of the muon $g-2$ discrepancy is comparable with the SM electroweak contributions \eqref{eq:gm2ew}, 
typical masses of the new particles are required to be $\lesssim m_W$ if the coupling is about $\alpha _2 = g^2 / 4 \pi $, 
where $g$ is the SU(2)$_L$ gauge coupling. 
However, no such particles have ever discovered yet.
On the other hand, when typical new particles masses are large enough to avoid current mass bound, 
the new contributions is naively too small to explain the muon $g-2$ discrepancy.
Therefore, new physics models which can explain the muon $g-2$ discrepancy should be categorized into the two type as follows:
\begin{enumerate}
\item[(1)] The models in which the new particles have small coupling with the muon ($\mathcal{O}(10^{-3})$) and small mass ($\mathcal{O}(1)\MeV$),
for example, Dark Photon model\cite{Fayet:2007ua, Pospelov:2008zw}.
\item[(2)] The models in which the new particles have large coupling with the muon  ($\mathcal{O}(1)$) and large mass ($\mathcal{O}(100)\GeV$),
for example, Supersymmetry.
\end{enumerate}

As we will see in Sec.~\ref{subsec:gm2ma}, SUSY contributions to the muon $g-2$ are proportional to $\tan \beta$, 
which is the ratio of the Higgs VEVs, $\tan \beta \equiv \langle H_u \rangle /  \langle H_d \rangle$.
If $\tan \beta $ is order of 10, $\mathcal{O}(10^{-9})$ contributions to the muon $g-2$ are realized by masses of $\mathcal{O}(100)\GeV$ smuons,
which is within kinematical reach of future experiments.
In this section, we summarize the MSSM contributions to the muon $g-2$.

%%%%%%%%%%%%%%%%%%%%%%%%%%%%%%%%%%%%%%%%%%%%%%%%%%%%%%%%%%%%%%%%%%%%%%%%%%%%%%
\subsection{Relevant interactions}
\label{subsec:interaction}

In general, particles coupled with muon affect the muon $g-2$.
In the MSSM, new contributions are induced by the superpartners of muon, 
muon neutrino, SU(2)$_L \times$U(1)$_Y$ gauge bosons, and Higgs bosons, respectively.
The corresponding interactions are those of gaugino-slepton-leptons.
In this section, we provide general relevant interactions including case of lepton flavor and CP violations.

%First, let us discuss leptons.
%We denote by $\ell _i$ the lepton mass eigenstates, 
%where the subscript $i$ ($i = 1,2,3$) represents the generation.
%As for neutrinos, we assume that they are massless since their masses are negligibly small in our calculations. 

First, let us discuss the sleptons.
We denote by $\widetilde{\ell }_{L. R}$ and $\widetilde{\nu }_{\ell }$ the superpartners of leptons $\ell _{L, R}$ and neutrinos $\nu _{\ell }$, respectively.
In this dissertation, we assume that the slepton mass matrices are universal among the flavors, 
such as in the gauge mediated SUSY-breaking models\cite{Giudice:1998bp}, 
while the Yukawa matrices are non-diagonal in the model basis.
After electroweak symmetry breaking, the mass matrix of the charged sleptons are written by
\beq
 \begin{pmatrix}
  \widetilde{\ell }^{\dagger}_L  & \widetilde{\ell }^{\dagger}_R \\
  \end{pmatrix} 
   \begin{pmatrix}
    m_{LL}^2 & m_{LR}^{2\dagger } \\
    m_{LR}^2 & m_{RR}^2,
  \end{pmatrix} 
  \begin{pmatrix}
  \widetilde{\ell }_L  \\
  \widetilde{\ell }_R \\
  \end{pmatrix} 
  , \label{eq:slepmass} 
\eeq
where the matrices $m^2_{LL, RR. LR}$ are written as
\beq
 m_{LL}^2 &=& \text{diag} (m_{\widetilde{e}_{LL}}^2, m_{\widetilde{\mu } _{LL}}^2, m_{ \widetilde{\tau}_{LL}}^2). \label{eq:slepll} \\
 m_{RR}^2 &=& \text{diag} (m_{\widetilde{e}_{RR}}^2, m_{\widetilde{\mu } _{RR}}^2, m_{ \widetilde{\tau}_{RR}}^2). \label{eq:sleprr} \\
 m_{LR}^2 &=& A_{\ell }  Y_{\ell } \langle H_d \rangle - Y_{\ell } \mu \langle H_u \rangle . \label{eq:sleplr}
\eeq
Here, $A_{\ell }$ is A-term for the sleptons, $\mu $ is the Higgsino mass parameter, 
and $Y_{\ell }$ is the lepton Yukawa matrix, 
which is diagonalized by two unitary matrices,  $U_L$, $U_R$ as 
\beq
 U_R Y_{\ell } U^{\dagger }_L &=& \text{diag} \left( Y^{\text{diag}}_e, Y^{\text{diag}}_{\mu } , Y^{\text{diag}}_{\tau} \right).  \label{eq:diagYukawa}
\eeq
Here, $U_L$, $U_R$ generically include mixings among the flavors and complex phases,
which cause LFV and CPV.
The explicit forms will be given in Sec.~\ref{subsec:non-univ}.
The squared masses of the left- and right-handed sleptons $m_{\widetilde{\ell }_{LL}}^2$ and $m_{\widetilde{\ell }_{RR}}^2$ $(\ell = e, \mu , \tau)$ 
are 
\beq
 m_{\widetilde{\ell }_{LL, RR}}^2 &=& m^2_{\widetilde{\ell }_{L, R}} + m^2_{\ell } + D_{\widetilde{\ell}_{L, R}}, \label{eq:slep2}
\eeq
where, $D_{\widetilde{\ell}_{L, R}}$ are D-terms, $D_{\widetilde{\ell} } = m^2_Z \cos 2 \beta (T_3^{\ell } - Q^{\ell } \sin ^2 \theta _W )$. 

We diagonalize the mass matrix of the sleptons by $6 \times 6$ unitary matrix $U^{\ell }$ as
\beq
 U^{\ell}  \begin{pmatrix}
    m_{LL}^2 & m_{LR}^{2\dagger } \\
    m_{LR}^2 & m_{RR}^2,
  \end{pmatrix}  U^{\ell \dagger} =  \text{(diagonal)}. \label{eq:slepdiag}
\eeq 
The mass eigenvalues are denoted by $m_{\widetilde{\ell} _X}$ ($X = 1, \cdots 6$)
and the mass eigenstates are written as 
$\widetilde{\ell}_X = U^{\ell}_{X,i} \widetilde{\ell}_{Li} + U^{\ell}_{X,i+3} \widetilde{\ell}_{Ri}  ~~(i = 1,2,3 )$. 

The mass matrix of sneutrinos is diagonal in the model basis.
Further, there are no right-handed sneutrinos in the MSSM.
Hence, in this dissertation, no sneutrino mixings between distinct flavors appear, except for Yukawa interactions.  

Then, we consider the charginos.
The mass matrix of the charginos is written as
\beq
\begin{pmatrix}
\overline{\widetilde{W}^-_R} & \overline{\widetilde{H}^-_{uR}} \\
\end{pmatrix}
\begin{pmatrix}
 M_2 & \sqrt{2} m_W \sin \beta  \\
 \sqrt{2} m_W \cos \beta & \mu \\  
 \end{pmatrix}
 \begin{pmatrix}
 \widetilde{W}^-_L \\
 \widetilde{H}^-_{dL} \\
 \end{pmatrix}.
\eeq
It is diagonalized by two unitary matrices, $O_{L}$, $O_{R}$.
We denote the eigenvalues by $m_{\widetilde{\chi} ^{\pm }_A}~(A=1,2)$. 
The mass eigenstates are called the charginos and are defined as 
\beq
 \widetilde{\chi }^-_{L} =  (O_L)
 \begin{pmatrix}
 \widetilde{W}^-_L \\
 \widetilde{H}^-_{dL} \\
 \end{pmatrix}, ~~~~
 \widetilde{\chi }^-_{R} = (O_R)
  \begin{pmatrix}
 \widetilde{W}^-_R \\
 \widetilde{H}^-_{uR} \\
 \end{pmatrix}. 
 \label{eq:charmasse}
\eeq
Then $\widetilde{\chi }^-_A =  \widetilde{\chi }^-_{AL} + \widetilde{\chi }^-_{AR}$ $(A = 1,2)$ forms a Dirac fermion.

Finally, we consider the neutralinos.
The mass term of the neutralinos is written as
\beq
\frac{1}{2} 
\begin{pmatrix}
\widetilde{B}_L & \widetilde{W}^0_L & \widetilde{H}^0_{dL} & \widetilde{H}^0_{uL} \\
\end{pmatrix}
\mathcal{M}_{N}
\begin{pmatrix}
 \widetilde{B}_L \\
 \widetilde{W}^0_L \\
 \widetilde{H}^0_{dL} \\
 \widetilde{H}^0_{uL} \\
\end{pmatrix} + \text{h.c.} ,\label{eq:neutmass} 
\eeq
where
\beq
 \mathcal{M}_N = \begin{pmatrix}
 M_1 & 0 & -m_Z \sin \theta _W \cos \beta  & m_Z \sin \theta _W \sin \beta  \\
 0 & M_2 & m_Z \cos \theta _W \cos \beta & -m_Z \cos \theta _W \sin \beta \\
 -m_Z \sin \theta _W \cos \beta  & m_Z \sin \theta _W \sin \beta & 0 & -\mu \\
 -m_Z \cos \theta _W \cos \beta & -m_Z \cos \theta _W \sin \beta & -\mu  & 0 \\
 \end{pmatrix}.  \label{eq:neutmassmat}
\eeq
The mass matrix $\mathcal{M}_N$ is diagonalized by unitary matrix $O_{N}$.
The mass eigenvalues are denoted by $m_{\widetilde{\chi} ^{0}_A}~(A=1,4)$ and are called the neutralinos.
They are defined as
\beq
 \widetilde{\chi }^0_{L} =  (O_N)
 \begin{pmatrix}
 \widetilde{B}_L \\
 \widetilde{W}^0_L \\
 \widetilde{H}^0_{dL} \\
 \widetilde{H}^0_{uL} \\
\end{pmatrix}. \label{eq:neutmasse}
\eeq
Then, $\widetilde{\chi }^0_A =  \widetilde{\chi }^0_{AL} + \widetilde{\chi }^0_{AR}$ $(A = 1, \cdots ,4)$ forms a Majorana fermion. 

We are ready to write the relevant interactions.
The interaction Lagrangian for gaugino-slepton-lepton is written as
\beq
 \mathcal{L}_{\text{int}} = \bar{\ell }_i \left( N^{R(e)}_{iAX} P_R + N^{L(e)}_{iAX}  P_L \right) \widetilde{\chi }^0_A \widetilde{\ell }_X  + 
 \bar{\ell }_i \left( C^{R(e)}_{iAX} P_R + C^{L(e)}_{iAX}  P_L \right) \widetilde{\chi }^{-}_A \widetilde{\nu }_{\ell _{X}} + \text{h.c.}, 
 \label{eq:intlep} 
\eeq
where the coefficients are 
\beq
C^{R(e)}_{iAX} &=&  -g \left( O^*_R \right) _{A1} \delta _{iX}, \label{eq:CRiAX} \\
 C^{L(e)}_{iAX} &=&  \left( Y_{\ell } \right) _{iX} \left( O^*_L \right) _{A2}. \label{eq:CLiAX} \\
 N^{R(e)}_{iAX} &=&  \frac{g}{\sqrt{2}} \left( \left( O_{N} \right)_{A1} \tan \theta _W + \left( O_{N} \right)_{A2}  \right) U^{l*}_{X,i} 
 - \left( Y_{\ell }\right) _{ij}  \left( O_N \right)_{A3} U^{l*}_{X,i+3} \label{eq:NRiAX} \\
 N^{L(e)}_{iAX} &=& -\sqrt{2} g \tan \theta _W \left( O^*_{N} \right)_{A1}  U^{l*}_{X,i+3} 
 - \left( Y_{\ell }\right) _{ij} \left( O^*_{N} \right)_{A3} U^{l*}_{X,j}.  \label{eq:NLiAX}
\eeq

%%%%%%%%%%%%%%%%%%%%%%%%%%%%%%%%%%%%%%%%%%%%%%%%%%%%%%%%%%%%%%%%%%

%%%%%%%%%%%%%%%%%%%%%%%%%%%%%%%%%%%%%%%%%%%%%%%%%%%%%%%%%%%%%%%%%%
\subsection{Formulae in mass eigenstates}
\label{subsec:gm2me}

%%%%%%%%%%%%%%%%%%%%%%%%%%%%%%% Diagram

\begin{figure}[t]
 \begin{center}\vspace{\baselineskip}
  \begin{fmffile}{feyns/char}
\begin{fmfgraph*}(125,70)
\fmfleft{d1,p1,d2,d3,d4}\fmfright{d5,p2,d6,d7,gc}
\fmf{fermion,label=$\mu$,l.s=right}{p1,x1}
\fmf{fermion,label=$\mu$,l.s=right}{x2,p2}
\fmf{dashes,tension=0.5,lab=$\widetilde{\nu }_{\mu }$,l.s=left}{x2,x1}
\fmf{phantom,tension=5}{gc,gb}
\fmf{phantom,tension=1}{gb,d1}
\fmfposition
\fmf{phantom,left,tag=1,lab=$\widetilde{\chi}^{\pm}$,l.s=left}{x1,x2}
\fmfipath{p[]}
\fmfiset{p1}{vpath1(__x1,__x2)}
\fmfi{photon}{point length(p1)/3*2 of p1 -- vloc(__gb)}
\fmfv{label=$\gamma$}{gb}
\fmfi{xgaugino}{p1}
\end{fmfgraph*}
 \end{fmffile}
 \hspace{10pt}
 \begin{fmffile}{feyns/neut}
\begin{fmfgraph*}(125,70)
\fmfleft{d1,p1,d2,d3,d4}\fmfright{d5,p2,d6,d7,gc}
\fmf{fermion,label=$\mu$,l.s=right}{p1,x1}
\fmf{fermion,label=$\mu$,l.s=right}{x2,p2}
\fmf{xgaugino,tension=0.5,lab=$\widetilde{\chi}^0$,l.s=left}{x2,x1}
\fmf{phantom,tension=5}{gc,gb}
\fmf{phantom,tension=1}{gb,d1}
\fmfposition
\fmf{phantom,left,tag=1,lab=$\widetilde{\mu }$,l.s=left}{x1,x2}
\fmfipath{p[]}
\fmfiset{p1}{vpath1(__x1,__x2)}
\fmfi{photon}{point length(p1)/3*2 of p1 -- vloc(__gb)}
\fmfv{label=$\gamma$}{gb}
\fmfi{dashes}{p1}
\end{fmfgraph*}
\end{fmffile}
\caption[The diagrams of the SUSY contributions to the muon $g-2$ (mass eigenstates).]
{The diagrams of the SUSY contributions to the muon $g-2$ in mass eigenstates. 
The left panel of the figure is called the chargino--muon sneutrino contribution (or chargino contribution), 
the right is the neutralino--smuon contribution (or neutralino contribution).}
\label{fig:gm2mass}
\end{center}
\end{figure}
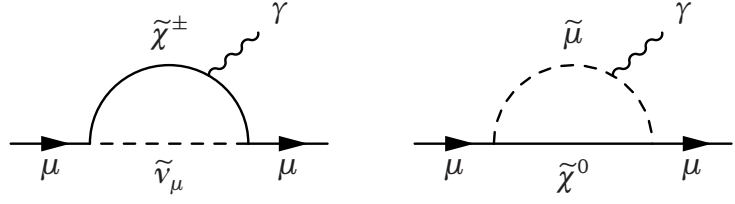

%%%%%%%%%%%%%%%%%%%%%%%%%%%%%%%

At leading order, the new contributions to the muon $g-2$ are induced by 
the chargino--muon sneutrino and the neutralino--smuon loop diagrams in Fig.~\ref{fig:gm2mass}.
Each diagram is called the chargino-muon sneutrino and the neutralino-smuon contribution, respectively.
Computing the Feynman diagrams by ordinary perturbation theory, 
we can obtain the SUSY contributions to the muon $g-2$ at one-loop level\footnote{
Strictly speaking, these results include the flavor/CP violating effects 
which are induced by off-diagonal elements of the slepton mass matrix.
Since they are subdominant contributions, we neglect them in this section.
These effects are discussed in Sec.~\ref{sec:LFVmssm} and \ref{subsec:non-univ}.} as\cite{Hisano:1995cp, moroi:1995yh}
\beq
\Delta a_{\mu }^{\widetilde{\chi }^{\pm}} &=& 
\frac{m_{\mu }}{48\pi ^2} \sum _{X=1}^3  \sum _{A=1}^2 \frac{1}{m^2_{\widetilde{\nu }_X}} 
\Bigg[
 \frac{m_{\mu }}{4}  \left(  \left|  C^{L(e)}_{2AX} \right| ^2  +  \left|  C^{R(e)}_{2AX} \right| ^2  \right) 
 F^C_{1} \left( \frac{m^2_{\widetilde{\chi }^{\pm}_A}}{m^2_{\widetilde{\nu }_{X}}} \right) \notag \\
 &&+m_{\widetilde{\chi }^{\pm}_A}  \Re \left[ C^{L(e)*}_{2AX} C^{R(e)}_{2AX}  \right] 
 F^C_{2} \left( \frac{m^2_{\widetilde{\chi }^{\pm}_A}}{m^2_{\widetilde{\nu }_{X}}} \right)  \Bigg],  \label{eq:gmc} \\
 \Delta a_{\mu }^{\widetilde{\chi }^0} &=& 
 -\frac{m_{\mu }}{48\pi ^2} \sum _{X=1}^6 \sum _{A=1}^4 \frac{1}{m^2_{\widetilde{\ell}_X}} 
 \Bigg[\frac{m_{\mu }}{4} 
 \left( \left|  N^{L(e)}_{2AX} \right| ^2  + \left|  N^{R(e)}_{2AX} \right| ^2  \right)
 F^N_{1} \left( \frac{m^2_{\widetilde{\chi }^0_A}}{m^2_{\widetilde{\ell }_{X}}} \right)   \notag \\
 &&+\frac{m_{\widetilde{\chi }^0_A}}{2}  \Re \left[N^{L(e)*}_{2AX} N^{R(e)}_{2AX} \right]
 F^N_{2}\left( \frac{m^2_{\widetilde{\chi }^0_A}}{m^2_{\widetilde{\ell }_{X}}} \right)   \Bigg]  , \label{eq:gmn} 
\eeq
where $F^C_{1,2}$ and $F^N_{1,2}$ are the loop functions.
\beq
F^C_1(x) &=& \frac{2}{(1-x)^4}\left[ 2 + 2x - 6x^2  +x^3 +6x \log x  \right] , \label{eq:fc1} \\
F^C_2(x) &=& \frac{3}{2(1-x)^3}\left[ -3 + 4x - x^2 -2  \log x  \right] , \label{eq:fc2} \\
F^N_1(x) &=& \frac{2}{(1-x)^4}\left[ 1 - 6x + 3x^2  + 2x^3 - 6x^2 \log x  \right] , \label{eq:fn1} \\
F^N_2(x) &=& \frac{3}{(1-x)^3}\left[ 1 - x^2  + 2x  \log x  \right]. \label{eq:fn2} 
\eeq

%%%%%%%%%%%%%%%%%%%%%%%%%%%%%%%%%%%%%%%%%%%%%%%%%%%%%%%%%%%%%%%%%%%%%%%%%%%%

\subsection{Formulae in gauge eigenstates}
\label{subsec:gm2ma}

%%%%%%%%%%%%%%%%%%%%%%%%%%%%%%%%%%%%%%%%%%%%%%% Diagram
\begin{figure}[t]
\vspace{\baselineskip}
\begin{center}
 \begin{fmffile}{feyns/gm2gauge}
 \begin{minipage}[t]{0.3\textwidth}
\begin{center}
   \begin{fmfgraph*}(125,70)
 \fmfleft{d1,p1,d2,d3,d4}\fmfright{d5,p2,d6,d7,gc}
 \fmf{fermion,label=$\mu _L$,l.s=right}{p1,x1}
 \fmf{fermion,label=$\mu _R$,l.s=left}{p2,x2}
 \fmf{dashes,tension=0.5,lab=$\widetilde{\nu }_{\mu}$,l.s=left}{x2,x1}
 \fmf{phantom,tension=5}{gc,gb}
 \fmf{phantom,tension=5}{gb,ga}
 \fmf{phantom,tension=1}{ga,d1}
 \fmfposition
 \fmf{phantom,left,tag=1,lab=$\widetilde{W}^{\pm}$~~~$\widetilde{H}^{\pm}$,l.s=left}{x1,x2}
 \fmfipath{p[]}
 \fmfiset{p1}{vpath1(__x1,__x2)}
 \fmf{photon}{gb,ga}
 \fmfv{label=$\gamma$}{gb}
 \fmfiv{d.sh=cross,d.size=5thick}{point length(p1)/2 of p1}
 \fmfi{plain}{p1}
   \end{fmfgraph*}
 \\(a)
\end{center} 
\end{minipage}
 \begin{minipage}[t]{0.3\textwidth}
\begin{center}
   \begin{fmfgraph*}(125,70)
 \fmfleft{d1,p1,d2,d3,d4}\fmfright{d5,p2,d6,d7,gc}
 \fmf{fermion,label=$\mu _L$,l.s=right}{p1,x1}
 \fmf{fermion,label=$\mu _R$,l.s=left}{p2,x2}
 \fmf{plain,tension=0.5,lab=$\widetilde{B}$,l.s=left}{x2,x1}
 \fmf{phantom,tension=5}{gc,gb}
 \fmf{phantom,tension=5}{gb,ga}
 \fmf{phantom,tension=1}{ga,d1}
 \fmfposition
 \fmf{phantom,left,tag=1,lab=$\widetilde{\mu }_L$~~~$\widetilde{\mu } _R$,l.s=left}{x1,x2}
 \fmfipath{p[]}
 \fmfiset{p1}{vpath1(__x1,__x2)}
 \fmf{photon}{gb,ga}
 \fmfv{label=$\gamma$}{gb}
 \fmfiv{d.sh=cross,d.size=5thick}{point length(p1)/2 of p1}
 \fmfi{dashes}{p1}
  \end{fmfgraph*}
 \\(b)
\end{center} 
\end{minipage}
 \begin{minipage}[t]{0.3\textwidth}
\begin{center}
   \begin{fmfgraph*}(125,70)
 \fmfleft{d1,p1,d2,d3,d4}\fmfright{d5,p2,d6,d7,gc}
 \fmf{fermion,label=$\mu _L$,l.s=right}{p1,x1}
 \fmf{fermion,label=$\mu _R$,l.s=left}{p2,x2}
 \fmf{plain,tension=0.5,lab=$\widetilde{W}^0$~~~~~~~$\widetilde{H}^0$,l.s=left}{x2,x1}
 \fmf{phantom,tension=5}{gc,gb}
 \fmf{phantom,tension=1}{gb,d1}
 \fmfposition
 \fmf{phantom}{x1,xx,x2}
 \fmf{phantom,left,tag=1,lab=$\widetilde{\mu } _L$,l.s=left}{x1,x2}
 \fmfipath{p[]}
 \fmfiset{p1}{vpath1(__x1,__x2)}
 \fmfi{photon}{point length(p1)/3*2 of p1 -- vloc(__gb)}
 \fmfv{label=$\gamma$}{gb}
 \fmfv{d.sh=cross,d.size=5thick}{xx}
 \fmfi{dashes}{p1}
  \end{fmfgraph*}
\\(c)
\end{center} 
\end{minipage}

\vspace{10pt}
 \begin{minipage}[t]{0.3\textwidth}
\begin{center}
   \begin{fmfgraph*}(125,70)
 \fmfleft{d1,p1,d2,d3,d4}\fmfright{d5,p2,d6,d7,gc}
 \fmf{fermion,label=$\mu _L$,l.s=right}{p1,x1}
 \fmf{fermion,label=$\mu _R$,l.s=left}{p2,x2}
 \fmf{plain,tension=0.5,lab=$\widetilde{B}$~~~~~~~$\widetilde{H}^0$,l.s=left}{x2,x1}
 \fmf{phantom,tension=5}{gc,gb}
 \fmf{phantom,tension=1}{gb,d1}
 \fmfposition
 \fmf{phantom}{x1,xx,x2}
 \fmf{phantom,left,tag=1,lab=$\widetilde{\mu } _L$,l.s=left}{x1,x2}
 \fmfipath{p[]}
 \fmfiset{p1}{vpath1(__x1,__x2)}
 \fmfi{photon}{point length(p1)/3*2 of p1 -- vloc(__gb)}
 \fmfv{label=$\gamma$}{gb}
 \fmfv{d.sh=cross,d.size=5thick}{xx}
 \fmfi{dashes}{p1}
  \end{fmfgraph*}
\\(d)
\end{center}
\end{minipage}
 \begin{minipage}[t]{0.3\textwidth}
\begin{center}
   \begin{fmfgraph*}(125,70)
 \fmfleft{d1,p1,d2,d3,d4}\fmfright{d5,p2,d6,d7,gc}
 \fmf{fermion,label=$\mu _L$,l.s=right}{p1,x1}
 \fmf{fermion,label=$\mu _R$,l.s=left}{p2,x2}
 \fmf{plain,tension=0.5,lab=$\widetilde{H}^0$~~~~~~~$\widetilde{B}$,l.s=left}{x2,x1}
 \fmf{phantom,tension=5}{gc,gb}
 \fmf{phantom,tension=1}{gb,d1}
 \fmfposition
 \fmf{phantom}{x1,xx,x2}
 \fmf{phantom,left,tag=1,lab=$\widetilde{\mu } _R$,l.s=left}{x1,x2}
 \fmfipath{p[]}
 \fmfiset{p1}{vpath1(__x1,__x2)}
 \fmfi{photon}{point length(p1)/3*2 of p1 -- vloc(__gb)}
 \fmfv{label=$\gamma$}{gb}
 \fmfv{d.sh=cross,d.size=5thick}{xx}
 \fmfi{dashes}{p1}
  \end{fmfgraph*}
\\(e)
\end{center} 
\end{minipage}
\end{fmffile}
\caption[The MSSM dominant contributions to the muon $g-2$ (gauge eigenstates).]
{The diagrams of the SUSY contributions to the muon $g-2$ in gauge eigenstates, 
where the diagrams (a)--(e) corresponds to Eqs.~\eqref{eq:WHsnu}--\eqref{eq:BHmuR}, respectively.
The diagram (a) comes from the chargino--muon sneutrino diagram, the diagrams (b)--(e) are from the neutralino--smuon diagram.}
\label{fig:gm2gauge}
\end{center}
\end{figure}
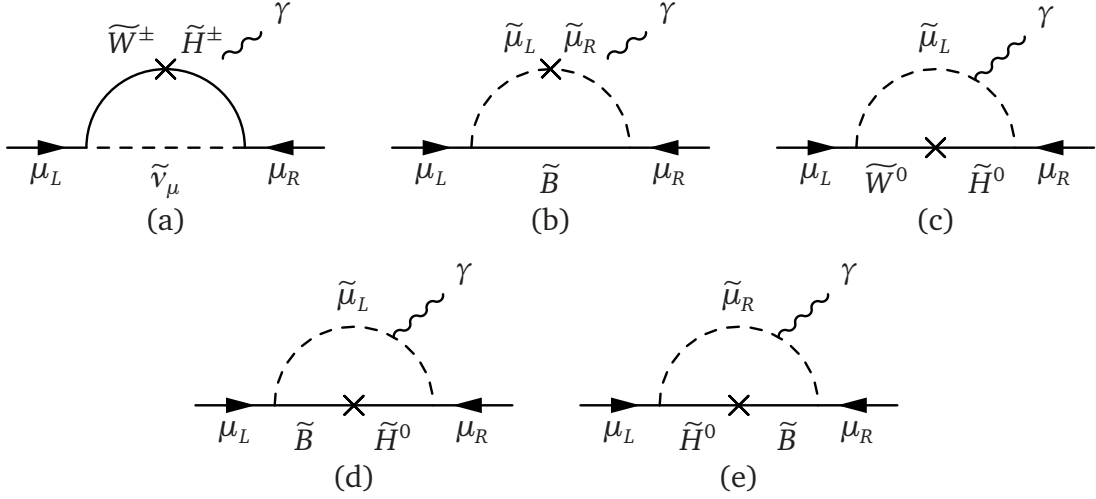

%%%%%%%%%%%%%%%%%%%%%%%%%%%%%%%%%%%%%%%%%%%%%%%

Even though the formulae in mass eigenstates \eqref{eq:gmc} and \eqref{eq:gmn} are useful for numerical calculations,
they are not suitable for understanding their dependence of the MSSM parameters.
The parameter dependences are hidden by the electroweak symmetry breaking which causes complicated mixings.
On the other hand, in weak eigenstates, the structure of one loop contributions become more transparent.
The expression in weak eigenstates are equivalent to the expansion of $m_W/m_{{\rm{soft}}}$, 
where $m_{{\rm{soft}}}$ represents SUSY-breaking masses and the Higgsino mass $\mu$.
At the leading order of $m_W/m_{{\rm{soft}}}$ and $\tan\beta$, 
they are evaluated as \cite{moroi:1995yh, Endo:2013bba}
\beq
  \Delta a_{\mu }(\widetilde{W}, \widetilde{H}, \widetilde{\nu}_\mu)
 &=& \frac{\alpha_2}{4\pi} \frac{m_\mu^2}{M_2 \mu} \tan\beta\cdot
f_C
 \left( \frac{M_2 ^2}{m_{\widetilde{\nu }}^2}, \frac{\mu ^2}{m_{\widetilde{\nu }}^2}  \right) , 
 \label{eq:WHsnu} \\
  \Delta a_{\mu }(\widetilde{\mu }_L, \widetilde{\mu }_R,\widetilde{B}) 
 &=& \frac{\alpha_Y}{4\pi} \frac{m_{\mu }^2 M_1 \mu}{m_{\widetilde{\mu }_L}^2 m_{\widetilde{\mu }_R}^2}  \tan \beta\cdot
 f_N \left( \frac{m_{\widetilde{\mu }_L}^2}{M_1^2}, \frac{m_{\widetilde{\mu }_R}^2}{M_1^2}\right), \label{eq:BmuLR} \\
 \Delta a_{\mu }(\widetilde{W}, \widetilde{H},  \widetilde{\mu}_L) 
 &=& - \frac{\alpha_2}{8\pi} \frac{m_\mu^2}{M_2 \mu} \tan\beta\cdot
 f_N
 \left( \frac{M_2 ^2}{m_{\widetilde{\mu }_L}^2}, \frac{\mu ^2}{m_{\widetilde{\mu }_L}^2} \right),  
 \label{eq:WHmuL}  \\ 
 \Delta a_{\mu }(\widetilde{B},\widetilde{H},  \widetilde{\mu }_L) 
  &=& \frac{\alpha_Y}{8\pi} \frac{m_\mu^2}{M_1 \mu} \tan\beta\cdot
 f_N 
 \left( \frac{M_1 ^2}{m_{\widetilde{\mu }_L}}, \frac{\mu ^2}{m_{\widetilde{\mu }_L}} \right),   
 \label{eq:BHmuL} \\ 
  \Delta a_{\mu }(\widetilde{B}, \widetilde{H},  \widetilde{\mu }_R) 
  &=& - \frac{\alpha_Y}{4\pi} \frac{m_{\mu }^2}{M_1 \mu} \tan \beta \cdot
  f_N \left( \frac{M_1 ^2}{m_{\widetilde{\mu }_R}^2}, \frac{\mu ^2}{m_{\widetilde{\mu }_R}^2} \right), \label{eq:BHmuR} 
\eeq
where $m_\mu$ is the muon mass, while $\alpha_Y$ and $\alpha_2$ are the fine structure constants of the U(1)$_Y$ and the SU(2)$_L$ gauge symmetries, respectively. 
They can be obtained from the Feynman diagrams shown in Fig.~\ref{fig:gm2gauge}.
The arguments in the left hand side of Eqs.~\eqref{eq:WHsnu}--\eqref{eq:BHmuR} represent the superparticles 
which propagate in each loop diagram.
If one of them decouples, the corresponding contribution is suppressed.
Eq.~\eqref{eq:WHsnu} comes from the chargino--muon sneutrino diagrams, 
and Eqs.~\eqref{eq:BmuLR}--\eqref{eq:BHmuR} are from the neutralino--smuon diagrams.

The loop functions are defined as\footnote{
The functions, $f_C$ and $f_N$, are reduced from the functions, $J_5$ and $I_4$, in Ref.~\cite{moroi:1995yh}.
}
\beq
f_C(x,y) &=& xy
\left[
\frac{5-3(x+y)+xy}{(x-1)^2(y-1)^2}
-\frac{2\log x}{(x-y)(x-1)^3}
+\frac{2\log y}{(x-y)(y-1)^3}
\right]\,,
\label{eq:fcma} \\
f_N(x,y) &=& xy
\left[
\frac{-3+x+y+xy}{(x-1)^2(y-1)^2}
+\frac{2x\log x}{(x-y)(x-1)^3}
-\frac{2y\log y}{(x-y)(y-1)^3}
\right]\,.\label{eq:fnma} 
\eeq
They are satisfy $0 \leq f_{C,N}(x,y) \leq 1$ and are monochromatically increasing for $x >0$ and $y > 0$.
When the superparticles have degenerate masses, they become $f_C(1,1) = 1/2$ and $f_N(1,1) = 1/6$.
 
\subsection{Structure of SUSY contributions}
\label{subsec:interpretations}

%%%%%%%%%%%%%%%%%%%%%%%%%%%%%%%%%%%%%%%%%%%%%%%% Table
\begin{table}[t]
\begin{center}
{\renewcommand\arraystretch{1.5}
  \begin{tabular}{|c|ccccc|c|} \hline
    Mass spectrum & (a)  & (b) & (c) & (d) & (e) & Note to solve the anomaly \\
     \hline \hline
    Degenerated case& \checkmark &  &  &  &  &  \\ \hline
    Large LR mixing & & \checkmark &  &  &  & Both $\widetilde{\mu }_{L,R}$ are light \\ \hline
    $\widetilde{W}$, $\tilde{\mu }_R$-decoupled & & &  & \checkmark &  &  \\ \hline
    $\widetilde{W}$, $\tilde{\mu }_L$-decoupled & &  &  &  &  \checkmark & $\mu  < 0$ for a positive contribution \\ \hline 
  \end{tabular}}
   \caption[Summary on the SUSY contributions to the muon $g-2$.]
   {Summary on the SUSY contributions to the muon $g-2$.
   The items (a)--(e) in the middle column correspond to the diagrams (a)--(e) in Fig. \ref{fig:gm2gauge}.
   The left column shows several mass spectrums. The check marks in the middle column denote the dominant contribution to the muon $g-2$ in the case of left column.}  
   \label{tab:gm2contr}
   \end{center}
\end{table}
%%%%%%%%%%%%%%%%%%%%%%%%%%%%%%%%%%%%%%%%%%%%%%%%
As discussed in Sec.~\ref{sec:gm2exp}, 
the muon $g-2$ has the discrepancy at more than $3\sigma$ levels between the experimental and the theoretical values.
In order to explain this discrepancy, the new contributions to the muon $g-2$ are required to be $\mathcal{O}(10^{-9})$ and positive.
In this section, we discuss the results obtained in Eqs.~\eqref{eq:WHsnu}--\eqref{eq:BHmuR} from above viewpoints.
The results are summarized in Table.~\ref{tab:gm2contr}.

Numerically, the SUSY contributions are evaluated as\cite{Endo:2013bba}
\beq
 \Delta a_{\mu }(\widetilde{W},\widetilde{H}, \widetilde{\nu}_{\mu} ) 
&\simeq  &
15 \times 10^{-9}
\left(\frac{\tan\beta}{10}\right)
\left(\frac{(100\GeV)^2}{M_2\mu}\right)
\left(\frac{f_C}{1/2}\right),  \label{eq:WHLnu_N}  \\
 \Delta a_{\mu }(\widetilde{\mu }_L, \widetilde{\mu }_R,\widetilde{B}) 
 &\simeq &
1.5 \times 10^{-9}
\left(\frac{\tan\beta}{10}\right)
 \left(\frac{(100\GeV)^2}{m_{\widetilde{\mu }_L}^2 m_{\widetilde{\mu }_R}^2/M_1\mu }\right)
\left(\frac{f_N}{1/6}\right). \label{eq:BLR_N} \\
 \Delta a_{\mu }(\widetilde{W},\widetilde{H}, \widetilde{\mu }_L) 
&\simeq &
-2.5 \times 10^{-9}
\left(\frac{\tan\beta}{10}\right)
\left(\frac{(100\GeV)^2}{M_2\mu}\right)
\left(\frac{f_N}{1/6}\right),  \label{eq:WHL_N}  \\
 \Delta a_{\mu }(\widetilde{B},  \widetilde{H},  \widetilde{\mu }_L) 
  &\simeq &
0.76\times 10^{-9}
\left(\frac{\tan\beta}{10}\right)
\left(\frac{(100\GeV)^2}{M_1\mu}\right)
\left(\frac{f_N}{1/6}\right), \label{eq:BHL_N} \\
 \Delta a_{\mu }(\widetilde{B}, \widetilde{H},  \widetilde{\mu }_R) 
 &\simeq &
-1.5\times 10^{-9}
\left(\frac{\tan\beta}{10}\right)
\left(\frac{(100\GeV)^2}{M_1\mu}\right)
\left(\frac{f_N}{1/6}\right), \label{eq:BHR_N}
\eeq
where Eqs.~\eqref{eq:WHsnu}--\eqref{eq:BHmuR} correspond to Eqs.~\eqref{eq:WHLnu_N}--\eqref{eq:BHR_N}, respectively.
As mentioned in Sec. \ref{subsec:gm2ma}, 
the arguments in the left hand side of Eqs.~\eqref{eq:WHLnu_N}--\eqref{eq:BHR_N} represent the superparticles which propagate in each loop diagram.
They are enhanced when $\tan \beta$ is large and $m_{\text{soft}}$ is small.
If relevant superparticles which propagate in each loop diagram have masses of $\mathcal{O}(100)\GeV$ and $\tan \beta = \mathcal{O}(10)$, 
they become $\mathcal{O}(10^{-9})$, which can explain Eq.\eqref{eq:g-2_deviation}.

Let us investigate parameter dependences of the SUSY contributions in detail.
There are two representative cases when the discrepancy of the muon $g-2$ is explained.
The first case is the one that the Wino--muon sneutrino contribution \eqref{eq:WHsnu} dominates the SUSY contributions.
This situation is realized when relevant superparticles have nearly degenerate mass spectrum, 
as can be seen in Eqs.~\eqref{eq:WHLnu_N}--\eqref{eq:BHR_N}.
In fact, the Wino--muon sneutrino contribution is dominant in wide SUSY models which explain the discrepancy of the muon $g-2$ 
with non-decoupling Higgsinos.

The second case is the one that the pure--Bino contribution \eqref{eq:BmuLR} is dominant.
As $\mu $ increases, it becomes relevant because it is proportional to $\mu $ including in left-right mixing of the smuon mass matrix.
The other contributions are suppressed because they are proportional to $1/ \mu $, which is from the Higgsino propagator.
In this case, the muon $g-2$ discrepancy is explained when both 
$m_{\widetilde{\mu }_L}$ and $m_{\widetilde{\mu }_R}$ are required to be $\mathcal{O}(100)\GeV$ and $\tan \beta = \mathcal{O}(10)$.

The other extreme cases also exist.
Here, we comment on two cases as follows, ($\widetilde{W}$, $\widetilde{\mu }_{R}$) and ($\widetilde{W}$, $\widetilde{\mu }_{L}$) decoupled cases, respectively.
First, when Wino and right-handed smuon are decoupled, 
the Bino--Higgsino--left-handed smuon contribution \eqref{eq:BHmuL} is relatively relevant, while the others are suppressed.
On the other hand, the Bino--Higgsino--right-handed smuon contribution \eqref{eq:BHmuR},
which gives positive contribution in $\mu < 0$ case, is relatively relevant,
when the Wino and the left-handed smuon are decoupled.
In these cases, since light smuon and large $\tan \beta $\footnote{
$m_{\widetilde{\mu }_{L,R}} \sim 100 \GeV$ and $\tan \beta = 40$--50 are required.} are required to explain the discrepancy of the muon $g-2$, 
the parameter regions are extremely limited.
In this dissertation, although these cases are also interesting, we will not discuss them anymore.

The Wino--Higgsino--left-handed smuon contribution \eqref{eq:WHmuL} has the same parameter dependence as the Wino--muon sneutrino one,
$M_2$, $\mu $, and $m_{\widetilde{\mu }_L}$,\footnote{
the muon sneutrino mass $m_{\widetilde{\nu }_{\mu }}$ is related to $m_{\widetilde{\mu }_L}$ by the SU(2)$_L$ symmetry as 
$m^2_{\widetilde{\nu }_{\mu }} = m^2_{\widetilde{\mu }_L} + m^2_W \cos 2 \beta $.} respectively.
As can be seen in Eq.~\eqref{eq:WHLnu_N} and \eqref{eq:WHL_N}, 
it can't be dominate in any parameter region.

Since the SUSY contributions are proportional to $1/m^2_{\text{soft}}$, 
all contributions \eqref{eq:WHsnu}--\eqref{eq:BHmuR} are rapidly suppressed, as both masses of the smuons increase.
In this case, no solution exists to explain the discrepancy of the muon $g-2$. \footnote{
Strictly speaking, when the left-right mixing of the smuons is extremely large, $\mathcal{O}(1)\TeV$ masses of the smuons is allowed.
We will discuss the case in Chapter.~\ref{chp:main}.}

So far, we discussed the SUSY contributions to the muon $g-2$ at one loop levels.
Some of higher order corrections are important in numerical calculations.
In this dissertation, the corrections that can be as large as or larger than $\mathcal{O}(10)\%$ are included
and will be discussed in Sec.~\ref{subsec:gm2vac}.

\section{LFV and EDM in the MSSM}
\label{sec:LFVmssm}
%!TEX root = ../Dthesis.tex

The MSSM, which explains the muon $g-2$ discrepancy, contains light superpartilces coupled with the muon.
In this case, Lepton Flavor Violation (LFV) and CP Violation (CPV)
which are sensitive to off-diagonal components of the slepton mass matrices, might be induced.
They are suppressed if the slepton mass matrices are universal among the flavors, such as in the gauge mediated SUSY-breaking models. 
However, even when the mass matrices are diagonal in the model basis, 
sizable FCNC and CPV are generically induced as long as the diagonal components are not equal to each others.
In fact, a lot of models have been proposed to explain the SM Yukawa couplings, 
and many of them predict non-diagonal Yukawa matrices in the model basis. 
Even if the sfermion mass matrices are diagonal in this basis, 
off-diagonal components are generated in the fermion mass eigenstate basis, 
which are obtained by rotating the mass matrices of the model basis with unitary matrices. 
In other words, the super GIM mechanism does not work generically unless the sfermion mass matrices are universal.

In this dissertation, we focus on muon FCNCs and electron EDM.
This section is summarized these formulae at leading order.
We do not consider the higher order corrections, i.g., Barr-Zee type diagrams, 
because the contributions at one-loop level are not suppressed and the heavy Higgs bosons are decoupled in our setup.   

\subsection{Formulae in $\Gamma (\ell ^-_j \rightarrow \ell ^-_i \gamma )$}
\label{subsec:LFVformula}

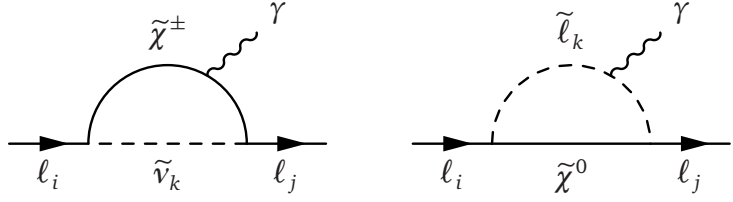
\begin{figure}[t]
 \begin{center}\vspace{\baselineskip}
  \begin{fmffile}{feyns/charLFV}
\begin{fmfgraph*}(125,70)
\fmfleft{d1,p1,d2,d3,d4}\fmfright{d5,p2,d6,d7,gc}
\fmf{fermion,label=$\ell _i$,l.s=right}{p1,x1}
\fmf{fermion,label=$\ell _j$,l.s=right}{x2,p2}
\fmf{dashes,tension=0.5,lab=$\widetilde{\nu }_{k}$,l.s=left}{x2,x1}
\fmf{phantom,tension=5}{gc,gb}
\fmf{phantom,tension=1}{gb,d1}
\fmfposition
\fmf{phantom,left,tag=1,lab=$\widetilde{\chi}^{\pm}$,l.s=left}{x1,x2}
\fmfipath{p[]}
\fmfiset{p1}{vpath1(__x1,__x2)}
\fmfi{photon}{point length(p1)/3*2 of p1 -- vloc(__gb)}
\fmfv{label=$\gamma$}{gb}
\fmfi{xgaugino}{p1}
\end{fmfgraph*}
 \end{fmffile}
 \hspace{10pt}
 \begin{fmffile}{feyns/neutLFV}
\begin{fmfgraph*}(125,70)
\fmfleft{d1,p1,d2,d3,d4}\fmfright{d5,p2,d6,d7,gc}
\fmf{fermion,label=$\ell _i$,l.s=right}{p1,x1}
\fmf{fermion,label=$\ell _j$,l.s=right}{x2,p2}
\fmf{xgaugino,tension=0.5,lab=$\widetilde{\chi}^0$,l.s=left}{x2,x1}
\fmf{phantom,tension=5}{gc,gb}
\fmf{phantom,tension=1}{gb,d1}
\fmfposition
\fmf{phantom,left,tag=1,lab=$\widetilde{\ell }_{k}$,l.s=left}{x1,x2}
\fmfipath{p[]}
\fmfiset{p1}{vpath1(__x1,__x2)}
\fmfi{photon}{point length(p1)/3*2 of p1 -- vloc(__gb)}
\fmfv{label=$\gamma$}{gb}
\fmfi{dashes}{p1}
\end{fmfgraph*}
\end{fmffile}
\caption[The diagrams of $\ell ^-_j \rightarrow \ell ^-_i \gamma $ decay.]
{The diagrams of $\ell ^-_j \rightarrow \ell ^-_i \gamma $ decay.
The left panel of the figure is the chargino--sneutrino contribution, 
the right is the neutralino--slepton contribution.}
\label{fig:LFVdiagram}
\end{center}
\end{figure}

We start to the effective Langangian \eqref{eq:eff_LFV}.
The diagrams which contribute to the $\ell ^-_j \rightarrow \ell ^-_i \gamma $ process have the same stricture as that of the muon $g-2$, 
as shown in Fig.~\ref{fig:LFVdiagram}.
Calculating the Feynman diagrams, the decay rate for $\ell ^-_j \rightarrow \ell ^-_i \gamma $ is evaluated as
\beq
 \Gamma (\ell ^-_j \rightarrow \ell ^-_i \gamma ) = \frac{\alpha}{4} m^5_{l_j} \left(  \left| A^L_C + A^L_N  \right| ^2 +  \left| A^R_C + A^R_N  \right| ^2  \right), \label{eq:drljtoijg}
\eeq
where $A^{L,R}_{C,N}$ are the amplitude which chargino--sneutrino and neutralino--slepton contribute to, and are given as
\beq
 A^L_{C} =  -\frac{1}{96\pi ^2} \sum _{X=1} ^3\sum _{A=1} ^2 
 \frac{1}{m^2_{\widetilde{\nu }_X}}\left[ \frac{1}{4} C^{L(e)}_{iAX} C^{L(e)*}_{jAX}
 F^C_{1}\left(\frac{m^2_{\widetilde{\chi}^{\pm }_A}}{m^2_{\widetilde{\nu}_{X}}} \right) + \frac{m_{\widetilde{\chi }^{\pm }_A}}{m_{\ell _j}} 
 C^{L(e)}_{iAX} C^{R(e)*}_{jAX} 
 F^C_{2}\left(\frac{m^2_{\widetilde{\chi}^{\pm }_A}}{m^2_{\widetilde{\nu }_X}} \right) \right], \label{eq:ALC}  \\
  A^L_{N} =  \frac{1}{96\pi ^2} \sum _{X=1} ^6\sum _{A=1} ^4  \frac{1}{m^2_{\widetilde{\ell }_X}}
  \left[ \frac{1}{4} N^{L(e)}_{iAX} N^{L(e)*}_{jAX} 
  F^N_{1}\left(\frac{m^2_{\widetilde{\chi}^0_A}}{m^2_{\widetilde{\ell }_X}} \right) 
  +\frac{1}{2} \frac{m_{\widetilde{\chi }^{0}_A}}{m_{l_j}} N^{L(e)}_{iAX} N^{R(e)*}_{jAX}
   F^N_{2}\left(\frac{m^2_{\widetilde{\chi}^0_A}}{m^2_{\widetilde{\ell }_X}} \right)
  \right], \label{eq:ALN} 
\eeq
and $ A^R_{C,N} =  A^{L}_{C,N}\big| _{L \to R}$.
The loop functions $F^{C,N}_{1,2}$ were given in Eqs.~\eqref{eq:fc1}--\eqref{eq:fn2}.

When $j=2$ and $i = 1$ (i.e.,  $\mu \to e \gamma $ decay), dominant decay mode of muons is $\mu ^- \rightarrow e^- \nu _{\mu } \bar{\nu }_e$. 
The branching ratio of $\mu \to e \gamma $ decay can be calculated in Eq.~\eqref{eq:mueg0}.

Once we can calculate Eqs.~\eqref{eq:ALC}, \eqref{eq:ALN}, 
$\mu \to e\bar{e}e$ and $\mu - e$ conversion processes are evaluated in Eqs.~\eqref{eq:muto3e} and \eqref{eq:mueconv},
if the dipole contributions are dominant.
In the MSSM, the dipole contributions are proportional to $\tan \beta $, 
while non-dipole contributions, which are from the dimension 6 operators, are not so.
Since large $\tan \beta $ is required to explain the discrepancy of the muon $g-2$, 
the dipole contributions are dominant in our assumption. 

Let us comment on higher order contributions, e.g., the Higgs-mediated FCNC\cite{Babu:1999hn}.
In supersymmetric models, the superpotential must be holomorphic.
Even though, one can diagonalize the mass matrices of fermion at tree level, 
off-diagonal components of Yukawa matrices are generated by radiative corrections, which pick up SUSY breaking effects as follows, 
\beq
  \left(\Delta_e\right)_{ij} \bar{\ell}_{Ri} \ell _{Lj} H^{0*}_u.  \label{eq:offYukawa}
\eeq
In this case, the basis that diagonalize fermion mass matrices do not accord with those of the Yukawa couplings of the Higgs fields, 
thus flavor-changing effects are induced at radiative levels.
The Higgs LFV effects contribute to the $\mu \to e \gamma $ decay through the diagrams called Barr-Zee type.
They are relevant in case where masses of superpartilcles are heavy,  while the CP-odd Higgs mass $m_A$ is relatively light.
In the main part of this dissertation, since we assume that only the Bino and the sleptons are light,
while the other superparticles are decoupled, the Higgs-mediated FCNCs are suppressed.

\subsection{Formulae in electron EDM}
\label{subsec:EDMformula}

The MSSM generically contains complex parameters, which cause CPV.  
Especially, CP phases in lepton sector contribute to the electric dipole moment (EDM) of the leptons.
The EDM can be calculated from the effective Lagrangian \eqref{eq:effH}.
The coefficient $F_3(0)$ can be evaluated in the same way as the muon $g-2$.
The one loop formulae for the electron EDM are given by
\begin{align}
\left\{ \frac{d_e}{e} \right\} _{\widetilde{\chi }^{\pm}} &= -\frac{1}{48\pi ^2} \sum _{X=1}^3 \sum _{A=1}^2 \frac{m_{\widetilde{\chi }^{\pm }_A}}{m^2_{\widetilde{\nu }_X}} \text{Im} \left( C^{L(e)*}_{1AX} C^{R(e)}_{1AX} \right) 
F^C_{2}\left(\frac{m^2_{\widetilde{\chi}^{\pm }_A}}{m^2_{\widetilde{\ell }_X}} \right) , \label{eq:eEDMc}  \\
\left\{ \frac{d_e}{e} \right\} _{\widetilde{\chi }^{0}} &= \frac{1}{96\pi ^2} \sum _{X=1}^6 \sum _{A=1} ^4 \frac{m_{\widetilde{\chi }^{0}_A}}{m^2_{\widetilde{\ell}_X}} \text{Im} \left( N^{L(e)*}_{1AX} N^{R(e)}_{1AX} \right) 
F^N_{2}\left(\frac{m^2_{\widetilde{\chi}^0_A}}{m^2_{\widetilde{\ell }_X}} \right), \label{eq:eEDMn}
\end{align} 
where Eqs.~\eqref{eq:eEDMc} and \eqref{eq:eEDMn} come from the chargino--sneutrino and the neutralino--slepton contributions, respectively.

Then, we discuss the CP violating phases without LFV, i.e., the slepton squared masses $m^2_{\widetilde{\ell }}$ are universal among the lepton flavor.
In this case,  the masses $m^2_{\widetilde{\ell }}$ must be real because of hermitically of Lagrangian.
Physical observables depend on CP violating phases of the following combinations\cite{Dugan:1984qf, Dimopoulos:1995kn}\footnote{
Strictly speaking, there are other conbinations, e.g., $\arg (M_i M^*_j)$ $(i \neq j)$.}
\beq
 \phi _i = \arg \left( M_i  \mu (B\mu )^* \right) , ~~~~\phi _{f} = \arg \left( A_f  \mu (B\mu )^* \right), \label{eq:phasemssm}
\eeq
where $M_i$ ($i = 1,2,3$) are gaugino masses, $A_f$ $(f = \ell , q)$ and $B\mu $ are coefficients of A- and  B-terms, respectively. 
Since these new phases do not generically vanish, and then lead to large EDMs, 
they are strongly limited by experiments. 
Although it is difficult to suppressed these phases in general SUSY models, 
we assume that they are vanished by some mechanisms, for simplicity. 

Even if $\phi _i , \phi _f \sim 0$ are realized, off-diagonal components of the lepton Yukawa matrices, 
which are transformed those of slepton mass matrices in mass eigenstates of leptons,   
could be new sources of CPVs, 
They are not suppressed if the slepton mass matrices are non-universal among the flavors because the super GIM mechanism does not work.
In Sec.~\ref{subsec:non-univ}, we will discuss sensitivities of these phases to by investigating constraints of the electron EDM.

\section{Higgs Coupling in the MSSM}
\label{sec:HiggsCmssm}
%!TEX root = ../Dthesis.tex

If the muon $g-2$ anomaly is due to the SUSY contributions, 
masses of smuons, electroweak gauginos, and higgsinos are required to be order of 100\GeV.
Further, when the slepton mass matrices are nearly universal among the flavors, staus should be light. 
Since staus have relatively large Yukawa coupling to the Higgs boson, 
the stau contribution to Higgs coupling to di-photon becomes sizable when the left-right mixing parameter of the staus is large.
In this section, we provide formulae for SUSY contributions to the Higgs coupling to di-photon.
Detailed analysis will be carried out in Chapter.~\ref{chp:main}.

The Higgs coupling to di-photon is composed of the contributions from SM particles and the MSSM.
Theoretically, $\kappa _{\gamma }$ is represented as
\beq
  \kappa_{\gamma} = 
  \frac{|\mathcal{M}_{\gamma\gamma}({\rm SM}) + \mathcal{M}_{\gamma\gamma}(\text{MSSM})|}{|\mathcal{M}_{\gamma\gamma}({\rm SM})|},
  \label{eq:kappa}
\eeq
where $\mathcal{M}_{\gamma \gamma }$ is related to the Higgs decay rate as
\beq
  \Gamma(h \to \gamma\gamma) = 
  \frac{\alpha^2m_h^3}{1024\pi^3} \left| \mathcal{M}_{\gamma\gamma} \right|^2.
\eeq

The right-hand side in Eq.~\eqref{eq:kappa} is dominated by one-loop contributions. 
The SM contributions $\mathcal{M}_{\gamma \gamma }(\text{SM})$ are composed of   
the electroweak gauge boson and the fermions at leading order. They are evaluated as\cite{Gunion:1989we}
\beq
 \mathcal{M}_{\gamma \gamma }(\text{SM}) = \frac{g_{hWW}}{m^2_W} A^h_1
 \left( \frac{m^2_h}{4m^2_W} \right) 
 + \sum _f \frac{2 g_{hff}}{m_f} N_{c,f} Q^2_f A^h_{\frac{1}{2}}
 \left( \frac{m^2_h}{4m^2_W} \right), \label{eq:kappasmcontrib}
\eeq
where the first term of the right hand side represents the contributions of the electroweak gauge boson, 
and the second terms are from fermions.
$m_h$ is the lightest (or SM-like) Higgs mass, $N_{c,f}$ and  $Q_f$ are the number of colors and the electric charge of particle $f$, respectively. 
The loop functions $A^h_1$ and $A^h_{\frac{1}{2}}$ are given by
\beq
 A^h_1 (x) &=& 2 + 3 x + 3x (2 - x) f(x), \notag \\
 A^h_{\frac{1}{2}} (x) &=& -2 x (1 + (1-x)) f(x), 
 \label{eq:loopgamma1}
\eeq
where the function $f(x)$ is 
\beq
 f(x) &=& \begin{cases}
                \arcsin ^2 \left( \sqrt{\frac{1}{x}} \right), ~~~~~~~~~~~~~~~ \text{if $x \geq 1$, } \\
                -\frac{1}{4} \left( \log \left( \frac{\eta _{+}}{\eta _{-}} \right) - i \pi \right) ^2, ~~~~\text{if $x < 1$,} 
          \end{cases}
          \label{eq:ffunc} \\
   \eta _{\pm } &\equiv& 1 \pm \sqrt{1 - x}. \label{eq:etafunc}  
\eeq
In $x \to \infty $ limit, the loop functions approach to $A^h_1 \to 7$, $A^h_{1/2} \to -3/4$, respectively.
The couplings $g_{hWW}$ and $g_{hff}$ are written as\footnote{
In the MSSM, there are additional mixing angle from the Higgs sector.
The Higgs couplings to the SM particles in the MSSM are expressed by
\beq
  g_{hWW} = \frac{g v}{\sqrt{2}} \sin (\beta - \alpha),~~~g_{hff} = \frac{m_f}{\sqrt{2}v} \frac{-\sin \alpha }{\cos \beta }, \label{eq:couplingMSSM}
\eeq
where $\alpha $ is mixing angle between the real part of the Higgs fields.
In this dissertation, we assume that the CP-odd Higgs $A$ is decoupled.
In this case, the Higgs couplings to SM particles approach to the SM values $\beta \simeq \alpha +  \frac{\pi }{2}$.
Therefore, we can neglect the MSSM contributions to the Higgs couplings to SM particles.}
\beq
 g_{hWW} = \frac{g v}{\sqrt{2}},~~~g_{hff} = \frac{m_f}{\sqrt{2}v}, \label{eq:couplingSM} 
\eeq
Numerically, the W boson contribution is leading one in the SM.
Among SM fermions, the top quark contribution is dominant because of large top Yukawa coupling.

The MSSM contributions $\mathcal{M}_{\gamma \gamma }(\text{MSSM})$ are composed of the sfermion $\mathcal{M}_{\gamma \gamma }(\widetilde{f})$, 
charged Higgs $\mathcal{M}_{\gamma \gamma }(H^{\pm}) $, and chargino contributions $\mathcal{M}_{\gamma \gamma }(\widetilde{\chi }^{\pm})$, respectively 
at the one-loop level.
They are evaluated as\cite{Gunion:1989we}
\beq 
\mathcal{M}_{\gamma \gamma }(\text{MSSM}) &=& \sum _{\widetilde{f}} \mathcal{M}_{\gamma \gamma }(\widetilde{f}) + \mathcal{M}_{\gamma \gamma }(H^{\pm}) 
+ \mathcal{M}_{\gamma \gamma }(\widetilde{\chi }^{\pm}), \notag \\
&=&  \sum _{\widetilde{f}} \frac{g_{h\widetilde{f}\widetilde{f}}}{m^2_{\widetilde{f}}} 
         N_{c,\widetilde{f}} Q^2_{\widetilde{f}} A^h_0 \left( \frac{m^2_h}{4m^2_{\widetilde{f}}} \right) 
         + \frac{g_{hH^+H^-}}{m^2_{H^{\pm}}} A^h_0 \left( \frac{m^2_h}{4 m^2_{H^{\pm}}} \right) \notag \\
         &&+ \sum _{A = 1,2} \frac{2 g_{h \widetilde{\chi }_A^{+} \widetilde{\chi }_A^{-}}  }{m_{\widetilde{\chi }^{\pm}_A} }
         A^h_{\frac{1}{2}} \left( \frac{m^2_h}{4 m^2_{\widetilde{\chi }^{\pm}_A}} \right)  
, \label{eq:kappamssmcontrib} 
\eeq 
The parameters $m_{\widetilde{f}}, m_{H^{\pm}}, m_{\widetilde{\chi }^{\pm}_A}$ are masses of the sfermions, charged Higgs bosons, and charginos, respectively. 
The loop function $A^h_0$ is given by
\beq
 A^h_0 (x) = x (1-x f(x)) . \label{eq:loopgamma2}
\eeq
It approaches to $A^h_0 \to -1/3$ in $x \to \infty $ limit.
The couplings $g_{h\widetilde{f}\widetilde{f}}, g_{hH^+H^-}, g_{h \widetilde{\chi }_A^{+} \widetilde{\chi }_A^{-}}$ are 
\beq
g_{h\widetilde{f}_{1,2}\widetilde{f}_{1,2}} &=&
  \frac{1}{2}(\delta m_{\widetilde{f}_{LL}}^2+\delta m_{\widetilde{f}_{RR}}^2) 
  \pm 
  \frac{1}{2}(\delta m_{\widetilde{f}_{LL}}^2-\delta m_{\widetilde{f}_{RR}}^2) \cos 2\theta_{\widetilde{f}} 
  \pm 
  \delta m_{\widetilde{f}{LR}}^2\, \sin 2\theta_{\widetilde{f}}, \notag \\
 g_{hH^+H^-}  &=& g \left(  m_W \sin (\beta - \alpha ) + \frac{m_Z \cos 2\beta }{2 \cos \theta _W} \sin (\alpha + \beta )\right), 
 \notag \\
 g_{h \widetilde{\chi }_A^{+} \widetilde{\chi }_A^{-}} &=&
 \frac{g}{\sqrt{2}} \left(- \left( O_L \right) _{i1} \left( O_R \right) _{2i} \sin \alpha  + \left( O_L \right) _{i2} \left( O_R \right) _{1i} \cos \alpha \right), 
 \label{eq:couplingMSSM2}
\eeq
where $O_{L,R}$ are the unitary matrices, which diagonalize the chargino mass matrix, as seen in Sec.~\ref{subsec:interaction}.
$\delta m_{\widetilde{f}_{LL,RR}}^2$ and $\delta m_{\widetilde{f}_{LR}}^2$ are defined as
\beq
  \delta m_{\widetilde{f}_{LL,RR}}^2 = 
  \frac{2}{v} (m_f^2 + D_{\widetilde{f}{L,R}}),~~~
  \delta m_{\widetilde{f}_{LR}}^2 = \frac{1}{v} m_{\widetilde{f}_{LR}}^2,
\eeq
where $D_{\widetilde{f}{L,R}}$ are D-terms, $D_{\widetilde{f}} = m^2_Z \cos 2 \beta \left( T^3_f  - Q_f \sin ^2 \theta _W \right)$, 
and $m_{\widetilde{f}_{LR}}^2$ is the left-right mixing of the sfermion mass matrices as 
\beq
  m_{\widetilde{f}_{LR}}^2 = 
  \frac{1}{2} (m_{\widetilde{f}_1}^2 - m_{\widetilde{f}_2}^2) 
  \sin 2 \theta_{\widetilde{f}}.
  \label{eq:MixingAngle}
\eeq
Here, $\theta _{\widetilde{f}}$ is mixing angles of the sfermions, defined as
\beq
  U_{\widetilde{f}} = 
  \begin{pmatrix}
  \cos\theta_{\widetilde{f}} & \sin\theta_{\widetilde{f}} \\
 -\sin\theta_{\widetilde{f}} & \cos\theta_{\widetilde{f}}
  \end{pmatrix}.
  \label{eq:UnitaryMatrix}
\eeq
The mass matrix of the sfermion is diagonalize by $U_{\widetilde{f}}$,
as $U_{\widetilde{f}} \mathcal{M}^2_{\widetilde{f}} U_{\widetilde{f}}^{\dagger} = \text{diag}(m^2_{\widetilde{f}_1}, m^2_{\widetilde{f}_2})$.
Here, we neglect effects of flavor/CP violations.
These contributions are expected to be suppressed since they are strongly constrained by lepton flavor experiments.

The contributions of the charged Higgs bosons and the charginos becomes relevant when these masses are order of 100\GeV.
For charged Higgs, the contributions are negligible since we assume that the CP-odd Higgs mass $m_A $ is decoupled.
This assumption is reasonable since none of them is discovered.
On the other hand, mass regions at several hundred \GeV ~for charginos are still remained.
As we will discuss in Sec.~\ref{sec:stau}, 
the chargino contributions to $\kappa _{\gamma }$ might become $\mathcal{O}(1)\%$.
In this dissertation, we take them as theoretical uncertainty.
Detailed estimation will be discussed in Sec.~\ref{sec:stau}.

%%%%%%%%%%%%%%%%%%%%%%%%%%%%%%%%%%%%%%%%%%% diagram
\begin{figure}[t]
 \begin{center}
  \begin{fmffile}{feyns/kappa}
\begin{fmfgraph*}(100,100)
 \fmfleft{h1}
 \fmfright{g1,g2}
 \fmf{dashes}{h1,v1}
  \fmf{dashes, left, tension=0.3, lab=${\widetilde{\tau}}_R$}{v1,v2}
  \fmf{dashes, left, tension=0.3, lab=${\widetilde{\tau}}_L$}{v2,v1}
  \fmf{phantom}{v2,gg1}
  \fmf{phantom}{v2,gg2}
 \fmf{photon}{gg1,g1}
 \fmf{photon}{gg2,g2}
\fmfv{decor.shape=cross, decor.size=10}{v2}
 \fmflabel{$h$}{h1}
 \fmflabel{$\gamma $}{g1}
 \fmflabel{$\gamma $}{g2}
 \end{fmfgraph*}
 \end{fmffile}
 \caption[Feynman diagram of the stau contribution the Higgs coupling to di-photon.]
 {Feynman diagram of the stau contribution to the Higgs coupling to di-photon.}
\label{fig:kappamssm}
\end{center}
\end{figure}
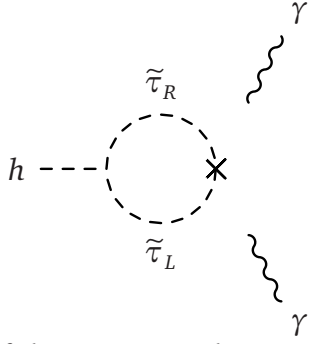
%%%%%%%%%%%%%%%%%%%%%%%%%%%%%%%%%%%%%%%% 
Then, let us consider the contributions of sfermions.
Light sfermions with large left-right mixing, i.e., the stop, sbottom, and stau, can provide large contributions to the Higgs coupling to di-photon\cite{Carena:2011aa}.
We assume that the colored superparticles are heavy enough to evade current LHC bound.
In our assumption, stau may have a large contribution to the Higgs coupling.
Fig.~\ref{fig:kappamssm} shows the leading stau contribution to Higgs coupling to di-photon amplitude in large left-right mixing region.
In this case, the leading stau corection is estimated as\cite{Kitahara:2012pb}
\beq
 \delta \kappa _{\gamma } \equiv \kappa _{\gamma } -1 \simeq - \sum _{i =1,2}
 \frac{0.05 \times \left(m^2_{\widetilde{\tau }_{LR}}\right)^2}{m^2_{\widetilde{\tau}_i} (m^2_{\widetilde{\tau}_1} - m^2_{\widetilde{\tau}_2})}, \label{eq:estimatekappa}
\eeq
where this approximation is applicable when the left-right mixing of the stau are extremely large (or $m^2_{\widetilde{\tau}_2} \gg m^2_{\widetilde{\tau}_1}$).
If $m^2_{\widetilde{\tau }_{LR}} \simeq \mathcal{O}(m^2_{\widetilde{\tau }})$, i.e., the off-diagonal component of the stau mass matrices is comparable to the diagonal parts, 
$\mathcal{O}(10)$\% correction to $\delta \kappa _{\gamma }$ can be realized.

However, extremely large left-right mixing might suffer from vacuum instability\cite{Rattazzi:1996fb,Casas:1995pd}.
As the left-right mixing increases, new charge-breaking minima become deeper, and our electroweak vacuum could decay them.
By requiring that the lifetime of our vacuum should be longer than the age of the Universe, the left-right mixing is bounded from above\cite{Hisano:2010re,Carena:2012mw}.
Hence, the stau contribution to the Higgs coupling to di-photon is also limited.
The upper limit is studied by several theoretical groups\cite{Sato:2012bf,Kitahara:2012pb,Carena:2012mw}, 
and becomes about 15\%.

In Chapter~\ref{chp:main}, we will study allowed stau mass region by the Higgs coupling to di-photon $\kappa _{\gamma }$, 
which is limited by the vacuum stability condition of stau--Higgs potential.
Then, future prospects for the stau based on above mass region are discussed.

\section{Current Status}
\label{sec:cstatus}
%!TEX root = ../Dthesis.tex

Supersymmetry, which provide a solution for the hierarchy problem, is required that the masses of superparticles are not so far above the electroweak scale, 
and has long been the leading candidate for new physics. 
This has motivated searches of superparticles at colliders, such as the Large Hadron Collider (LHC).
Particularly, the main searches has been for strong-interactiong superparticles (colored superparticles), such as gluinos and squarks,
because of large production cross section.
They are produced through $p p \to \widetilde{g}\widetilde{g}, \widetilde{g}\widetilde{q}, \widetilde{q}\widetilde{q}$.
Although these limit depend on the decays, masses ($m_{\widetilde{g}}$, $m_{\widetilde{q}}$) is strongly limited by current LHC data, 
in case assuming some of simplified models (or decay modes). 

%%%%%%%%%%%%%%%%%%%%%%%%%%%%%%%%%%%%%%%%%%%%% Figure
\begin{figure}[t]
 \begin{center}
 \includegraphics[width=15cm]{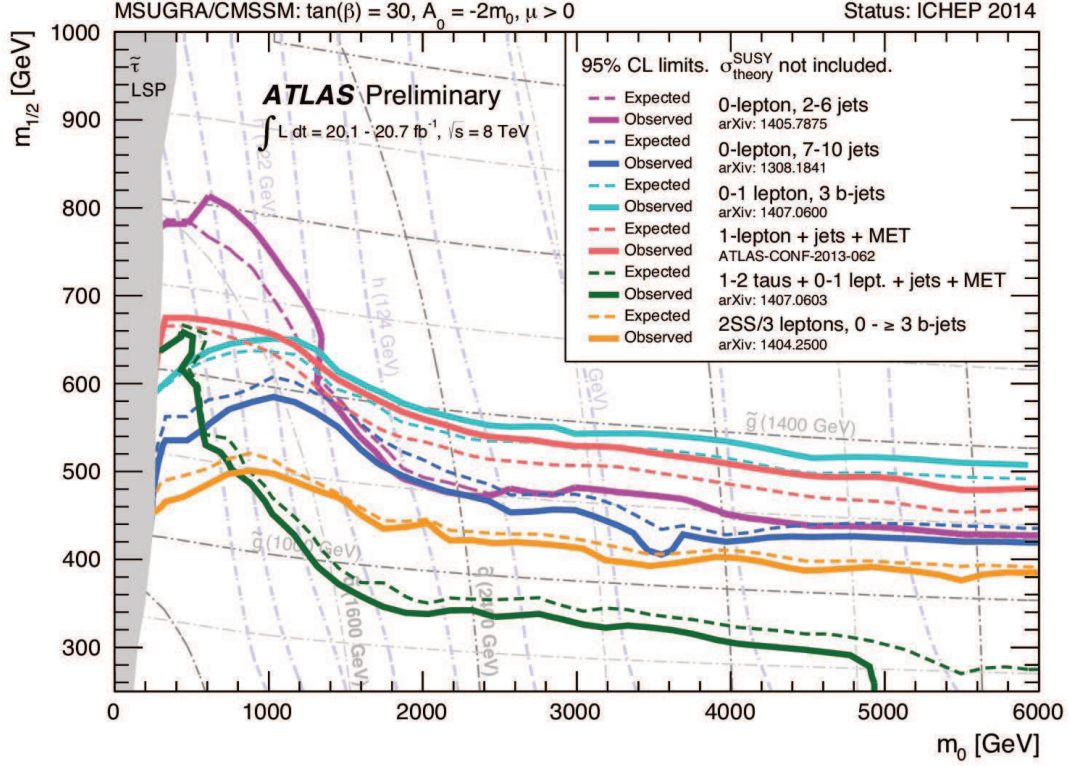}
 \end{center}
 \caption[Current status for mSUGRA.]
 {
Constraints on mSUGRA/CMSSM from ATLAS collaboration at LHC with $\mathcal{L} \simeq 20\invfb$ and $\sqrt{s} = 8\TeV$.
$A_0 = -2m_0$, $\tan \beta = 30$, $\mu > 0$ are taken to realize the Higgs mass $m_h \simeq 126\GeV$.
In this figure, ($m_0$, $m_{1/2}$) plane of mSUGRA is limited from the jets + $\Slash{E}_{T}$.
This figure is from Ref.~\cite{Mitrevski:1955518}.
}
\label{fig:mSUGRAconstraint}
\end{figure}
%%%%%%%%%%%%%%%%%%%%%%%%%%%%%%%%%%%%%%%%%%%%%

One of the benchmark models is the framework of minimal supergravity (mSUGRA), also knew as the constrained MSSM (CMSSM).
It has 5 parameters, a unified scalar mass $m_0$, a unified gaugino mass $m_{1/2}$, 
a unified trilinear scalar coupling $A_0$, the ratio of Higgs VEVs $\tan \beta $, and one discrete choice, the sign of $\mu $.
Fig.~\ref{fig:mSUGRAconstraint} shows summary on constraints in mSUGRA/CMSSM.
This results imply $m_{\widetilde{g}} \gtrsim 1.4\TeV$ and $m_{\widetilde{q}} \gtrsim 1.6\TeV$\cite{Mitrevski:1955518}.

Current mass bound of gluinos and squarks, $m_{\widetilde{g}, \widetilde{q}} \gtrsim 1\TeV$, is consistent with the Higgs mass of 126\GeV.
In the MSSM, the Higgs mass at tree level is about the $Z$ boson mass, $m^{2}_h \leq m^2_Z \cos ^2 2\beta$, 
because the quartic terms in the Higgs potential is dominated by gauge couplings. 
This contradicts the LEP bound\cite{Barate:2003sz}, $m_h \geq 114.4\GeV$.

The relation to the Higgs mass receives quantum corrections.
Relevant contributions come from top and stop loops.
If the corrections are comparable to the tree level contribution, the Higgs mass of about 126\GeV ~can be realized.
At one-loop order and taking in the decoupling limit ($m_A \gg m_Z$), the Higgs mass is evaluated as\cite{Okada:1990vk, Ellis:1990nz, Haber:1990aw, Ellis:1991zd, Okada:1990gg}
\beq
 m^2_h &\simeq&  m^2_Z \cos ^2 2\beta + \frac{3}{4\pi ^2} Y^2_t  \sin ^2\beta \left[ m^2_t \log \frac{m_{\widetilde{t}_1}m_{\widetilde{t}_2}}{m^2_t} 
 + \cos ^2 \theta _{\widetilde{t}} \sin ^2 \theta _{\widetilde{t}} 
 \left( m^2_{\widetilde{t}_2} -  m^2_{\widetilde{t}_1} \right) \log \frac{m^2_{\widetilde{t}_2}}{m^2_{\widetilde{t}_1}} \right. \notag \\
 &&+   \left.   \cos ^4 \theta _{\widetilde{t}} \sin ^4 \theta _{\widetilde{t}}  
 \left\{  \left( m^2_{\widetilde{t}_2} -  m^2_{\widetilde{t}_1} \right) ^2 - 
 \frac{1}{2}  \left( m^4_{\widetilde{t}_2} -  m^4_{\widetilde{t}_1} \right) \log \frac{m^2_{\widetilde{t}_2}}{m^2_{\widetilde{t}_1}}   \right\} / m^2_t \right] \notag \\
 &\sim& m^2_Z \cos ^2 2\beta + \frac{3}{4\pi ^2} Y^2_t  \sin ^2\beta \left[ m^2_t \log \frac{M^2_S}{m^2_t} 
 + \frac{X^2_t}{M^2_S} \left( 1- \frac{X^2_t}{12 M^2_S}\right) + \cdots \right] , \label{eq:mssmhiggsmass}
\eeq
where $M_S = \sqrt{m_{\widetilde{t}_1} {m_{\widetilde{t}_2}}}$ and $X_t = A_t - \mu \cot \beta$.

From Eq.~\eqref{eq:mssmhiggsmass},\footnote{
In fact, in order to evaluate the Higgs mass precisely, it is also important to take into account higher order corrections.
However, qualitative consequence does not change.} it is found that there are two possibilities to enhance the Higgs boson mass as follows, 
\begin{itemize}
\item $M_S \gg 1\TeV$, $X_t \simeq 0$ (Heavy scalar top)\cite{Okada:1990vk, Ellis:1990nz, Haber:1990aw, Ellis:1991zd} 
\item $M_S \simeq 1\TeV$, $X_t \simeq \sqrt{6} M_S$ (Maximal mixing)\cite{Okada:1990gg} 
\end{itemize}
Anyway, $M_S \gtrsim 1\TeV$ is favored to explain the Higgs mass, and is consistent with current mass bound of colored superparticles.

These two phenomenological requirements, (A) the mass bound of colored superparticles is $\gtrsim 1\TeV$, (B) the Higgs boson mass is $126\GeV$, respectively, 
naively conflict with  (C) the indication of the muon $g-2$ anomaly where superparticles have a mass of $\mathcal{O}(100)\GeV$, as seen in Sec.~\ref{sec:gm2mssm}.
In fact, some of the representative SUSY-breaking mediation mechanism such as mSUGRA and minimal GMSB cannot satisfy (A)--(C) simultaneously\cite{Endo:2011gy}.
There are two possibilities to solve above inconsistency as follows, 
\begin{enumerate}
\item[(i)] 
Model building (extending MSSM).  
\item[(ii)]
Model-independent approach. Particularly, we consider a split mass spectrum of the superparticles as
\beq
 m_{\widetilde{g}}, m_{\widetilde{q}} \gg m_{\widetilde{ \ell }}, m_{\widetilde{W}}, m_{\widetilde{B}}, m_{\widetilde{H}}, \label{eq:splitmassspec}
\eeq
\end{enumerate}
where the squarks are heavy enough to explain the Higgs boson mass and satisfy the current LHC limit, 
while the non-colored superparticles are light to solve the muon $g-2$ anomaly.

In the case (i), some theoretical groups have proposed extended models to explain the muon $g-2$ and the Higgs mass simultaneously.
For example, an extension of MSSM with vector-like supermultiplets\footnote{
Other extensions are additional U(1)$^{\prime}$ gauge symmetry imposed on the Higgs fields\cite{Endo:2011gy}, 
a large LR mixing of the stops by introducing a Higgs-messenger mixing\cite{Evans:2011bea, Evans:2012hg}, 
splitting F-terms between colored and non-colored messengers\cite{Ibe:2012qu}, and so on.
} 
was considered to increase the Higgs boson mass by extra contributions\cite{Moroi:1992zk, Babu:2004xg,Babu:2008ge, Martin:2009bg,Asano:2011zt}, 
while explaining the discrepancy to the muon $g-2$\cite{Endo:2011mc, Endo:2011xq}.
Then, the collider phenomenology were studied in Ref.~\cite{Endo:2012cc}.

%%%%%%%%%%%%%%%%%%%%%%%%%%%%%%%%%%%%%%%%%%%%% Figure
\begin{figure}[t]
 \begin{center}
 \includegraphics[width=8cm]{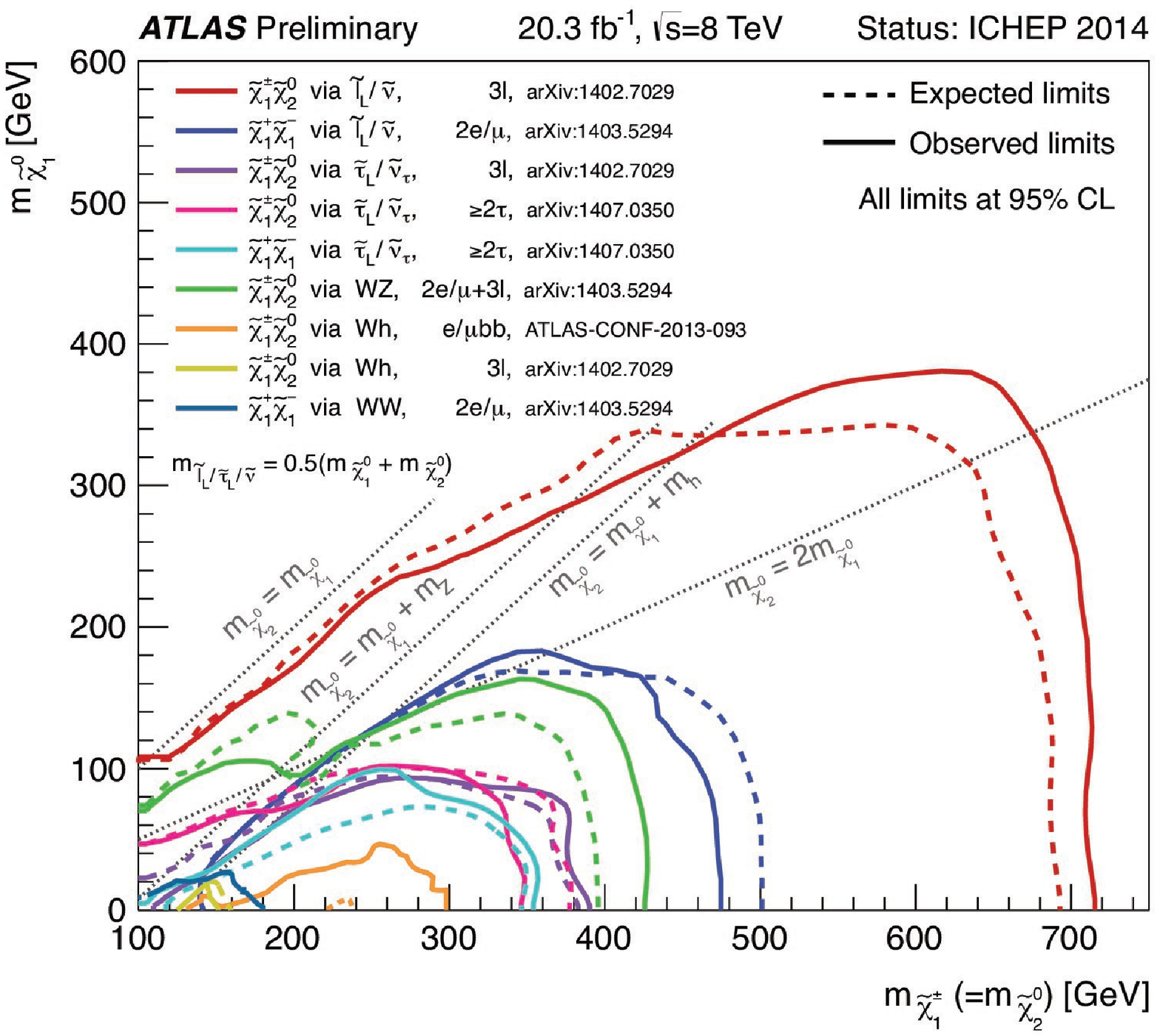}
  \includegraphics[width=8cm]{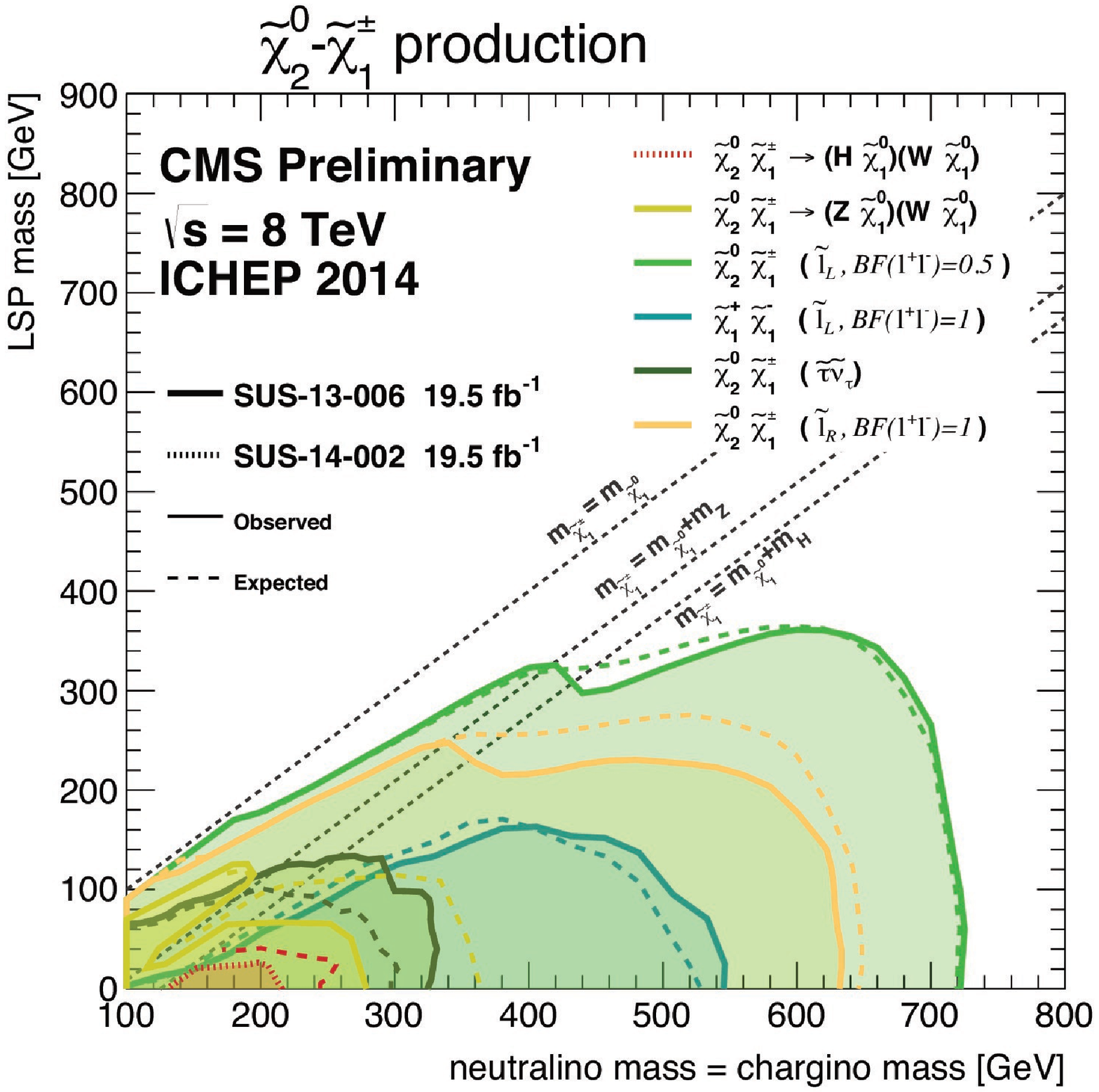}
 \end{center}
 \caption[Mass bound of the electroweak gauginos.]
 {
 Mass bound of the electroweak gauginos, which are produced through $p p \to \widetilde{\chi}^{\pm, 0} \widetilde{\chi } ^{\pm, 0}$,  
 by the ATLAS (left) and the CMS (right) collaborations, respectively. 
 These figures are from Refs.~\cite{Santoyo-Castillo:1952963,CMS:summary14}.
}
\label{fig:EWinoconstraint}
\end{figure}
%%%%%%%%%%%%%%%%%%%%%%%%%%%%%%%%%%%%%%%%%%%%%
The scenario (ii) forces us to change strategy of superparticle search.
The standard SUSY searches are based on productions of colored superparticles as mentioned in the beginning of this section.
In this case, they are not always promising because the colored superparticles could be away from the LHC reach. 
On the contrary, the searches for weak interacting superparticles (non-colored superparticles), 
such as electroweak gauginos, Higgsinos, and sleptons, become important.
If the muon $g-2$ anomaly is true, non-colored superparticles couple to the muon should have masses of $\mathcal{O}(100)\GeV$, 
and are expected to be produced at LHC, through direct productions such as 
$p p \to \widetilde{\chi}^{\pm, 0} \widetilde{\chi } ^{\pm, 0}, \widetilde{\ell} \widetilde{\ell}$. 

Now, let us review the searches for electroweak gauginos (chagrinos and neutralinos). 
The ATLAS\cite{Aad:2014nua, Aad:2014vma} and the CMS\cite{Khachatryan:2014qwa} collaborations respectively 
reported results of their searches for above direct productions.
The mass bound of the electroweak gauginos is shown in Fig.~\ref{fig:EWinoconstraint}.
The left panel of Fig.~\ref{fig:EWinoconstraint} is result of ATLAS collaboration, the right is from CMS.
Although the results strongly depend on the mass spectrum of non-colored superparticles, 
the mass regions for the superparticles are beginning to be limited by current LHC data.
%\footnote{
%Note that this is somewhat misunderstanding result.
%Constraints of non-colored superparticles strongly depend on their mass spectrum.
%For example, the red line is assumed that $m_{\widetilde{\ell }} = (m_{\widetilde{\chi }_1} + m_{\widetilde{\chi }_2}) /2$.
%In this case, since the signal multilepton + $\Slash{E}_{T}$ is enhanced. it is strongly limited.
%In general SUSY models, the mass limits of non-colored superparticles are still weak.}.

%%%%%%%%%%%%%%%%%%%%%%%%%%%%%%%%%%%%%%%%%%%%% Figure
\begin{figure}[t]
 \begin{center}
 \includegraphics[width=10cm]{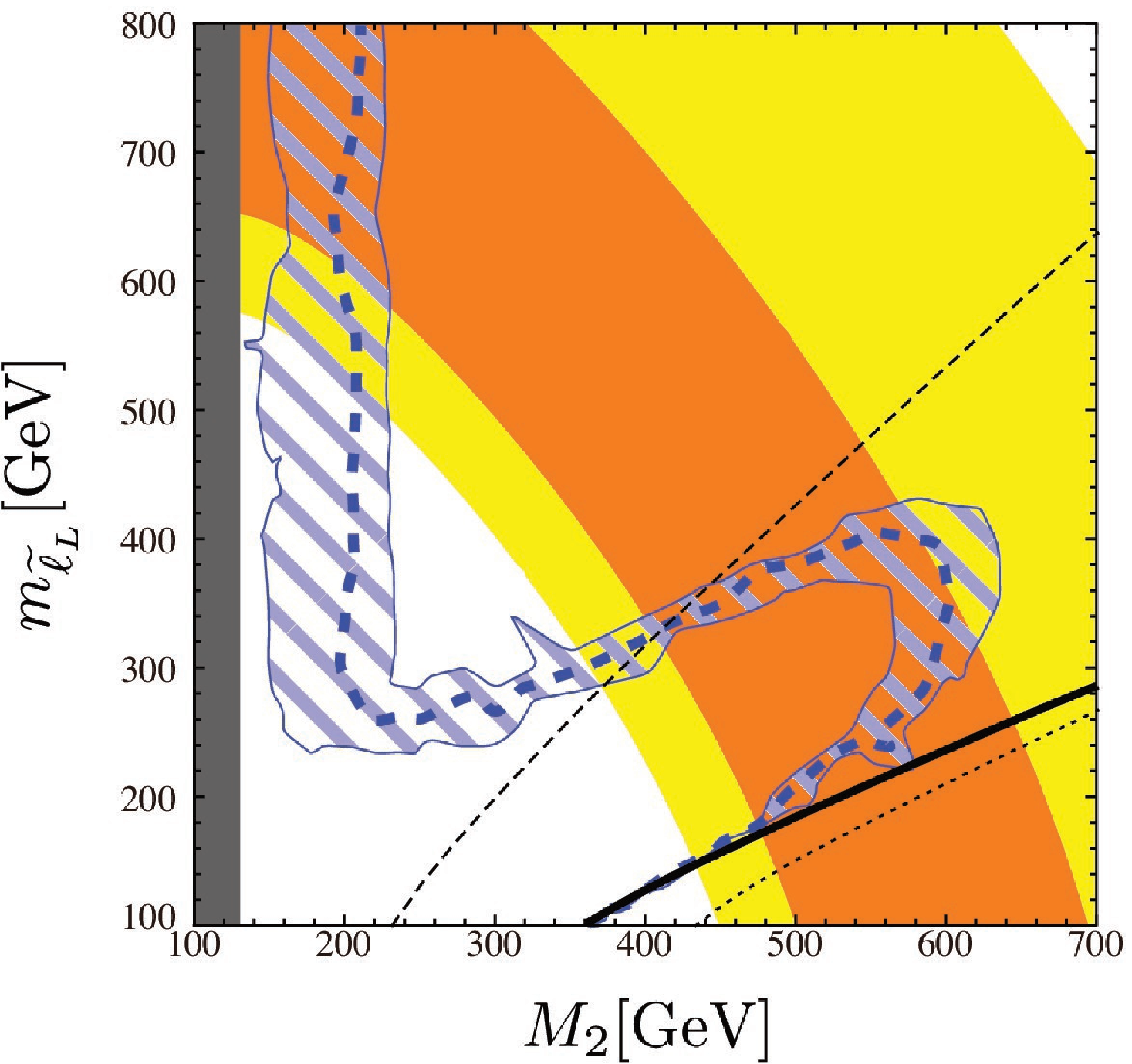}
 \end{center}
 \caption[Current LHC bounds on the SUSY $g-2$ explanations.]
 {
 Current LHC bounds on the SUSY $g-2$ explanations\cite{Endo:2013bba}, 
 by the production $p p \to \widetilde{\chi }^{\pm }_1 \widetilde{\chi }^{0}_2 \to 3 \ell + \Slash{E}_T$ 
 in the data of $\int \mathcal{L} = 13.0\invfb$\cite{ATLAS-CONF-2012-154}.  
 The parameters are taken $\mu = M_2 = 2M_1$, $m_{\widetilde{\ell }_R} = 3\TeV$, and $\tan \beta =40$, respectively.
 In the orange (yellow) regions, the muon $g-2$ discrepancy is explained at $1\sigma$ ($2\sigma$) level, 
 where $1\sigma$ ($2\sigma$) mean that the SUSY contribution to the muon $g-2$ are $(26.1 \pm 8.0) \times 10^{-10}$ $((26.1 \pm 16.0) \times 10^{-10} )$, 
 as will see in Sec.~\ref{subsec:gm2vac}. 
 The blue dotted line represents the LHC exclusion limit at 95\% CL.
 Including 30\% errors of Parton Distribution Function (PDF) and scale uncertainties, the limit distribute in the blue hatched regions.
 The LSP is $\widetilde{\chi }^0_1$ in the regions above the black thick lines, while the sneutrino is lightest below them.
 Sleptons are lighter than $\widetilde{\chi }^0_1$ ($\widetilde{\chi }^0_2$) below the black dotted (dashed) lines, respectively.
 We find that the muon $g-2$ regions are partly excluded by current LHC data.}
\label{fig:EWgm2}
\end{figure}
%%%%%%%%%%%%%%%%%%%%%%%%%%%%%%%%%%%%%%%%%%%%%

LHC phenomenology of the superparticles which are relevant for the muon $g-2$ has been also studied in Ref.~\cite{Endo:2013bba}.
 \footnote{
 This is previous our study. For detailed discussion, see Ref.~\cite{Endo:2013bba}.}
 Fig.~\ref{fig:EWgm2} shows the LHC exclusion limits with the blue dotted lines at 95\% CL, 
 where the SUSY contributions to the muon $g-2$ are dominated by chargino-muon sneutrino diagrams.
 When the slepton are heavier than the Bino, while are lighter than the Wino, 
 a signal of 3 leptons plus large missing transverse energy is enhanced.
 Thus, it is found that the muon $g-2$ regions are partly excluded by current LHC data, and 
 it has been shown that almost all parameter regions are expected to be probed in near future by LHC at $14\TeV$.
 
However, the searches in Ref.~\cite{Endo:2013bba} rely on the assumption that Wino is light.
This assumption is not always necessary to explain the muon $g-2$ discrepancy, 
when the SUSY contributions is mainly from neutralino-smuon diagrams in Fig.~\ref{fig:gm2mass} of Sec.~\ref{sec:gm2mssm}.
Among the neutralino contributions, Bino--smuon diagram as Fig.~\ref{fig:gm2gauge} (b) of Sec.~\ref{sec:gm2mssm} 
is proportional to the left-right mixing of the slepton.
When only the Bino and the sleptons are light, and the left-right mixing is large, the Bino-smuon contribution is enhanced.
Eventually, we notice that the models in which only the Bino and the sleptons are light, are "minimal"  SUSY models 
to solve the muon $g-2$ anomaly.

If the Wino is heavy, collider searches differ significantly from those in Ref.~\cite{Endo:2013bba}, 
thus it is important to investigate the phenomenology of the "minimal" SUSY models to explain the muon $g-2$ discrepancy.
In the next chapter, we will study phenomenology of this "minimal" models, 
where only the Bino and the sleptons are light.

\chapter{Prospects for Slepton Searches}
\label{chp:main}
%!TEX root = ../Dthesis.tex

This chapter is the main part of this dissertation and is based on \cite{Kitahara:2013lfa, Endo:2013lva, Endo:2014pja}.
In Sec.~\ref{sec:foundation}, we introduce an effective theory in which only the Bino and the sleptons are light, 
while the other superparticles are decoupled.
In this setup, the left-right mixing becomes large in order to solve the muon $g-2$ anomaly.
Such a large mixing spoils the stability the  electroweak symmetry breaking vacuum. 
We also discuss the vacuum meta-stability condition of the Higg-slepton potential 
and the slepton mass bound.

Searches for sleptons depend on the mass spectrum of the sleptons.
We will show the left-right mixing of the slepton is constrained by the vacuum meta-stability condition 
of the slepton in Sec.~\ref{subsec:vacuum}.
Then, it is found that the slepton mass is bounded from above in order to explain the muon $g-2$.
Since the most severe constraint is obtained from the stau stability condition because of large tau Yukawa coupling, 
we consider the two following mass spectrum, 
\begin{enumerate}
\item[(1)] Universal mass spectrum (Sec.~\ref{subsec:universal}, \ref{sec:stau})
\beq
 m_{\widetilde{e}} = m_{\widetilde{\mu }} = m_{\widetilde{\tau }}. \label{eq:univmass}
\eeq
\item[(2)] Non-universal mass spectrum (Sec.~\ref{subsec:non-univ})
\beq
 m_{\widetilde{e}} = m_{\widetilde{\mu }} \ll  m_{\widetilde{\tau }}. \label{eq:nonunivmass}
\eeq
\end{enumerate}

In the Universal case (1), we will show that smuons are limited to be $m_{\widetilde{\mu }_1} \lesssim 500\GeV$ to solve the muon $g-2$ anomaly.
This is within kinematical reach of the LHC and/or ILC sensitivity.
The searches is discussed in Sec.~\ref{subsec:universal}.
Furthermore, light staus also affect the Higgs coupling to di-photon $\kappa _{\gamma }$,
whose sensitivity is expected to be improved by HL-LHC and ILC.
Once the excess from the SM prediction of $\kappa _{\gamma }$ is observed, 
we can estimate not only the stau contribution to $\kappa _{\gamma }$, but also their masses and mixing.
In Sec.~\ref{sec:stau}, we discuss the stau searches in ILC.

On the other hand, the vacuum condition is relaxed when staus are much larger than other sleptons.
In this case, smuon masses of $\mathcal{O}(1)\TeV$ are allowed to solve the muon $g-2$ anomaly, 
and outside the kinematical reach of HL-LHC and ILC.
However, such a non-universal mass spectrum generically induces sizable lepton flavor violation and the CP violation. 
In Sec.~\ref{subsec:non-univ}, we discuss the sensitivity to the future lepton flavor experiments.
%Finally, we summarize in Sec.~\ref{sec:summary} 

\section{Foundation}
\label{sec:foundation}
%!TEX root = ../Dthesis.tex

\subsection{Setup}
\label{subsec:setup}
%!TEX root = ../Dthesis.tex

In this chapter, we consider a low-energy effective theory, in which only the following superparticles are light,
\beq
  \tilde B,~\tilde{\ell}_L,~\tilde{\ell}_R.
  \label{eq:particles}
\eeq
Here, $\tilde{\ell}$ denotes the selectron and the smuon (and the stau, depending on the mass spectrum).
The model parameters are as follows, 
\beq
M_1, m_{\tilde\ell_L}^2, m_{\tilde\ell_R}^2, m_{\tilde\ell_{LR}}^2.
\label{eq:parameters}
\eeq
Here, $M_1$ is the Bino mass. 
On the other hand, $m_{\tilde\ell_L}^2$ and $m_{\tilde\ell_R}^2$ are soft SUSY-breaking masses of the left- and right-handed sleptons, respectively. 
$m_{\tilde\ell_{LR}}^2$ is off-diagonal components of the slepton mass matrices. 
This is the "minimal" SUSY model to solve the muon $g-2$ anomaly.

All colored superparticles are set to be very heavy. In fact, none of them have been discovered at LHC. 
The Higgs boson mass of $126\GeV$ favors the scalar top masses to be $\Order(10-100)\TeV$,
if the trilinear coupling of the top squark is suppressed. 
Similarly, the heavy Higgs bosons of the two Higgs doublets are assumed to be heavy for simplicity. 
In this dissertation, all of them are considered to be decoupled, 
although the results of this dissertation are almost independent on their mass spectrum.\footnote{
The non-decoupling effect of the heavy superparticles appears the Bino coupling with the smuons as will see in Sec.~\ref{subsec:gm2vac}.
They yield correction of 5--10\% to the SUSY one-loop contribution to the muon $g-2$.
In this dissertation, masses of the heavy superparticles are set to 30\TeV.}

In this dissertation, we consider the case that the Higgsino mass is large.
The SUSY contributions to the muon $g-2$ are dominated by 
the Wino--Higgsino--muon sneutrino and/or the Bino--smuon contributions.
In our setup, the Bino-smuon contribution is dominant in order to explain the discrepancy of the muon $g-2$.
It is enhanced by the left-right mixing of the smuon, which is determined by the muon Yukawa coupling, 
the Higgsino mass $\mu $, and $\tan \beta$.
Since too large $\tan \beta $ spoils perturbativity of the down-type Yukawa interaction, $\mu $ is favored to be large.  
Therefore, we focus on the large Higgsino mass region.

When $M_2$ and $\mu $ $> m_{\widetilde{\ell }_L}$, the chargino--muon sneutrino contribution to the muon $g-2$ is proportional to $1/ M_2 \mu $, 
where $M_2$ is the Wino mass.
Thus, it is suppressed by large ($M_2$, $\mu $) case.
The Winos are also supposed to be decoupled for simplicity.\footnote{
Note that the setup of this dissertation is meaningful when the Wino is decoupled.
When $M_2 \lesssim 1\TeV$, the chargino--muon sneutrino contribution can be effective, 
and cannot be neglected.
Therefore, the analysis dramatically changes due to large theoretical uncertainty in this case\cite{Endo:2013bba}.
In this dissertation, we set to $M_2 = 30\TeV$ in order to suppress the chargino--muon sneutrino contribution. 
}
 
Before proceeding, let us comment on the left-right mixing of the slepton. It includes the scalar trilinear coupling of the slepton, 
$A_\ell$, as well as $\mu$ and $\tan\beta$. 
The Bino--smuon contribution to the muon $g-2$ could be enhanced by $A_\mu$ with $\mu\tan\beta$ kept small. 
However, this requires $A_\mu$ to be extraordinary large. If the trilinear coupling is universal among the matter scalar fermions, this implies extremely large trilinear coupling for the stop sector, resulting in either too large Higgs boson mass or rapid decays of our electroweak vacuum into charge/color breaking vacua. In this dissertation, $A_\ell $ is set to be zero for simplicity, and the left-right mixing of the smuon is determined by $\mu$ and $\tan\beta$.

\subsection{Muon $g-2$}
\label{subsec:gm2vac}
%!TEX root = ../Dthesis.tex

%%%%%%%%%%%%%%%%%%%%%%%%%%%%%%%%%%%%%%%%%%%%% Figure
\begin{figure}[tbp]
 \begin{center}
 \includegraphics[width=10cm]{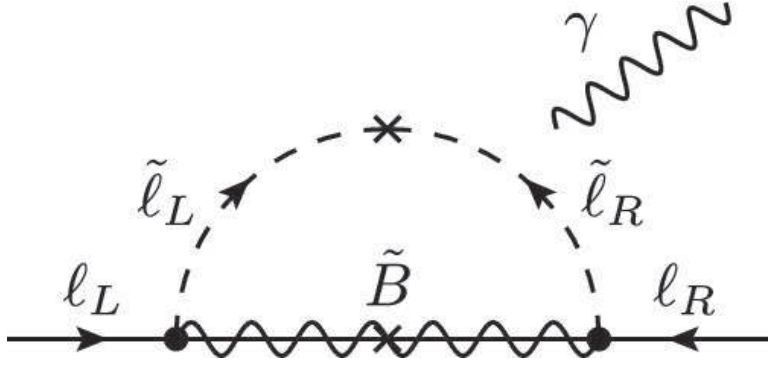}
 \end{center}
 \caption[The Bino--smuon contribution to the magnetic dipole operators.]
 {The Bino--smuon contribution to the magnetic dipole operators.}
 \label{fig:bino}
\end{figure}
%%%%%%%%%%%%%%%%%%%%%%%%%%%%%%%%%%%%%%%%%%%%%

In this section, we discuss the SUSY contribution to the muon $g-2$ in the minimal SUSY model.
The feynman diagram for the Bino--smuon contribution to the muon $g-2$ is shown in Fig.~\ref{fig:bino}, where the lepton $\ell = \mu $. 
In the mass insertion approximation, it is estimated as
\beq
 a_{\mu}({\rm SUSY}) 
 &=& 
 -(1 + \delta^{\rm 2loop}) \frac{\alpha_Y}{4\pi}
 \frac{ m_\mu M_1 m_{\widetilde\mu_{LR}}^2 }{m^2 _{\widetilde{\mu}_{L}}m^2 _{\widetilde{\mu}_{R}}}\,
 f_N \left( \frac{m^2_{\widetilde{\mu}_{L}}}{M^2_1},\frac{m^2 _{\widetilde{\mu}_{R}}}{M^2_1} \right)
 \notag \\ 
 &=& 
 \frac{1 + \delta^{\rm 2loop}}{1 + \Delta_\mu} \frac{\alpha_Y}{4\pi}
 \frac{ m_{\mu }^2  M_1 \mu }{m^2 _{\widetilde{\mu}_{L}}m^2 _{\widetilde{\mu}_{R}}} \tan \beta 
 \cdot f_N \left( \frac{m^2_{\widetilde{\mu}_{L}}}{M^2_1},\frac{m^2 _{\widetilde{\mu}_{R}}}{M^2_1} \right)
 \notag \\
 &\simeq&
1.5\times 10^{-9} ~\frac{1 + \delta^{\rm 2loop}}{1 + \Delta_\mu}
\left(\frac{\tan\beta}{10}\right)
\left(\frac{(100\GeV)^2}{m_{\widetilde{\mu }_L}^2 m_{\widetilde{\mu }_R}^2/M_1\mu }\right)
\left(\frac{f_N}{1/6}\right),
\label{eq:gminus2bino}
\eeq
at the leading order of $\tan \beta$. The loop function $f_N(x,y)$ is defined as Eqs.~\eqref{eq:fnma}.
It is noticed that the contribution is proportional to the left-right mixing of the slepton, $m_{\widetilde\ell_{LR}}^2$.
\beq
m_{\widetilde\ell_{LR}}^2 
= - Y_\ell\, v_u \mu
= - \frac{m_\ell}{1+\Delta_\ell} \mu \tan\beta,
\label{eq:LRmixing}
\eeq
where $v_u$ is the vacuum expectation value (VEV) of the up-type Higgs field $H_u$, 
and $\tan \beta$ is the ratio of Higgs VEVs, $\tan \beta \equiv \langle H_u \rangle /  \langle H_d \rangle $.
The parameter $\Delta_\ell$ is a correction to the lepton Yukawa coupling constant\cite{Marchetti:2008hw}.
It appears when the lepton Yukawa coupling in the MSSM is matched to the physical lepton mass $m_{\ell}$ or the lepton Yukawa coupling in the SM.
In the low-energy effective theory, it is evaluated as
\beq
 \Delta _{\ell } \simeq \frac{\alpha _Y}{4\pi } M_1 \mu \tan \beta \cdot I \left( M^2_1, m^2_{\widetilde{\ell}_L}, m^2_{\widetilde{\ell}_R} \right),
 \label{eq:Yukawacorr}
\eeq
where the diagrams including Higgsino propagator are neglected, because they are suppressed in large $\mu $ regions. 
Also, no-$\tan \beta$ enhanced terms are discarded.
The loop function $I(a,b,c)$ is defined as
\beq
 I(a,b,c) = - \frac{ab \log \left(\frac{a}{b}\right) + bc \log \left(\frac{b}{c}\right) +ca \log \left( \frac{c}{a} \right) }{(a-b)(b-c)(c-a)}.
 \label{eq:Yukawacorrloop}
\eeq
In particular, when $\mu \tan \beta $ is very large, 
$\Delta _{\mu }$ becomes as large as or larger than $\mathcal{O}(0.1 - 1)$.

The correction $\delta^{\rm 2 loop}$ denotes leading contributions of higher order corrections. 
It is estimated as
\begin{align}
1 + \delta^{\rm 2 loop} = 
\left(1 - \frac{4\alpha}{\pi} \ln \frac{m_{\rm soft}}{m_\mu} \right)
\left[1 + \frac{1}{4\pi}
  \left( 2\alpha_Y\Delta b + \frac{9}{4}\alpha_2 \right) \ln \frac{M_{\rm soft}}{m_{\rm soft}} 
\right].
\label{eq:2-loop}
\end{align}
In the right-hand side, the first bracket is QED corrections to the muon $g-2$\cite{vonWeitershausen:2010zr}.
They can be estimated by renormalization group contributions the magnetic dipole operator from the (slepton or Bino) soft mass scale $m_{\text{soft}}$ to the muon mass scale.
In the numerical analysis, $m_{\rm soft}$ is chosen to be the smuon mass, and this correction is $\sim 10\%$. 
Non-logarithmic terms evaluated in Ref.~\cite{vonWeitershausen:2010zr} are found to be very small in our parameter regions. 

The second bracket in Eq.\eqref{eq:2-loop} is corrections to the Bino couplings with the smuons.\footnote{
Ref.~\cite{Fargnoli:2013zda} also studied non-decoupling two-loop contributions from sfermions,
which partially overlaps with our discussion. Our results are consistent with those of Ref.~\cite{Fargnoli:2013zda}.}
When SUSY is unbroken, the gaugino coupling is equal to the gauge coupling constant thanks to SUSY. 
This equality is violated after heavy superparticles are decoupled at a scale $M_{\rm soft}$.
The Bino--muon--smuon interactions are
\begin{align}
 \mathcal{L}_{\text{int}} = -\frac{1}{\sqrt{2}}\widetilde{g}_L\, \overline{\widetilde{B}} \mu _L\, \widetilde{\mu }^*_L + \sqrt{2} \widetilde{g}_R\, \overline{\widetilde{B}} \mu _R\, \widetilde{\mu}_R^* + {\rm h.c.},
\end{align}
where the coefficients are the Bino couplings as
\begin{align}
\widetilde g_L &= g_Y + \delta \widetilde g_L \simeq 
g_Y 
\left[1 + \frac{1}{4\pi}
  \left( \alpha_Y\Delta b + \frac{9}{4}\alpha_2 \right) \ln \frac{M_{\rm soft}}{m_{\rm soft}} 
\right], \label{eq:gL} \\
\widetilde g_R &= g_Y + \delta \widetilde g_R \simeq
g_Y 
\left[1 + \frac{\alpha_Y}{4\pi}
  \Delta b\, \ln \frac{M_{\rm soft}}{m_{\rm soft}} 
\right]. \label{eq:gR}
\end{align}
For derivation, see Appendix \ref{app:bino}. Here, the terms of $\alpha_Y$ are corrections to the Bino self-energy, 
which are SUSY analog of the oblique corrections~\cite{Nojiri:1996fp, Nojiri:1997ma, Cheng:1997sq, Cheng:1997vy, Katz:1998br}.
They are called the super-oblique corrections.
The U(1)$_Y$ gauge coupling constant, $g_Y$, is evaluated at $m_{\rm soft}$.
The corrections are represented by the difference between the beta functions of the U(1)$_Y$ gauge and Bino couplings. 
In the minimal model, SM particles and light sleptons contribute to the beta functions. 
Thus, the coefficient becomes $\Delta b = 41/6 - n_{\rm slepton}$, 
where $n_{\rm slepton}$ is number of the generations of light sleptons; for instance, 
$n_{\rm slepton} = 3$ if all the sleptons are light, and $n_{\rm slepton} = 2$ when the staus are decoupled. 
On the other hand, the term of $\alpha_2$ is non-oblique corrections after the Wino decoupled.
In case that the gluinos are decoupled, similar calculation is evaluated in Ref.~\cite{Hikasa:1995bw}.  
The weak boson couples only to the left-handed (s)leptons at the Bino--muon--smuon vertices. 
Since the Bino--smuon contribution to the muon $g-2$ includes both $\widetilde g_L$ and $\widetilde g_R$, 
the loop correction \eqref{eq:2-loop} is obtained. 
It yields a correction of $5-10\%$ for $M_{\rm soft} = 10-100\TeV$ with $m_{\rm soft} \sim 100\GeV$. 
In our analysis, only the logarithmic terms are considered.
Non-logarithmic terms are expected to be suppressed. 

%In Eq.~\eqref{eq:2-loop}, we include the corrections that can be as large as or larger than $\mathcal{O}(10)$\%.
%Other two loop contributions are unknown or expected to be small.
%See Refs.~\cite{Heinemeyer:2003dq,Heinemeyer:2004yq,Feng:2008nm,Feng:2008cn,Feng:2009gn,Stockinger:2006zn}
In Eq.~\eqref{eq:2-loop}, we include the corrections that can be as large as or larger than $\mathcal{O}(10)\%$.
Other 2-loop corrections are unknown or expected to be small.
The SUSY corrections to the SM one-loop diagram is calculated in 
Refs.~\cite{Heinemeyer:2003dq, Heinemeyer:2004yq, Feng:2008nm,Feng:2008cn,Feng:2009gn}.
When the superparticles except those of \eqref{eq:parameters} are decoupled, these contributions are negligibly small.
Other corrections that have not yet been estimated include electroweak and SUSY two-loop contributions to SUSY one-loop diagrams. 
They might provide $\delta a_{\mu } \sim 10^{-10}$, according to Ref.~\cite{Stockinger:2006zn}.
In addition, non-logarithmic corrections to $\delta ^{\text{2loop}}$ could be a few percents of the SUSY one-loop contributions, 
similarly to the discussions in Ref.~\cite{Stockinger:2006zn}.

We comment on other one-loop contributions to the muon $g-2$. 
The Bino--Higgsino--smuon contributions \eqref{eq:BHmuL}, \eqref{eq:BHmuR} can be $\lesssim \Order(10^{-10})$ 
when the Higgsinos are relatively light due to the vacuum stability bound of the stau-Higgs potential in Sec.~\ref{subsec:vacuum}. 
We include these contributions in the numerical analysis for completeness.\footnote{
This contribution can dominate the SUSY contributions to the muon $g-2$, when $\mu$ is small while decoupling the Wino. 
Since they are enhanced only by $\tan\beta$, superparticles are required to be light to explain \eqref{eq:g-2_deviation}. They are detectable in colliders. 
In particular, the Higgsino production can be significant. 
}
On the other hand, the chargino--muon sneutrino contributions \eqref{eq:WHmuL}, 
which are dominated in the case of nearly degenerate mass spectrum, are less than $\Order(10^{-11})$ for $M_2 > 10\TeV$, i.e., negligible.
We do not include them in the numerical analysis.

Summarizing the above, the theoretical uncertainty of the SUSY contribution to the muon $g-2$ can be estimated at $\mathcal{O}(10)$\%.
It is smaller than the current experimental and theoretical (SM) uncertainties.
Here and hereafter, we understand that the results of this dissertation have uncertainty of $\mathcal{O}(10)$\%, 
which depends on the heavy mass scale $M_{\text{soft}}$, 
and unknown higher-order corrections.
In this dissertation, we set to $M_{\text{soft}} = 30\TeV$, for simplicity.  

%%%%%%%%%%%%%%%%%%%%%%%%%%%%%%%%%%%%%%%%%%%%% Figure
\begin{figure}[thb]
 \begin{center}
 \includegraphics[width=8cm]{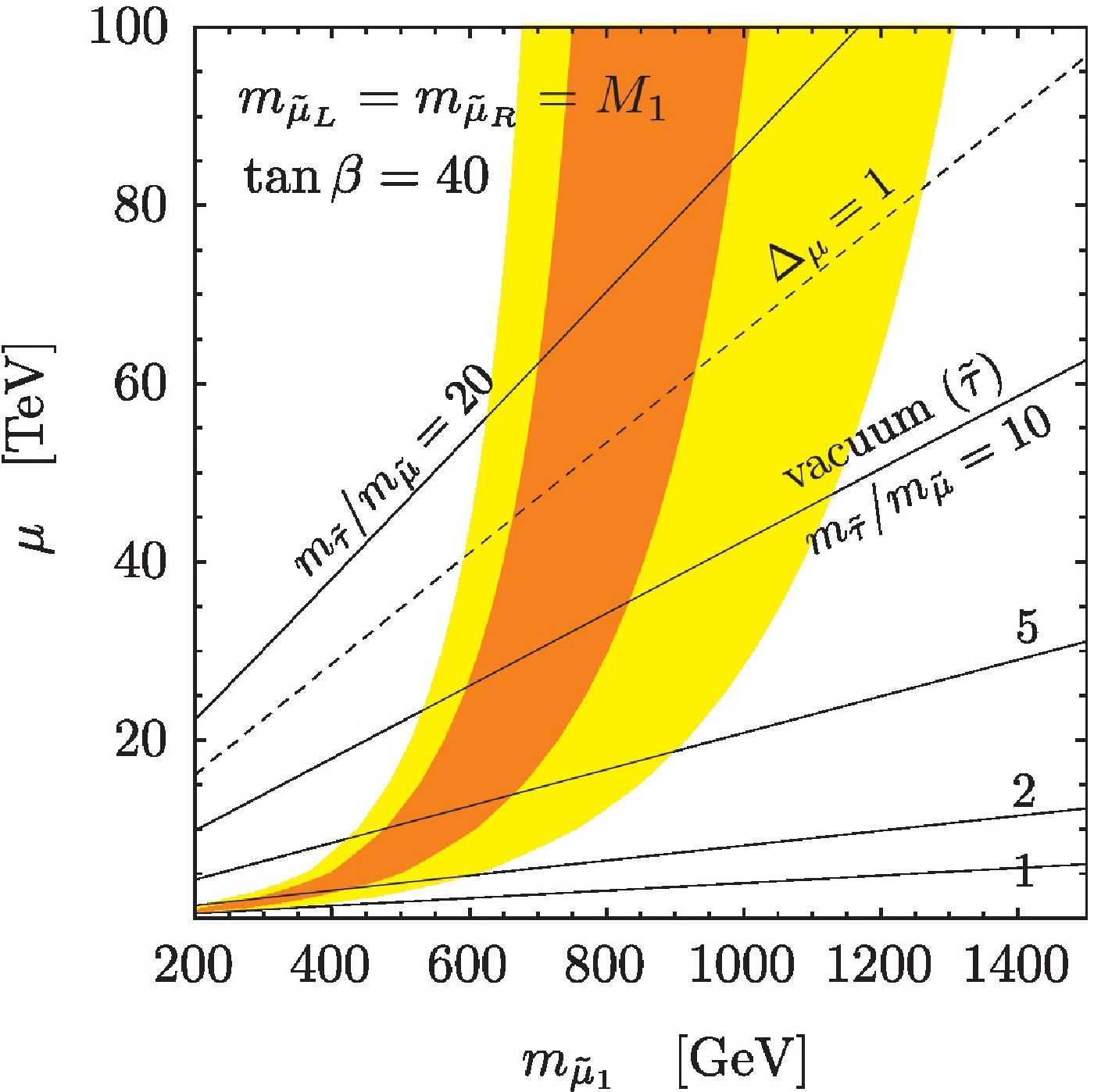}\hspace*{2mm}
 \includegraphics[width=8cm]{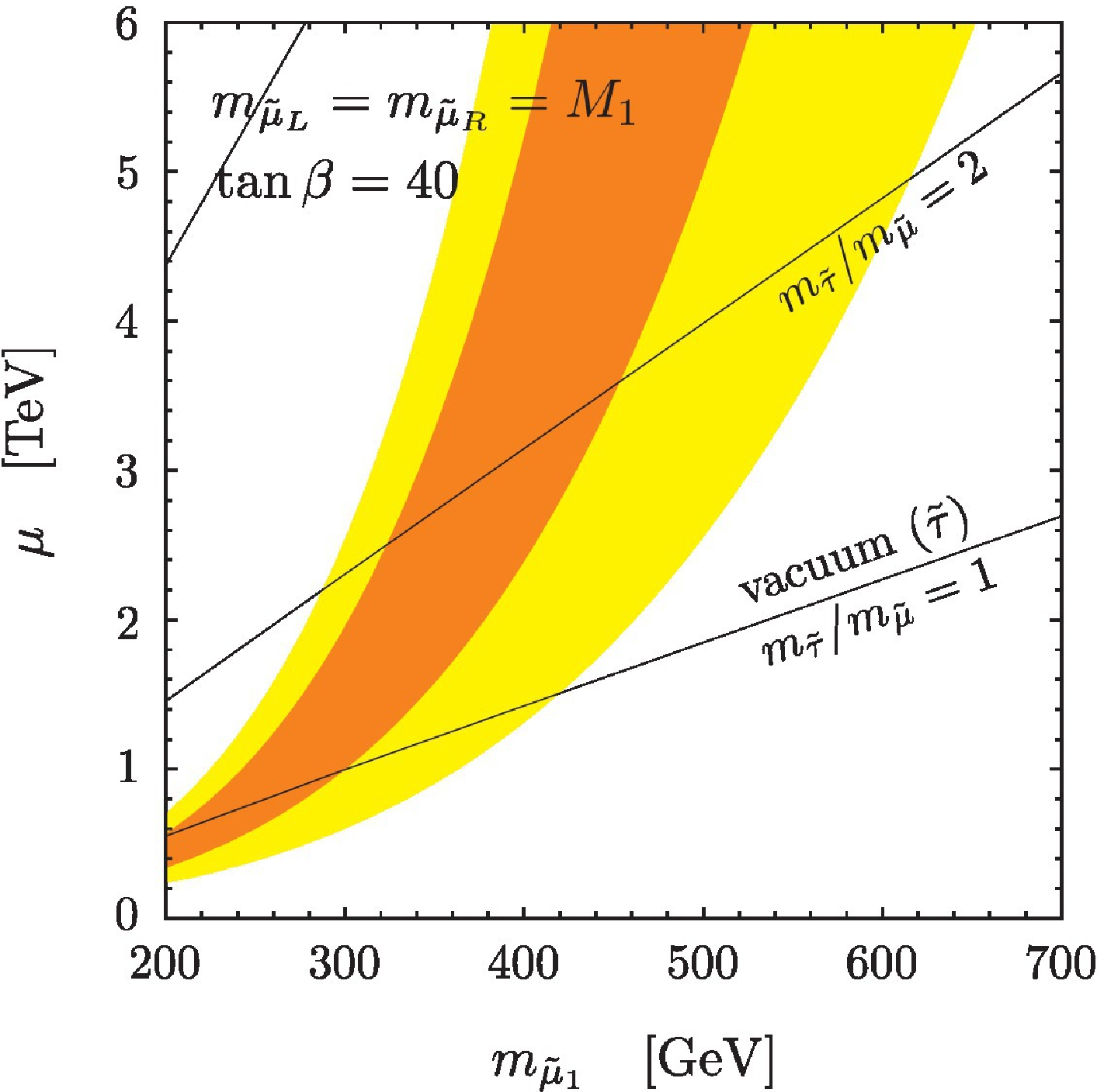}
 \end{center}
 \caption[The SUSY contributions to the muon $g-2$ and the vacuum stability bounds.]
 {
The SUSY contributions to the muon $g-2$ and the vacuum stability bounds are shown. 
In the orange (yellow) regions, the muon $g-2$ discrepancy \eqref{eq:g-2_deviation} is explained at the $1 \sigma$ $(2 \sigma)$ level, 
where $1\sigma$ ($2\sigma$) mean that the SUSY contribution to the muon $g-2$ are $(26.1 \pm 8.0) \times 10^{-10}$ $((26.1 \pm 16.0) \times 10^{-10} )$. 
The black solid lines represent the upper bound on $\mu$ from the vacuum stability bound of the stau-Higgs potential. 
The stau masses are set to be $m_{\widetilde{\tau}}/m_{\widetilde{\mu}}=1, 2, 5, 10$ and 20 from bottom to top.
Above the black dashed line, $\Delta_{\mu}$ becomes larger than unity. 
The parameters are $M_1 = m_{\widetilde\mu_L} = m_{\widetilde\mu_R}$, $\tan\beta = 40$ and $M_{\rm soft} = 30\TeV$. 
In the right panel, a part of the parameter region of the left panel is magnified.
}
\label{fig:gminus2_stausmubound}
\end{figure}
%%%%%%%%%%%%%%%%%%%%%%%%%%%%%%%%%%%%%%%%%%%%%

In Fig.~\ref{fig:gminus2_stausmubound}, we show contours of the SUSY contributions to the muon $g-2$. 
The horizontal and vertical axises are the lightest smuon mass $m_{\widetilde\mu_1}$ and $\mu$, respectively. 
The parameters are set as $M_1 = m_{\widetilde\mu_L} = m_{\widetilde\mu_R}$, $\tan\beta = 40$ and $M_{\rm soft} = 30\TeV$. 
In the right panel of Fig.~\ref{fig:gminus2_stausmubound},  a part of the parameter region of the left panel is showed.
In the orange (yellow) regions, 
the SUSY contribution to the muon $g-2$ becomes $(26.1 \pm 8.0) \times 10^{-10}$ 
$((26.1 \pm 16.0) \times 10^{-10})$.
In this dissertation, we define these regions as $1\sigma $ $(2\sigma ) $ regions.
% and call that the muon $g-2$ discrepancy is explained at the $1\sigma $ $(2 \sigma )$ level. 
Hereafter, we will use the above definition	without notice.
From Fig.~\ref{fig:gminus2_stausmubound}, it is found that they are enhanced by large $\mu$, and the smuon masses can be $1\TeV$ for $\mu=\Order(10-100)\TeV$. 
This is contrasted to the chargino--muon sneutrino contributions to the muon $g-2$, where $\mu$ is favored to be small~\cite{Endo:2013bba}. 
On the other hand, detailed dependences on the superparticle mass spectrum are determined by the loop function \eqref{eq:fnma} and the vacuum stability condition.  
They will be discussed in Sec.~\ref{subsec:vacuum}.

\subsection{Vacuum Stability}
\label{subsec:vacuum}
%!TEX root = ../Dthesis.tex

As shown in Sec.~\ref{subsec:gm2vac}, 
the Bino--smuon contribution to the muon $g-2$ is enhanced by a large left-right mixing of the smuon. 
However, too large left-right mixing spoils the stability of the electroweak symmetry breaking vacuum (EW vacuum). 
In this section, we discuss the vacuum stability condition of the sleptons and slepton mass region using the condition.
%This section is based on Refs.~\cite{Kitahara:2013lfa, Endo:2013lva}. 

Let us consider the scalar potential for the neutral component of the up-type Higgs, $H_u$, the left-handed slepton, $\widetilde{\ell}_L$, 
and the right-handed slepton, $\widetilde{\ell}_R$.
It is given as\cite{Kitahara:2013lfa}
\beq
 V &=& (m_{H_u}^2 + \mu ^2) | H_u | ^2 + m_{\widetilde{\ell}_L }^2 | \widetilde{\ell}_L |^2  + m_{\widetilde{\ell }_R}^2 | \widetilde{\ell}_R | ^2
       - (Y_{\ell } \mu H_u ^{*}  \widetilde{\ell}_L^{*}  \widetilde{\ell }_R + {\rm h.c.}) + Y_{\ell }^2 | \widetilde{\ell}_L^{*}  \widetilde{\ell}_R|^2 \nonumber \\
     &&+ \frac{g^2}{8} (|  \widetilde{\ell}_L  |^2 + |H_u|^2)^2 +\frac{g^2_Y}{8} (|  \widetilde{\ell}_L  | ^2 - 2 | \widetilde{\ell}_R| ^2 - |H_u| ^2) ^2
       + \frac{g^2 + g^2_Y}{8}\delta _H |H_u| ^4 , 
       \label{eq:potential}
\eeq
where $m^2_{H_u}$ is the soft SUSY-breaking mass of the up-type Higgs.
When $\tan \beta$ and the CP-odd Higgs mass, $m_A$ are large, 
the terms including the down-type Higgs, $H_d$ can be neglected.\footnote{
Authors in Ref.~\cite{Carena:2012mw} analyzed the scalar potential including the down-type Higgs field for completeness. 
As $\tan \beta $ approaches to 1, the effects of the down-type Higgs become relevant.
However, we will not consider such small $\tan \beta $ region since it is difficult to explain the muon $g-2$ anomaly.}
Further, only the renormalizable terms are taken into account in this dissertation. 
Higher-dimensional terms depend on the superparticles that decouple at $M_{\rm soft}$, which is mentioned in Sec.~\ref{subsec:gm2vac}. 
Since $M_{\rm soft}$ is very large, their contributions to the vacuum meta-stability condition are considered to be small.

We take account of the RG-improved term in the scalar potential, which is the last term in Eq.~\eqref{eq:potential}.
The leading correction for $\delta _H$ is represented as 
\beq
 \delta _H \simeq  \frac{3}{2\pi ^2} \frac{Y^2_t}{g^2 + g^{\prime 2} }
 \log \frac{\sqrt{m^2_{\widetilde{t}_1} m^2_{\widetilde{t}_2} }}{m^2_t} , \label{eq:corrpotential}
\eeq
where $Y_t$ is the top Yukawa coupling, and $m_t$ and $m_{\widetilde{t}}$ are the top and stop masses, respectively.
The Higgs mass is enhanced if $\delta _H$ is sizable.
In our analysis, we set $\delta _H \sim 1$ to reproduce the Higgs mass of 126\GeV. 

The coefficient of the quartic terms in the scalar potential, $Y^2_{\ell }$ is somewhat important.
At the tree level, $Y_{\ell } = m_{\ell }/ (v \cos \beta ) \simeq (m_{\ell }/ v) \tan \beta $ in a large $\tan \beta $ limit.
As mentioned in Eq.~\eqref{eq:LRmixing}, It is modified at the one loop level as follows:
\beq
 Y_{\ell } = \frac{m_{\ell }}{v \cos \beta } \frac{1}{1 + \Delta _{\ell }} \simeq \frac{m_{\ell }}{v} \frac{\tan \beta }{1 + \Delta _{\ell }}. \label{eq:leptonYukawa}
\eeq
The dependence on $Y_{\ell }$ to the vacuum stability condition will be discussed later.

Expanding $H_u$ around the EW vacuum, $H_u = v_u + \frac{1}{\sqrt{2}} h_u$, 
we have
\beq
 V &=& \frac{ g^2 + g^{2}_Y }{4}(1 + \delta _H ) v^2_u h^2_u 
 + (\widetilde{\ell }^*_L, \widetilde{\ell }^*_R )
 \begin{pmatrix}
  m_{\widetilde{\ell}_{LL} }^2   & -\big| m_{\widetilde\ell_{LR}}^2 \big| \\
   -\big| m_{\widetilde\ell_{LR}}^2 \big| & m_{\widetilde{\ell}_{RR} }^2   \\
  \end{pmatrix}
  \begin{pmatrix}
  \widetilde{\ell}_L  \\
  \widetilde{\ell}_R  \\
  \end{pmatrix} \notag \\
  && -\left( \frac{1}{\sqrt{2}v_u} \big| m_{\widetilde\ell_{LR}}^2 \big|  h_u \widetilde{\ell}_L^{*} \widetilde{\ell }_R + {\rm{h.c.}} \right)
  + \frac{g^2 - g^{2}_Y}{2\sqrt{2}} v_u h_u  | \widetilde{\ell}_L |^2
  + \frac{ g^{2}_Y}{2\sqrt{2}} v_u h_u  | \widetilde{\ell}_R |^2 \notag \\
  && + \frac{g^2 + g^{2}_Y}{32} \left( 8\sqrt{2} v_u h^3_u   + h^4_u\right) 
  + \frac{g^2 + g^{2}_Y}{8}  | \widetilde{\ell}_L |^4 + \frac{g^{2}_Y}{2}  | \widetilde{\ell}_R |^4  \notag \\
  && + \left( Y_{\ell }^2  - \frac{1}{2} g^{2}_Y  \right) | \widetilde{\ell}_L^{*}  \widetilde{\ell}_R|^2 
  + \frac{g^2 -g^{2}_Y }{8} h^2_u | \widetilde{\ell}_L |^2 + \frac{g^{2}_Y}{4} h^2_u | \widetilde{\ell}_R |^2.
  \label{eq:potentialew}
 \eeq
where $m_{\widetilde\ell_{LL, RR}}^2$ are the squired masses of the left- and right-handed sleptons as seen in \eqref{eq:slep2} of Sec.~\ref{subsec:interaction}, 
and $m_{\widetilde\ell_{LR}}^2$ is the left-right mixing of the sleptons defined in Eq.~\eqref{eq:LRmixing}.
 
The scalar potential has the ordinary electroweak symmetry breaking minimum (EW vacuum), $h_u = \widetilde{\ell }_L = \widetilde{\ell }_R = 0$.
As  the trilinear coupling of the sleptons and the Higgs boson, $m_{\widetilde\ell_{LR}}^2$ increases,
disastrous charge-breaking minima (CB vacuum), $\widetilde{\ell }_L \neq 0$ or $\widetilde{\ell }_R \neq 0$ in the scalar potential become deeper, 
and our EW vacuum could decay them by quantum tunneling effects.
By requiring that the lifetime of the EW vacuum should be longer than the age of the Universe, 
the mixing $m_{\widetilde\ell_{LR}}^2$ is bounded from above.

The vacuum transition rate from the false vacuum (EW vacuum) to the true vacuum 
(CB vacuum) can be evaluated by semiclassic technique\cite{Coleman:1977py, Callan:1977pt}.
In this technique, the imaginary part of energy of the false vacuum corresponds to the transition rate (see Appendix \ref{app:vacuum}).
The vacuum transition rate per unit space-time volume is evaluated as follows, 
\beq
 \Gamma / V = A \cdot e^{-B} , \label{eq:decayrate}
\eeq 
where the prefactor $A$ is the fourth power of the typical energy scale in the potential,
and expected to be roughly $(100\GeV)^4$.
Although it is hard to precisely calculate $A$, the decay rate is not sensitive to it. 
On the other hand, the Euclidean action which is evaluated on the bounce solution, 
$B$,  is important to estimate the decay rate. It is defined as\cite{Hisano:2010re}
\beq
 B = S_E[\bar{\phi }(r)] - S_E[ \phi ^f], \label{eq:bounce}
\eeq
where $S_E$ is Euclidean action, $\bar{\phi}$ is a bounce configuration, and $\phi ^f$ is field values at the false vacuum.
When $\Gamma / V$ is  smaller than the fourth power of the present Hubble parameter $H_0 = 1.5 \times 10^{-42} \GeV$,
(i.e. the lifetime of the false vacuum is longer than the age of the Universe), 
the false vacuum becomes unstable.
The vacuum meta-stability condition is approximately given as $B \gtrsim 400$
%\footnote{
%The meta-stability condition has only logarithmic dependence of $H_0$ and $A$.
%Thus, even if the value of $H_0$ or $A$ change $\mathcal{O}(10)\%$, 
%the change of this condition is less than one percent.} 
if we assume $A \sim (100\GeV)^4$.

The public package  \texttt{CosmoTransitions 1.0.2}\cite{Wainwright:2011kj},\footnote{
In order to evaluate the bounce solution and the $B$ numerically, 
we used modules \texttt{tunneling1D.py} and \texttt{pathDeformation.py}, and set the parameter $\alpha = 3$ in the modules.} 
which is the numerical package to analyze cosmological phase transitions for single and multiple scalar fields,
is used for the numerical evaluation of the Euclidean action $B$.
The scalar potential is analyzed at the zero temperature.
The vacuum mata-stability conditions correspond to constraint of $m_{\widetilde\ell_{LR}}^2$.
The fitting formula of the condition is obtained as \cite{Kitahara:2013lfa, Endo:2013lva}
\begin{align}
\left|m_{\widetilde\ell_{LR}}^2 \right| 
&\leq
\eta_\ell 
\bigg[
1. 01  \times 10^2 \GeV \sqrt{m_{\widetilde{\ell}_L} m_{\widetilde{\ell}_R}} 
+ 1.01 \times 10^2 \GeV  (m_{\widetilde{\ell}_L} + 1.03 m_{\widetilde{\ell}_R}) 
\notag \\
-2.27 \times 
&
10^4\GeV^2
+ \frac{2.97 \times 10^6\GeV^3}{m_{\widetilde{\ell}_L} + m_{\widetilde{\ell}_R}} 
- 1.14 \times 10^8\GeV^4 
  \left( \frac{1}{m^2_{\widetilde{\ell}_L}} +  \frac{0.983}{m^2_{\widetilde{\ell}_R}} \right) 
\bigg]. 
\label{eq:kyb}
\end{align}
The fitting formula reproduces results of \texttt{CosmoTransitions} at better than the 1\% level. 
Note that an asymmetry of the coefficients of $m_{\widetilde{\ell }_L}$ and $m_{\widetilde{\ell }_R}$ is due to the  D-term potential, 
i.e. the terms including $g^2 $ and $g^2_Y$ in Eq.~\eqref{eq:potential},
although the effects of this asymmetry are negligible in numerical analysis.

Let us compare Eq.~\eqref{eq:kyb} to the result in Ref.~\cite{Hisano:2010re}.
The meta-stability condition of the stau-Higgs potential was evaluated as\cite{Hisano:2010re}
\beq
\left|  \mu \tan \beta \right| &<& 213.5 \sqrt{m_{\widetilde{\tau }_L} m_{\widetilde{\tau }_R} } 
 - 17.0 (m_{\widetilde{\tau }_L} + m_{\widetilde{\tau }_R}) \notag \\
 &+& 45.2 \times 10^{-2} \GeV ^{-1} (m_{\widetilde{\tau }_L} - m_{\widetilde{\tau }_R})^2 
 - 1.30 \times 10^4 \GeV,  \label{eq:hisanobound}
\eeq
by numerical calculation. 
It is valid in the regions of $m_{\widetilde{\tau }_L} $, $m_{\widetilde{\tau }_R} \leq 600\GeV$, 
and fails as either the left- or right-handed slepton mass becomes extremely large.
The fitting formula \eqref{eq:kyb} is not only consistent with Eq.~\eqref{eq:hisanobound} in light slepton mass regions, 
but also applicable when masses of the sleptons are $\mathcal{O}(1)\TeV$.

%%%%%%%%%%%%%%%%%%%%%%%%%%%%%%%%%%%%%%%%%%%%% Figure
\begin{figure}[t]
 \begin{center}
 \includegraphics[width=8cm]{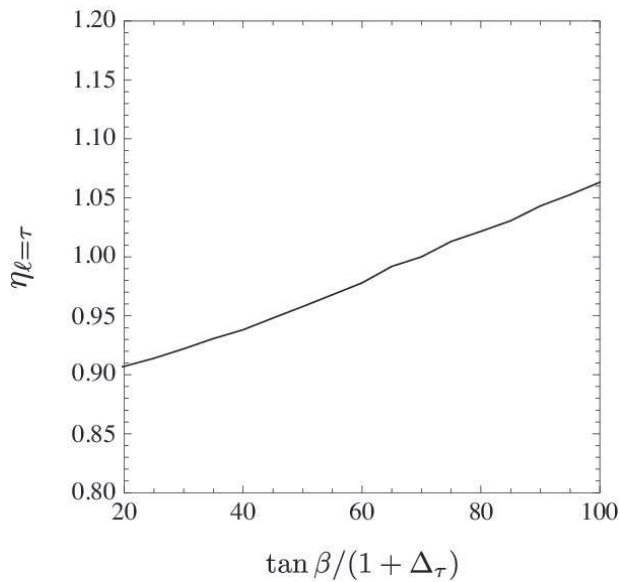}
 \caption[$\tan \beta$ dependence of the vacuum stability condition.]
 {$\tan \beta$ dependence of the vacuum stability condition of the stau-Higgs potential.
 Here, we set to $m_{\widetilde{\tau }_L} = m_{\widetilde{\tau }_R} = 300\GeV.$}
  \label{fig:bounddep}
 \end{center}
 \end{figure}
%%%%%%%%%%%%%%%%%%%%%%%%%%%%%%%%%%%%%%%%%%%%%
The fitting formula (\ref{eq:kyb}) is universal for all the slepton flavors up to corrections, $\eta_\ell \sim 1$. 
Given the soft scalar masses, the meta-stability condition is independent of the flavor except through Yukawa interactions in the quartic terms of the scalar potential. 
Since the quartic terms are dominated by gauge interactions, $\eta_\ell$ changes little around unity. It also depends on $\tan\beta$ when the Yukawa coupling is large. 
In practice, $\eta_\tau = 1$ is set for the stau with $\tan\beta/(1+\Delta_\tau) = 70$\cite{Hisano:2010re, Kitahara:2013lfa}. 
Numerical result of $\eta_\tau$ for the stau with various $\tan\beta$ is found in Fig.~\ref{fig:bounddep}. 
Here, although the stau mass is taken to be $m_{\widetilde{\tau }_L} = m_{\widetilde{\tau }_R} = 300\GeV$, 
the result is (most) independent of the stau mass. 
For instance, it is $\eta_\tau \simeq 0.94$ for $\tan\beta/(1+\Delta_\tau) = 40$. On the other hand, we obtain $\eta_\ell \simeq 0.88$ for the smuons and the selectrons, which is independent of the flavor and $\tan\beta$ because of small Yukawa couplings\cite{Endo:2013lva}.

The theoretical uncertainty of the fitting formula (\ref{eq:kyb}) is expected to be at most $\mathcal{O}(10)$\%.
First, the meta-stability condition corresponding to $B$ has only logarithmic dependence of $H_0$ and $A$.
Thus, even if the value of $H_0$ or $A$ change $\mathcal{O}(10)\%$, the change of the condition is less than one percent.
Then, thermal corrections can change the transition rate.
They could be significant (but $\lesssim \mathcal{O}(10)\%$) when the staus are light, for instance, 
for $m_{\widetilde{\tau }_1} \lesssim 200 \GeV$\cite{Endo:2010ya, Endo:2011uw} 
in the scalar potential for the SM Higgs and the lightest stau. 
Although the corrections in case of three-dimensional potential are not discussed, 
it is expected that the results are almost unchanged, 
since the trilinear coupling of the lightest stau with the Higgs boson is more important for the vacuum meta-stability condition. 
For more precise evaluation, detailed analysis is needed.

The most severe constraint on $\mu\tan\beta$ is obtained from the stability condition of the stau-Higgs potential. 
This is because the left-right mixing is proportional to the Yukawa coupling. 
In Fig.~\ref{fig:gminus2_stausmubound}, we show the upper limits of $\mu \tan \beta $ for $\tan\beta = 40$. 
Combined with the muon $g-2$, the smuon masses are bounded from above for given stau masses. 
For $m_{\widetilde e} = m_{\widetilde\mu} = m_{\widetilde\tau}$, the lightest smuon is limited to be $m_{\widetilde{\mu}_1}\lesssim 300~(420)\GeV$ at the $1\sigma$ ($2\sigma$) level of the muon $g-2$.

%%%%%%%%%%%%%%%%%%%%%%%%%%%%%%%%%%%%%%%%%%%%% Figure
\begin{figure}[t]
 \begin{center}
 \includegraphics[width=8cm]{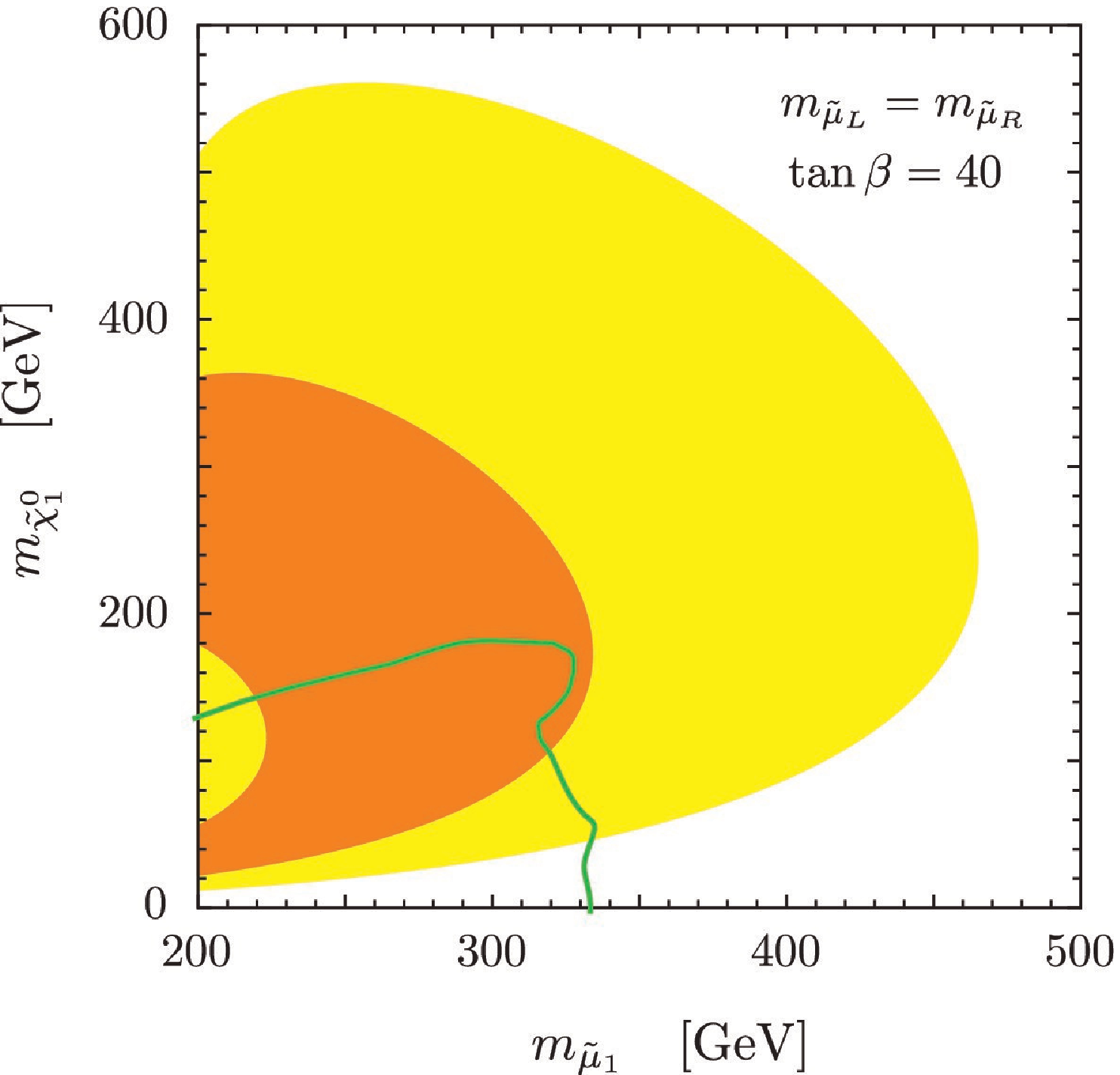}
 \caption[The SUSY contributions to the muon $g-2$ as a function of the lightest smuon mass, $m_{\widetilde\mu_1}$, and the lightest neutralino mass, $m_{\widetilde\chi^0_1}$.]{The SUSY contributions to the muon $g-2$ as a function of the lightest smuon mass, $m_{\widetilde\mu_1}$, and the lightest neutralino mass, $m_{\widetilde\chi^0_1}$.
 In the orange (yellow) regions, the muon $g-2$ discrepancy \eqref{eq:g-2_deviation} is explained at the $1 \sigma$ $(2 \sigma)$ level.  The left-right mixing is maximized under the vacuum stability condition. 
 The parameters are set to $m_{\widetilde\ell_L} = m_{\widetilde\ell_R}$, $\tan\beta = 40$ and $M_{\rm soft} = 10\TeV$. The stau soft masses are equal to those of the selectrons and smuons.
 The region below the green line is excluded by LHC.}
 \label{fig:MSL-M1}
 \end{center}
\end{figure}
%%%%%%%%%%%%%%%%%%%%%%%%%%%%%%%%%%%%%%%%%%%%%

The SUSY contributions to the muon $g-2$ depend on superparticle mass spectra mainly through the loop function \eqref{eq:fnma}. In Fig.~\ref{fig:MSL-M1}, we show contours of the muon $g-2$ as a function of smuon and neutralino masses. 
In the orange (yellow) regions, the muon $g-2$ discrepancy is explained at the $1\sigma$ ($2\sigma$) level. 
The left-right mixing is maximized with satisfying the vacuum stability condition. 
Here, the stau soft mass is supposed to be degenerate with that of the smuon, $m_{\widetilde e} = m_{\widetilde\mu} = m_{\widetilde\tau}$. 
It is found that the lightest smuon mass can be $330~(460)\GeV$ at the $1\sigma$ ($2\sigma$) level of the muon $g-2$, when the lightest neutralino mass, $m_{\widetilde\chi^0_1}$, is almost a half of $m_{\widetilde{\mu}_1}$. Note that $m_{\widetilde\chi^0_1}$ is almost equal to $M_1$. On the other hand, $m_{\widetilde\chi^0_1} \simeq 560\GeV$ is realized for $m_{\widetilde{\mu}_1} \simeq 250\GeV$ at the $2\sigma$ level of the muon $g-2$. 

For a fixed value of the lightest smuon mass, 
the SUSY contributions to the muon $g-2$ is maximized when the left-handed smuon mass, 
$m_{\widetilde\mu_L}$ is equal to the right-handed smuon mass, $m_{\widetilde\mu_R}$. 
Eq.~\eqref{eq:gminus2bino} depends on the left- and right-handed smuon masses as $f_N(x,y)/xy$ with $x = m_{\widetilde\mu_L}^2/M_1^2$ and $y = m_{\widetilde\mu_R}^2/M_1^2$. When $xy$ is fixed, it is maximized for $x=y$ and rapidly decreases for $x \neq y$. On the other hand, the vacuum stability condition \eqref{eq:kyb} is relaxed for $m_{\widetilde\ell_L} \neq m_{\widetilde\ell_R}$. Since the former dependence is stronger than the latter, the SUSY contributions to the muon $g-2$ become largest when $m_{\widetilde\mu_L}$ is equal to $m_{\widetilde\mu_R}$ for given $M_1$ and $\tan\beta$. In other words, the orange/yellow regions in Fig.~\ref{fig:MSL-M1}, i.e., the parameter regions favored by the muon $g-2$, shrink toward smaller superparticle masses for $m_{\widetilde\mu_L} \neq m_{\widetilde\mu_R}$.

%%%%%%%%%%%%%%%%%%%%%%%%%%%%%%%%%%%%%%%%%%%%% Figure
\begin{figure}[t]
 \begin{center}
 \includegraphics[width=8cm]{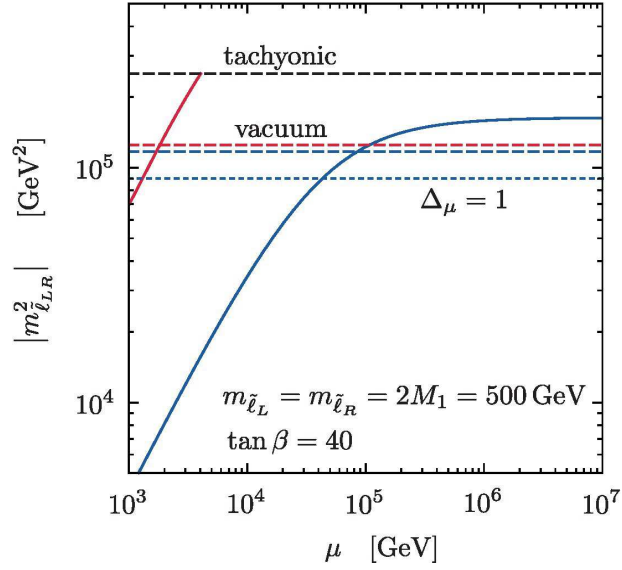}
 \caption[The left-right mixing of the smuon and the stau.]
 {The left-right mixing of the smuon (blue solid) and the stau (red solid) as a function of $\mu$.
 The blue (red) dashed line represents the vacuum meta-stability bound with $\eta = 0.88 ~(0.94)$.
 Above the black dashed line, sleptons become tachyonic. 
 Below the blue dotted line, $\Delta_{\mu} < 1$ is satisfied.
 The parameters are $m_{\widetilde\ell_L} = m_{\widetilde\ell_R} = 2M_1 = 500\GeV$, $\tan\beta = 40$ and $M_{\rm soft} = 30\TeV$.}
  \label{fig:LRmixing}
 \end{center}
 \end{figure}
%%%%%%%%%%%%%%%%%%%%%%%%%%%%%%%%%%%%%%%%%%%%%
The vacuum stability condition is relaxed as staus become heavy.
When they are decoupled, the constraint almost disappear, and extremely large $\mu $ is allowed.
As $\mu $ increases, the left-right mixing \eqref{eq:LRmixing} approaches to
\beq
 m^2_{\widetilde{\ell }_{LR}} = \frac{m_{\ell }}{1 + \Delta _{\ell }} \mu \tan \beta \to \frac{4\pi }{\alpha _Y} \frac{m_{\ell }}{M_1}
  \frac{1}{I(M^2_1, m^2_{\widetilde{\ell}_L},m^2_{\widetilde{\ell}_R})} ~~~~(\text{for } \mu \to \infty).
  \label{eq:limitbound}
\eeq
In Fig.~\ref{fig:LRmixing}, the left-right mixings of the smuon and the stau, $|m_{\widetilde\ell_{LR}}^2|$, are plotted as a function of $\mu$, by the blue and red solid lines, respectively. Since $m_{\widetilde\ell_{LR}}^2$ is proportional to the Yukawa coupling, $|m_{\widetilde\tau_{LR}}^2|$ is larger than $|m_{\widetilde\mu_{LR}}^2|$. When the staus are light, $\mu$ is bounded from above by the stau stability condition (red dashed line). For heavier staus, the red dashed line (as well as the stau tachyonic bound or the black dashed line) is lifted up, and larger $\mu$ is allowed. 

%%%%%%%%%%%%%%%%%%%%%%%%%%%%%%%%%%%%%%%%%%%%% Figure
\begin{figure}[t]
 \begin{center}
  \includegraphics[width=8cm]{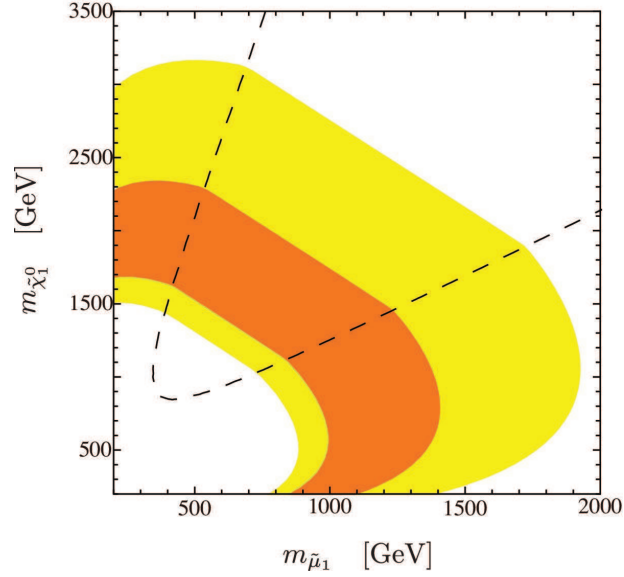}
 \caption[Mass bound for smuon]
 {Mass bound for smuon when the staus are decoupled.
 In orange (yellow) regions, the muon $g-2$ discrepancy is explained at $1\sigma$ ($2\sigma$) level.
 The parameters are $m_{\widetilde{\mu }_L}= m_{\widetilde{\mu }_R}$, $\tan \beta = 40$, and $M_{\text{soft}} = 30\TeV$.
 The left-right mixing is limited by the vacuum stability condition of the smuon-Higgs potential.
 On the black line, $m^{2, \text{vac}, \mu \to \infty}_{\widetilde{\mu }_{LR}} / m^{2, \text{vac}}_{\widetilde{\mu }_{LR}} = 1$ is satisfied.}
  \label{fig:masslimit}
 \end{center}
 \end{figure}
%%%%%%%%%%%%%%%%%%%%%%%%%%%%%%%%%%%%%%%%%%%%%
Even if the staus are decoupled, the left-right mixing is limited by the vacuum stability condition of the smuon-Higgs potential.
Fig.~\ref{fig:masslimit} shows contours of the muon $g-2$ under the smuon stability condition, .
In the outside of the black-dashed line, the constraint \eqref{eq:kyb} is more severe than Eq.~\eqref{eq:limitbound}, 
i.e., $m^{2, \text{vac}, \mu \to \infty}_{\widetilde{\mu }_{LR}} > m^{2, \text{vac}}_{\widetilde{\mu }_{LR}}$ is satisfied for given smuon masses.
We show that the lightest smuon mass can be as large as 1.4 (1.9)$\TeV$ at $1\sigma$ ($2\sigma$) level of the muon $g-2$.
Since it is very weak, this bound does not appear in the parameter range of Fig.~\ref{fig:gminus2_stausmubound}.

In $\mu \to \infty$ limit, the situation is essentially the same as the soft Yukawa coupling models~\cite{Borzumati:1999sp,Crivellin:2010ty}, 
where the Yukawa couplings at tree level equal to zero, and the fermion masses are generated by only radiative corrections.
In this dissertation, the vacuum stability condition is analyzed seriously and 
is found to restrict the SUSY contributions to the muon $g-2$, 
resulting in tighter bound on the smuon mass than Ref.~\cite{Crivellin:2010ty}.
 
Note that $\Delta_\ell$ is larger than unity when $\mu$ is very large. 
Then, the radiative correction exceeds the tree level contribution to the lepton mass. For $\Delta_\ell \gg 1$, 
the Yukawa coupling constant is suppressed by $\Delta_\ell$, as observed by Eq.~\eqref{eq:LRmixing}. 
This may be disfavored by a naturalness argument~\cite{Crivellin:2010gw}. In Fig.~\ref{fig:gminus2_stausmubound}, 
$\Delta_\mu$ becomes larger than unity above the black dashed line, and above the blue dotted line in Fig.~\ref{fig:LRmixing}.

So far, we discussed that the slepton masses are bounded from above by the vacuum meta-stability condition and the muon $g-2$.
Furthermore, when the slepton masses are almost same among lepton flavor, 
the stau mass region might be also detected by the Higgs coupling to di-photon $\kappa _{\gamma }$. 
In the next subsection, we discuss the stau mass region.

\subsection{Stau Mass Region}
\label{subsec:kg}
%!TEX root = ../Dthesis.tex

We defined the minimal SUSY model, 
where only the Bino and the sleptons are light, while the other superparticles are decoupled in Sec.~\ref{subsec:setup}.
If sleptons have nearly degenerate mass spectrum, 
staus also become light and the stau contribution to the Higgs coupling to di-photon $\kappa_{\gamma}$ might be sizable.
Maximal value of $\kappa _{\gamma }$ (or the decay rate of $\Gamma (h \to \gamma \gamma )$) by the vacuum meta-stability condition have been discussed 
by some theoretical groups\cite{Kitahara:2012pb, Sato:2012bf,Carena:2012mw}.
In contrast, we discuss the stau mass region where the Higgs coupling  $\kappa_{\gamma}$ is deviated from SM prediction 
in the viewpoint of the vacuum meta-stability, then classify the stau contribution.
Our study is expected to provide some hints whether the deviation is from the stau contribution or not, if an excess of $\kappa _{\gamma }$ is observed.

In this section, we discuss the stau mass region based on the Higgs coupling $\kappa _{\gamma }$ and the vacuum meta-stability condition of the stau-Higgs potential. 
The staus are characterized by the three parameters as follows, 
\beq
 m_{\widetilde{\tau}_1}, m_{\widetilde{\tau}_2}, \theta _{\widetilde{\tau }},
 \label{eq:parameters_s}
\eeq
where the left-right mixing angle $\theta _{\widetilde{\tau}}$ is associated with the off-diagonal element of the stau mass matrix as
(cf. Sec.~\ref{sec:HiggsCmssm})
\beq
   m_{\widetilde{\tau }_{LR}}^2 = 
  \frac{1}{2} (m_{\widetilde{\tau }_1}^2 - m_{\widetilde{\tau}_2}^2) 
  \sin 2 \theta_{\widetilde{\tau}}.
  \label{eq:staumix}
\eeq
Here, $m_{\widetilde{\tau }_1} < m_{\widetilde{\tau }_2}$ is assumed.
The stau contribution to the Higgs coupling is determined, once the parameters \eqref{eq:parameters_s} are given. 
They are constrained by the vacuum meta-stability condition. 

% Figure %%%%%%%%%%%%%%%%%%%%%%%%%%%%%%%%
\begin{figure}[t]
 \begin{center}
 \includegraphics[width=8cm]{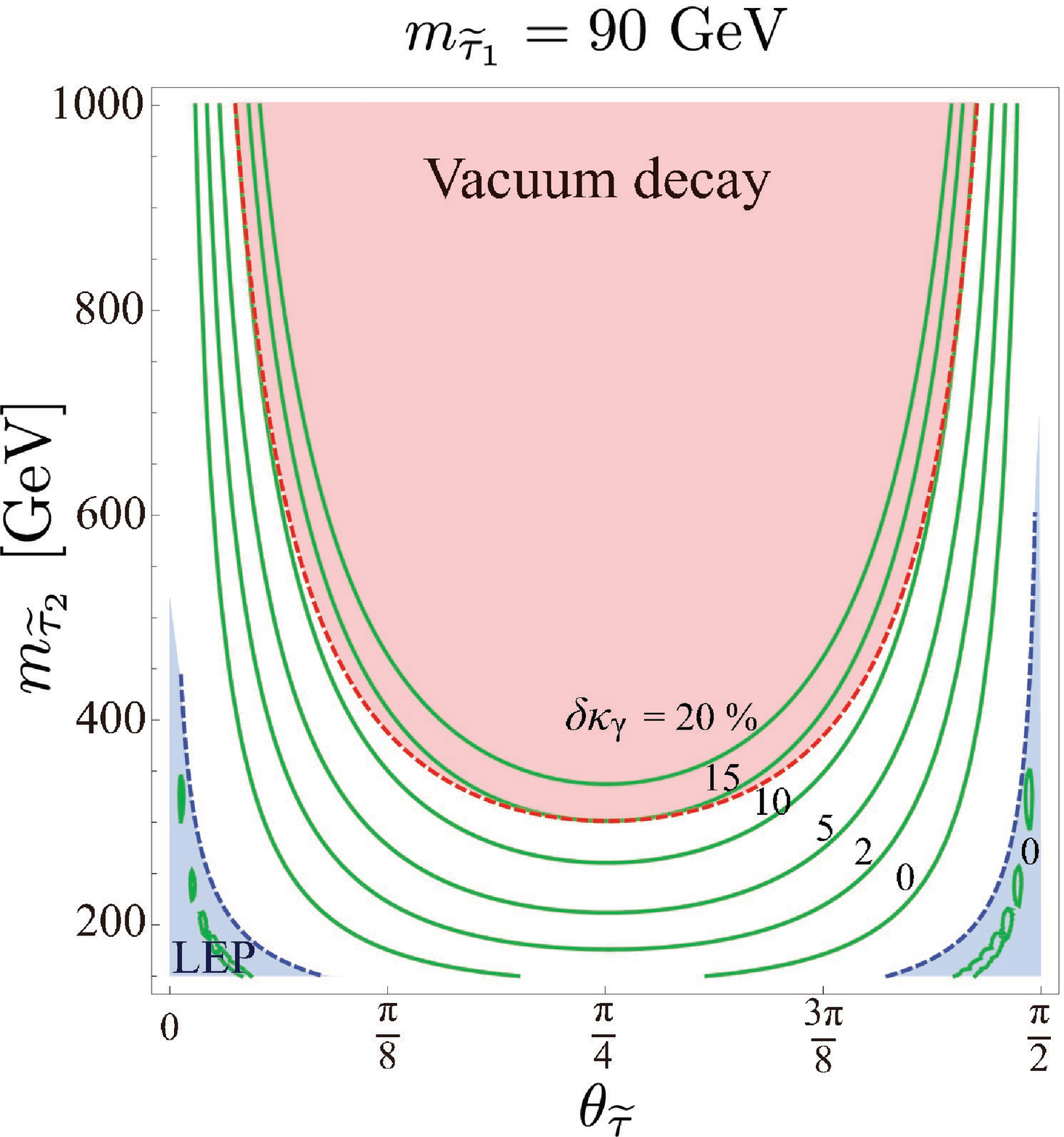} \hspace*{2mm}
 \includegraphics[width=8cm]{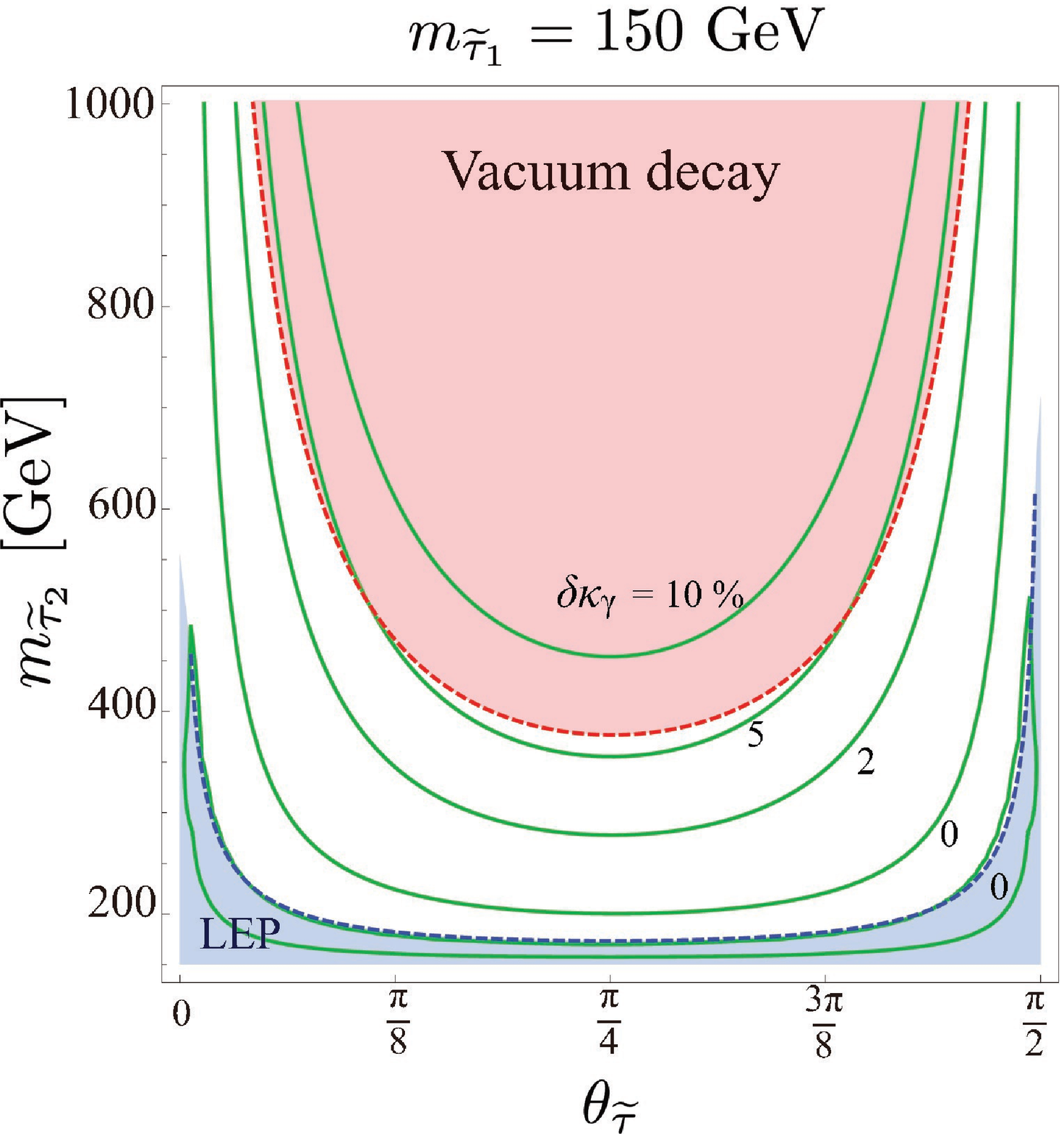} 
 \end{center}
 \caption[Contours of $\delta \kappa_{\gamma}$ as a function of $\theta _{\widetilde{\tau}}$ and $m_{\widetilde{\tau }_2}$.]
 {Contours of $\delta \kappa_{\gamma}$ are shown by the green solid lines. 
The lightest stau mass is $90\GeV$ in the left panel and $150\GeV$ in the right. 
Here, we set to $\tan\beta = 20$ and $A_\tau = 0$. 
The red regions  are excluded by the vacuum meta-stability condition \eqref{eq:kyb}. 
The blue regions are excluded by the chargino search at LEP experiment.}
 \label{fig:mass_angle}
\end{figure}
%%%%%%%%%%%%%%%%%%%%%%%%%%%%%%%%%%%%%
In Fig.~\ref{fig:mass_angle}, we show contours of $\delta \kappa_{\gamma} = \kappa _{\gamma} -1$ as a function of $\theta_{\widetilde\tau}$ and $m_{\widetilde\tau_2}$, 
where the green solid lines are contours of $\delta \kappa_{\gamma}$ for given $m_{\widetilde\tau_1}$. 
The stau contribution depends on $\theta_{\widetilde\tau}$ and is maximized when $\sin 2\theta_{\widetilde\tau}$ is close to unity ($\theta_{\widetilde\tau} \sim \pi/4$) for fixed $m_{\widetilde\tau_1}$ and $m_{\widetilde\tau_2}$. 
Also, $\delta \kappa_{\gamma}$ is enhanced by larger $m_{\widetilde\tau_2}$.
On the other hand, if $\widetilde\tau_2$ is extremely heavy ($\widetilde\tau_2 = \mathcal{O}(1)\TeV$), the stau contribution to $\kappa_{\gamma}$ becomes insensitive to $m_{\widetilde\tau_2}$ and controlled by $m_{\widetilde\tau_1}$ and $\theta_{\widetilde\tau}$.

In Fig.~\ref{fig:mass_angle}, the red regions are excluded by the vacuum meta-stability condition. 
Eq.~\eqref{eq:kyb} gives an upper bound on $m_{\widetilde\tau_{LR}}^2$ for given $m_{\widetilde\tau_1}$ and $m_{\widetilde\tau_2}$. 
Combined with Eq.~\eqref{eq:staumix}, 
the mixing angle $\theta_{\widetilde\tau}$ is constrained as a function of $m_{\widetilde\tau_1}$ and $m_{\widetilde\tau_2}$.
When $m_{\widetilde\tau_2}$ is small, the angle is not limited by the vacuum meta-stability condition, 
and $\delta \kappa_{\gamma}$ is maximized when $\sin 2\theta_{\widetilde\tau} = 1$ is satisfied. 
It is found that $\delta \kappa_{\gamma}$ becomes largest just below the red region with $\sin 2\theta_{\widetilde\tau} = 1$ in each panel of Fig.~\ref{fig:mass_angle}.
On the other hand, the vacuum meta-stability condition constrains the stau mixing angle for large $m_{\widetilde\tau_2}$.
The maximal value of $\delta \kappa_{\gamma}$ decreases, as $m_{\widetilde\tau_2}$ increases.
When $\widetilde\tau_2$ is extremely heavy, the vacuum meta-stability condition becomes insensitive to $m_{\widetilde\tau_2}$ and determined by $m_{\widetilde\tau_1}$. 
This is because, in the decoupling limit, $\widetilde\tau_2$ does not contribute to the field configuration of the bounce solution to derive the vacuum meta-stability condition. 
The maximal value of $\delta \kappa_{\gamma}$ is determined by $m_{\widetilde\tau_1}$.

In the analysis, we choose $\tan\beta = 20$, $A_\tau = 0$ and $M_2 = 500\GeV$, where $M_2$ is the Wino mass.\footnote{
We consider the chargino contribution as theoretical error in this analysis.
Detailed estimation for the error will be discussed in Sec.~\ref{subsec:discussion}.}
The stau contribution to $\kappa_{\gamma}$ and the vacuum meta-stability condition are almost independent of them, once $m_{\widetilde{\tau}_{LR}}^2$ is given. 
Rather, they are included in the definition of $m_{\widetilde{\tau}_{LR}}^2$ in association with the Higgsino mass parameter $\mu$. 
For fixed $m_{\widetilde{\tau}_{LR}}^2$, $\mu$ becomes smaller, as $\tan\beta$ increases. 
When the charginos are light, they can affect the Higgs coupling \cite{Carena:2011aa,Batell:2013bka}. 
Their contribution to $\kappa_{\gamma}$ is taken into account for completeness.
It is at most a few percents in the vicinity of the blue region and much less than 1\% around the red region in Fig.~\ref{fig:mass_angle}.
The blue region is already excluded by LEP experiment\cite{CharginoLEP}, where the lightest chargino mass is less than $104\GeV$.

% Figure %%%%%%%%%%%%%%%%%%%%%%%%%%%%%%%%
\begin{figure}[tb]
\begin{center}
 \includegraphics[width=15cm]{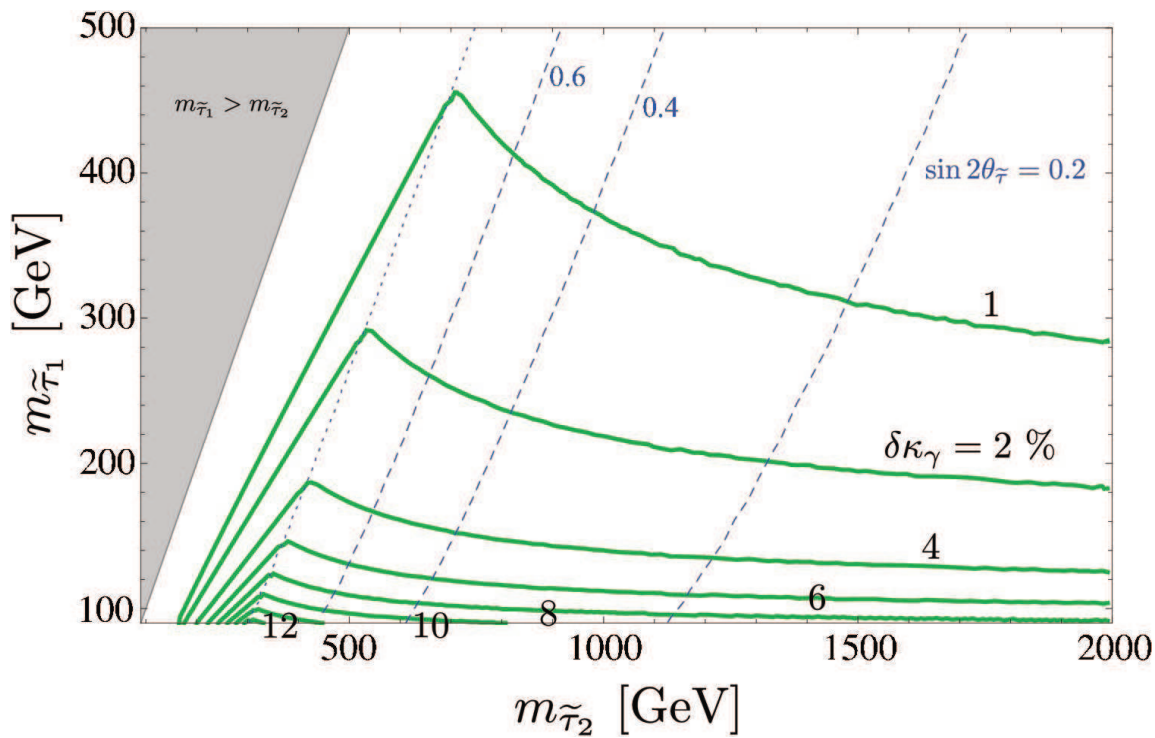}  
 \end{center}
 \caption[Contours of $\delta \kappa _{\gamma }$ as a function of $m_{\widetilde{\tau }_1}$ and $m_{\widetilde{\tau }_2}$.]
 {Contours of $\delta \kappa_{\gamma}$ are shown by the green solid lines.
At each point, $\delta \kappa_{\gamma}$ is maximized under the vacuum meta-stability condition.
Here, $\tan\beta = 20$, $A_\tau = 0$ and $\pi/4 \leq\theta_{\widetilde{\tau}} < \pi/2$ are taken. 
The blue dashed lines are contours of $\sin 2 \theta_{\widetilde{\tau}}$.
In the left region of the blue dotted line (the leftmost blue line), $\sin 2 \theta_{\widetilde{\tau}} = 1$ is satisfied.}
 \label{fig:StauMass}
\end{figure}
%%%%%%%%%%%%%%%%%%%%%%%%%%%%%%%%%%%%%

Next, let us study the stau mass region. 
In Fig.~\ref{fig:StauMass}, we show contours of $\delta \kappa_{\gamma}$ by the green solid lines. 
Here, $\delta \kappa_{\gamma}$ is maximized with satisfying the vacuum meta-stability condition \eqref{eq:kyb}
at each $(m_{\widetilde\tau_1},m_{\widetilde\tau_2})$. 
Each contour is composed of the two regions. 
In the left region of the peak, i.e. small $m_{\widetilde\tau_2}$ region, $\sin 2\theta_{\widetilde\tau} = 1$ is satisfied. 
The stau contribution to $\kappa_{\gamma}$ is enhanced when $m_{\widetilde\tau_2}$ is larger, as explained above.  
On the other hand, in the right region of the peak, $\theta_{\widetilde\tau}$ is limited by the vacuum meta-stability condition. 
Here, $\sin 2\theta_{\widetilde\tau}$ is less than unity. 
Contours of $\sin 2\theta_{\widetilde\tau}$ are shown by the blue dashed lines in Fig.~\ref{fig:StauMass}. 
As already found in Fig.~\ref{fig:mass_angle}, $\kappa_{\gamma}$ is enhanced, when $m_{\widetilde\tau_2}$ is smaller. 

In the Figs.~\ref{fig:StauMass}, we also choose $\tan\beta = 20$, $A_\tau = 0$ and $M_2 = 500\GeV$. 
The results are almost independent of them except for the region in the vicinity of $m_{\widetilde\tau_1} = m_{\widetilde\tau_2}$.
When $m_{\widetilde\tau_1}$ is very close to $m_{\widetilde\tau_2}$, the Higgsinos become light, because the stau left-right mixing parameter tends to be small (see Eq.\eqref{eq:staumix}).
Then, the charginos can contribute to $\kappa_{\gamma}$. 
Otherwise, their contribution is negligible in Fig.~\ref{fig:StauMass}. 
In the figure, it is also supposed that the lightest stau is mainly composed of the right-handed component, $\pi/4 \leq\theta_{\widetilde{\tau}} < \pi/2$.
As mentioned above, the stau mass region in Fig.~\ref{fig:StauMass} is almost insensitive to this choice. 

Currently, the measured values of $\kappa_{\gamma}$ at LHC are consistent with the SM prediction. 
The uncertainties are 15\% (ATLAS) \cite{Aad:2013wqa} and 25\% (CMS) \cite{CMS:yva}. 
As found in Fig.~\ref{fig:StauMass}, they are not precise enough to probe the stau contribution for $m_{\widetilde\tau_1} > 100\GeV$.
In future, the sensitivity will be improved very well, as mentioned in Sec.~\ref{sec:kappaexp}.
It is expected that LHC measures $\kappa_{\gamma}$ at about 7\% and 5\% for the luminosities, $300\invfb$ and $3000\invfb$, respectively with $\sqrt{s}=14\TeV$ \cite{ATLAS:2013hta,CMS:2013xfa}. 
If the measurement of $\textrm{Br}(h \to \gamma \gamma)/\textrm{Br}(h \to ZZ^{\ast})$ at HL-LHC is combined with the measurements of the Higgs couplings at ILC, it was argued that the uncertainty of $\kappa_{\gamma}$ can be reduced to be about 2\% ($1\sigma$) at $250\GeV$ ILC with $\mathcal{L} = 250\invfb$ \cite{Peskin:2013xra}. 
If the luminosity is accumulated up to $2500\invfb$ at $1\TeV$ ILC, it has been estimated that the accuracy of $\kappa_{\gamma}$ can be better than 1\% \cite{Peskin:2013xra}. 

It is noteworthy that, once an excess of $\kappa_{\gamma}$ is measured, the mass region of staus are determined from Fig.~\ref{fig:StauMass}.
From the joint analysis of $250\GeV$ ILC and HL-LHC, $\delta \kappa_{\gamma}$ is expected to be measured with the uncertainty of $2\%$ at the $1\sigma$ level.
If $\delta \kappa_{\gamma}$ is measured to be larger than 4\%, the upper bound is obtained as $m_{\widetilde\tau_1} < 200\GeV$.\footnote
{
The vacuum meta-stability condition determines the upper bound. If thermal transitions are taken into account, the constraint could be more severe especially when the stau is light \cite{Endo:2011uw}.
}  
Such a stau can be discovered at $500\GeV$ ILC. 
In fact, the stau is detectable up to $230\GeV$ at ILC with $\sqrt{s}=500\GeV$ and $\mathcal{L} = 500\invfb$ \cite{Baer:2013vqa}.
On the other hand, if $\delta \kappa_{\gamma}$ is measured to be $2\%$ ($1\%$), the stau mass is predicted to be less than $290\GeV$ ($460\GeV$). 
This is within the kinematical reach of $1\TeV$ ILC.
Therefore, if the stau contribution to $\kappa_{\gamma}$ is large enough to be measurable, the stau is predicted to be discovered at ILC.\footnote
{
Although the stau mass region could also be accessed by LHC, future sensitives of the stau searches have not been known. 
In particular, ILC is superior when the stau mass is degenerate with the Bino mass. 
} 

Let us mention the case when the heaviest stau is very heavy. 
In contrast to the lightest stau $\widetilde\tau_1$, 
the heaviest stau $\widetilde\tau_2$ can be decoupled with $\kappa_{\gamma}$ enhanced and the vacuum meta-stability condition satisfied.
In Fig.~\ref{fig:StauMass}, $\delta \kappa_{\gamma}$ is insensitive to $m_{\widetilde\tau_2}$ and determined by $m_{\widetilde\tau_1}$ for very large $m_{\widetilde\tau_2}$.
In the limit, $\mathcal{M}_{\gamma\gamma}(\widetilde\tau)$ is determined only by $m_{\widetilde\tau_1}$ and $g_{h\widetilde\tau_{1}\widetilde\tau_{1}}$. 
The vacuum meta-stability condition of $g_{h\widetilde\tau_{1}\widetilde\tau_{1}}$ is independent of $m_{\widetilde\tau_2}$ and approximately proportional to $m_{\widetilde\tau_1}$ \cite{Endo:2011uw}. 
Since the loop function $A_0^h (x_{\widetilde\tau_1})$ is insensitive to $m_{\widetilde\tau_1}$ for $m_{\widetilde\tau_1}\gtrsim 100\GeV$, $\mathcal{M}_{\gamma\gamma}(\widetilde\tau)$ is almost scaled by $1/m_{\widetilde\tau_1}$, when the heaviest stau is decoupled. 
Thus, the excess of $\kappa_{\gamma}$ is explained by a light stau. As found in Fig.~\ref{fig:StauMass}, the upper bound on $m_{\widetilde\tau_1}$ for larger $m_{\widetilde\tau_2}$ is more severe than that for smaller $m_{\widetilde\tau_2}$. Such a light stau can be discovered at ILC.

We discussed that once the deviation from the SM prediction of $\kappa _{\gamma }$ were observed, 
the stau mass region were determined by the vacuum meta-stability condition.
In Sec.~\ref{sec:stau}, we will study stau searches for ILC based on the stau mass region.

\section{Selectron/Smuon}
\label{sec:slepton}
%!TEX root = ../Dthesis.tex

\subsection{Overview}

In Sec.~\ref{subsec:gm2vac} and \ref{subsec:vacuum}, 
we showed that when the muon $g-2$ anomaly \eqref{eq:g-2_deviation} is solved by the Bino--smuon contribution \eqref{eq:gminus2bino}, 
soft masses of the Bino and the smuon are bounded from above by the vacuum meta-stability condition \eqref{eq:kyb}. 
The result is summarized in Tab.~\ref{tab:summary}. 
In this section, we study experimental status and future prospects to search for such SUSY models. 
The mass bounds depend on the slepton mass spectrum. 
When the stau is degenerate with the selectron and the smuon, a tight constraint is imposed by the vacuum meta-stability condition of the staus. 
In Sec.~\ref{subsec:universal}, it will be discussed that the limit is strong enough for the superparticles to be detectable directly in colliders. 
On the other hand, if the staus are heavier or decoupled, the meta-stability bound is relaxed. 
The superparticle masses can exceed the collider sensitivities. In Sec.~\ref{subsec:non-univ}, 
it will be argued that such hierarchical mass spectrum can be probed by LFV and CPV.

%%%%%%%%%%%%%%%%%%%%%%%%%%%%%%%%%%%%%%%% Table
\begin{table}[t]
\begin{center}
  \begin{tabular}{l|cccc} \hline
    Mass spectrum & Smuon & Vacuum  & LHC/ILC & LFV/EDM \\ \hline \hline
    $m_{\tilde e} = m_{\tilde\mu} = m_{\tilde\tau}$ (Sec.~\ref{subsec:universal}) & $< 330/460\GeV$ & $\tilde\tau$ & \checkmark &  \\  
    $m_{\tilde e} = m_{\tilde\mu} < m_{\tilde\tau}$ (Sec.~\ref{subsec:non-univ}) & $< 1.4/1.9\TeV$  & $\tilde\tau$ or $\tilde\mu$  &  & \checkmark \\ \hline
  \end{tabular}
   \caption[Summary of smuon searches.]{Summary of searches. The lightest smuon mass is restricted to explain the muon $g-2$ discrepancy at the $1\sigma/2\sigma$ level. The left-right mixing is limited by the vacuum stability condition of the stau-- or smuon--Higgs potential, depending on the stau masses. 
 The check mark shows promising experimental searches.
 Models of the universal mass spectrum can be tested by LHC/ILC, while those of the non-universal spectrum predict large LFV/EDM.
}
   \label{tab:summary}
\end{center}
\end{table}

%%%%%%%%%%%%%%%%%%%%%%%%%%%%%%%%%%%%%%%%

%%%%%%%%%%%%%%%%%%%%%%%%%%% Subsection
\subsection{Universal Slepton Mass}
\label{subsec:universal}
%!TEX root = ../Dthesis.tex

In this section, we study collider searches for the SUSY models with the universal slepton mass spectrum. 
The left- (right-) handed sleptons have a common soft SUSY-breaking mass,
\begin{align}
m_{\widetilde e_L} = m_{\widetilde\mu_L} = m_{\widetilde\tau_L},~~~
m_{\widetilde e_R} = m_{\widetilde\mu_R} = m_{\widetilde\tau_R}.
\end{align}
Then, the vacuum meta-stability condition from the stau--Higgs potential restricts slepton masses up to $330-460\GeV$ to solve the muon $g-2$ anomaly. This is within the reach of the LHC or ILC sensitivity. 
In fact, some of the parameter regions have already been excluded by current LHC data, as shown later. 

Collider signatures depend on the LSP. 
In the case of universal soft slepton masses, 
either the lightest neutralino or the lightest stau become LSP.\footnote{
The lightest stau is lighter than sneutrinos when $\mu$ is large. 
}
In the latter case, the stau is likely to be long-lived.\footnote{
The stau could decay in detectors, for instance, through R-parity violations. The SUSY signatures depend on decay channels. Such cases are not discussed here. 
}
The (meta-) stable staus leave charged tracks in detectors. 
%Such signatures have been studied by ATLAS~\cite{ATLAS-CONF-2013-058} and CMS~\cite{Chatrchyan:2013oca}. 
Such signatures have been studied by ATLAS~\cite{ATLAS:2014fka} and CMS~\cite{Chatrchyan:2013oca}.
The CMS constraint on the cross section of the stau direct production provides the 95\% CL exclusion limit, $m_{\widetilde\tau_1} > 339\GeV$. 
If this is imposed in addition to the vacuum stability condition, 
all the parameter regions of the long-lived stau are excluded in the case of $m_{\widetilde\tau_1} \leq m_{\widetilde\chi^0_1}$. 

%%%%%%%%%%%%%%%%%%%%%%%%%%%%%%%%%%%%%%%%%%%%% Figure
\begin{figure}[t]
 \begin{center}
 \includegraphics[width=8cm]{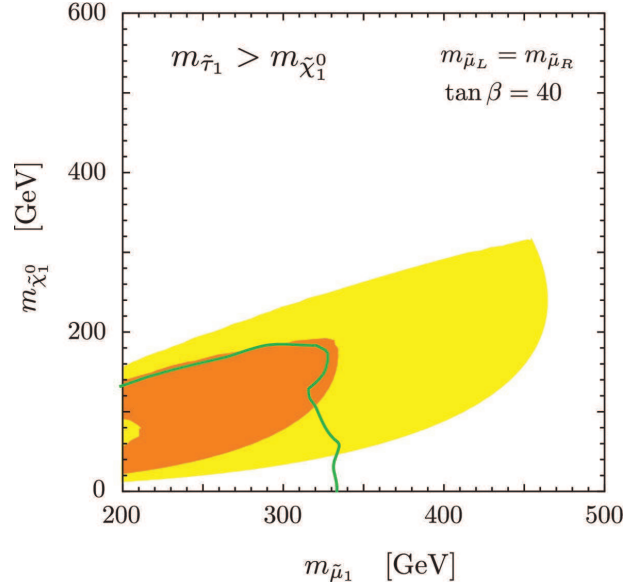}
 \caption[The SUSY contributions to the muon $g-2$ and the vacuum stability bounds in $m_{\widetilde\tau_1} > m_{\widetilde\chi^0_1}$.]
 {The SUSY contributions to the muon $g-2$ as a function of the lightest smuon, $m_{\widetilde{\mu }_1}$, 
 and the lightest neutralino, $m_{\widetilde{\chi }^0_1}$.
 In the orange (yellow) regions, the discrepancy of the muon $g-2$ \eqref{eq:g-2_deviation} is explained at the 1$\sigma $ (2$\sigma$) level.
 The left-right mixing is maximized under the vacuum stability condition \eqref{eq:kyb}.
 The parameters are $m_{\widetilde{\ell }_L} = m_{\widetilde{\ell }_R}$, $\tan \beta = 40$, and $M_{\text{soft}} = 10\TeV$.
 Further, the condition that the neutralino LSP, $m_{\widetilde\tau_1} > m_{\widetilde\chi^0_1}$ is imposed.
 All the parameter regions of the long-lived stau ($m_{\widetilde\tau_1} \leq m_{\widetilde\chi^0_1}$) are excluded by the ATLAS and CMS (long-lived stau search).
 The region below the green line is excluded by search for direct smuon production at ATLAS.}
 \label{fig:M1smu_plane}
 \end{center}
 \end{figure}
%%%%%%%%%%%%%%%%%%%%%%%%%%%%%%%%%%%%%%%%%%%%%
 
\begin{figure}[tbp]
 \begin{center}
 \includegraphics[width=8cm]{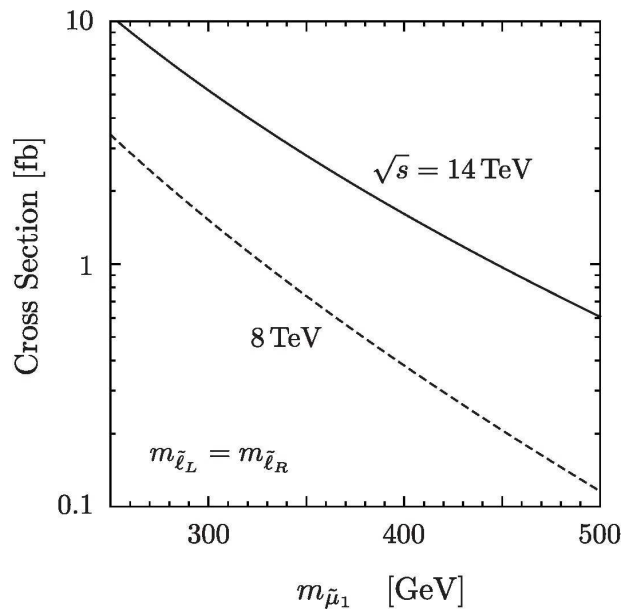}
 \caption[The total cross section of the smuon productions at LHC with $\sqrt{s} = 8\TeV$ and $14\TeV$.]
 {The total cross section of the smuon productions at LHC with $\sqrt{s} = 8\TeV$ (dashed) and $14\TeV$ (solid). The parameters satisfy $m_{\widetilde\ell_L} = m_{\widetilde\ell_R} = 2M_1$. The left-right mixing is maximized under the vacuum stability condition with $\tan\beta=40$.}
 \label{fig:LHC_smuon}
 \end{center}
 \end{figure}

In Fig.~\ref{fig:M1smu_plane}, we show the SUSY contributions to the muon $g-2$ when the left-right mixing is maximized under the two conditions; (i) the vacuum stability and (ii) the neutralino LSP, $m_{\widetilde\tau_1} > m_{\widetilde\chi^0_1}$. Then, SUSY signature is (opposite-sign same-flavor) di-lepton with large missing transverse energy. First, selectrons and smuons are produced by collisions. They subsequently decay into the lightest neutralino and a partner lepton. 
%This signature was studied by ATLAS~\cite{ATLAS-CONF-2013-049} and CMS~\cite{CMS-PAS-SUS-13-006}. 
This signature was studied by ATLAS~\cite{Aad:2014vma} and CMS~\cite{Khachatryan:2014qwa}. 
In particular, we show the 95\% CL exclusion limit for $m_{\widetilde{\mu}_L} = m_{\widetilde{\mu}_R}$ 
by ATLAS in Fig.~\ref{fig:M1smu_plane} (and in Fig.~\ref{fig:MSL-M1}).
Here, the region below the green line is excluded. 
In detail, the left-right mixing is negligible in the ATLAS analysis. 
However, the left- and right-handed smuons maximally mix with each other in Fig.~\ref{fig:M1smu_plane}. 
The total cross section of the smuon productions decreases by 10\% compared to the ATLAS analysis. 
Since it is sufficiently small, the exclusion limit is considered to be almost the same as the ATLAS result. 
As a result, we find that almost all the $1\sigma$ parameter region of the muon $g-2$ is already excluded.\footnote{
In the ATLAS analysis, $m_{\widetilde{e}} = m_{\widetilde{\mu}}$ is assumed. 
%The selectron production provides almost the same constraint as the smuon~\cite{ATLAS-CONF-2013-049}. 
The selectron production provides almost the same constraint as the smuon~\cite{Aad:2014vma}. 
Thus, if selectrons are decoupled, the mass bound becomes weaker. However, LFV/CPV constraints are severe, as will be mentioned in Sec.~\ref{subsec:non-univ}.
}
The sensitivity will be improved by the upgrade of the energy and the luminosity. 
Fig.~\ref{fig:LHC_smuon} shows the total cross section of the smuon productions at $\sqrt{s}=8\TeV$ and $14\TeV$. The cross section is estimated at the leading order.
For instance, it becomes $1\,{\rm fb}$ for the lightest smuon mass of $330\GeV$ at $\sqrt{s}=8\TeV$, which corresponds to the current LHC bound. The same cross section is obtained for $450\GeV$ at $14\TeV$.
Studies for the future sensitivity are required. 

%%%%%%%%%%%%%%%%%%%%%%%%%%%%%%%%%%%%%%%%%%%%% Figure
%%%%%%%%%%%%%%%%%%%%%%%%%%%%%%%%%%%%%%%%%%%%%

In the neutralino LSP case, the stau productions have also been studied in LEP and LHC. 
If their mass difference is larger than $15\GeV$, the stau mass is constrained to be $\gtrsim 81.9\GeV$ by LEP~\cite{Beringer:1900zz}. 
%LHC is still ineffective to search for di-tau events from the direct stau productions~\cite{ATLAS-CONF-2013-028}. 
LHC is still ineffective to search for di-tau events from the direct stau productions~\cite{Aad:2014yka}. 
These limits are so sufficiently weak that Fig.~\ref{fig:M1smu_plane} does not change even if they are imposed. 

%%%%%%%%%%%%%%%%%%%%%%%%%%%%%%%%%%%%%%%%%%%%% Table
\begin{table}[tbp]
\begin{center}
\begin{tabular}{|c|ccc|ccccccc|c|}
\hline
&
$m_{\widetilde\ell}$ & $M_1$ & $\mu$ & 
$m_{\widetilde{e}_1}$ & $m_{\widetilde{e}_2}$ & 
$m_{\widetilde{\mu}_1}$ & $m_{\widetilde{\mu}_2}$ & 
$m_{\widetilde{\tau}_1}$ & $m_{\widetilde{\tau}_2}$ & 
$m_{\widetilde{\chi}^0_1}$ 
& $\Delta a_\mu$ 
\\ \hline
A &
300 & 200 & 756 &
303 & 304 &
298 & 309 &
199 & 380 &
199 & 16.1
\\
$\rm{A^{\prime}}$&
300 & 200 & 699 &
303 & 304 &
299 & 308 &
209 & 375 &
199 & 14.6
\\
B &
470 & 250 & 1680 &
472 & 472 &
465 & 479 &
329 & 581 &
250 & 10.2
\\
C &
340 & 160 & 1138 &
343 & 343 &
336 & 350 &
199 & 442 &
160 & 18.0
\\ \hline
\end{tabular}
\caption[Model parameters at several sample point.]
{Model parameters and mass spectra at several model points in Fig.~\ref{fig:M1smu_plane}. The masses are in units of GeV, and the muon $g-2$ is scaled by $10^{-10}$. Here, $m_{\widetilde\ell}$ denotes $m_{\widetilde\ell_L} = m_{\widetilde\ell_R}$, and $\tan\beta=40$ is set.}
\label{tab:mass}
\end{center}
\end{table} 
%%%%%%%%%%%%%%%%%%%%%%%%%%%%%%%%%%%%%%%%%%%%%

In Tab.~\ref{tab:mass}, superparticle mass spectra are listed for several points in Fig.~\ref{fig:M1smu_plane}. The lightest neutralino mass is close to the Bino mass, since $\mu$ is very large. 
Given $m_{\widetilde\ell} \equiv m_{\widetilde\ell_L} = m_{\widetilde\ell_R}$, the slepton masses are hierarchical except for those of the selectrons due to a large left-right mixing. The lightest stau mass is closest to the neutralino mass among the sleptons. In Fig.~\ref{fig:M1smu_plane}, 
the point A (A') is around the upper side of the contour of the muon $g-2$. 
In the vicinity of this side, $m_{\widetilde\tau_1}$ has the closest value to $m_{\widetilde\chi^0_1}$ in the allowed range. 
If $m_{\widetilde\tau_1} > m_{\widetilde\chi^0_1}$ is imposed, 
it satisfies $m_{\widetilde\tau_1} = m_{\widetilde\chi^0_1}$ to maximize the SUSY contributions to the muon $g-2$ around the upper side of the contour. 
Above it, the regions are excluded by the long-lived stau search, 
or the muon $g-2$ becomes too small. 
On the other hand, if the mass difference between the stau and the neutralino is assumed to be larger than $\delta M$, 
they satisfy $m_{\widetilde\tau_1}=m_{\widetilde\chi^0_1}+\delta M$  in the vicinity of the upper side. 
At the point A', $\delta M = 10\GeV$ is imposed, and $m_{\widetilde\tau_1}$ is larger than $m_{\widetilde\chi^0_1}$ by $10\GeV$. 
In contrast, the points B and C are away from this region. 
B (C) is close to the maximal end point of the lightest smuon mass which explains the muon $g-2$ at the $2\sigma$ ($1\sigma$) level. 
Here, the left-right mixing is determined by the vacuum meta-stability condition of the stau. 

%%%%%%%%%%%%%%%%%%%%%%%%%%%%%%%%%%%%%%%%%%%%% Figure
\begin{figure}[t]
 \begin{center}
 \includegraphics[width=8cm]{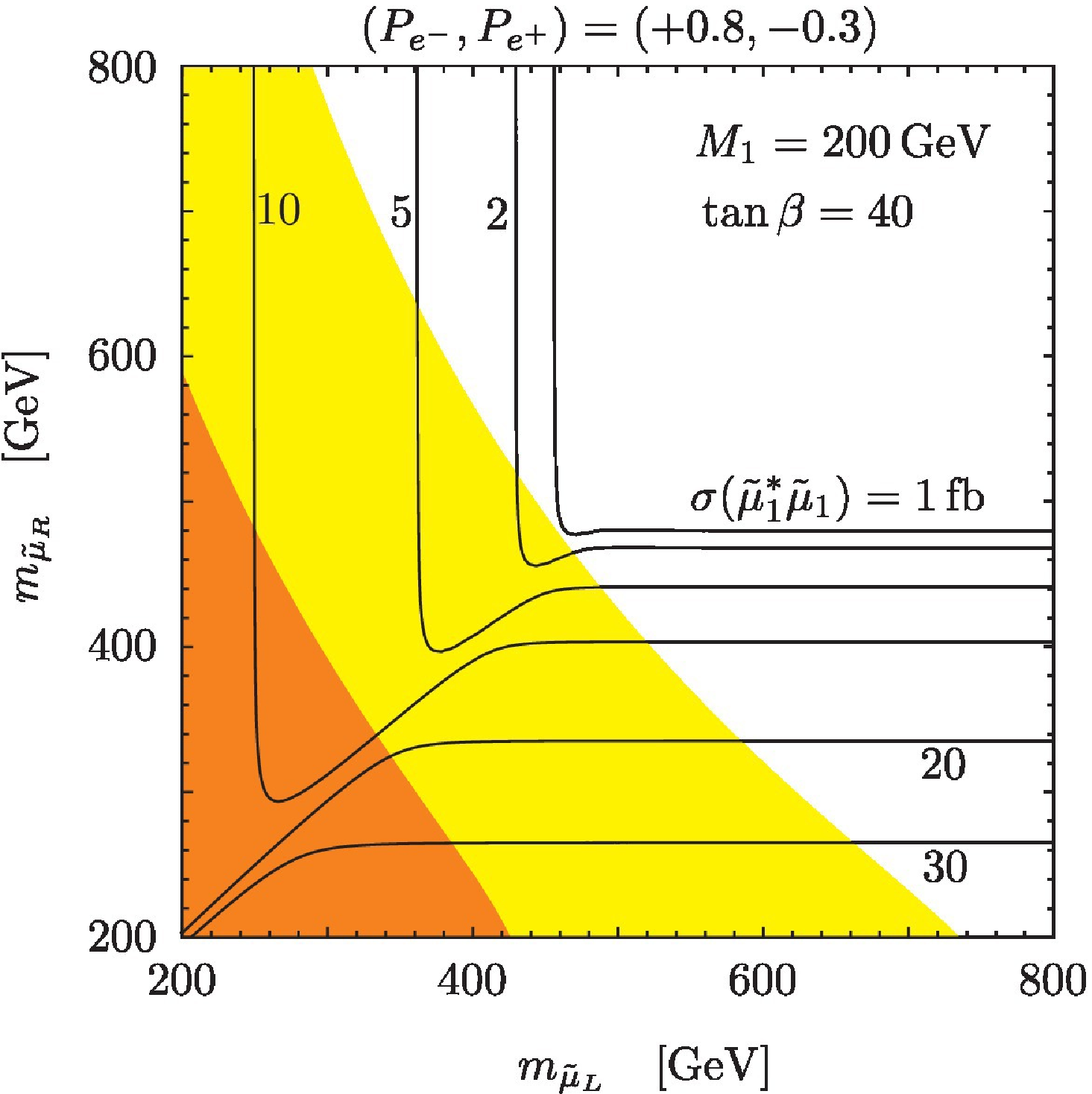}\hspace*{5mm}
 \includegraphics[width=8cm]{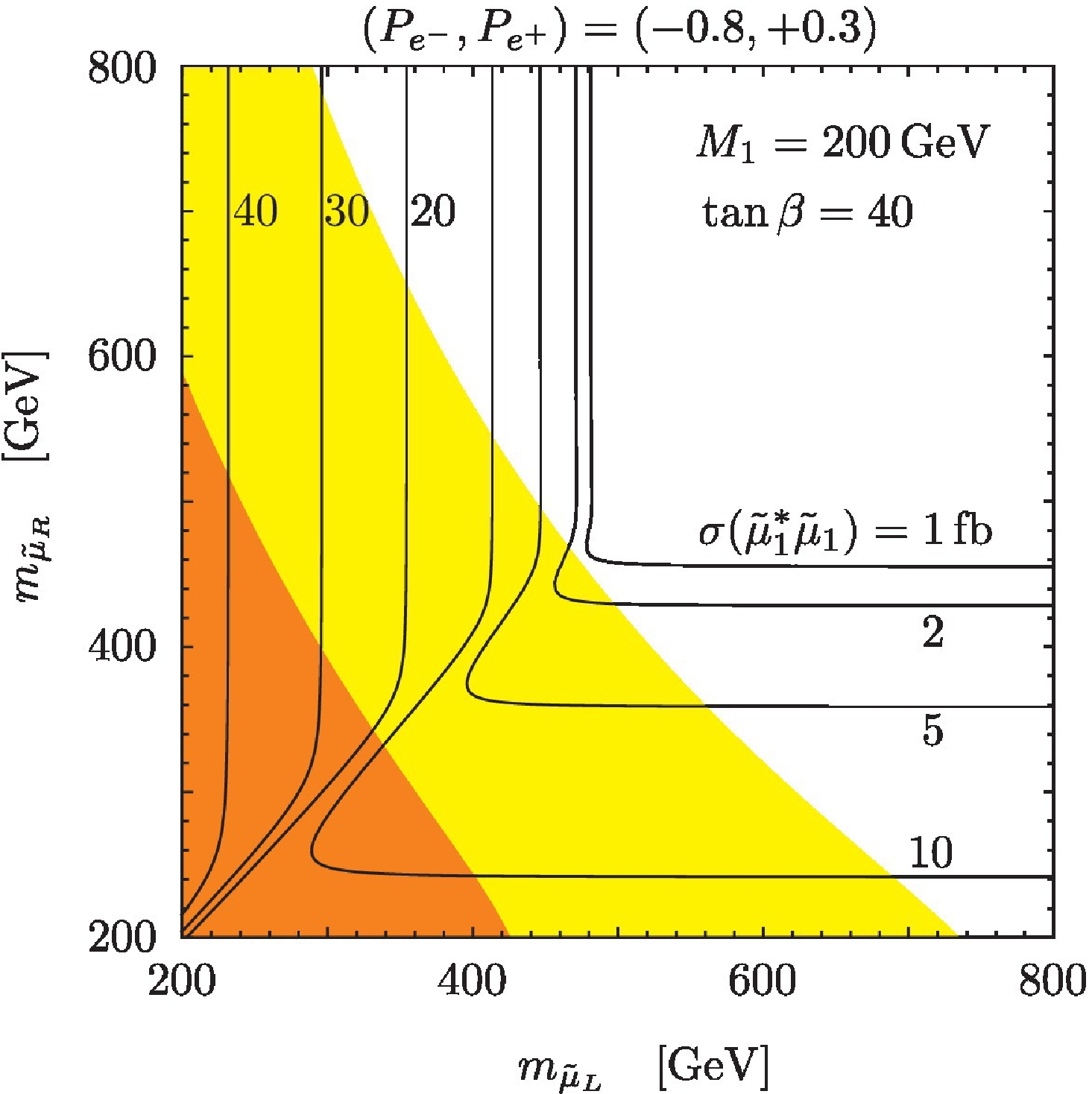}
  \end{center}
 \caption[The production cross section of the lightest smuon at ILC with $\sqrt{s} = 1\TeV$. ]
 {The production cross section of the lightest smuon at ILC with $\sqrt{s} = 1\TeV$. The beams are assumed to be polarized at 80\% $(e^-)$ and 30\% $(e^+)$. The left-right mixing is maximized under the vacuum stability condition with $\tan\beta=40$. The $1\sigma$ ($2\sigma$) regions of the muon $g-2$ are also shown by the orange (yellow) bands. The parameters are $M_1=200\GeV$ and $M_{\rm soft} = 10\TeV$.} 
  \label{fig:SmuonProdILC}
\end{figure}
%%%%%%%%%%%%%%%%%%%%%%%%%%%%%%%%%%%%%%%%%%%%%

Linear colliders of $e^+e^-$ (ILC) are very useful (see e.g., Ref.~\cite{Baer:2013cma,Baer:2013vqa}). 
They can provide rich informations of the models. 
Moreover, they are superior to LHC when the slepton masses are close to that of the lightest neutralino~\cite{Martyn:2004jc}. 
In the orange/yellow regions of Fig.~\ref{fig:M1smu_plane}, selectrons, smuons and staus can be produced at the linear colliders. 
In Fig.~\ref{fig:SmuonProdILC}, we show production cross sections of the lightest smuon for the left- and right-handed smuon masses. 
Here, the production cross section of the lightest smuon are \cite{Freitas:2003yp}\footnote{
The cross section is calculated at the leading order in Fig.~\ref{fig:SmuonProdILC}. It can be enhanced by $\sim 10\%$ near the mass threshold~\cite{Freitas:2003yp}. 
} 
\beq
\sigma(e^+e^-\to\widetilde\mu_1\widetilde\mu_1) &=& 
\frac{8\pi\alpha^2}{3s} \lambda^{\frac{3}{2}} 
\bigg[
  c_{11}^2 \frac{\Delta_Z^2}{\sin^42\theta_W}(P_{-+}L^2 + P_{+-}R^2)
\nonumber\\&&
  + \frac{1}{16} (P_{-+} + P_{+-})
  + c_{11}\frac{\Delta_Z}{2\sin^22\theta_W}(P_{-+}L + P_{+-}R)
\bigg].
\label{eq:CrossSectionsmuon}
\eeq
and it is assumed that the collision energy is $\sqrt{s} = 1\TeV$ at ILC, and the beams are polarized at 80\% for the electron and 30\% for the positron. 
The parameters are defined in Appendix \ref{app:slepprod}.
The left-right mixing is maximized under the vacuum stability condition. The other parameters are $M_1=200\GeV$ and $\tan\beta=40$. 
It is found that the cross section is larger than 20 (1)$\invfb$ for the $1\sigma$ ($2\sigma$) parameter region of the muon $g-2$. It is expected that the smuons can be discovered in almost all the parameter region that are kinematically allowed (cf.~\cite{Baer:2013vqa}).

It is possible to measure masses of the smuon and the neutralino from event distributions and the cross section~\cite{Tsukamoto:1993gt,Nojiri:1996fp}. Also, the chirality structure of the smuons could be determined by the beam polarization. We assume $(P_{e^-},P_{e^+}) = (+0.8,-0.3)$ in the left panel of Fig.~\ref{fig:SmuonProdILC}, and $(-0.8,+0.3)$ in the right panel. The cross section of a chiral smuon is sensitive to the polarization, because the productions proceed by the s-channel $\gamma/Z$ exchanges. 
On the other hand, when $m_{\widetilde\mu_L}=m_{\widetilde\mu_R}$, the cross section is insensitive to the beam polarization. Since the left-right mixing is large, the left- and right-handed smuons are maximally mixed with each other. Both of them can be produced, and the mass difference is expected to be measured, for instance, by measuring threshold productions. At the model points in Tab.~\ref{tab:mass}, the difference is $\gtrsim 10\GeV$, which is much larger than the uncertainty of the mass measurement at ILC, $\Order(10-100)\MeV$~\cite{Baer:2013cma}. It is emphasized that the smuon productions are clean and direct signatures of the SUSY contributions to the muon $g-2$. 

%%%%%%%%%%%%%%%%%%%%%%%%%%%%%%%%%%%%%%%%%%%%% Figure
\begin{figure}[t]
 \begin{center}
 \includegraphics[width=8cm]{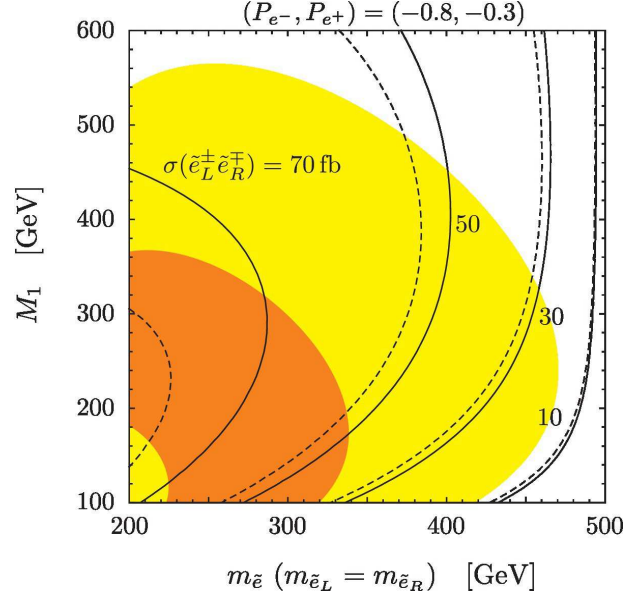}
  \end{center}
 \caption[The production cross section of $\widetilde e^+_R \widetilde e^-_L$ and $\widetilde e^+_L \widetilde e^-_R$ at ILC with $\sqrt{s} = 1\TeV$.]
 {The production cross section of $\widetilde e^+_R \widetilde e^-_L$ and $\widetilde e^+_L \widetilde e^-_R$ at ILC with $\sqrt{s} = 1\TeV$. The parameters are same as Fig.~\ref{fig:SmuonProdILC}, but $M_1$ is varied with $m_{\widetilde e_L}=m_{\widetilde e_R} \equiv m_{\widetilde e}$. The corrections to the Bino coupling due to decoupling of heavy superparticles are included (discarded) in the black solid (dashed) lines.} 
  \label{fig:SelectronProdILC}
\end{figure}
%%%%%%%%%%%%%%%%%%%%%%%%%%%%%%%%%%%%%%%%%%%%%

The selectron productions proceed not only by the s-channel $\gamma/Z$ exchanges, but also by the t-channel Bino exchange. The latter contribution can enhance the cross section and provide individual information in addition to mass measurements of the selectron and the neutralino. The productions, $e^+_L e^-_L \to \widetilde e^+_R \widetilde e^-_L$ and $e^+_R e^-_R \to \widetilde e^+_L \widetilde e^-_R$, proceed by the t-channel. In Fig.~\ref{fig:SelectronProdILC}, 
we show their total cross sections by the black solid lines (For the formulae of the cross section, see Appendix \ref{app:slepprod}.).
Here, the parameters are $m_{\widetilde e_L} = m_{\widetilde e_R}$, and the beam polarization is 80\% for the electron and 30\% for the positron. It is found that the cross sections are $\Order(10)\,{\rm fb}$ in the muon $g-2$ parameter regions. They are larger than those of the smuon in Fig.~\ref{fig:SmuonProdILC}. Importantly, this channel is useful to measure the Bino--electron--selectron couplings~\cite{Nojiri:1996fp,Nojiri:1997ma}. They are deviated from the U(1)$_Y$ gauge coupling constant by decoupling heavy superparticles, as discussed in Sec.~\ref{subsec:gm2vac}. In Fig.~\ref{fig:SelectronProdILC}, the cross sections are estimated with (without) $\delta\widetilde g_L$ and $\delta\widetilde g_R$~(see Eqs. (\ref{eq:gL}) and (\ref{eq:gR})). The results are shown by the solid (dashed) lines. It is found that 
the corrections enhance the cross section by $8-10\%$ for $M_{\rm susy} = 10\TeV$. This is measurable at ILC~\cite{Nojiri:1996fp,Nojiri:1997ma}. 

Cross sections of $e^+_L e^-_R \to \widetilde e^+_R \widetilde e^-_R$ and $e^+_R e^-_L \to \widetilde e^+_L \widetilde e^-_L$ also include the t-channel contribution of the Bino. 
In particular, the former cross section is enhanced well compared to those solely by the s-channel $\gamma/Z$ exchanges. It can be $\gtrsim 100\,{\rm fb}$ when $M_1$ is relatively small in the muon $g-2$ parameter regions. However, in the latter process, the t-channel contribution interferes destructively with the s-channel contribution. On the other hand, the cross section of $e^+_L e^-_R \to \widetilde e^+_R \widetilde e^-_R$ differs by $5-6\%$ between the cases with and without $\delta\widetilde g_L$ and $\delta\widetilde g_R$. This is smaller than the above channels. 

Let us comment on the dark matter in the present model. The lightest neutralino is a candidate of the dark matter. The neutralino relic abundance becomes consistent with the measured cold dark matter abundance~\cite{Hinshaw:2012aka,Ade:2013zuv} by the stau co-annihilation. The mass difference between the stau and the neutralino is required to be $5-10\GeV$. This corresponds to the model points in the vicinity of the upper side of the muon $g-2$ contours in Fig.~\ref{fig:M1smu_plane}, including the points A or A'. 
%This region has not been excluded by the studies of the di-lepton signature at LHC~\cite{ATLAS-CONF-2013-049, CMS-PAS-SUS-13-006}. 
This region has not been excluded by the studies of the di-lepton signature at LHC~\cite{Aad:2014vma, Khachatryan:2014qwa}. 
Masses of the selectrons and the smuons are too close to that of the neutralino for LHC. 
ILC is superior to study the region~\cite{Baer:2013cma,Baer:2013vqa}. Selectrons and smuons as well as staus can be produced, as discussed above. 

Finally, we comment on the Higgs coupling to di-photon.
As seen in \ref{subsec:kg}, when the stau is light and the left-right mixing is large, the coupling  $\kappa _{\gamma }$ can be sizably enhanced.
Comparing with the muon $g-2$ parameter regions in Fig.~\ref{fig:M1smu_plane}, 
the stau contribution $\delta \kappa _{\gamma }$ becomes $5-20\%$ $(2-5\%)$ in $1\sigma $ ($2\sigma$) region.
The latter is comparable with or larger than the future sensitivity to $\kappa _{\gamma }$ using joint analysis of HL-LHC and ILC. 
If the deviation $\delta \kappa _{\gamma }$ is observed by future collider experiments and can be explained by the stau contribution, 
we will get more strong evidence for the SUSY. 
The stau searches are discussed in Sec.~\ref{sec:stau}.

\subsection{Non-Universal Slepton Mass}
\label{subsec:non-univ}
%!TEX root = ../Dthesis.tex

%%%%%%%%%%%%%%%%%%%%%%%%%%%%%%%%%%%%%%%%%%%%% Figure
\begin{figure}[t]
 \begin{center}
 \includegraphics[width=8cm]{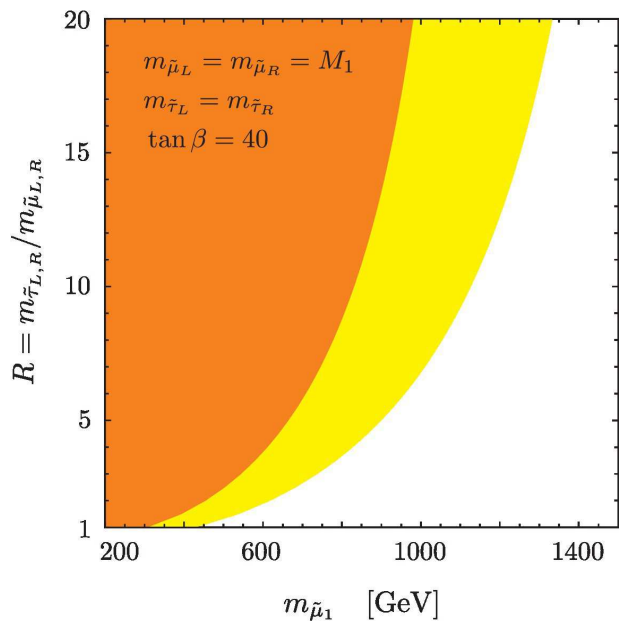}
 \caption[The lower bound on the stau mass to satisfy the vacuum stability condition of the stay.]
 {The lower bound on the stau mass to satisfy the vacuum stability bound of the stau--Higgs potential. In the orange (yellow) region, the SUSY contribution to the muon $g-2$ can explain the discrepancy \eqref{eq:g-2_deviation} at the $1\sigma$ $(2\sigma)$ level. The parameters are $m_{\tilde{\mu}_L} = m_{\tilde{\mu}_R}= M_1$, $m_{\tilde{\tau}_L} = m_{\tilde{\tau}_R}$, $\tan\beta = 40$ and $M_{\rm soft} = 30\TeV$. }
  \label{fig:boundR}
 \end{center}
 \end{figure}
%%%%%%%%%%%%%%%%%%%%%%%%%%%%%%%%%%%%%%%%%%%%%

In this section, we discuss the case without slepton mass universality. The slepton soft SUSY-breaking masses satisfy 
\begin{align}
m_{\tilde e_L} = m_{\tilde\mu_L} < m_{\tilde\tau_L},~~~
m_{\tilde e_R} = m_{\tilde\mu_R} < m_{\tilde\tau_R}.
\label{eq:non-univ}
\end{align}
As discussed in Sec.~\ref{subsec:vacuum}, 
the vacuum meta-stability bound of the stau--Higgs potential is relaxed when the staus are heavy, and the smuon masses are allowed to be larger in order to solve the muon $g-2$ anomaly. In other words, the stau masses are required to be large relative to the smuons, when the smuons are heavy. 
In Fig.~\ref{fig:boundR}, we show the lower bound on the stau mass. 
In the orange (yellow) regions, the muon $g-2$ anomaly is solved at the $1\sigma$ ($2\sigma$) level, while the vacuum meta-stability constraint is avoided. 
The ratios of the stau and smuon masses are defined as
\begin{align}
R_L \equiv \frac{m_{\tilde{\tau}_L}}{m_{\tilde{\mu}_L}},~~~
R_R \equiv \frac{m_{\tilde{\tau}_R}}{m_{\tilde{\mu}_R}}.
\end{align}
The parameters are $M_1 = m_{\tilde{\mu}_L} = m_{\tilde{\mu}_R}$, $m_{\tilde{\tau}_L} = m_{\tilde{\tau}_R}$, $\tan \beta = 40$ and $M_{\rm soft} = 30\TeV$. The vertical axis is $R \equiv R_L = R_R$. If the staus are decoupled, the left-right mixing is bounded either by Eq.~\eqref{eq:limitbound} or the vacuum meta-stability bound of the smuon--Higgs potential, as discussed in Sec.~\ref{subsec:vacuum}. The smuon masses can be $1.4\TeV$ $(1.9\TeV)$ for the $1\sigma$ $(2\sigma)$ level of the muon $g-2$. Even though it becomes difficult to produce such sleptons at LHC or ILC, non-universal slepton masses generically cause problems of too large lepton flavor violations (LFV) and CP violations (CPV).

When the muon $g-2$ anomaly is solved by the SUSY contributions, 
LFV and lepton EDM are generically sizable. 
In this section, the following setup is considered. In the model basis, the slepton mass matrices are diagonal among the flavors, 
\begin{align}
(m_{\tilde\ell_L}^2)_{ij} &= \text{diag}\left( m_{\tilde e_L}^2, m_{\tilde\mu_L}^2, m_{\tilde\tau_L}^2 \right), \\
(m_{\tilde\ell_R}^2)_{ij} &= \text{diag}\left( m_{\tilde e_R}^2, m_{\tilde\mu_R}^2, m_{\tilde\tau_R}^2 \right). 
\end{align}
Furthermore, $m_{\tilde e_L} = m_{\tilde\mu_L}$ and $m_{\tilde e_R} = m_{\tilde\mu_R}$ are imposed. 
Otherwise, LFV becomes too large, as shown later. The Yukawa matrices are generally non-diagonal in the model basis. The mass eigenstate basis of the charged leptons is obtained by the left- and right-handed unitary matrices, $U_L, U_R$, as Eq.~\eqref{eq:diagYukawa}.
In this dissertation, we assume that the unitary matrices are represented as
\begin{equation}
U_{L,R} = \exp 
\left[
\begin{pmatrix}
 0 & (\delta_{L,R})_{12} & (\delta_{L,R})_{13} \\
 -(\delta_{L,R})^{* }_{12} & 0 & (\delta_{L,R})_{23} \\
-(\delta_{L,R})^{* }_{13}  & -(\delta_{L,R})^{*}_{23} & 0 \\
\end{pmatrix} 
\right].
\label{eq:unitary}
\end{equation}
Non-vanishing mixings, $(\delta_{L,R})_{ij}$, induce LFV and EDM, as long as the slepton masses are non-universal.\footnote{
Similar setup has been studied in Ref.~\cite{Endo:2010fk}, where squarks of the first two generations are light.
}
On the other hand, we assume that quark FCNC and CPV are suppressed by heavy colored superparicles in this dissertation.

The magnetic dipole contributions to LFV are represented in Eq.~\eqref{eq:eff_LFV}, 
as mentioned in Sec.~\ref{subsec:TDM}.
In our setup, the Wilson coefficients, $A^L_{ij}$ and $A^R_{ij}$, are dominated by the Bino--slepton contributions, similarly to the muon $g-2$. 
In the mass insertion approximation, they are estimated as \cite{Cho:2001hx, Endo:2013lva}
\beq
A^L_{ij} &=& (1 + \delta^{\rm 2loop}) \frac{\alpha_Y}{8\pi}
\frac{M_1\mu\tan\beta}{m_{\ell_j}} \sum_{a,b=1, 2, 3} 
\Big[ U_R \Big]_{ib} \Big[ M_\ell \Big]_{ba} \Big[ U_L^\dagger \Big]_{aj}\,
F_{a,b}, 
\label{eq:LFVaml} \\ 
A^R_{ij} &=& (1 + \delta^{\rm 2loop}) \frac{\alpha_Y}{8\pi}
\frac{M_1\mu\tan\beta}{m_{\ell_j}} \sum_{a,b=1, 2, 3}
\Big[ U_L \Big]_{ia} \Big[ M_\ell^\dagger \Big]_{ab} \Big[ U_R^\dagger \Big]_{bj}\,
F_{a,b}.
\label{eq:LFVamr} 
\eeq
The two loop factor $\delta^{\rm 2loop}$ is found in Eq.~\eqref{eq:2-loop}. The loop function $F_{a,b}$ is defined as
\begin{align}
F_{a,b} = 
\frac{1}{m_{\tilde\ell_{La}}^2 m_{\tilde\ell_{Rb}}^2} 
f_N \left( \frac{m_{\tilde\ell_{La}}^2}{|M_1|^2}, \frac{m_{\tilde\ell_{Rb}}^2}{|M_1|^2} \right).
\end{align}
The muon FCNCs are most sensitive to the non-universal slepton mass and the non-diagonal Yukawa matrix.
In this case, the coefficients $A^{L.R}_{12}$ is important.
The unitarity matrices in $A^R_{12}$ are expanded as
\begin{align}
\sum_{a,b=1, 2, 3} 
\Big[ U_L \Big]_{1a} \Big[ M_\ell^\dagger \Big]_{ab} \Big[ U_R^\dagger \Big]_{b2}\, F_{a,b}
&=
- \frac{m_\mu}{1+\Delta_\mu} (\delta_L)_{12} \left(F_{1,2} - F_{2,2}\right) 
\\ 
&~~~ +\frac{m_\tau}{1+\Delta_\tau} (\delta_L)_{13} (\delta_R)_{23}^* 
\left(F_{1,2} - F_{1,3} - F_{3,2} + F_{3,3}\right), 
\notag 
\end{align}
at the leading order of $(\delta_L)_{ij}$, $(\delta_R)_{ij}$ and $m_\mu/m_\tau$. 
Here and hereafter, we set $m_e = 0$, for simplicity. 
Similarly, $A^L_{12}$ is obtained by replacing $L \leftrightarrow R$. In the last term, $F_{1,2}$ is dominant when the staus are heavy, i.e., $F_{1,2} \gg F_{1,3}, F_{3,2}, F_{3,3}$ for $m_{\tilde e_L}, m_{\tilde\mu_L} \ll m_{\tilde\tau_L}$ and $m_{\tilde e_R}, m_{\tilde\mu_R} \ll m_{\tilde\tau_R}$. On the other hand, the right-hand side vanishes when the slepton masses are universal, as expected from the super GIM mechanism. In particular, when the sleptons are degenerate in the first two generations, the first term becomes zero due to $F_{1,2}=F_{2,2}$. Otherwise, the muon LFV is induced at the order of $(\delta_L)_{12}$.

The decay rate of $\mu \to e \gamma $ is evaluated by Eq.~\eqref{eq:mueg0}.
It is compared with the SUSY contribution to the muon $g-2$. In the non-universal slepton mass spectrum, it is represented by $A^L_{ij}$ and $A^R_{ij}$ as
\beq
a_\mu({\rm SUSY})
&=& m_\mu^2\, {\rm Re} \left[ A^L_{22} + A^R_{22} \right] \notag \\
&=& (1 + \delta^{\rm 2loop}) \frac{\alpha_Y}{4\pi} 
m_\mu M_1\mu\tan\beta 
\left[ 
\frac{m_\mu}{1+\Delta_\mu} F_{2,2} + \kappa
\right]. 
\label{eq:gmin2_non-univ}
\eeq
This is same as Eq.~\eqref{eq:gminus2bino} up to a correction $\kappa$, 
which is represented as
\begin{align}
\kappa = \frac{m_\tau}{1+\Delta_\tau} {\rm Re} \left[ (\delta_L)_{23} (\delta_R)_{23}^* \right]
\left(F_{2,2} - F_{2,3} - F_{3,2} + F_{3,3}\right) + \cdots. 
\end{align} 
Here, the omitted terms are suppressed by higher order of $(\delta_L)_{ij}$, $(\delta_R)_{ij}$ or $m_\mu/m_\tau$. 
If the slepton mass matrices are universal, $\kappa $ vanishes.  
For the mass spectrum \eqref{eq:non-univ} with $R_L, R_R \gg 1$, the ratio is
\begin{align}
\frac{{\rm Br}(\mu \to e \gamma)}{a_\mu({\rm SUSY})^2} \simeq 
\frac{1}{\Gamma_{\rm tot}} \frac{\alpha m_{\mu}}{16}\,
\big|\delta_{13}\delta_{23}\big|^2
\left( \frac{m_{\tau}}{m_{\mu}} \frac{1 + \Delta_{\mu}}{1 + \Delta_{\tau}} \right)^2,
\label{eq:muegammag-2}
\end{align}
where $\Gamma_{\rm tot}$ is the total decay rate of muon, and the mixing is defined as
\begin{align}
\big|\delta_{13}\delta_{23}\big|^2 \equiv 
\big|(\delta_R)_{13}(\delta_L)_{23}\big|^2 + \big|(\delta_L)_{13}(\delta_R)_{23}\big|^2.
\label{eq:mixing}
\end{align}
It is independent of superparticle mass spectra except through $\Delta_{\mu}$ and $\Delta_{\tau}$. Thus, when the muon $g-2$ discrepancy \eqref{eq:g-2_deviation} is explained by the SUSY contributions, $\mu \to e\gamma$ is induced by the non-universal slepton mass sizably. It is important that the decay is not suppressed by heavy slepton masses, for instance, $m_{\tilde\mu_1} = 1.4\TeV$ or $1.9\TeV$ in Tab.~\ref{tab:summary}, for given SUSY contributions to the muon $g-2$. 

%%%%%%%%%%%%%%%%%%%%%%%%%%%%%%%%%%%%%%%%%%%%%
\begin{figure}[t]
\begin{center}
 \includegraphics[width=8cm]{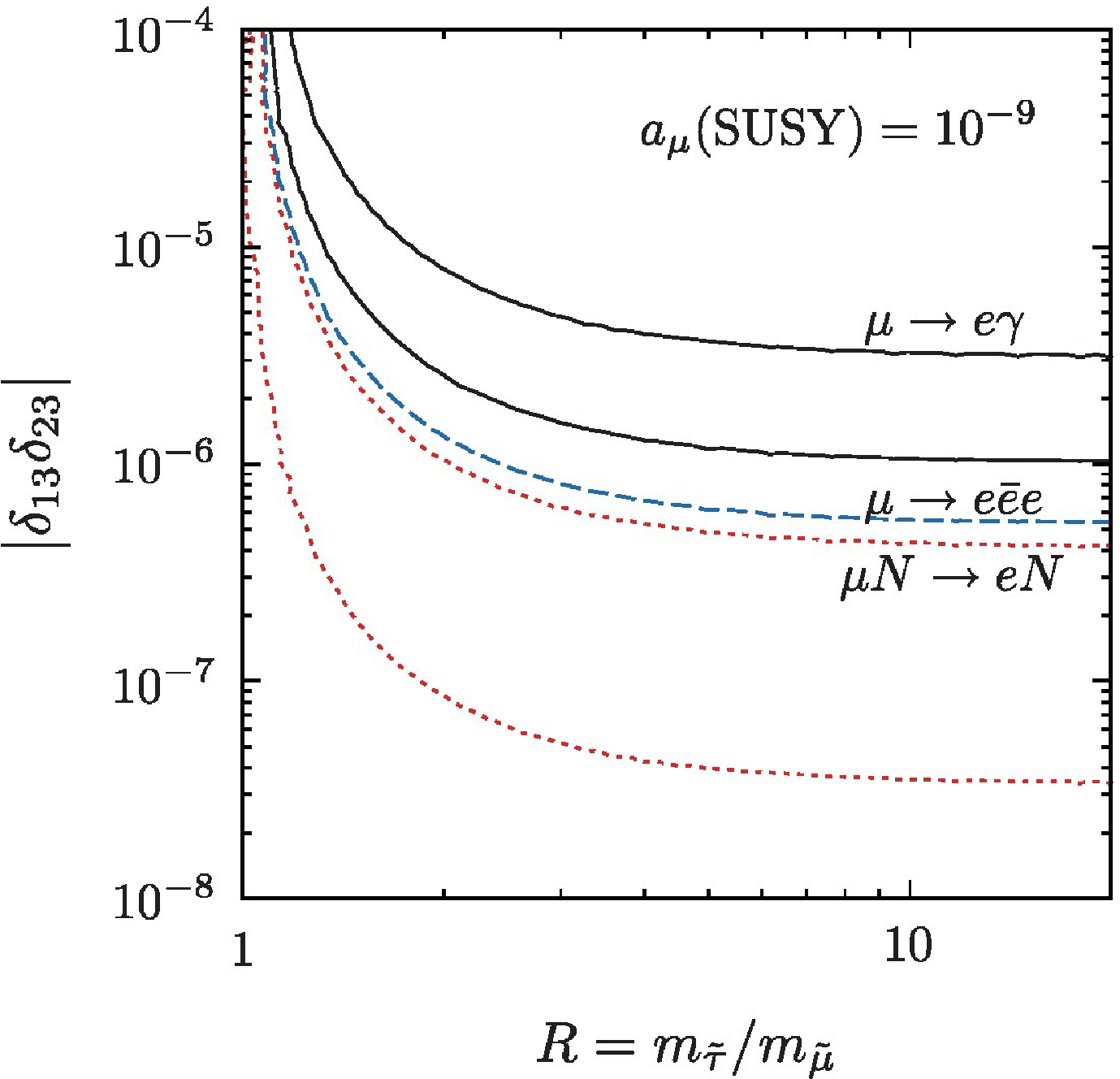}\hspace*{2mm}
 \includegraphics[width=8cm]{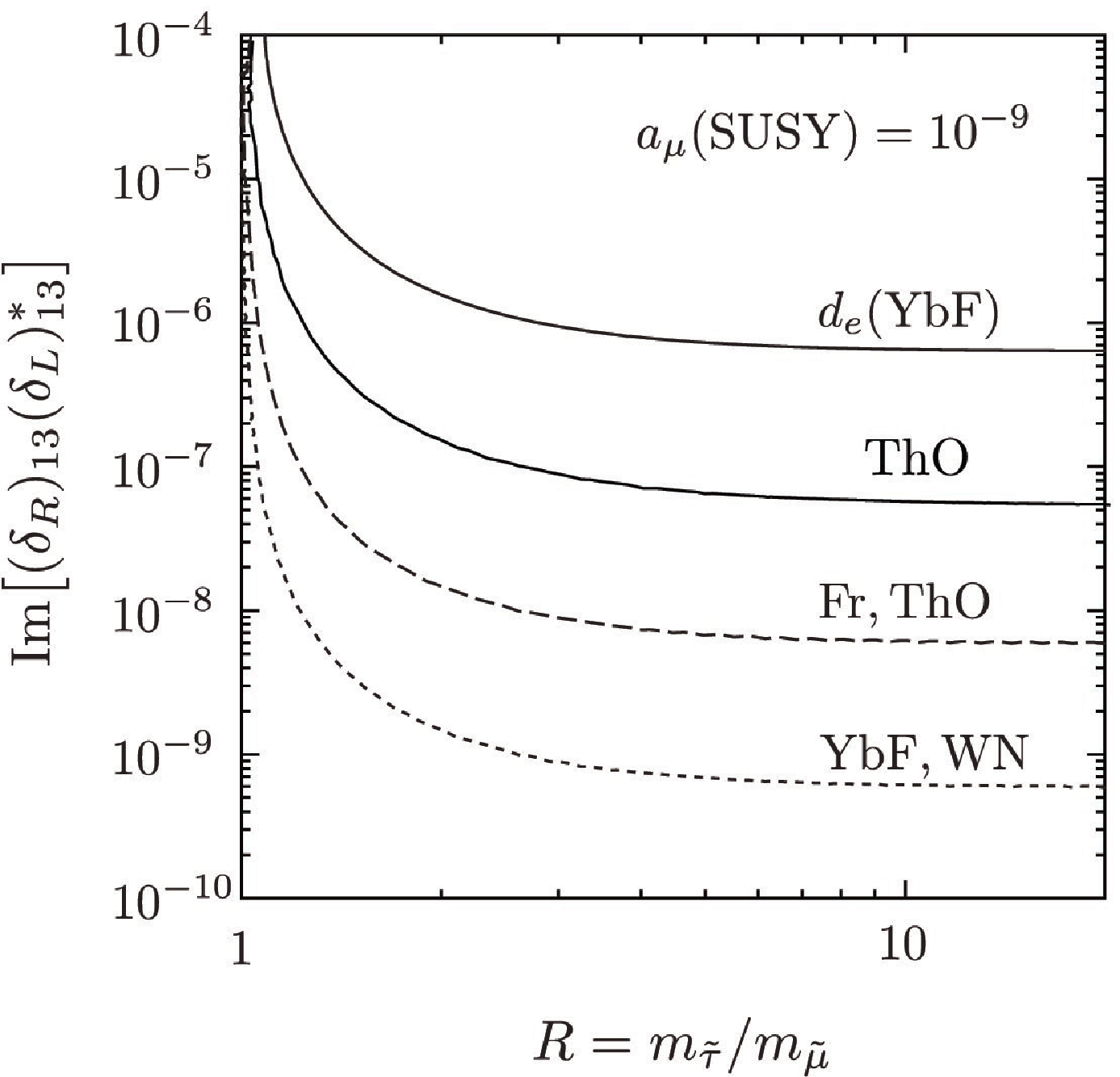}
 \caption[Contours of ${\rm Br}(\mu \to e \gamma)$ and $d_e$ for $a_\mu({\rm SUSY}) = 10^{-9}$ with $M_{\rm soft} = 30\TeV$. ]
 {Contours of ${\rm Br}(\mu \to e \gamma)$ (left) and $d_e$ (right) for $a_\mu({\rm SUSY}) = 10^{-9}$ with $M_{\rm soft} = 30\TeV$. In the left panel, the upper black solid line is the current bound by MEG, ${\rm Br}(\mu \to e \gamma) < 5.7 \cdot 10^{-13}$ at 90\% CL~\cite{Adam:2013mnn}. The lower one is the sensitivity of the MEG upgrade, ${\rm Br}(\mu \to e \gamma) = 6 \cdot 10^{-14}$~\cite{Baldini:2013ke}. The blue dashed line is the Mu3e sensitivity, ${\rm Br}(\mu \to e\bar ee) = 10^{-16}$~\cite{Blondel:2013ia}. The $\mu-e$ conversion is expected to probe down to $R_{\mu e} = 3 \cdot 10^{-17}$ (upper red dotted) by COMET/Mu2e~\cite{Kuno:2013mha,Abrams:2012er} and $2 \cdot 10^{-19}$ (lower red dotted) by PRISM/PRIME~\cite{Kuno:2012pt}. 
 In the right panel, the black solid line is the bounds by using YbF and ThO, $|d_e| < 1.05 \cdot 10^{-27}\,e{\rm cm}$ and $8.7\times 10^{-29}e$cm at 90\% CL~\cite{Hudson:2011zz,Baron:2013eja}. The sensitivity is planned to be improved: $|d_e| = 10^{-29}\,e{\rm cm}$ by Fr or ThO~\cite{Sakemi:2011zz,Vutha:2009ux,Campbell:2013ota}, and $10^{-30}\,e{\rm cm}$ by YbF or WN~\cite{Kara:2012ay,Kawall:2011zz}.
 Here, $m_{\tilde\ell_L} = m_{\tilde\ell_R}$ is assumed.
}
 \label{fig:LFVCPV}
\end{center}
\end{figure}
%%%%%%%%%%%%%%%%%%%%%%%%%%%%%%%%%%%%%%%%%%%%%

In the left panel of Fig.~\ref{fig:LFVCPV}, contours of ${\rm Br}(\mu \to e \gamma)$ are shown. Here, $m_{\tilde\ell_L} = m_{\tilde\ell_R}$ is assumed. The SUSY contributions to the muon $g-2$ is fixed to be $a_\mu({\rm SUSY}) = 1 \times 10^{-9}$ with $M_{\rm soft} = 30\TeV$. The small corrections are taken at $M_1 = m_{\tilde{\mu}} = 400\GeV$ and $\tan\beta = 40$ as a reference, though the result is almost independent of them. In the figure, the contours correspond to the current limit and future sensitivities of experiments,
\begin{itemize}
\item the current limit of the MEG experiment, ${\rm Br}(\mu \to e \gamma) < 5.7 \times 10^{-13}$ at 90\% CL~\cite{Adam:2013mnn} (upper black solid line in the figure).
\item the sensitivity of the MEG upgrade, ${\rm Br}(\mu \to e \gamma) = 6 \times 10^{-14}$~\cite{Baldini:2013ke} (lower black solid line).
\item the sensitivity of the Mu3e experiment, ${\rm Br}(\mu \to e\bar ee) = 10^{-16}$ at Phase II~\cite{Blondel:2013ia} (blue dashed line).
\item the sensitivity of the COMET experiment, $R_{\mu e} = 3 \times 10^{-17}$ at Phase II~\cite{Kuno:2013mha} (upper red dotted line). The Mu2e experiment has a similar sensitivity~\cite{Abrams:2012er}.
\item the proposal of the PRISM/PRIME project, $R_{\mu e} = 2 \times 10^{-19}$~\cite{Kuno:2012pt} (lower red dotted line).
\end{itemize}
Note that $\mu \to e\bar ee$ and $\mu-e$ conversion experiments have better sensitivity in future than those of $\mu \to e \gamma$. In particular, the latter experiment has low (accidental) backgrounds. On the other hand, the current constraint and future sensitivities of the tau LFV's are weaker than those of the muon, though they are also induced by the non-universal slepton masses with finite $(\delta_L)_{23}$ and $(\delta_R)_{23}$. 

From the figure, it is found that the LFV decay rate increases rapidly as the staus become heavier than  the smuons, $R > 1$. When the staus are decoupled, $R \gg 1$, the mixing $|\delta_{13}\delta_{23}|$ in Eq.~\eqref{eq:mixing} is limited to be smaller than $3 \times 10^{-6}$ by MEG for $a_\mu({\rm SUSY}) = 1 \times 10^{-9}$. For instance, when the smuon and selectron masses are larger than $1\TeV$, the stau are required to be heavier than $7\TeV$ ($R > 7$) to avoid the vacuum stability constraint, according to Fig.~\ref{fig:boundR}. If the lepton Yukawa matrix is related to the quark sector, e.g., by the GUT relation, it is naively expected to be $|\delta_{13}\delta_{23}| \sim V_{ub} V_{cb} \sim 10^{-4}$. This already exceeds the above limit. Thus, non-universal slepton mass spectra are tightly constrained by LFV. In future, if sleptons are neither discovered at LHC nor ILC, the model is expected to be probed by LFV. Otherwise, the SM Yukawa matrices are tightly limited in the model basis, or when flavor models are constructed.

The flavor off-diagonal components of the Yukawa matrices are sources of the CP violations, when the slepton mass matrices are non-universal. Similarly to the correction $\kappa$ of the muon $g-2$ in Eq.~\eqref{eq:gmin2_non-univ}, 
the electron EDM is induced as
\begin{align}
\frac{d_e}{e}
&=  \frac{m_e}{2}\, {\rm Im} \left[ A^L_{11} - A^R_{11} \right] \notag \\
&=  (1 + \delta^{\rm 2loop}) \frac{\alpha_Y}{8\pi} 
M_1\mu\tan\beta \notag \\
&~~~ \times
\bigg[
\frac{m_\mu}{1+\Delta_\mu} 
{\rm Im} \left[ (\delta_R)_{12} (\delta_L)_{12}^* \right]
\left(F_{1,1} - F_{1,2} - F_{2,1} + F_{2,2}\right)
\notag \\
&~~~~~~ + 
\frac{m_\tau}{1+\Delta_\tau} 
{\rm Im} \left[ (\delta_R)_{13} (\delta_L)_{13}^* \right]
\left(F_{1,1} - F_{1,3} - F_{3,1} + F_{3,3}\right) 
+ \cdots \bigg].
\label{eq:EDM}
\end{align}
Here, we assume $\arg(M_1\mu\tan\beta) =0$.
The omitted terms are suppressed by orders of $(\delta_L)_{ij}$, $(\delta_R)_{ij}$ or $m_\mu/m_\tau$. In the last term, $F_{1,1} \gg F_{1,3}, F_{3,1}, F_{3,3}$ is obtained when the staus are heavy. On the other hand, the right hand side vanishes when the slepton masses are universal, because the complex phases can be rotated away. Comparing Eq.~\eqref{eq:EDM} with Eq.~\eqref{eq:gmin2_non-univ}, we obtain
\begin{align}
\frac{d_e / e}{a_\mu({\rm SUSY})} \simeq 
\frac{1}{2 m_{\mu}}
{\rm{Im}}[ (\delta_R )_{13}  (\delta_L)^{ \ast}_{13} ]\,
\frac{m_{\tau}}{m_\mu}\frac{1 + \Delta_{\mu}}{1 + \Delta_{\tau}},
\label{eq:EDMg-2}
\end{align}
for the mass spectrum \eqref{eq:non-univ} with $R_L, R_R \gg 1$. It is independent of superparticle mass spectra except through $\Delta_{\mu}$ and $\Delta_{\tau}$. Thus, if the muon $g-2$ anomaly is solved by the SUSY contributions, EDM becomes sizable by the non-universal slepton mass. Similarly to LFV, this is valid even for large selectron and smuon masses.

In the right panel of Fig.~\ref{fig:LFVCPV}, contours of the electron EDM are shown. Here, the SUSY contributions to the muon $g-2$ is fixed to be $a_\mu({\rm SUSY}) = 1 \times 10^{-9}$ with $M_{\rm soft} = 30\TeV$. The result is almost independent of superparticle mass spectra except for small corrections, $\Delta_\mu$ and $\Delta_\tau$. The following data are used,
\begin{itemize}
\item the limit with the YbF molecule, $|d_e| < 1.05 \times 10^{-27}\,e{\rm cm}$ at 90\% CL~\cite{Hudson:2011zz} (upper solid line in the figure).
\item the current limit with the ThO molecule, $|d_e| < 8.7 \times 10^{-29}\,e{\rm cm}$ at 90\% CL~\cite{Baron:2013eja} (lower solid line in the figure).
\item the sensitivity with the Fr atom, $|d_e| = 10^{-29}\,e{\rm cm}$~\cite{Sakemi:2011zz} (dashed line). The experiment with the ThO molecule could have a similar sensitivity by accumulating data, $|d_e| = 1 \times 10^{-28}/\sqrt{({\rm day})}\,e{\rm cm}$~\cite{Vutha:2009ux,Campbell:2013ota}.
\item the sensitivity with the YbF molecule, $|d_e| = 10^{-30}\,e{\rm cm}$~\cite{Kara:2012ay} (dotted line). The experiment with the WN ion can probe down to, $|d_e| = 10^{-30}\,e{\rm cm}/{\rm day}$, where the systematic limit is at the level of $10^{-31}\,e{\rm cm}$~\cite{Kawall:2011zz}.
\end{itemize}
From the figure, it is found that EDM is sensitive to ${\rm{Im}}[ (\delta_R )_{13} (\delta_L)^*_{13} ]$ when the staus are heavier than the selectrons. The current experimental limit puts a constraint, ${\rm{Im}}[ (\delta_R )_{13} (\delta_L)^*_{13} ] < 6 \times 10^{-8}$ for $R \gg 1$ and $a_\mu({\rm SUSY}) = 1 \times 10^{-9}$. The sensitivity will be improved very well. The mixing will be able to be probed at the level of $10^{-10}$. Thus, if sleptons are neither discovered at LHC nor ILC in future, the model can be sensitively probed by EDM as well as LFV. If no signal will be observed, the CP violating phase must be suppressed very tightly in order to explain the muon $g-2$ anomaly by the SUSY contributions.

%%%%%%%%%%%%%%%%%%%%%%%%%%%

\section{Stau}
\label{sec:stau}
%!TEX root = ../Dthesis.tex

So far, we studied the prospects for the selection/smuon searches.
In this section, we discuss the stau searches.
If the staus have masses of $\mathcal{O}(100) \GeV$ and large left-right mixing,
the Higgs coupling to the di-photon $\kappa _{\gamma }$ is deviated from the SM prediction. 
Such a large left-right mixing is constrained by the vacuum meta-stability condition of the stau-Higgs potential, 
as mentioned in Sec~\ref{subsec:vacuum}.
Then, we discussed that once the deviation from the SM prediction of $\kappa _{\gamma }$ were observed, 
the mass region for the staus were determined by the condition in Sec.~\ref{subsec:kg}.

Once the stau is discovered at ILC, its properties including the mass are determined. 
Particularly, it is important to measure the mixing angle of the stau $\theta_{\widetilde\tau}$. 
When $\sin 2\theta_{\widetilde\tau}$ is sizable, the angle can be measured at ILC \cite{Nojiri:1996fp,Bechtle:2009em,Boos:2003vf,Endo:2013xka}. 
As observed in Fig.~\ref{fig:StauMass}, it is likely to be sizable to enhance $\kappa_{\gamma}$. 
In particular, if $\sin 2\theta_{\widetilde\tau}$ is large enough to be measurable, the heaviest stau is likely to be light. 
Thus, it may be possible to discover the heaviest stau and measure its mass at ILC.
Then, the stau contribution to $\kappa_{\gamma}$ can be reconstructed by using the measured masses and mixing angle. 
This provides a direct test whether the stau contribution is the origin of the deviation of $\kappa_{\gamma}$. 
On the other hand, the heaviest stau is not always discovered at the early stage of ILC, even if the stau mixing angle is measured. 
If $\theta_{\widetilde\tau}$ as well as $m_{\widetilde\tau_1}$ is measured, $m_{\widetilde\tau_2}$ may be estimated in order to explain the excess of $\kappa_{\gamma}$.
In this section, we will study the reconstruction of the stau contribution to $\kappa_{\gamma}$. 
The mass of the heaviest stau and theoretical uncertainties will also be discussed.

\subsection{Reconstruction}
\label{subsec:reconst}

If both of $\widetilde\tau_{1}$ and $\widetilde\tau_{2}$ are measured, the stau contribution to $\kappa_{\gamma}$ can be reconstructed. The contribution is determined by the parameters in Eq.~\eqref{eq:parameters_s}. In this subsection, 
we assume the situation that both of $\widetilde\tau_{1}$ and $\widetilde\tau_{2}$ are observed by ILC.
Then, we discuss how and how accurately the model parameters are measured at ILC, and consequently the stau contribution to $\kappa_{\gamma}$ is reconstructed.

%%%%%%%%%%%%%%%%%%%%%%%%%%%%%%%%%%%%%
\begin{table}[t]
\centering
    \caption[Model parameters at our sample point.]
    {Model parameters at our sample point. In addition, $\tan\beta = 5$ and $A_\tau = 0$ are set, though the results are almost independent of them.}
    \label{table:ModelPoint}
\vspace{0.5em}
    \begin{tabular}{l|cccc|c}
      \hline\hline
      {Parameters} & 
      $m_{\widetilde{\tau}_1}$ & 
      $m_{\widetilde{\tau}_2}$ & 
      $\sin 2\theta_{\widetilde{\tau}}$& 
      $m_{\widetilde{\chi}^0_1}$ & 
      $\delta \kappa_{\gamma}$\\
      \hline
      {Values} & 
      100\GeV & 230\GeV & 0.92 & 90\GeV & 
      3.6\% \\
      \hline\hline
    \end{tabular}
\end{table}
%%%%%%%%%%%%%%%%%%%%%%%%%%%%%%%%%%%%%

Let us first specify a model point to quantitatively study the accuracies. 
In Tab.~\ref{table:ModelPoint}, the stau masses, the stau mixing angle, and the Bino (-like neutralize) mass are shown. 
The point is not so far away from the SPS1a' benchmark point \cite{AguilarSaavedra:2005pw}, where ILC measurements have been studied (see e.g., Ref.~\cite{Bechtle:2009em}). 
The stau mixing angle is chosen to enhance the Higgs coupling as $\delta \kappa_{\gamma} = 3.6\%$. 
The staus masses are within the kinematical reach of ILC at $\sqrt{s}=500\GeV$. 
The point is consistent with the vacuum meta-stability condition and the current bounds from LHC and LEP. 
The most tight bound on the stau mass has been obtained at LEP as $m_{\widetilde{\tau}_1} > 81.9\GeV$ at 95\% CL \cite{Beringer:1900zz}. 
%On the other hand, LHC constraints are still weak \cite{ATLAS:2013yla}.
On the other hand, LHC constraints are still weak \cite{Aad:2014yka}.
The other superparticles are assumed to be decoupled for simplicity. 
In particular, $\tan\beta = 5$ and $A_\tau = 0$ are chosen, where the Higgsino masses are about 2.2\TeV.

In order to reconstruct the stau contribution to $\kappa_{\gamma}$, it is required to measure the stau masses and the mixing angle.
At ILC, staus are produced in $e^+e^-$ collisions and decay into the tau and the Bino.
The stau masses are measured by studying the endpoints of the tau jets. 
In Ref.~\cite{Bechtle:2009em}, the mass measurement has been studied in detail at SPS1a'. 
It is argued that the mass can be measured at the accuracy of about $0.1\GeV$ ($6\GeV$) for $\widetilde\tau_1$ ($\widetilde\tau_2$).
Here, $\sqrt{s}=500\GeV$ and ${\cal L} = 500\invfb$ are assumed for ILC. 
The mass resolution may be improved by scanning the threshold productions \cite{Grannis:2002xd,Baer:2013cma}.
The accuracy could be $\sim 1\GeV$ for $m_{\widetilde{\tau}_2} = 206\GeV$.
Since the model parameters of our sample point are not identical to those of SPS1a', the mass resolutions may be different from those estimated at SPS1a'. 
For instance, the production cross section of staus becomes different, while the SUSY background is negligible in our sample point. 
Profile of the tau jets depends on the masses of the staus and the Bino. 
In this dissertation, instead of analyzing the Monte Carlo simulation, we simply adopt the mass resolution,\footnote
{
The resolutions estimated in Ref.~\cite{Bechtle:2009em} depend on the uncertainty of the measured Bino mass. The Bino mass can be measured very precisely at ILC by the productions of selectrons or smuons \cite{Baer:2013vqa}, though they are irrelevant for $\kappa_{\gamma}$ and the vacuum meta-stability condition.
}
\begin{equation}
\Delta m_{\widetilde{\tau}_1} \sim 0.1\GeV,~
\Delta m_{\widetilde{\tau}_2} \sim 6\GeV.
\label{eq:ErrorMass}
\end{equation}

% Figure %%%%%%%%%%%%%%%%%%%%%%%%%%%%%%%%
\begin{figure}[t]
\begin{center}
 \includegraphics[width=8cm]{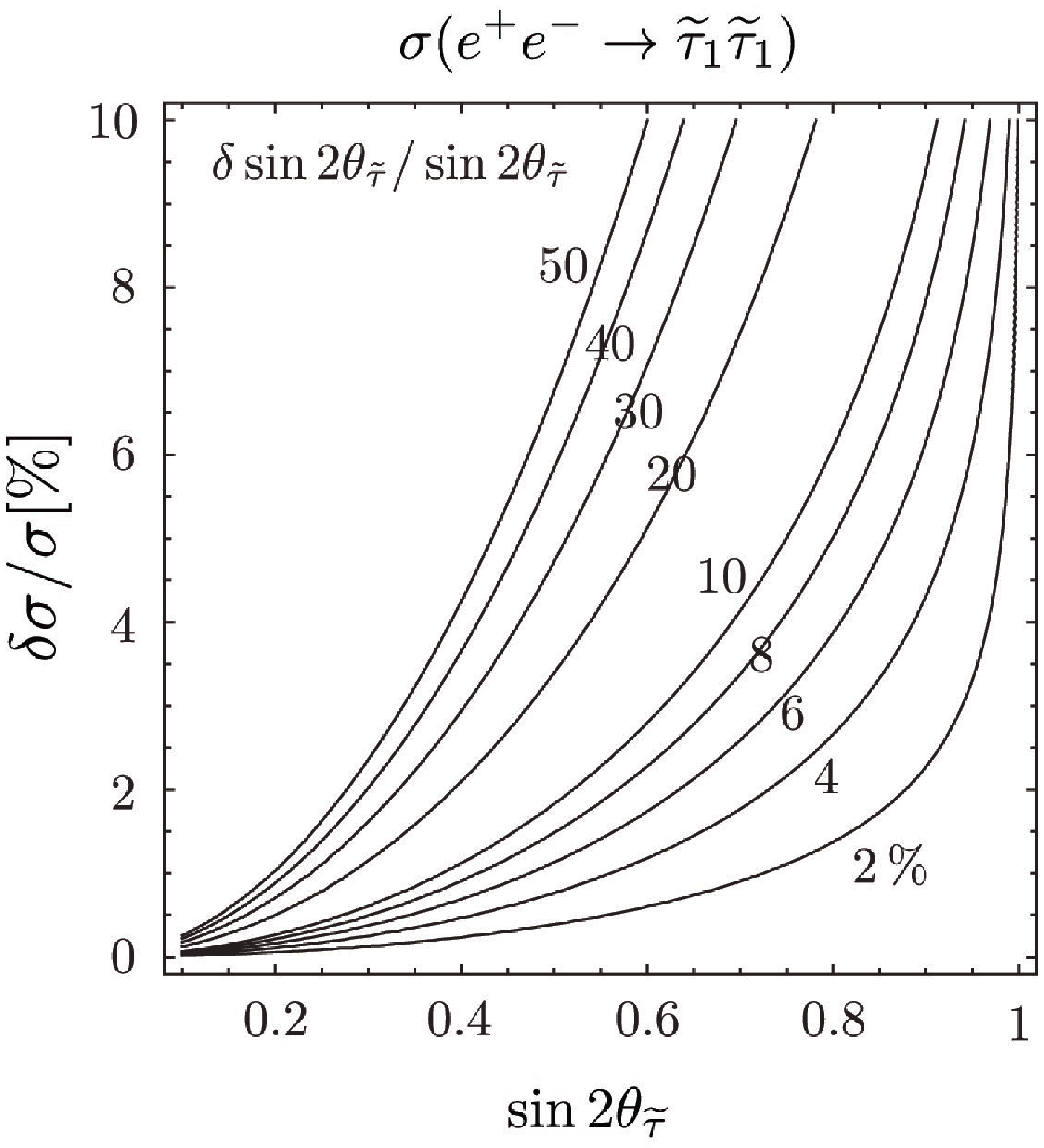}\hspace*{2mm}
 \includegraphics[width=8cm]{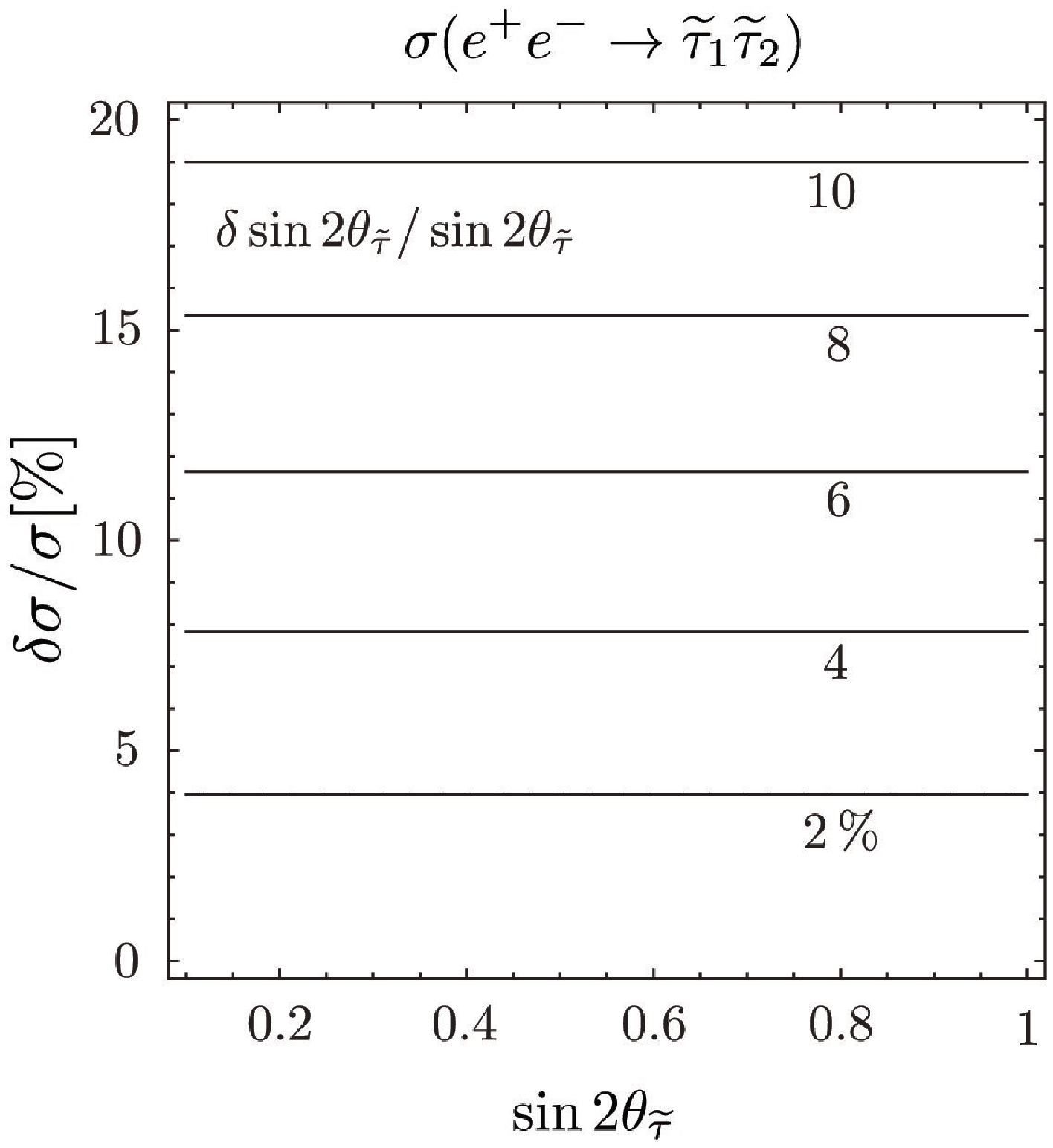}
 \end{center}
 \caption[Contours of $\delta \sin 2\theta_{\widetilde\tau}/\sin 2\theta_{\widetilde\tau}$.]
 {Contours of $\delta \sin 2\theta_{\widetilde\tau}/\sin 2\theta_{\widetilde\tau}$ determined by the measurement of the production cross section of a pair of $\widetilde\tau_1$ (left) and that of $\widetilde\tau_1$ and $\widetilde\tau_2$ (right). 
 Uncertainties from the mass resolutions are not taken into account.}
 \label{fig:StauMixingAngle}
\end{figure}
%%%%%%%%%%%%%%%%%%%%%%%%%%%%%%%%%%%%%

Next, let us discuss the measurement of the stau mixing angle, $\theta_{\widetilde{\tau}}$.
Several methods have been studied for ILC. 
For instance, the polarization of the tau which is generated at the stau decay has been studied in Ref.~\cite{Nojiri:1996fp,Boos:2003vf,Bechtle:2009em}.
The angle can also be extracted from the production cross section of a pair of the lightest stau \cite{Boos:2003vf}. 
Note that accuracies of these angle measurement depend on the model point, i.e., the input value of $\theta_{\widetilde\tau}$. 

In order to study the accuracy of the stau mixing angle at our sample point, let us investigate the production cross section of the lightest stau by following the procedure in Ref.~\cite{Endo:2013xka}.
The production cross section of the lightest stau is given by \cite{Boos:2003vf}  (cf, Appendix \ref{app:slepprod})
\beq
\sigma(e^+e^-\to\widetilde\tau_1\widetilde\tau_1) &=& 
\frac{8\pi\alpha^2}{3s} \lambda^{\frac{3}{2}} 
\bigg[
  c_{11}^2 \frac{\Delta_Z^2}{\sin^42\theta_W}(P_{-+}L^2 + P_{+-}R^2)
\nonumber\\&&
  + \frac{1}{16} (P_{-+} + P_{+-})
  + c_{11}\frac{\Delta_Z}{2\sin^22\theta_W}(P_{-+}L + P_{+-}R)
\bigg],
\label{eq:CrossSection}
\eeq
at the tree level, where the parameters are $\lambda = 1-4 m_{\widetilde\tau_1}^2/s$, $\Delta_Z = s/(s-m_Z^2)$, and $c_{11} = [(L+R) + (L-R) \cos 2\theta_{\widetilde{\tau}}]/2$ with $L=-1/2+\sin^2\theta_W$ and $R=\sin^2\theta_W$. 
The beam polarizations are parameterized as $P_{\mp\pm} = (1 \mp P_{e-}) (1 \pm P_{e+})$.
In the bracket, the first and second terms come from the s-channel exchange of the Z boson and the photon, respectively. The last term is induced by the interference of them.
The dependence on the stau mixing angle originates in the Z boson contribution. 

Since Eq.~\eqref{eq:CrossSection} is a function of the stau mass and mixing angle, $\theta_{\widetilde\tau}$ is determined by measuring the cross section and the stau mass.
If the errors are summed in quadrature, the accuracy of the stau mixing angle is estimated as
\beq
\left( \Delta \sin 2\theta_{\widetilde\tau } \right) ^2 = 
 \left( \frac{\partial  \sin 2\theta_{\widetilde\tau }}{\partial  \sigma (\widetilde{\tau}_1)} \right) ^2 \left( \Delta  \sigma (\widetilde{\tau}_1)\right) ^2 +
 \left( \frac{\partial  \sin 2\theta_{\widetilde\tau }}{\partial  m _{\widetilde{\tau }_1}} \right) ^2 \left( \Delta m _{\widetilde{\tau }_1} \right) ^2 , 
 \label{eq:errorangleestimate}
\eeq
where $\sigma(\widetilde\tau_1) = \sigma(e^+e^-\to\widetilde\tau_1\widetilde\tau_1)$. 
In Fig.~\ref{fig:StauMixingAngle}, we show contours of the uncertainty of the stau mixing angle, $\delta \sin 2\theta_{\widetilde\tau}/\sin 2\theta_{\widetilde\tau}$.
In the left panel, the angle is determined from the production cross section of the lightest stau. 
The accuracy is sensitive to the input value of $\sin 2\theta_{\widetilde\tau}$ and $\delta\sigma(\widetilde\tau_1)/\sigma(\widetilde\tau_1)$.
In contrast, the uncertainty from the mass resolution of $\widetilde\tau_1$ in Eq.~\eqref{eq:ErrorMass} is negligible.
The accuracy of $\sin 2\theta_{\widetilde\tau}$ becomes better for larger $\sin 2\theta_{\widetilde\tau}$.
If the stau contributes to $\kappa_{\gamma}$ sizably, $\sin 2\theta_{\widetilde\tau}$ is likely to be large, as observed in Fig.~\ref{fig:StauMass}. 
Thus, the mixing angle is expected to be measured well.
At the sample point, where $\sin 2\theta_{\widetilde\tau} = 0.92$, $\delta \sin 2\theta_{\widetilde\tau}/\sin 2\theta_{\widetilde\tau}$ is estimated to be better than 10\%, if the cross section is measured as precisely as $\delta\sigma(\widetilde\tau_1)/\sigma(\widetilde\tau_1) < 10\%$. 
At ILC, it is argued that the production cross section can be measured at the accuracy of about 3\%, according to the analysis in Ref.~\cite{Bechtle:2009em} at SPS1a'.
If $\delta\sigma(\widetilde\tau_1)/\sigma(\widetilde\tau_1) \sim 3\%$ is applied to our sample point, the accuracy is estimated to be
\begin{equation}
\Delta \sin 2\theta_{\widetilde\tau}/\sin 2\theta_{\widetilde\tau} \sim 2\%.
\label{eq:ErrorAngle}
\end{equation}

From Eqs.~\eqref{eq:ErrorMass} and \eqref{eq:ErrorAngle}, the accuracy of the reconstruction of the stau contribution to $\kappa_{\gamma}$ is estimated. 
If the errors are summed in quadrature, the uncertainty is estimated as
\beq
 \left( \Delta\kappa_{\gamma} \right) ^2 = 
 \left( \frac{\partial \kappa _{\gamma }}{\partial  m _{\widetilde{\tau }_1}} \right) ^2 \left( \Delta m _{\widetilde{\tau }_1} \right) ^2  + 
 \left( \frac{\partial \kappa _{\gamma }}{\partial  m _{\widetilde{\tau }_2}} \right) ^2 \left( \Delta m _{\widetilde{\tau }_2} \right) ^2 + 
 \left( \frac{\partial \kappa _{\gamma }}{\partial  \sin 2 \theta _{\widetilde{\tau }}} \right) ^2 \left( \Delta \sin 2 \theta _{\widetilde{\tau }} \right) ^2. 
 \label{eq:errorkappaestimate}
\eeq
Substituting Eqs.~\eqref{eq:ErrorMass} and \eqref{eq:ErrorAngle} for Eq.~\eqref{eq:errorkappaestimate}, the uncertainty is obtained as
\beq
\Delta\kappa_{\gamma} \sim 0.5\%,
\label{eq:ErrorKappa}
\eeq
at the sample point, where $\delta \kappa_{\gamma} = 3.6\%$. 
Note that the uncertainty of the measurement of $\kappa_{\gamma}$ is 1--2\% from HL-LHC and ILC, as mentioned above.
Since the reconstruction error is comparable to or smaller than that of the measured $\kappa_{\gamma}$, it is possible to check whether the stau is the origin of the excess of the Higgs coupling $\kappa_{\gamma}$. 
It is emphasized that this is a direct test of the stau contribution to $\kappa_{\gamma}$. 

In Eq.~\eqref{eq:ErrorKappa}, the error is dominated by the uncertainties of the heaviest stau mass and the stau mixing angle. 
The former may be reduced by scanning the threshold of the stau productions, as mentioned above. 
For instance, if we adopt $\Delta m_{\widetilde{\tau}_2} \sim 1\GeV$ as implied in Ref.~\cite{Grannis:2002xd,Baer:2013cma}, 
the error becomes $\Delta\kappa_{\gamma} \sim 0.3\%$. 
On the other hand, the latter uncertainty may be improved by studying the production cross section of $\widetilde\tau_1$ and $\widetilde\tau_2$ \cite{Bechtle:2009em}. 
The cross section is given by \cite{Boos:2003vf} (or see Appendix \ref{app:slepprod})
\beq
\sigma(e^+e^-\to\widetilde\tau_1\widetilde\tau_2) &=& 
\frac{8\pi\alpha^2}{3s} \lambda^{\frac{3}{2}} 
\bigg[
  2 \times c_{12}^2 \frac{\Delta_Z^2}{\sin^42\theta_W}(P_{-+}L^2 + P_{+-}R^2) \bigg], 
\label{eq:CrossSection12}
\eeq
where  $\lambda = [1- (m_{\widetilde{\tau }_1} + m_{\widetilde{\tau }_2})^2 / s] [1- (m_{\widetilde{\tau }_1} - m_{\widetilde{\tau }_2})^2 / s]$, 
and $c_{12} = c_{21} =  (L-R) \sin 2\theta_{\widetilde{\tau}} /2$.
Since $e^+e^-\to\widetilde\tau_1\widetilde\tau_2$ proceeds by the s-channel exchange of the Z-boson, its cross section is proportional to $\sin^2 2\theta_{\widetilde\tau}$.
Thus, it is very sensitive to the stau mixing angle, and further, the accuracy is independent of the model point, once the error of the production cross section is given. 
In the right panel of Fig.~\ref{fig:StauMixingAngle}, we show contours of $\delta \sin 2\theta_{\widetilde\tau}/\sin 2\theta_{\widetilde\tau}$,
where, the angle is extracted from $\sigma(e^+e^-\to\widetilde\tau_1\widetilde\tau_2)$.
It is found that the accuracy is independent of the input $\sin 2\theta_{\widetilde\tau}$. 
Here, uncertainties from the mass resolution are neglected. 
In particular, if the mass resolution of $\widetilde\tau_2$ is large, the accuracy of the mixing angle becomes degraded. 
Unfortunately, the accuracy of the measurement of $\sigma(e^+e^-\to\widetilde\tau_1\widetilde\tau_2)$ has not been analyzed for ILC. 
Since $\sin 2\theta_{\widetilde\tau}$ is likely to be large to enhance $\kappa_{\gamma}$, the cross section can be sizable.
At the sample point, it is estimated to be about 6\,\text{fb} for $\sqrt{s}=500\GeV$ with $(P_{e-},P_{e+})=(-0.8,0.3)$.
It is necessary to study this production process in future.

Let us comment on the $\tan\beta$ dependence.
At the sample point, we set to $\tan\beta = 5$. 
Although the stau contribution to $\kappa_{\gamma}$ includes $\tan\beta$, once the stau masses and mixing angle are measured, the reconstruction of the Higgs coupling is almost insensitive to it. 
This is because the stau left-right mixing parameter $m_{\widetilde\tau_{LR}}^2$ is 
determined by the measured masses and mixing angle through Eq.~\eqref{eq:staumix}.
One can check that, even if $\tan\beta$ is varied, the accuracy \eqref{eq:ErrorKappa} is almost unchanged. 

Finally, we comment on the muon $g-2$.
In this section, we discussed our sample point, where $m_{\widetilde{\tau }_1} = 100\GeV$, $m_{\widetilde{\tau }_2} = 230\GeV$, $\sin 2 \theta _{\widetilde{\tau }}= 0.92$, 
$m_{\widetilde{\chi }^0_1} = 90\GeV$, and $\tan \beta =5$, respectively.
The soft SUSY breaking masses can be estimated as $m_{\widetilde{\tau}_L} = 145\GeV$, $m_{\widetilde{\tau}_R} = 291\GeV$,\footnote{
In this case, we consider $0 \leq \theta _{\widetilde{\tau }} < \frac{\pi }{4}$ region.} and 
the SUSY contribution to the muon $g-2$ is $a_{\mu }(\text{SUSY}) \sim 1.8 \times 10^{-9}$,  
if the sleptons masses are universal among the flavors. Here, we set to $M_{\text{soft}}=30\TeV$ 
and assume that the left-right mixing of the smuon is $m_{\widetilde{\mu }_{LR}} = (m_{\mu } /m _{\tau }) \times m_{\widetilde{\tau }_{LR}}$.
It is found that our sample point is explained the muon $g-2$ discrepancy at 1$\sigma $ level 
and is not excluded by dilepton searche \cite{ATLAS:2014fka, Chatrchyan:2013oca}.
The mass difference between $m_{\widetilde{\mu}_L}$ and $m_{\widetilde{\chi }^0_1}$ are $\sim 50\GeV$, 
which is much larger than the uncertainty of the mass measurement by smuon productions at ILC.
Therefore, the smuon productions are expected to provide direct signatures in this scenario.

%%%%%%%%%%%%%%%%%%%%%%%%%%%%%%%%%%%%%%%%%%%%%%%%%%%%%%%%%%%%%%%%%%%%%%%%%%%%%%%%%%%%%%%%
\subsection{Discussions} 
\label{subsec:discussion}

Let us discuss miscellaneous prospects for the stau searches and theoretical uncertainties which have not been mentioned so far.

First of all, let us consider the situation when the heaviest stau is not discovered at the early stage of ILC, i.e., at $\sqrt{s}=500\GeV$. 
As found in Sec.~\ref{subsec:kg}, if the excess of $\kappa_{\gamma}$ is measured at this stage, the lightest stau is already discovered. 
Then, it is possible to determine the stau mixing angle by measuring the production cross section of the lightest stau, 
as long as $\sin 2\theta_{\widetilde\tau}$ is sizable as observed in Fig.~\ref{fig:StauMixingAngle}.
From the measurements of $m_{\widetilde{\tau}_1}$, $\theta_{\widetilde{\tau}}$ and $\kappa_{\gamma}$, the mass of the heaviest stau $m_{\widetilde{\tau}_2}$ is determined. 
The predicted mass could be tested at the next stage of ILC, e.g., $\sqrt{s}=1\TeV$. 

In order to demonstrate the procedure, let us consider a model point with $m_{\widetilde{\tau}_1}=150\GeV$, $m_{\widetilde{\tau}_2}=400\GeV$ and 
$\sin 2\theta_{\widetilde\tau}=0.91$. 
At the point, the deviation from the SM prediction of the Higgs coupling is $\delta \kappa_{\gamma}=5.6\%$.
At the early stage of ILC, it is expected that $\kappa_{\gamma}$ is determined with the uncertainty $\Delta\kappa_{\gamma} \sim 2\%$, and the lightest stau is measured with $\Delta m_{\widetilde{\tau}_1} \sim 0.1\GeV$ and $\delta\sigma(\widetilde\tau_1)/\sigma(\widetilde\tau_1) \sim 3\%$. 
The stau mixing angle is extracted from the cross section as 
\beq
\Delta \sin 2\theta_{\widetilde\tau}/\sin 2\theta_{\widetilde\tau} \sim 2.5\%, 
\eeq 
as observed in Fig.~\ref{fig:StauMixingAngle}.
Since $\delta \kappa_{\gamma}$ is a function of $m_{\widetilde{\tau}_1}$, $m_{\widetilde{\tau}_2}$ and $\theta _{\widetilde{\tau}}$, 
the uncertainty of the heaviest stau is estimated as
\beq
 \left( \Delta m_{\widetilde{\tau }_2}\right) ^2 = 
 \left( \frac{\partial m_{\widetilde{\tau }_2}} {\partial  m _{\widetilde{\tau }_1}} \right) ^2 \left( \Delta m _{\widetilde{\tau }_1} \right) ^2  + 
 \left( \frac{\partial m_{\widetilde{\tau }_2}} {\partial  \sin 2 \theta _{\widetilde{\tau }}} \right) ^2 \left( \Delta \sin 2 \theta _{\widetilde{\tau }} \right) ^2 +
 \left( \frac{\partial m_{\widetilde{\tau }_2}} {\partial  \kappa_{\gamma}} \right) ^2 \left( \Delta  \kappa_{\gamma} \right) ^2,
 \label{eq:errorstaumassestimate}
\eeq
thus the mass of the heaviest stau is determined with the accuracy 
\beq
\Delta m_{\widetilde{\tau}_2} \sim 53\GeV,
\eeq
where, the largest uncertainty comes from the measurement of the Higgs coupling.
If the error is reduced to be $\Delta\kappa_{\gamma} \sim 1\%$ due to reductions of the HL-LHC systematic uncertainties (see Ref.~\cite{Peskin:2013xra}), $\Delta m_{\widetilde{\tau}_2} \sim 26\GeV$ is achieved. 
Such a prediction can be checked at ILC with $\sqrt{s}=1\TeV$.
This result would be helpful for choosing the beam energy to search for the heaviest stau at ILC.
Once $\widetilde{\tau}_2$ is discovered, the stau contribution to the Higgs coupling can be reconstructed as Sec.~\ref{subsec:reconst}. 

When the sleptons have universal mass spectrum, the left- and right-handed slepton masses can be estimated by 
$m_{\widetilde{\ell}_L} = 345\GeV$, $m_{\widetilde{\ell}_R} = 250\GeV$, respectively.\footnote{
We consider $\frac{\pi }{4} \leq \theta _{\widetilde{\tau }} < \frac{\pi }{2}$ region.}
If the lightest neutralino are much lighter than the smuons,  they are expected to be probed at LHC with $\sqrt{s} = 13$--$14\TeV$.
The SUSY contributions to the muon $g-2$ is calculated as $a_{\mu }(\text{SUSY}) \sim 2.0 \times 10^{-9}$, 
and is explained the discrepancy of the muon $g-2$ at $1\sigma $ level, 
where the parameters are taken to $M_1 = 140\GeV$, and $M_{\text{soft}} = 30\TeV$. 
Combining the information of the Higgs coupling with that of the muon $g-2$, 
we may detect effects of sleptons more definitely.

The uncertainty of the prediction of the heaviest stau mass depends on the model point, especially the stau mixing angle. 
If $m_{\widetilde{\tau}_2}$ is larger, $\sin 2\theta_{\widetilde\tau}$ is likely to be smaller, as expected from Fig.~\ref{fig:StauMass}. The measurement of the stau mixing angle, then, suffers from a larger uncertainty, and it becomes difficult to determine the mass of the heaviest stau. 

Next, let us mention extra contributions to the Higgs coupling from other SUSY particles. 
So far, they are suppressed because those particles are supposed to be heavy.
However, if the chargino, the stop or the sbottom is light, its contribution can be sizable \cite{Carena:2011aa,Batell:2013bka}. 
These particles are searched for effectively at (HL-) LHC (see e.g., Ref.~\cite{ATLAS:2013hta,CMS:2013xfa}).\footnote
{
On the other hand, it is possible to determine $\tan\beta$ by studying decays of staus, neutralinos or charginos at ILC, if the Higgsinos are light \cite{Boos:2003vf,Baer:2013cma}.
}
If none of them is discovered, their masses are bounded from below, and upper limits on their contributions to $\kappa_{\gamma}$ are derived.\footnote
{
If extra SUSY particles such as charginos are discovered, their contributions to $\kappa_{\gamma}$ may be reconstructed.
}
These extra contributions should be taken into account as a theoretical (systematic) uncertainty in the analysis of $\delta \kappa_{\gamma}$.

% Figure %%%%%%%%%%%%%%%%%%%%%%%%%%%%%%%%
\begin{figure}[t]
\begin{center}
 \includegraphics[width=8cm]{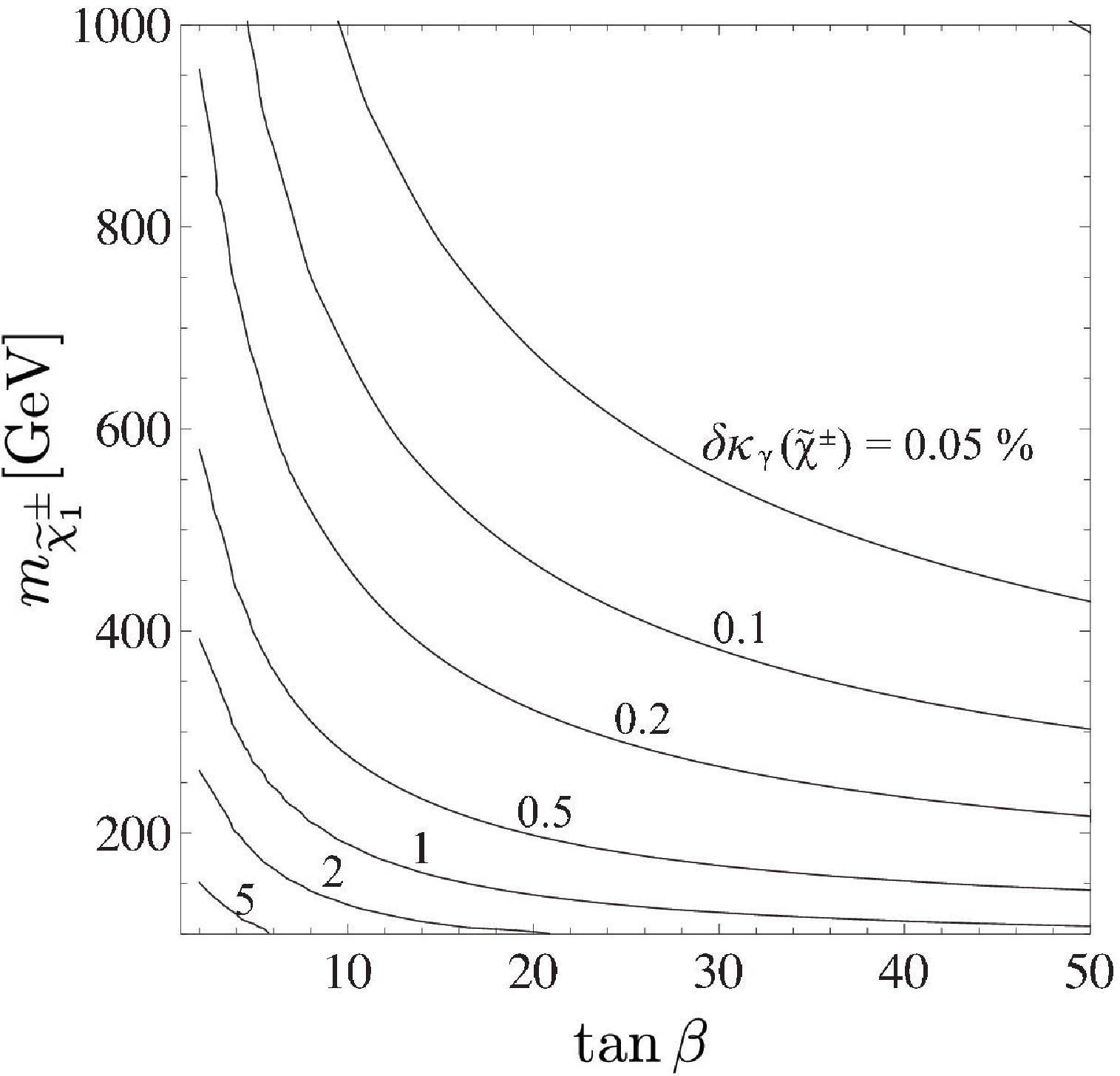}\hspace*{2mm}
 \includegraphics[width=8cm]{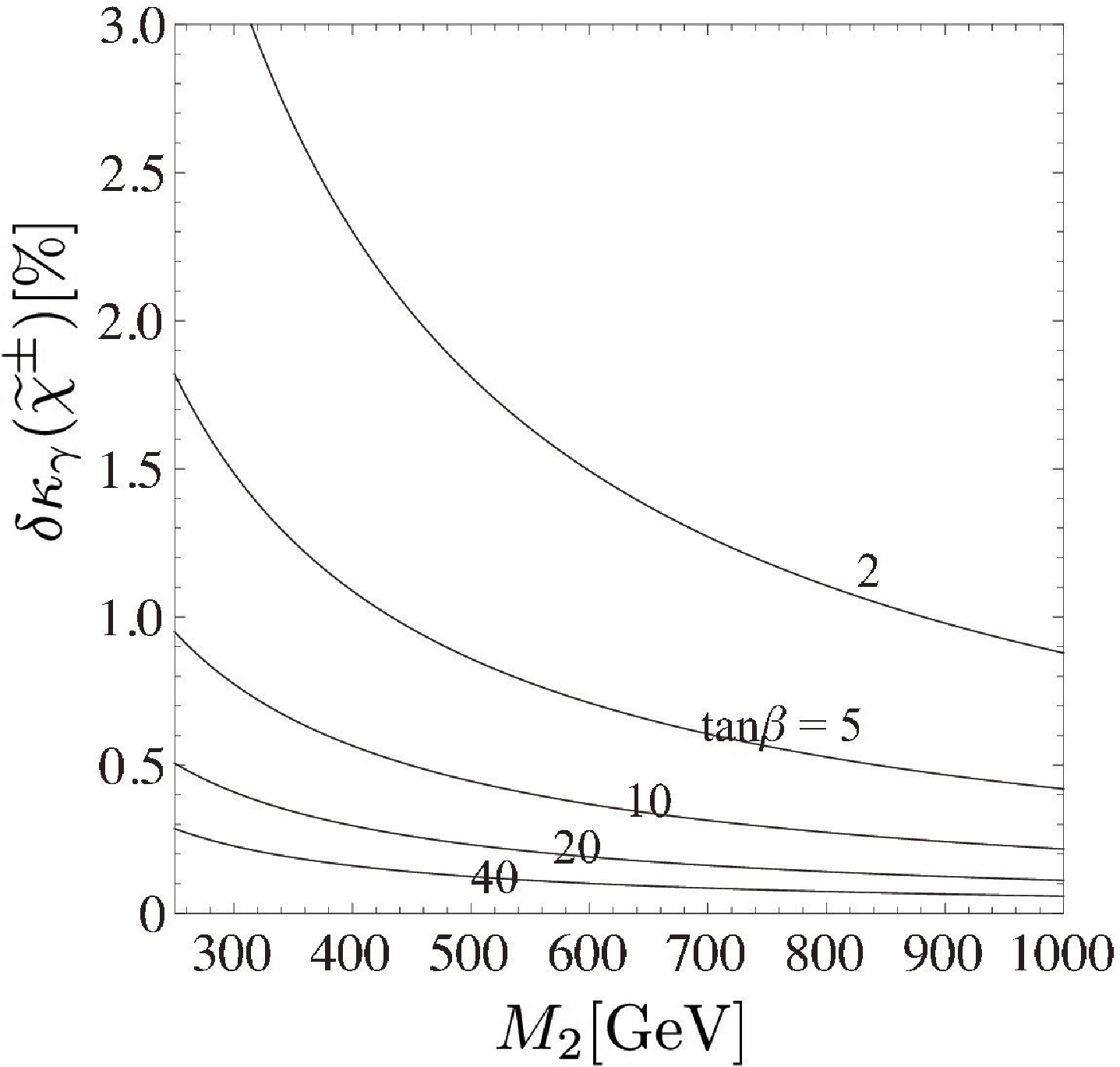}
 \end{center}
 \caption[Chargino contribution to $\delta \kappa_{\gamma}$.]
 {Chargino contribution to the Higgs coupling $\delta \kappa_{\gamma}$.
 The left panel shows contours of the chargino contribution to $\delta \kappa_{\gamma}(\widetilde\chi^\pm)$, 
 where the Wino mass $M_2$ is equal to the Higgsino mass $\mu$.  
 In the right panel, the chargino contribution $\delta \kappa_{\gamma}(\widetilde\chi^\pm)$ is displayed for various $\tan\beta$,  where $\mu_H$ is fixed to be $250\GeV$.
}
 \label{fig:Chargino}
\end{figure}
%%%%%%%%%%%%%%%%%%%%%%%%%%%%%%%%%%%%%

We discuss the chargino contribution.
Since the mass bounds on the charginos are still weak, it can be as large as the theoretical uncertainty.
At one loop level, the formula was given in \eqref{eq:kappamssmcontrib} of Sec.~\ref{sec:HiggsCmssm}.
When the charginos are heavier than the electroweak gauge bosons, it is approximated as\cite{Batell:2013bka}
\beq
\mathcal{M}_{\gamma\gamma}(\widetilde\chi^{\pm}) \simeq 
\frac{4}{3}
\frac{g^2 v \sin 2 \beta}{M_2 \mu - \frac{1}{4} g^2 v^2 \sin 2 \beta}.
\label{eq:MgammaChargino}
\eeq
Since $\sin 2 \beta = 2 \tan \beta / (1 + \tan ^2 \beta )$, 
it is found that it is suppressed in large $\tan \beta $, 
thus is relevant when $\tan \beta = \mathcal{O}(1)$.

The left panel of Fig.~\ref{fig:Chargino} shows contours of $\delta \kappa _{\gamma } (\widetilde{\chi }^{\pm})$ 
as a function of $\tan \beta$  and the lightest chargino mass $m_{\widetilde{\chi }^{\pm}_1}$.
Here, we take $M_2 = \mu $, for simplicity.
The contribution decrease as $m_{\widetilde{\chi }^{\pm}_1}$ or $\tan \beta$ increases.
If the charginos are constrained to be heavier than $600\GeV$ ($1\TeV$), these contribution to $\kappa_{\gamma}$ is estimated to be smaller than $0.5\%$ ($0.2\%$) for $\tan\beta > 2$.
%\footnote{The chargino contribution is unlikely to dominate the contributions to $\delta \kappa_{\gamma}$, unless it is very light and $\tan\beta$ is small.}
This is considered to be a theoretical uncertainty. 
In addition, if either of the Wino or the Higgsino is decoupled, $\delta \kappa_{\gamma}(\tilde\chi^\pm)$ is suppressed.
Such a feature is observed in the right panel of Fig.~\ref{fig:Chargino}, where $\mu_H$ is fixed to be $250\GeV$ for various $\tan\beta$.

The stop and the sbottom can also contribute to $\kappa _{\gamma }$ sizably\cite{Carena:2011aa}. 
In addition, they simultaneously modify the Higgs coupling to di-gluon.
The coupling $\kappa _g$ is expected to be measured precisely at ILC at (sub) percent level\cite{Peskin:2013xra, Asner:2013psa}.
Hence, if deviations are discovered in $\kappa _g$ as well as $\kappa _{\gamma}$, 
it is interesting to study the contributions of stop or sbottom.

%\section{Summary}
%\label{sec:summary}

\chapter{Conclusion}
\label{chp:summary}
%!TEX root = ../Dthesis.tex

Supersymmetry is one of the most motivated candidates of new physics.
If the muon $g-2$ anomaly is solved by SUSY, some of superparticles are relatively light.
In this dissertation, we concentrated on the "minimal" SUSY model in which 
only the Bino and the slepton are light, while the other superparticles are decoupled.
In this model, the muon $g-2$ discrepancy is explained by the Bino-smuon diagram.
It is enhanced by large left-right mixing $m^2_{\widetilde{\ell }_{LR}}$.
We have investigated various phenomena in the minimal model.

The analyses were classified by the slepton mass spectrum. 
When the mass spectrum is universal among the flavors, 
we found that the vacuum meta-stability of the stau-Higgs potential restricts the smuon masses tightly. 
They are predicted to be within 330 (460) $\GeV$ at the $1\sigma$ ($2\sigma$) level of the muon $g-2$. 
It was shown that part of the parameter region is already excluded by LHC, 
and argued that such sleptons are expected to be probed at LHC or ILC in future.

If the staus are (much) heavier than the smuons, the vacuum stability bound of the staus is relaxed. 
In this mass spectrum, the smuon masses are limited by the vacuum mata-stability condition of the smuon-Higgs potential. 
It was found that they are less than 1.4 (1.9) $\TeV$ at the $1\sigma$ ($2\sigma$)  level of the muon g − 2. 
Such slepton masses exceed the LHC/ILC reach. 
Instead, the non-universal slepton mass spectrum generically predicts too large LFV and EDM. 
They originate in the non-diagonal SM Yukawa matrices in the model basis. 
Although the prediction depends on the flavor models, LFV and/or EDM is very likely to be sizable in wide models. 
If no sleptons are discovered at LHC or ILC, the model is expected to be probed by LFV and EDM sensitively.

In the universal mass spectrum case, the stau constribution to the Higgs coupling to di-photon $\kappa _{\gamma }$ becomes large.
The coupling will be measured at the percent levels by the joint analysis of HL-LHC and ILC.
In this dissertation, we considered a situation that an excess of  $\kappa _{\gamma }$ is measured in HL-LHC and ILC.
First, we studied the stau mass region based on the vacuum meta-stability of the stau-Higgs potential.
Consequently, we showed that the lightest stau $m_{\widetilde{\tau } _1}$ is predicted to be lighter than about 200\GeV, 
if the excess is measured to be larger than 4\% 
in the early stage of $\sqrt{s} = 500\GeV$ ILC.
Further, it was found that $m_{\widetilde{\tau } _1}$ is limited to be less than 290--460\GeV, if the excess is measured to be 1--2\% 
by accumulating luminosity at $\sqrt{s} =1\TeV$ ILC.
Such stau is within the kinematical reach of ILC.
Therefore, we concluded that the stau contribution to $\kappa _{\gamma }$ can be probed by discovering the stau, 
if the excess is measured in the future experiments, and if it originates in the stau contribution.

Once the stau is discovered at ILC, its properties are determined precisely. 
In this dissertation, we also studied the reconstruction of the stau contribution to $\kappa _{\gamma }$ by using the information which is available at ILC. 
It was estimated that the contribution can be reconstructed at $\sim 0.5\%$ at the sample point, which is comparable to or smaller than the measured value of the Higgs coupling. 
Thus, it is possible to test directly whether the excess originates in the stau contribution. 
Here, the measurement of the stau mixing angle is crucial. We also argued that, 
if the stau mixing angle is measured at the early stage of ILC, 
it is also possible to predict the heaviest stau mass, even when the heaviest stau is not yet discovered at the moment. 
Therefore, the stau contribution to $\kappa _{\gamma }$ can be probed not only by discovering the lightest stau, but also by studying the stau properties.

%Discoveries of new physics are the next target after the discovery of the Higgs boson. 
%The measurement of the Higgs couplings to di-photon is one of the hopeful channels to search for the new physics. 
%The stau contribution to the Higgs coupling could be probed or tested in future colliders by following the analysis in this letter.

%\section{Discussion}

%\TODO{stability : quantum correction}
%\TODO{14 LHC, ILC, collider simulation -> sensitivity }
%\TODO{Model Building}
%\TODO{Cosmology}
%\TODO{Other scenarios}

In conclusion, we studied the minimal SUSY models that explain the muon $g-2$ discrepancy with the Bino-smuon contributions. 
It was shown that they are expected to be probed by LHC/ILC and LFV/EDM complementarily in future.

%%%%%%%%%%%%%%%%%%%%%%%%%%%%%

%%%%%%%%%%%%%%%%%%%%%%%%%%%%% Acknowledgement
\chapter*{Acknowledgement}
%!TEX root = ../Dthesis.tex

Author firstly acknowledges Japan Society for the Promotion of Science (JSPS) 
and an Advanced Leading Graduate Course for Photon Science (ALPS) for fellowship.
The works by Author which are cited this dissertation were supported by JSPS KAKENHI Grant No. 25-10486, 
and supported by ALPS grant.

\vspace{20pt}
Author would like to express his sincerest gratitude to his supervisor Prof.~Koichi Hamaguchi 
for various instructive suggestions, stimulating discussion and collaborations, hearty encouragements and continuous support.

Author is also obliged to Dr.~Endo for advice, especially his patient answering to many questions.
Without his encouragement and support, this dissertation would not have materialized.

Author would like to show his gratitude to the collaborators of his works, 
Dr.~Endo, Prof.~Hamaguchi, Dr.~Sho Iwamoto, and Mr.~Teppei Kitahara.
Author was much helped by Mr.~Kitahara's works.

\vspace{10pt}
Author also thanks to this dissertation committee members for their advice and supports; 
Prof.~Yutaka Matsuo (the chief), Prof.~Masahiro Ibe, Prof.~Masahiro Kawasaki, Prof.~Tomio Kobayashi, and Prof.~Satoru Yamashita.

\vspace{20pt}
Finally, Author expresses his sincere gratitude to his family, colleagues, and friends. 
Especially, Author would like to offer his special thanks to his wife, Ms.~Suzuka Yoshinaga for a grateful support about his private life.

%%%%%%%%%%%%%%%%%%%%%%%%%%%%%

%%%%%%%%%%%%%%%%%%%%%%%%%%%%% Appendix
\appendix
\chapter{Vacuum Decay}
\label{app:vacuum}
%!TEX root = ../Dthesis.tex

In this appendix, we briefly review the vacuum decay.
The vacuum state is stable if the vacuum expectation value of scalar potential is at a true minimum.
However, when it is at a local minimum which is higher than the true minimum, 
the false vacuum becomes meta-stable and could decay into the true vacuum by quantum tunneling effect.
The vacuum transition from the false vacuum to the true vacuum can be evaluated by semiclassical technique\cite{Coleman:1977py, Callan:1977pt}.

The vacuum decay rate comes from the imaginary part of energy of the false vacuum state. 
In general , the energy of the ground state $E_0$ is estimated by path integral method in Euclidean space-time as follows,
\beq
 E_0 = - \lim _{T \to \infty} \frac{1}{T} \log \left(\int \left[ \mathcal{D} \phi  \right] \exp (- S_E[\phi ]), \right) \label{eq:pathint}
\eeq
Here, $S_E[\phi ]$ is an Euclidean action, and $\phi $ denotes a scalar field in a theory.
In this dissertation, $\phi _1= h_u$, $\phi _2 = \widetilde{\ell }_L$, and  $\phi _3 = \widetilde{\ell }_R$, respectively. 
The integral \eqref{eq:pathint} is over all fields satisfying the conditions as
\beq
 \lim _{T\to \infty} \phi \left(  \overrightarrow{x}, \pm \frac{T}{2} \right) = \phi ^f,  \label{eq:bouncecond}
\eeq
where, $\phi ^f$ is a field configuration of the false vacuum.
For a stable state, $E_0$ is real.
When the state is unstable,  a non-zero imaginary part appears in $E_0$, because such unstable state is not eigenstate of the Hamiltonian.
The decay rate $\Gamma $ corresponds to the imaginary part as 
\beq
 \Gamma = -2\cdot \text{Im} E_0 . \label{eq:decayrate}
\eeq

The path integral is dominated by bounce configurations, 
which is a stationary point of the action and satisfy the boundary condition \eqref{eq:bouncecond}.
We assume an $O(4)$ symmetric action to look for the bounce solutions as
\beq
 S_E[\phi (r)] =  2 \pi ^2 \int dr r^3 \left[  \frac{1}{2}\left( \frac{d h_u}{dr} \right) ^2  +  \sum _{i = 2} ^3 \left( \frac{d\phi _i}{dr} \right) ^2  + V(\phi )\right] , 
\eeq
where $r$ is radial coordinate in four-dimensional spacetime, 
and $V(\phi )$ is the potential energy. 
The equation of motion are evaluated as 
\beq
 &&\frac{d^2 h_u}{dr^2} + \frac{3}{r} \frac{d h_u}{dr} = \frac{\partial V(\phi )}{\partial h_u}, \label{eq:eombounce1}  \\
 &&2\frac{d^2 \phi _i}{dr^2} + \frac{6}{r} \frac{d\phi _i}{dr} = \frac{\partial V(\phi )}{\partial \phi _i} ~~~~(i =2,3), \label{eq:eombounce2}
\eeq
where the boundary conditions are $\lim _{r \to \infty} \phi _i(r) = \phi _i^f$ and $\frac{d\phi _i}{dr} (0) = 0$.
The solutions in \eqref{eq:eombounce1} and  \eqref{eq:eombounce2} are the bounce solutions $\bar{\phi }_i$.
In this dissertation, they are calculated by numerical calculations.

\chapter{Bino Coupling}
\label{app:bino}
%!TEX root = ../Dthesis.tex

In this appendix, we briefly summarize the non-decoupling corrections to the Bino couplings with the smuons.
The Bino-muon-smuon interactions are 
\beq
 \mathcal{L}_{\text{int}} = -\frac{1}{\sqrt{2}}\widetilde{g}_L\, \overline{\widetilde{B}} \mu _L\, \widetilde{\mu }^*_L + \sqrt{2} \widetilde{g}_R\, \overline{\widetilde{B}} \mu _R\, \widetilde{\mu}_R^* + {\rm h.c.}, 
 \label{eq:appbino}
\eeq
where $\widetilde{g}_{L,R}$ are the Bino couplings.

The non-decoupling SUSY-breaking effects are understood in term of renormalization group equations.
Above the heavy superpartner scale $M_{\text{soft}}$, SUSY is exact symmetry.
Thus, the relation $\widetilde{g}_{L,R} = g_Y$ is maintained.
Below the  scale $M_{\text{soft}}$, the heavy superparticles are decoupled.
The light particle (SM fermions, light sleptons) loops still renormalize the gauge boson wave function.
On the other hand, the heavy superparticles loops decouple gauge boson and gaugino wave function renormalization, respectively.
Therefore, SUSY is broken in gauge sector, and U(1)$_Y$ gauge coupling $g_Y$ and Bino coupling $\widetilde{g}_{L,R}$ start to evolve differently.

First, we derive the coefficients $\Delta b$ in \eqref{eq:gL} and \eqref{eq:gR} of Sec.~\ref{subsec:gm2vac}. 
At one loop, the solution of the RGE of the U(1)$_Y$ gauge coupling and the Bino coupling are\cite{Cheng:1997sq}
\beq
 \frac{1}{g^2_Y(m_\text{soft})} &=&  \frac{1}{g^2_Y(M_\text{soft})} + \frac{\beta _{g_Y}}{8\pi ^2} \log \frac{M_{\text{{soft}}}}{m_{\text{soft}}}, \label{eq:solgauge} \\
 \frac{1}{\widetilde{g}^2_{L,R}(m_\text{soft})} &=& 
 \frac{1}{\widetilde{g}^2_{L,R}(M_\text{soft})} + \frac{\beta _{\widetilde{g}}}{8\pi ^2} \log \frac{M_{\text{{soft}}}}{m_{\text{soft}}}, \label{eq:solbino}
\eeq
where $\beta _{g_Y}$ and $\widetilde{g}$ are beta function of U(1)$_Y$ gauge and Bino coupling, respectively. 
Since $\widetilde{g}^2_{L,R}(M_\text{soft}) = g^2_Y(M_\text{soft})$ at boundary, 
the ratio of \eqref{eq:solgauge} and \eqref{eq:solbino} becomes 
\beq
 \frac{\widetilde{g}_{L,R}(m_\text{soft})}{g_Y(m_\text{soft})} \simeq 1 + \frac{g^2_Y(m_\text{soft})}{16 \pi ^2} (\beta _{g_Y} - \beta _{\widetilde{g}}) 
 \log \frac{M_{\text{{soft}}}}{m_{\text{soft}}}, 
 \label{eq:ratiobino}
\eeq
where $m_{\text{soft}}$ is the light superparticle scale.
Then, let us define 
\beq
 a_f =  N_{\text{SU(2)}_L} \times N_C \times Y^2_f, \label{eq:coeffalpha} 
\eeq
where $N_{\text{SU(2)}_L}$ is a dimension of SU(2)$_L$ representation, $N_C$ is a color number, and $Y_f$ is $U(1)_Y$ hypercharge of a particle $f$, respectively.    
For example, $a_Q = 2 \times 3 \times (1/6)^2$ for the left-handed quark doublet.
In our setup, the matter content is just SM fermions plus $n_{\text{slepton}}$ sleptons.
The beta function of U(1)$_Y$ gauge and Bino coupling is estimated as\footnote{
For the full MSSM,  coefficients or the beta functions are
\beq
 \beta _{g_Y} &=& \frac{2}{3} \left[ 3 \times (\alpha _Q + \alpha _U + \alpha _D + \alpha _L + \alpha _E) + 2 \alpha _H \right]  \notag \\
 &&+ \frac{1}{3} \left[ 3 \times (\alpha _Q + \alpha _U + \alpha _D + \alpha _L + \alpha _E) + 2 \alpha _H \right] 
 = 11 , \notag \\ 
 \beta _{\widetilde{g}} &=& 3 \times (\alpha _Q + \alpha _U + \alpha _D + \alpha _L + \alpha _E) + 2 \alpha _H = 11, \notag 
\eeq
thus, $\widetilde{g}_{L,R} = g_Y$ is maintained.}
\beq
 \beta _{g_Y} &=& \frac{2}{3} \left[ 3 \times (\alpha _Q + \alpha _U + \alpha _D + \alpha _L + \alpha _E) + 2 \alpha _H \right]  \notag \\
 &&+ \frac{1}{3} \left[n_{\text{slepton}} \left( \alpha _L + \alpha _E \right) +  \alpha _H \right] 
 = \frac{41}{6} + \frac{1}{2} n_{\text{slepton}},  \\
  \beta _{\widetilde{g}} &=& n_{\text{slepton}} \left(  \alpha _L + \alpha _E \right) = \frac{3}{2} n_{\text{slepton}}.
\eeq
Thus the coefficient $\Delta b$ is evaluated as
\beq
 \Delta b \equiv \beta _{g_Y} - \beta _{\widetilde{g}} = \frac{41}{6}  - n_{\text{slepton}}.
\eeq

Then, we consider $\mathcal{O}(\alpha _2)$ correction due to Wino decoupling.
As for logarithmic correction, 
it can be also understood $\mathcal{O} (\alpha _2)$ RGEs for the Bino coupling with the left-handed smuon $\widetilde{g}_L$.
If SUSY is exact, the Wino and the $W$ boson contributions to the Bino coupling $\widetilde{g}_L$ cancel each other at $\mathcal{O}(\alpha _2)$.
When the Winos are decoupled, only the $W$ boson contribution remains,   
whereas, the U(1)$_Y$ gauge coupling $g_Y$ does not run at $\mathcal{O}(\alpha _2)$.
Therefore, below the Wino mass scale $M_{\text{soft}}$, the relation $\widetilde{g}_L = g_Y$ does not hold.
Up to $\mathcal{O}(\alpha _2)$, renormalization group equation for the Bino coupling $\widetilde{g}_L$ is evaluated as (cf. Ref.~\cite{Hikasa:1995bw})
\beq
 \frac{d\widetilde{g}^{-2}_L}{d \log \mu } =  \frac{\alpha _2}{2\pi } \frac{9}{4} \widetilde{g}^{-2}_L - \frac{\beta _{\widetilde{g}}}{8\pi ^2}, \label{eq:winorge}
\eeq
where the first term of right hand side corresponds to the $W$ boson contribution.
Solving \eqref{eq:winorge} and substituting \eqref{eq:solgauge} for the solution of \eqref{eq:winorge}, 
we can obtain Eq.\eqref{eq:gL} in Sec.~\ref{subsec:gm2vac}.

\chapter{Slepton Production}
\label{app:slepprod}
%!TEX root = ../Dthesis.tex

In this appendix, we briefly summarize the total cross section of slepton production at $e^+ e^-$ collider.
Smuons and Staus are produced via s-channel $\gamma / Z $ exchanges. 
The production cross section is given by \cite{Boos:2003vf}
\beq
\sigma(e^+e^-\to\widetilde\ell_i\widetilde\ell_j) &=& 
\frac{8\pi\alpha^2}{3s} \lambda^{\frac{3}{2}} 
\bigg[
  c_{ij}^2 \frac{\Delta_Z^2}{\sin^42\theta_W}(P_{-+}L^2 + P_{+-}R^2)
\nonumber\\&&
  + \delta _{ij} \frac{1}{16} (P_{-+} + P_{+-})
  + \delta _{ij} c_{ij}\frac{\Delta_Z}{2\sin^22\theta_W}(P_{-+}L + P_{+-}R)
\bigg],
\label{eq:AppCrossSection}
\eeq\
where we denote $\ell = \mu$, $\tau$, and the parameters are defined as
\beq
 \lambda &=&  [1- (m_{\widetilde{\ell }_1} + m_{\widetilde{\ell }_2})^2 / s] [1- (m_{\widetilde{\ell }_1} - m_{\widetilde{\ell  }_2})^2 / s], \\
 \Delta Z &=& \frac{s}{s-m^2_Z}, \\
 c_{11/22} &=& \frac{1}{2}[(L+R) \pm (L-R) \cos 2\theta_{\widetilde{\ell}}], \\
 c_{12} &=& c_{21} =  \frac{1}{2}(L-R) \sin 2\theta_{\widetilde{\ell}}, \\
 L&=&-1/2+\sin^2\theta_W, \\
 R&=&\sin^2\theta_W.
\eeq
The beam polarizations are parameterized as $P_{\mp\pm} = (1 \mp P_{e-}) (1 \pm P_{e+})$.

Selectrons can be produced not only via s-channel $\gamma / Z $ exchanges, 
but also by the t-channel Bino exchange.
In this dissertation, we neglect the left-right mixing of the selectrons because of small Yukawa coupling.
The polarized cross section is evaluated as \cite{Boos:2003vf}
\beq
\sigma (e^+_{R, L} e^-_{L, R} \to \widetilde{e} ^+_{L, R}\widetilde{e}^- _{L,R}) &=& \frac{2\pi \alpha ^2}{3s} \lambda ^{\frac{3}{2}} \left[  1 + g_{L, R}^2 \Delta _Z \right] ^2 
+ \frac{16\pi \alpha ^2}{s} \sum _{A=1}^{4} \sum _{B=1}^{4} \left| X_{(L,R) A}  \right|^2 \left| X_{(L,R) B}  \right|^2 h^{AB} \notag  \\
&&+ \frac{8\pi \alpha ^2}{s} \sum _{A=1}^{4}  \left| X_{(L,R) A}  \right|^2 \left[  1 + g_{L, R}^2 \Delta _Z \right] f^{A}, \label{eq:sele1}\\
\sigma (e^+_{L, R} e^-_{R, L} \to  \widetilde{e} ^+_{L, R}\widetilde{e}^- _{L,R} ) &=& 
\frac{2\pi \alpha ^2}{3s} \lambda ^{\frac{3}{2}} \left[  1 + g_{L, R} g_{R, L} \Delta _Z \right] ^2, \label{eq:sele2} \\
\sigma (e^+_{L} e^-_{L} \to  \widetilde{e} ^+_{R}\widetilde{e}^- _{L} ) &=& \frac{16\pi \alpha ^2}{s} 
\sum _{A=1}^{4} \sum _{B=1}^{4} X_{LA}X^{*}_{RA} X_{RB}X^{*}_{LB} H^{AB}, \label{eq:sele3} \\
\sigma (e^+_{R} e^-_{R} \to  \widetilde{e} ^+_{L}\widetilde{e}^- _{R} ) &=&\sigma (e^+_{L} e^-_{L} \to  \widetilde{e} ^+_{R}\widetilde{e}^- _{L} ) , \label{eq:sele4}
\eeq
 where
\beq
  g_L &=& \frac{-1 + 2 \sin ^2 \theta _W}{2 \cos \theta _W}, ~~~ g_R = \frac{\sin \theta _W}{\cos \theta _W}, \\
 f^A &=& \Delta _A \beta - \frac{\Delta ^2_A  - \beta ^2}{2} \log \frac{\Delta _A  + \beta }{\Delta _A  - \beta } , \\
 h^{AB} &=& \begin{cases}
                        - 2 \beta + \Delta _A  \log \frac{\Delta _A  + \beta }{\Delta _A  - \beta }  ~~~(A = B)\\
                        \frac{f^B - f^A}{\Delta _A -\Delta _B}~~~~~~~~~~~~~~~~~~~~~~~(A \neq B)
                     \end{cases}, \\
  G ^{AB}_{\pm } &=& \frac{2}{s} \frac{m_{\widetilde{\chi }^0_A} m_{\widetilde{\chi }^0_B}}{\Delta _A  \pm \Delta _B}
  \left[ 
  \log \frac{\Delta _B  + \beta }{\Delta _B  - \beta } \pm \log \frac{\Delta _A  + \beta }{\Delta _A  - \beta }
  \right] , \\
   H^{AB} &=& \begin{cases}
                     \frac{4\beta }{s} \frac{m^2_{\widetilde{\chi }^0_A}}{\Delta ^2_A  - \beta ^2}~~~~~~~(A = B) \\
                     G^{AB}_{-} ~~~~~~~~~~~~~~(A \neq B) 
                  \end{cases}
 \eeq
For the case of $\widetilde{e}_{L,R} \widetilde{e}_{L,R} $ pairs, the parameter $\Delta _A$ is given by 
 \beq
  \beta = 1 - \frac{4 m^2_{\widetilde{e}_{L,R}}}{s}, ~~~
 \Delta _A = \frac{2}{s} \left( m^2_{\widetilde{e}_{L, R} } - m^2_{\widetilde{\chi }^0_A} \right) -1,
 \eeq
 while for $\widetilde{e}_{L} \widetilde{e}_{R}$ pairs
  \beq
  \beta =  [1- (m_{\widetilde{e}_L} + m_{\widetilde{e}_R})^2 / s] [1- (m_{\widetilde{e}_L} - m_{\widetilde{e}_R})^2 / s] , ~~~
 \Delta _A = \frac{1}{s} \left( m^2_{\widetilde{e}_L} + m^2_{\widetilde{e}_R} - 2 m^2_{\widetilde{\chi }^0_A} \right) -1.
 \eeq
The matrices $X_{(L,R)A}$ are given by 
 \beq
  X_{LA} =\frac{1}{\sqrt{2}} \widetilde{g}_L (O_N) _{A1} , ~~~X_{RA} =\sqrt{2} \widetilde{g}_R (O_N) _{A1}  ,  \eeq
 where $O_N$ is the unitary matrices which dianonalize the neutralino mass matrices.

%%%%%%%%%%%%%%%%%%%%%%%%%%%%% Reference
\bibliographystyle{utphys}
\bibliography{ref}

%%%%%%%%%%%%%%%%%%%%%%%%%%%%%

\end{document}